\documentclass[11pt]{article}
\pdfoutput=1
\usepackage{amsmath,amssymb,epsfig,amsfonts}
\usepackage{graphicx}
\usepackage{pdflscape}
\usepackage{subfig}
\usepackage[usenames, dvipsnames]{color}
\usepackage{adjustbox}
 \usepackage{multirow}

\usepackage{hyperref}
\usepackage[dvipsnames]{xcolor}
\usepackage{cite}
\usepackage{multirow}
\usepackage{slashed}
\usepackage{verbatim} 
\usepackage{enumitem}
\usepackage{rotating}
\usepackage{xcolor}
\usepackage{multirow}
\usepackage{bm}
\usepackage[normalem]{ulem}
\usepackage{cleveref}
\usepackage{booktabs} 
\usepackage{tikz-cd} 
\usepackage[dvipsnames]{xcolor}

\usepackage[most]{tcolorbox}

\definecolor{ourcolorforheader}{named}{YellowOrange}

\usepackage[fleqn,tbtags]{mathtools}

\graphicspath{ {figures/} }

\definecolor{purp}{rgb}{0.53, 0.37, 0.8}

\usepackage{tikz}
\usepackage{diagbox}
\usetikzlibrary{positioning}
\usetikzlibrary{calc}
\usetikzlibrary{decorations.pathreplacing,calligraphy}

\usepackage{xstring}
\usetikzlibrary{decorations.pathmorphing} 
\usetikzlibrary{decorations.markings} 
\usetikzlibrary{arrows} 
\usetikzlibrary{shapes} 
\usetikzlibrary{matrix} 
\usetikzlibrary{positioning} 
\usepackage[english]{babel} 
\usepackage[autostyle]{csquotes}
\usepackage{pifont}
\usetikzlibrary{shapes.multipart}

\tikzset{->-/.style={decoration={
  markings,
  mark=at position .5 with {\arrow{>}}},postaction={decorate}}}

\newcommand{\bea}{\begin{eqnarray}}
\newcommand{\eea}{\end{eqnarray}}
\newcommand{\be}{\begin{equation}}
\newcommand{\ee}{\end{equation}}
\newcommand{\ba}{\begin{aligned}}
\newcommand{\ea}{\end{aligned}}
\newcommand{\bit}{\begin{itemize}}
\newcommand{\eit}{\end{itemize}}
\newcommand{\ben}{\begin{enumerate}}
\newcommand{\een}{\end{enumerate}}

\newcommand{\id}{\text{id}}

\newcommand{\Triv}{\text{Triv}}

\newcommand{\Neu}{\text{Neu}}
\newcommand{\Dir}{\text{Dir}}

\newcommand{\Bsym}{\mathfrak{B}_{\text{sym}}}
\newcommand{\Bphys}{\mathfrak{B}_{\text{phys}}}

\newcommand{\lb}{\left(}
\newcommand{\rb}{\right)}

\newcommand{\wt}{\widetilde}
\newcommand{\wh}{\widehat}

\newcommand{\Z}{{\mathbb Z}}

\newcommand{\Q}{{\mathbb Q}}

\newcommand{\cA}{\mathcal{A}}

\newcommand{\cC}{\mathcal{C}}
\newcommand{\cD}{\mathcal{D}}
\newcommand{\cE}{\mathcal{E}}
\newcommand{\cF}{\mathcal{F}}

\newcommand{\cL}{\mathcal{L}}
\newcommand{\cM}{\mathcal{M}}

\newcommand{\cO}{\mathcal{O}}

\newcommand{\cS}{\mathcal{S}}

\newcommand{\cZ}{\mathcal{Z}}

\renewcommand{\S}{\mathsf{S}}

\newcommand{\D}{\mathsf{D}}

\newcommand{\fT}{\mathfrak{T}}
\newcommand{\fB}{\mathfrak{B}}

\renewcommand{\Vec}{\mathsf{Vec}}
\newcommand{\Rep}{\mathsf{Rep}}

\renewcommand{\dim}{\text{dim}}

\newcommand{\TwoVec}{2\mathsf{Vec}}
\newcommand{\TwoRep}{2\mathsf{Rep}}

\renewcommand{\Q}{{\bm{Q}}}

\newcommand{\phys}{\text{phys}}

\newcommand{\I}{\text{I}}
\newcommand{\II}{\text{II}}
\newcommand{\diag}{\text{diag}}
\allowdisplaybreaks
\usepackage{physics}
\usepackage{makecell}

\usepackage{tikz-cd}

\begin{document}

\baselineskip=18pt  
\numberwithin{equation}{section}  
\allowdisplaybreaks  

\thispagestyle{empty}

\vspace*{0.8cm} 
\begin{center}
{{\Huge  Gapped Phases in (2+1)d with\\
\bigskip
Non-Invertible Symmetries: Part II 
}}

\vspace*{0.8cm}
Lakshya Bhardwaj$^1$,   Sakura Sch\"afer-Nameki$^2$,
Apoorv Tiwari$^3$,
Alison Warman$^2$

\vspace*{.4cm} 
{\it 
${}^1$~School of Mathematics, University of Birmingham, \\
Watson Building, Edgbaston, Birmingham B15 2TT, UK \\
${}^2$~Mathematical Institute, University of Oxford, \\
Andrew Wiles Building,  Woodstock Road, Oxford, OX2 6GG, UK\\
${}^3$~Center for Quantum Mathematics at IMADA, Southern Denmark University, \\
Campusvej 55, 5230 Odense, Denmark
}

\vspace*{0cm}
 \end{center}
\vspace*{0.4cm}

\noindent
We use the  Symmetry Topological Field Theory (SymTFT) to systematically characterize gapped phases in 2+1 dimensions with categorical symmetries. 
The SymTFTs that we consider are (3+1)d Dijkgraaf-Witten (DW) theories for finite groups $G$, whose gapped boundaries realize all so-called 
``All Bosonic type" fusion 2-category symmetries. In \cite{Bhardwaj:2024qiv} we provided the general framework and studied the case where $G$ is an abelian group. 
In this work we focus on the case of non-Abelian $G$.
Gapped boundary conditions play a central role in the SymTFT construction of symmetric gapped phases.
These fall into two broad families: minimal and non-minimal boundary conditions, respectively.
The first kind corresponds to boundaries on which all line operators are obtainable as boundary projections of bulk line operators.
The symmetries on such boundaries include (anomalous) 2-groups and 2-representation categories of 2-groups.
Conversely non-minimal boundaries contain line operators that are intrinsic to the boundaries.
The symmetries on such boundaries correspond to fusion 2-categories where modular tensor categories intertwine non-trivially with the above symmetry types.
We discuss in detail the generalized charges of these symmetries and their condensation patterns that give rise to a zoo of rich beyond-Landau gapped phases.
Among these are phases that exhibit novel patterns of symmetry breaking wherein the symmetry broken vacua carry distinct kinds of topological order.
We exemplify this framework for the case where $G$ is a dihedral group.

\newpage

\tableofcontents

\section{Introduction and Summary}

The classification and characterization of phases of matter protected by (generalized) global symmetries, are some of the key fundamental questions in theoretical physics.
This addresses fundamental issues both in condensed matter and high energy physics wherein such infra-red (IR) phases can describe the low-energy physics of lattice models, and of quantum field theories (QFTs) alike. 

 The structure of gapped phases in 2+1 dimensions is particularly interesting due to the possibility of a vast landscape of topological orders \cite{Wen:1989iv, Rowell:2007dge, Wen:2015qgn}.
Moreover the existence of topologically ordered phases in theories with additional (0-form) symmetries give rise to symmetry enriched topological orders \cite{Barkeshli:2014cna, Teo:2015xla, Teo:2015lpm, Teo:2017tjt,mesaros2013classification, lu2016classification, Essin:2013rca}, where global symmetries intertwine with topological properties of these systems in non-trivial ways. 
In addition to being of fundamental theoretical importance, such phases also have many applications ranging from the understanding of fractional excitations in quantum spin liquids \cite{Savary:2016ksw} to topological quantum computation \cite{Kitaev:1997wr, Nayak:2008zza}.

 The advent of higher categorical symmetries \cite{Kaidi:2021xfk, Choi:2021kmx, Bhardwaj:2022yxj} (see \cite{Schafer-Nameki:2023jdn, Shao:2023gho} for reviews) has revitalized the study of phases of matter.
Generalized symmetries provide powerful and systematic constraints on phases of matter \cite{Komargodski:2020mxz, Gaiotto:2017yup, Bhardwaj:2023idu} and can explain topologically ordered phases from a uniform symmetry perspective.
Moreover, an understanding of these symmetries enables the identification of numerous new stable phases, which will be exemplified throughout this work.
Additionally generalized symmetries facilitate the characterization of phases in terms of order parameters, in particular local and extended operators that form generalized charges of generalized symmetries \cite{Lin:2022dhv, Bhardwaj:2023wzd, Bhardwaj:2023ayw, Bartsch:2023pzl, Bartsch:2022ytj}.

 In this paper we continue the analysis of gapped phases in 2+1d started in \cite{Bhardwaj:2024qiv}, for so-called all-Bosonic type fusion 2-category symmetries. The main distinction in this paper compared to Part I is that we extend the analysis to include fusion 2-categories that are related by generalized gauging to non-Abelian finite group 0-form symmetries (as opposed to the analysis in Part I that was focused on Abelian groups). {This is not merely a technical upgrade, but as we will see, leads to a whole zoo of new, interesting gapped phases with categorical symmetries, distinct from the ones that can be realized in the abelian case. }

 The setup as explained in Part I \cite{Bhardwaj:2024qiv} is to study gapped phases in the presence of fusion 2-category symmetries using the Symmetry Topological Field Theory (SymTFT) \cite{Ji:2019jhk, Gaiotto:2020iye, Lichtman:2020nuw, Apruzzi:2021nmk, Freed:2022qnc}  approach to categorical phases, referred to as the ``Categorical Landau Paradigm" \cite{Bhardwaj:2023fca}, see \cite{Chatterjee:2022tyg, Moradi:2022lqp, Wen:2023otf, Moradi:2023dan, Huang:2023pyk, Bhardwaj:2023bbf, Bhardwaj:2024qrf, Bhardwaj:2024ydc, Choi:2024rjm, Apte:2022xtu} for related constructions in group and categorical symmetries and gapless phases. 
The SymTFT provides a powerful and economical way to fully characterize the universal properties of gapped phases with higher categorical symmetries.
The key technical input to study gapped phases in the presence of categorical symmetries is an understanding of the topological defects and boundary conditions \footnote{In (2+1)d systems with (3+1)d SymTFT the question of boundary conditions of DW theories were discussed e.g. in 
 \cite{Chen:2015gma, Chen:2017zzx, Bullivant:2020xhy,Luo:2022krz,  Ji:2022iva, Zhao:2022yaw} and in high energy physics in 
\cite{Bergman:2020ifi, vanBeest:2022fss}.} of the SymTFT . 
A general result \cite{Decoppet:2024htz} divides fusion 2-categories into two classes:
\bit
\item All Bosonic type (AB-type)
\item Emergent Fermionic type (EF-type) \footnote{We stress that EF-type fusion 2-categories are still bosonic fusion 2-categories describing symmetries of bosonic 3d theories, and should not be confused with the notion of fermionic fusion 2-categories that would describe symmetries of fermionic 3d theories.}
\eit
In this paper as in Part I, we study gapped phases in theories with AB-type fusion 2-categories. 
AB type of fusion 2-categories are related by generalized gaugings to possibly anomalous finite group 0-form symmetries.
Their associated SymTFTs are 4d Dijkgraaf-Witten (DW) theories \cite{Dijkgraaf:1989pz} based on a finite (possibly non-abelian) group $G$, along with a twist valued in the group cohomology $\pi\in H^4(G,U(1))$. 
We further restrict our study to fusion 2-categories for which the SymTFTs are untwisted DW theories, i.e. have $\pi=0$.

The salient new features can be summarized as follows: 
although the methodology is similar to the Abelian case, the non-Abelian case is significantly richer due to existence of (non-condensation) non-invertible topological defects in the non-abelian DW theory.
Consequently, the resulting symmetries are more interesting and stabilize a more intricate set of gapped phases in 2+1d. 
The symmetries include 2-groups and their 2-categories of 2-representations.
In addition, it includes symmetry types that are obtained by intertwining modular tensor categories with the above mentioned symmetries.
The first step in identifying the categorical symmetries that can be realized from a given SymTFT, is to understand the classification of gapped boundary conditions (BCs). 
For the 4d SymTFT, this was done in \cite{Bhardwaj:2024qiv, Bullimore:2024khm, Xu:2024pwd}.
These boundary conditions fall into two categories, using the terminology in \cite{Bhardwaj:2024qiv}, will be denoted by minimal and non-minimal.
Furthermore we can generate these by starting with the canonical Dirichlet boundary condition $\mathfrak{B}_{\Dir}$, which realizes the $G$ 0-form symmetry, possibly with anomaly via a procedure of stacking and gauging.

\begin{table}
$$
\begin{array}{|c|c|}\hline
\quad \phantom{.} \text{Minimal BC} \quad \phantom{.}& \quad \text{Symmetry} \quad \phantom{.}
\cr \hline
\hline 
\hline
\text{Dirichlet} & \quad   \TwoVec_{S_3}:S_3 \text{ 0-form symmetry }  \quad \phantom{.}
\cr
\hline
\text{Neumann}(\Z_3) & \TwoVec_{\mathbb{G}^{(2)}}:\text{2-group }\mathbb{G}^{(2)} =  \Z_3^1 \rtimes \Z_2^{(0)}\,  
\cr
\hline
\text{Neumann}(\Z_2) &  \quad \TwoRep(\mathbb{G}^{(2)}):\text{2-representations of the 2-group }\mathbb G^{(2)} (\Rep(\mathbb G^{(2)}))\quad \phantom{.}
 
\cr
\hline
\text{Neumann}(S_3) & \quad \TwoRep(S_3) :\text{2-representations of }S_3\quad \phantom{.}
\cr
\hline
\end{array}
$$
\caption{Summary of 2-fusion categorical symmetries on minimal boundary conditions (BCs) of the 3+1d $S_3$ Dijkgraaf-Witten theory.\label{tab:S3Syms}} 
\end{table}

The minimal and non-minimal BCs are obtained as follows
\begin{itemize}
\item {\bf Minimal Gapped Boundary Conditions:} stacking of $\mathfrak{B}_{\Dir}$ with an SPT and gauging a non-anomalous sub-group $H$. 
These boundary conditions have the property that every non-trivial line on the boundary is obtained via the projection of a bulk line. 
Equivalently, there is a topological local operator that connects every boundary line to a bulk line.
The web of symmetries related to a non-Abelian group via minimal gaugings was studied in detail in \cite{Bhardwaj:2022maz}. 
\item {\bf Non-minimal Gapped Boundary Conditions:} These are constructed via stacking of $\mathfrak{B}_{\Dir}$ with non-trivial $G$-topological order and gauging a non-anomalous sub-group $H$.
In contrast to minimal boundary conditions, these boundaries contain lines that are intrinsic to the boundary, i.e., they are not obtained via the projection of a bulk line.
\end{itemize}
These gapped BCs play the role of both the symmetry and physical boundary in the construction of gapped phases following the SymTFT approach -- we will review this in section \ref{Sec:generalframework}. 
We then construct the SymTFT sandwich compactifications with a fixed symmetry boundary, focusing on minimal BCs, but including physical BCs that are  non-minimal.

The main examples we will focus on are the simplest finite groups $G =S_3, D_8$, the permutation group on 3 elements and the dihedral group (symmetry of a square). For the most exciting phases, we will also include a discussion for general $D_{2n}$. We will here exemplify this with the $G=S_3$ case.
The starting point is the SymTFT which is the DW theory for $G$, and we first classify the minimal gapped BCs. 
These will in particular play the role of symmetry boundaries. In this case there are four types of minimal BC (upto choices of discrete torsion) of the $S_3$ DW theory in 3+1d, which are either Dirichlet, Neumann for $\Z_2$, for $\Z_3$ or $S_3$. These correspond in turn, if taken to be the symmetry boundaries to those shown in table \ref{tab:S3Syms}.

In table~\ref{tab:obstsalat} we have summarized  possible minimal gapped phases. 
Note that the $H$ Neumann boundary conditions have an additional choice of discrete torsion $\omega\in H^3(H,U(1))$.
These choices only affect the properties of topological defects ending on these boundaries, and therefore do not affect the symmetry categories associated with these boundaries.
These choices do however affect the structure of the gapped phase obtained upon compactifying the SymTFT.
Each entry in table~\ref{tab:obstsalat} yields a 2+1d gapped phase with a particular symmetry breaking/preserving pattern. 
In particular for the invertible symmetry categories corresponding to group like and 2-group like symmetries, we find various phases with interesting patterns of 0-form and 1-form symmetry breaking.
The preserved 0-form symmetries realize potential SPTs while the 1-form symmetry breaking patterns are characterized by the topological datum (fusion, topological spins and associators) of the lines charged under the 1-form symmetries.
We also find 1-form symmetry preserving or confining phases.

For the  non-invertible symmetry category describing 2-representations of the 2-group, among others, we find a particularly noteworty phase in which the symmetry is completely broken.
This phase has the remarkable feature that {\bf one of its vacua is topologically trivial while the other is topologically ordered} and realizes the ground state of toric code or double semion topological order.
Such a possibility is forbidden in the case of invertible symmetry categories.
We have dubbed this the ``superstar phase" (marked with $\star$ in table \ref{tab:obstsalat}).
We encounter similar phases for all examples and in fact can generalize these to the symmetry categories $\TwoRep(\Z_n^{(1)}\rtimes \Z_2^{(0)})$ which is related to the $D_{2n}$ 0-form symmetry via generalized gauging.
Similar examples also recently appeared in the context of lattice models \cite{Choi:2024rjm, Gorantla:2024ptu}.

\begin{table}
$$
\begin{array}{|c||c|c|c|c|}\hline\rule[-1em]{0pt}{2.8em} 
\text{Sym}\backslash\Bphys &  {\rm Dir} &\Neu(\Z_3),p  &\Neu (\Z_2),t &  \Neu (S_3),(p,t) \cr \hline \hline \rule[-1em]{0pt}{2.8em} 
 \TwoVec_{S_3}  & 
 \makecell[c]{\text{$S_3$ SSB} \\
 {\scriptsize\text{Sec. } \ref{sec:papaya1}}
 }
& 
\makecell[c]{\text{$\Z_2$ SSB $\boxtimes$ $\Z_3$ SPT}\\
 {\scriptsize\text{Sec. } \ref{sec:papaya3}} 
 }
& 
\makecell[c]{\text{$\Z_3$ SSB $\boxtimes$ $\Z_2$ SPT}\\
 {\scriptsize\text{Sec. } \ref{sec:papaya2}} 
 }
 &
 \makecell[c]{\text{$S_3$ SPT} \\
 {\scriptsize\text{Sec. } \ref{sec:papaya4}} 
 }
\cr\hline\rule[-1em]{0pt}{2.8em} 
  \TwoVec_{\mathbb{G}^{(2)}} 
  & 
\makecell[c]{\text{$\Z^{(0)}_2$ SSB $\boxtimes$  $\Z_3^{(1)}$ Triv} \\ 
 {\scriptsize \text{Sec. } \ref{sec:pineapple1}}
 }
 &
\makecell[c]{\text{$\Z_3^{(1)}\rtimes \Z_2^{(0)}$ SSB} \\
 {\scriptsize \text{Sec. } \ref{sec:pineapple2} }
 }
 &
 \makecell[c]{\text{$\Z_3^{(1)}\rtimes \Z_2^{(0)}$ SPT} \\
   {\scriptsize\text{Sec. } \ref{sec:pineapple3}}
   }
   &
\makecell[c]{\text{$\Z_3$ DW} \\
   {\scriptsize \text{Sec. } \ref{sec:pineapple4}} 
   }
\cr \hline\rule[-1em]{0pt}{2.8em} 
 \TwoRep (\mathbb{G}^{(2)}) 
 & 
 \makecell[c]{\text{${\TwoRep(\mathbb{G}^{(2)})\over \Z_2^{(1)}}$ SSB} \\
\text{$\boxtimes$ $\Z_2^{(1)}$ Triv} \\
 {\scriptsize\text{Sec. } \ref{sec:apple1}} 
 }
&
\makecell[c]{\text{$\TwoRep(\mathbb{G}^{(2)})$ SPT }\\
\text{$\boxtimes$ $\Z_2^{(1)}$ Triv}\\
{\scriptsize \text{Sec. } \ref{sec:apple2}}} &
\makecell[c]{\text{$\Z_2$ DW $\boxplus$ $\Z_2^{(1)}$ Triv $\star$}\\
  {\scriptsize \text{Sec. } \ref{sec:apple3}}} &
  \makecell[c]{ \text{$\Z_2$ DW}\\  {\scriptsize\text{Sec. } \ref{sec:apple4} }}
\cr \hline \rule[-1em]{0pt}{2.8em} 
 \TwoRep (S_3) & 
   \makecell[c]{\text{Triv} \\
   {\scriptsize\text{Sec. } \ref{sec:orange1} }
   }
&
   \makecell[c]{ \text{$\Z_3 \text{DW} \boxtimes \Z_2^{(1)}$ Triv }\\
   {\scriptsize \text{Sec. } \ref{sec:orange2}} } &
   \makecell[c]{\text{$\Z_2 \text{DW} \boxtimes \Z_3^{(1)}$ Triv } \\
    {\scriptsize\text{Sec. } \ref{sec:orange3}}} &
    \makecell[c]{ \text{DW $S_3$}\\
   {\scriptsize \text{Sec. } \ref{sec:orange4} }}
\cr \hline
\end{array}
$$
\caption{Phase-Map for minimal BC for the SymTFT given by $S_3$ DW theory in 3+1d: the vertical axis describes the four minimal symmetry boundaries, whereas the horizontal the physical, gapped boundary choices. Each phase is characterized in terms of a symmetry breaking pattern and SPT/TO. The 2-group $\mathbb{G}^{(2)}= \Z_3^{(1)} \rtimes \Z_2^{(0) }$. }
\label{tab:obstsalat}
\end{table}

 In addition, we also study non-minimal phases obtained from non-minimal gapped physical boundaries. 
These host phases such as symmetry enriched topological orders and non-minimal symmetry broken phases for invertible or non-invertible 1-form symmetries.
There is also a {\bf non-invertible version of the  ``superstar phase"} which hosts a topological order $\fT$ in one vacuum and its $\Z_2$ gauged version $\fT/\Z_{2}$ in the second vacuum.
The general analysis for non-minimal gapped phases is carried out for the minimal $S_3$ symmetry web while the non-minimal extensions of the superstar phase are discussed for the 2-representation 2-group symmetries in the $D_{2n}$ symmetry webs.

 In conclusion, this paper provides a complete systematic analysis of gapped phases with (all bosonic type) fusion 2-category symmetries. Clearly the next step is to extend this to gapless phases, which are second-order phase transitions between these gapped phases. 
For studies of gapless phases in 2+1d from the SymTFT perspective some works have appeared in 
\cite{Antinucci:2024ltv, Wen:2024qsg} and will appear shortly in \cite{RuiWenGapless, GaplessPaper}. In tandem a development of lattice models for these gapped phases and phase transitions will appear in \cite{LatticePaper}.

\section{General SymTFT Framework for  Gapped BCs and Phases}
\label{Sec:generalframework}

The SymTFT provides a framework to systematically characterize gapped and gapless phases for a given categorical symmetry. Here we will focus on fusion 2-categories (of so-called all bosonic type). The first step in any SymTFT analysis is to characterize the gapped boundary conditions (BCs). Each symmetry corresponds to a fixed BC, $\Bsym$, and for gapped phases, the physical BC $\Bphys$ identifies each  gapped phase.

\subsection{SymTFTs for Fusion 2-Categories}

Let us recall a few basics of the SymTFT construction for fusion 2-categories (see Part I \cite{Bhardwaj:2024qiv} for more details). 
The SymTFT is compactified either on a semi-infinite interval (the quiche) or an interval (sandwich):
\be
\begin{split}
\label{QuicheSandwich}
\begin{tikzpicture}
 \begin{scope}[shift={(0,0)},scale=0.6] 
\draw [cyan, fill=cyan!80!red, opacity =0.5]
(0,0) -- (0,4) -- (2,5) -- (5,5) -- (5,1) -- (3,0)--(0,0);
\draw [black, thick, fill=white,opacity=1]
(0,0) -- (0, 4) -- (2, 5) -- (2,1) -- (0,0);
\draw [cyan, thick, fill=cyan, opacity=0.2]
(0,0) -- (0, 4) -- (2, 5) -- (2,1) -- (0,0);
\draw[line width=1pt] (1,2.5) -- (3,2.5);
\draw[line width=1pt] (3,2.5) -- (4.2,2.5);
\fill[red!80!black] (1,2.5) circle (3pt);
\node at (2,5.4) {$\Bsym$};
\draw[dashed] (0,0) -- (3,0);
\draw[dashed] (0,4) -- (3,4);
\draw[dashed] (2,5) -- (5,5);
\draw[dashed] (2,1) -- (5,1);
\draw [cyan, thick, fill=cyan, opacity=0.1]
(0,4) -- (3, 4) -- (5, 5) -- (2,5) -- (0,4);
\draw [black, thick, dashed]
(3,0) -- (3, 4) -- (5, 5) -- (5,1) -- (3,0);
\node[below, red!80!black] at (1,2.5) {$\cE_{d-1}$};
\node  at (2.5, 3) {$\Q_{d}$};
\node at (2.5, -1) {Quiche};
\end{scope}
\end{tikzpicture}    
\qquad \qquad \qquad 
\begin{tikzpicture}
 \begin{scope}[shift={(0,0)},scale=0.6] 
\draw [cyan, fill=cyan!80!red, opacity =0.5]
(0,0) -- (0,4) -- (2,5) -- (5,5) -- (5,1) -- (3,0)--(0,0);
\draw [black, thick, fill=white,opacity=1]
(0,0) -- (0, 4) -- (2, 5) -- (2,1) -- (0,0);
\draw [cyan, thick, fill=cyan, opacity=0.2]
(0,0) -- (0, 4) -- (2, 5) -- (2,1) -- (0,0);
\draw[line width=1pt] (1,2.5) -- (3,2.5);
\draw[line width=1pt,dashed] (3,2.5) -- (4,2.5);
\fill[red!80!black] (1,2.5) circle (3pt);
\draw [fill=blue!40!red!60,opacity=0.2]
(3,0) -- (3, 4) -- (5, 5) -- (5,1) -- (3,0);
\fill[red!80!black] (1,2.5) circle (3pt);
\fill[red!80!black] (4,2.5) circle (3pt);
\draw [black, thick, opacity=1]
(3,0) -- (3, 4) -- (5, 5) -- (5,1) -- (3,0);
\node at (2,5.4) {$\Bsym$};
\node at (5,5.4) {$\Bphys$};
\draw[dashed] (0,0) -- (3,0);
\draw[dashed] (0,4) -- (3,4);
\draw[dashed] (2,5) -- (5,5);
\draw[dashed] (2,1) -- (5,1);
\draw [cyan, thick, fill=cyan, opacity=0.1]
(0,4) -- (3, 4) -- (5, 5) -- (2,5) -- (0,4);
\node[below, red!80!black] at (.8, 2.4) {$\cE_{d-1}$};
\node[below, red!80!black] at (3.8, 2.4) {$\wt{\cE}_{d-1}$};
\node  at (2.5, 3) {$\Q_{d}$};
\node  at (6.7, 2.5) {$=$};
\draw [fill=blue!60!green!30,opacity=0.7]
(8,0) -- (8, 4) -- (10, 5) -- (10,1) -- (8,0);
\draw [black, thick, opacity=1]
(8,0) -- (8, 4) -- (10, 5) -- (10,1) -- (8,0);
\fill[red!80!black] (9,2.5) circle (3pt);
\node[below, red!80!black] at (8.8, 2.4) {${\cO}_{d-1}$};
\end{scope}
\node at (1, -0.6) {Sandwich};
\end{tikzpicture}  
\end{split}
\ee
We sometimes will show the same figures in a projection to 2d. For instance for line defects we will draw dashed lines:  
\be
\begin{split}
\begin{tikzpicture}
\begin{scope}[shift={(0,0)}]
\draw [cyan,  fill=cyan] 
(0,0) -- (0,2) -- (2,2) -- (2,0) -- (0,0) ; 
\draw [white] (0,0) -- (0,2) -- (2,2) -- (2,0) -- (0,0)  ; 
\draw [very thick] (0,0) -- (0,2) ;
\node[above] at (0,2) {$\Bsym$}; 
\draw [thick, dashed](0,1) -- (2,1);
\node[left] at (0,1) {$\cE_0^{R, i}$};
\draw [thick, fill=black] (0,1) ellipse (0.05 and 0.05);
\node[] at (1,1.3) {$\Q_{1}^{\bm{R}}$};
\end{scope}
\end{tikzpicture}    
\qquad \qquad 
\begin{tikzpicture}
\begin{scope}[shift={(0,0)}]
\draw [cyan,  fill=cyan] 
(0,0) -- (0,2) -- (2,2) -- (2,0) -- (0,0) ; 
\draw [white] (0,0) -- (0,2) -- (2,2) -- (2,0) -- (0,0)  ; 
\draw [very thick] (0,0) -- (0,2) ;
 \draw [very thick] (2,0) -- (2,2) ;
\node[above] at (0,2) {$\Bsym$}; 
 \node[above] at (2,2) {$\Bphys$}; 
\draw [thick, dashed](0,1) -- (2,1);
\node[left] at (0,1) {$\cE_0^{R, i}$};
 \draw [thick, fill=black] (2,1) ellipse (0.05 and 0.05);
\draw [thick, fill=black] (0,1) ellipse (0.05 and 0.05);
\node [right] at (2,1) {$\wt{\mathcal{E}}_0^{R, i}$};
\node at (1,1.3) {$\Q_{1}^{\bm{R}}$};
\end{scope}
\end{tikzpicture}  
\end{split}
\ee
Similarly solid lines will denote surfaces $\Q_2$, with boundaries $\cE_{1}$ and $\wt{\cE}_{1}$.

 A prominent role in our analysis is played by the topological defects of these SymTFTs -- shown in the last figure in terms of $\Q_d$. 
We can characterize the defects according to their dimensions $d= 0,\cdots, 3$. 
The 3-dimensional topological operators are all condensation defects as they can be obtained by condensing line (1-dimensional) and surface (2-dimensional) operators on a 3-dimensional world-volume. 
The topological defects of dimensions 1 and 2 can either end on the boundaries, giving rise to 0 and 1-dimensional defects -- shown as $\cE$ in (\ref{QuicheSandwich}) --  or alternatively, they project to 1 and 2 dimensional defects. More generally they can give rise to so-called L-shaped configurations, that result in non-genuine operators on the boundary, which we will discuss shortly.

 The topological defects of dimensions less than equal to two form a braided fusion 2-category $\cZ(\TwoVec_{G})$ that is identified as the center of the fusion 2-category $\TwoVec_{G}$ of $G$-graded 2-vector spaces. This 2-category admits a decomposition of the following form 
\be\label{2Center}
\cZ(\TwoVec_{G}) = \boxplus_{[g]}~\TwoRep (H_g)\,,
\ee
where $[g]$ are conjugacy classes in $G$,  $H_{g}$ is the centralizer of a representative element $g$ in $[g]$ and $\TwoRep (H_g)$ is the fusion 2-category formed by 2-representations of $H_g$.
All the surface defects in $\TwoRep (H_g)\subset\cZ(\TwoVec_{G})$ component are related to each other via condensations of line operators living inside these surface operators. We will thus focus our studies on a representative element in each such component, corresponding to the trivial 2-representation, denoting the corresponding surface operator as
\be
\Q_2^{[g]}\,.
\ee
The identity surface operator is part of this subset of operators and is obtained by setting $[g]=[\id]$.

 The line operators living on $\Q_2^{[g]}$ form the fusion category $\Rep(H_g)$ of representations of $H_g$. In particular, for $[g]=[\id]$ we have line operators forming $\Rep(G)$. These are in fact the genuine topological line operators of the 4d SymTFT and simple line operators of this type are labeled as
\be
\Q_1^R\,,
\ee
for irreducible representations $R$ of $G$.

\subsection{Gapped Boundary Conditions: Minimal}
The topological boundary conditions (BCs) of such a 4d SymTFT can be divided into two classes: minimal and non-minimal. The minimal BCs are those for which every topological line defect of the boundary arises by projecting a topological line defect of the bulk. Equivalently, this means that there exist topological local operators on the boundary that connect each boundary line to some bulk line. The non-minimal BCs are those which have additional boundary lines not satisfying this property.

 All minimal BCs are related by gaugings (upto stacking with an invertible 3d TQFT). The Dirichlet BC for the 4d $G$ DW theory is minimal as the symmetry carried by it is $G$ 0-form symmetry which is generated by surface operators (there are no line operators involved). We label the Dirichlet boundary condition as
\begin{equation}
    \fB_{\Dir}\,.
\end{equation}
The other minimal BCs are obtained from Dirichlet by gauging some subgroup $H$ of $G$, possibly with discrete torsion $\tau\in H^3(H,U(1))$. 
We label the boundary conditions thus obtained as
\begin{equation}
    \fB_{\Neu(H),\tau}\,.
\end{equation}
Indeed, the lines on $\fB_{\Neu(H),\tau}$ are all Wilson lines for this gauging and are valued in the braided fusion category $\Rep(H)$, and all these lines are obtained from the $\Rep(G)$ valued bulk lines by the functor
\be
    \kappa:\quad\Rep(G)\to\Rep(H)\,,
\ee
that restricts the $G$ action on a $G$ representation to the subgroup $H$.
Note that $\fB_{\Neu(H),\tau}$ for $H=1$ is $\fB_\Dir$.
Non-minimal BCs can also be obtained by a procedure involving gauging starting from the Dirichlet BC. In this procedure we first stack $\fB_\Dir$ with a $G$ symmetric 3d TQFT and then gauge a subgroup $H$ of the diagonal $G$ symmetry of the combined system, with a discrete torsion $\tau$.

\subsubsection{Dirichlet Boundary Condition}
\label{sec:Dir}

We can discuss the structure of some specific minimal BCs in some detail.
The Dirichlet BC has the property that all genuine lines can end on this boundary
\be
\Q_1^{R} \Bigg|_{\Dir} = \cE_0^{R, i}\,, \quad  i=1, \cdots , \dim (R) \,. 
\ee
Each line has $\dim (R)$ many linearly independent local operators as endpoints, labeled by $\cE_0$ above.
We will depict this by 
\be
\begin{split}
\begin{tikzpicture}
 \begin{scope}[shift={(0,0)},scale=0.8] 
\draw [cyan, fill=cyan!80!red, opacity =0.5]
(0,0) -- (0,4) -- (2,5) -- (5,5) -- (5,1) -- (3,0)--(0,0);
\draw [black, thick, fill=white,opacity=1]
(0,0) -- (0, 4) -- (2, 5) -- (2,1) -- (0,0);
\draw [cyan, thick, fill=cyan, opacity=0.2]
(0,0) -- (0, 4) -- (2, 5) -- (2,1) -- (0,0);
\draw[line width=1pt] (1,2.5) -- (3,2.5);
\draw[line width=1pt] (3,2.5) -- (4.2,2.5);
\fill[red!80!black] (1,2.5) circle (3pt);
\node at (2,5.4) {$\fB_\Dir$};
\draw[dashed] (0,0) -- (3,0);
\draw[dashed] (0,4) -- (3,4);
\draw[dashed] (2,5) -- (5,5);
\draw[dashed] (2,1) -- (5,1);
\draw [cyan, thick, fill=cyan, opacity=0.1]
(0,4) -- (3, 4) -- (5, 5) -- (2,5) -- (0,4);
\draw [black, thick, dashed]
(3,0) -- (3, 4) -- (5, 5) -- (5,1) -- (3,0);
\node[below, red!80!black] at (.5, 3.2) {$\cE_0^{R,i}$};
\node  at (2.5, 3) {$\Q_{1}^{R}$};
\node  at (9, 2.5) {$i=1\,,\dots\,, {\rm dim}(R)\,.$};
\end{scope}
\end{tikzpicture}    
\end{split}
\ee
The bulk topological surfaces give rise to the boundary surfaces generating the $G$  0-form symmetry, by projecting them parallel to the boundary, where they are generically reducible and split into the elements of the conjugacy class 
\be\label{VecSym}
\Q_2^{[g]}\Bigg|_{{\Dir}} = \bigoplus_{g\in [g]} D_2^g \,.
\ee
The set of $D_2^{g}$, $g\in G$, generate the 0-form symmetry group $G$ associated to $\fB_\Dir$.
We can also consider L-shaped configurations, which will be important when discussing non-genuine line operators which take the following form: 
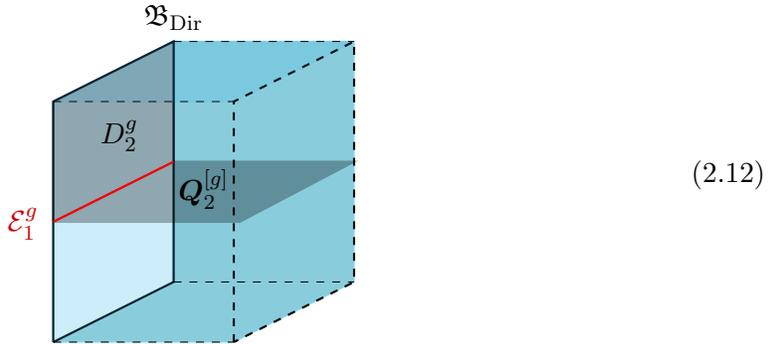
\be\label{LShaped}
\begin{split}
\begin{tikzpicture}
 \begin{scope}[shift={(0,0)},scale=0.8] 
\draw [cyan, fill=cyan!80!red, opacity =0.5]
(0,0) -- (0,4) -- (2,5) -- (5,5) -- (5,1) -- (3,0)--(0,0);
\draw [black, thick, fill=white,opacity=1]
(0,0) -- (0, 4) -- (2, 5) -- (2,1) -- (0,0);
\draw [cyan, thick, fill=cyan, opacity=0.2]
(0,0) -- (0, 4) -- (2, 5) -- (2,1) -- (0,0);
\draw [black, thick, fill=black, opacity=0.3]
(0.,2) -- (0., 4) -- (2, 5) -- (2,3) -- (0,2);
\draw [black, thick, fill=black, opacity=0.3]
(0.,2) -- (3.1, 2) -- (5., 3) -- (2, 3) -- (0.,2);
\draw [black, thick, dashed]
(3,0) -- (3, 4) -- (5, 5) -- (5,1) -- (3,0);
\node at (2,5.4) {$\fB_\Dir$};
\draw[dashed] (0,0) -- (3,0);
\draw[dashed] (0,4) -- (3,4);
\draw[dashed] (2,5) -- (5,5);
\draw[dashed] (2,1) -- (5,1);
\draw [cyan, thick, fill=cyan, opacity=0.1]
(0,4) -- (3, 4) -- (5, 5) -- (2,5) -- (0,4);
\draw [thick, red] (0.0, 2.) -- (2, 3) ;
\node[below, red!80!black] at (-0.5, 2.4) {$\cE_1^g$};
\node  at (2.5, 2.55) {$\Q_{2}^{[g]}$};
\node[above] at (1.1, 3) {$D_2^{g}$};
\end{scope}
\end{tikzpicture}    
\end{split}
\ee
There is exactly one line operator $\cE_1^g$ between $D_2^g$ on boundary and the bulk operator $\Q_2^{[g]}$.
\be
    \Q_{2}^{[g]}\Bigg|_{{\Dir}}=\left\{(D_2^{g},\cE_1^{g})\,,\quad g\in [g]\right\}\,.
\ee
Additionally there are SymTFT bulk condensation defects whose projections on $\fB_{\Dir}$ can be deduced from the projections of the non-condensation defects.
In the rest of the paper we will mostly avoid detailing the properties of the condensation defects unless necessary.

Note that the surfaces $D_2^g$ generate precisely a $G$ 0-form symmetry, which is why this BC when used as the symmetry boundary gives rise to $\TwoVec_G$.

\subsubsection{Other Minimal Boundary Conditions}
\label{sec:Minies}

For a Neumann type minimal BC $\fB_{\Neu(H),\tau}$, we have line operators valued in $\Rep(H)$ on the boundary. Given a representation $R$ of $G$ and a representation $R'$ of $H$, we can have L-shaped configurations of the type: 
\be
\begin{split}\label{L1Shaped}
\begin{tikzpicture}
 \begin{scope}[shift={(0,0)},scale=0.8] 
\draw [cyan, fill=cyan!80!red, opacity =0.5]
(0,0) -- (0,4) -- (2,5) -- (5,5) -- (5,1) -- (3,0)--(0,0);
\draw [black, thick, fill=white,opacity=1]
(0,0) -- (0, 4) -- (2, 5) -- (2,1) -- (0,0);
\draw [cyan, thick, fill=cyan, opacity=0.2]
(0,0) -- (0, 4) -- (2, 5) -- (2,1) -- (0,0);
\draw[line width=1pt] (1,2.5) -- (3,2.5);
\draw[line width=1pt] (1,2.5) -- (1,4.5);
\draw[line width=1pt] (3,2.5) -- (4.2,2.5);
\fill[red!80!black] (1,2.5) circle (3pt);
\node at (2,5.4) {$\fB_{\Neu(H),\tau}$};
\draw[dashed] (0,0) -- (3,0);
\draw[dashed] (0,4) -- (3,4);
\draw[dashed] (2,5) -- (5,5);
\draw[dashed] (2,1) -- (5,1);
\draw [cyan, thick, fill=cyan, opacity=0.1]
(0,4) -- (3, 4) -- (5, 5) -- (2,5) -- (0,4);
\node[below, red!80!black] at (.8, 2.4) {$\cE_0^{R,R'}$};
\draw [black, thick, dashed]
(3,0) -- (3, 4) -- (5, 5) -- (5,1) -- (3,0);
\node  at (2.5 , 3) {$\Q_{1}^{R}$};
\node  at (0.5, 3.3) {$D_{1}^{R'}$};
\node  at (5.5, 2.5) {$,$};
\end{scope}
\end{tikzpicture}    
\end{split}
\ee
with the number of local operators arising between the two lines being the number of copies of $R'$ arising in the decomposition of $R$ as an $H$-representation.

 The boundary surface operators fall into condensation classes labeled by double $H$-cosets.
A bulk surface operator of type $\Q_2^{[h]}$ for $h\in H$ can end on the boundary. 
More generally a bulk surface operator $\Q_2^{[g]}$ can have multiple ends on $\fB_{\Neu(H),\tau}$.
These are given by the splitting of $G$ conjugacy classes into $H$-conjugacy classes via the equivalence relation $g_1\sim g_1'$ if $g_1=hg_1'h^{-1}$ for some $h$ in $H$.
For example, let some $G$ conjugacy class $[g]$ split as 
\begin{equation}
    [g]=\oplus_{i} [g_i]_{H}\,,
\end{equation}
into $H$ conjugacy classes. 

Correspondingly the bulk surface operator $\Q_2^{[g]}$ has the following ends on $\fB_{\Neu(H),\tau}$
\begin{equation}
    \Q_2^{[g]}\Bigg|_{\Neu(H),\tau}=\bigoplus_{g_i\notin H}(D_2^{[g_i]_H}\,, \cE_1^{[g_i]_H})
    \oplus \bigoplus_{g_i\in H} \cE_1^{[g_i]_H}\,.
\end{equation}
Which is depicted as
\be
\begin{split}
\begin{tikzpicture}
 \begin{scope}[shift={(0,0)},scale=0.8] 
\draw [cyan, fill=cyan!80!red, opacity =0.5]
(0,0) -- (0,4) -- (2,5) -- (5,5) -- (5,1) -- (3,0)--(0,0);
\draw [black, thick, fill=white,opacity=1]
(0,0) -- (0, 4) -- (2, 5) -- (2,1) -- (0,0);
\draw [cyan, thick, fill=cyan, opacity=0.2]
(0,0) -- (0, 4) -- (2, 5) -- (2,1) -- (0,0);
\draw [black, thick, fill=black, opacity=0.3]
(0.,2) -- (3.1, 2) -- (5., 3) -- (2, 3) -- (0.,2);
\node at (2,5.4) {$\fB_{\Neu(H),\tau}$};
\draw[dashed] (0,0) -- (3,0);
\draw[dashed] (0,4) -- (3,4);
\draw[dashed] (2,5) -- (5,5);
\draw[dashed] (2,1) -- (5,1);
\draw [black, thick, dashed]
(3,0) -- (3, 4) -- (5, 5) -- (5,1) -- (3,0);
\draw [cyan, thick, fill=cyan, opacity=0.1]
(0,4) -- (3, 4) -- (5, 5) -- (2,5) -- (0,4);
\draw [thick, red] (0.0, 2.) -- (2, 3) ;
\node[below, red!80!black] at (-0.8, 2.4) {$\cE_1^{[g_i]_H}$};
\node  at (2.5, 2.55) {$\Q_{2}^{[g]}$};
\node[above] at (5.5, 2.55) {$,$};
\node[below] at (2, -0.5) {$(g_i\in H)$};
\end{scope}
\begin{scope}[shift={(7,0)},scale=0.8] 
\draw [cyan, fill=cyan!80!red, opacity =0.5]
(0,0) -- (0,4) -- (2,5) -- (5,5) -- (5,1) -- (3,0)--(0,0);
\draw [black, thick, fill=white,opacity=1]
(0,0) -- (0, 4) -- (2, 5) -- (2,1) -- (0,0);
\draw [cyan, thick, fill=cyan, opacity=0.2]
(0,0) -- (0, 4) -- (2, 5) -- (2,1) -- (0,0);
\draw [black, thick, fill=black, opacity=0.3]
(0.,2) -- (0., 4) -- (2, 5) -- (2,3) -- (0,2);
\draw [black, thick, fill=black, opacity=0.3]
(0.,2) -- (3.1, 2) -- (5., 3) -- (2, 3) -- (0.,2);
\node at (2,5.4) {$\fB_{\Neu(H),\tau}$};
\draw[dashed] (0,0) -- (3,0);
\draw[dashed] (0,4) -- (3,4);
\draw[dashed] (2,5) -- (5,5);
\draw[dashed] (2,1) -- (5,1);
\draw [black, thick, dashed]
(3,0) -- (3, 4) -- (5, 5) -- (5,1) -- (3,0);
\draw [cyan, thick, fill=cyan, opacity=0.1]
(0,4) -- (3, 4) -- (5, 5) -- (2,5) -- (0,4);
\draw [thick, red] (0.0, 2.) -- (2, 3) ;
\node[below, red!80!black] at (-0.8, 2.4) {$\cE_1^{[g_i]_H}$};
\node  at (2.5, 2.55) {$\Q_{2}^{[g]}$};
\node[above] at (1.1, 3) {$D_2^{[g_i]_H}$};
\node[above] at (5.5, 2.55) {$.$};
\node[below] at (2, -0.5) {$(g_i\notin H)$};
\end{scope}
\end{tikzpicture}    
\end{split}
\ee
The structure of these ends can be deduced straightforwardly by gauging $H$ on the boundary $\fB_{\Dir}$ starting from the configurations in \eqref{LShaped}.
The lines $\cE_1^{[g_i]_H}$ descend from the twisted sector line
\begin{equation}
    \bigoplus_{g\in [g_i]_H}\cE_1^g\,,
\end{equation}
on $\fB_{\Dir}$ upon such a gauging.
The surfaces $D_2^{[g_i]_H}$ on $\fB_{\Neu(H),\tau}$ are invertible or non-invertible depending on whether the order of the conjugacy class $[g_i]_H$ is 1 or greater than 1 respectively.
The fusion rules (upto connect components) of these surfaces can be deduced from the product structure on the $H$-conjugacy classes. 
The relation to $H$-double cosets is due to the fact that the number of 
$H$-conjugacy classes into which a $G$-conjugacy class splits is equal to the number of double cosets $HgH$ that intersect the $G$-conjugacy class.
Additionally, some higher junctions of topological defects may carry charges under other defects. 
These reflect in the anomaly structure of the associated 2-fusion category of defects on this boundary.
There are also associators of the lines $\cE^{[g_i]_H}$ for $g_i\in H$ that are determined by $\tau\in H^3(H,U(1))$.
We discuss such boundaries for $G=S_3$ and $D_8$ in detail in other parts of the paper.

 For purely Neumann BCs having $H=G$, all topological surfaces on the boundary are condensations of $\Rep(G)$ lines on the boundary, and each bulk surface ends along the boundary and admits only a single line operator living at the end.
The associator of these lines is determined by $\tau\in H^3(G,U(1))$.
The symmetry carried by $\fB_{\Neu(G),\tau}$ is
\be
\TwoRep(G) \,,
\ee
which is precisely the fusion 2-category formed by condensation surfaces for $\Rep(H)$ lines.

We summarize the possible minimal BC for a 4d DW $G$ theory in table \ref{tab:MinBCTab}, including the associated symmetries that they would correspond to if chosen as $\Bsym$. We also provide all the topolotical defects that can end (have Dirichlet BCs) in each case, as well as the 2d projection of the SymTFT picture.

\begin{table}
$$
\begin{array}{|c|c|c|}\hline
\text{BC, Symmetry} & \Q_p\text{ with Dirichlet BCs} & \text{SymTFT Quiche} \cr \hline \hline 
{\rm Dir}: \TwoVec_{G} &
\begin{cases}
     \Q_2^{[\id]}\,,\\ 
 \bigoplus_{R \in \Rep(G)} \dim (R) \Q_1^{R} 
\end{cases}
&
\begin{tikzpicture}[baseline]
\begin{scope}[shift={(0,-1)}]
\draw [cyan,  fill=cyan] 
(0,0) -- (0,2) -- (2,2) -- (2,0) -- (0,0) ; 
\draw [white] (0,0) -- (0,2) -- (2,2) -- (2,0) -- (0,0)  ; 
\draw [very thick] (0,0) -- (0,2) ;
\node[above] at (0,2) {$\fB_{\rm Dir}$}; 
\draw [thick, dashed](0,1) -- (2,1);
\draw [thick, fill=black] (0,1) ellipse (0.05 and 0.05);
\node[] at (1,1.3) {$\Q_{1}^{\bm{R}}$};
\node at (0, -0.1) {};
\end{scope}
\end{tikzpicture}
\cr\hline
\Neu(N),t\ N \trianglelefteq G:\  \TwoVec_{\mathbb{G}^{(2)}} & 
\begin{cases}
     \bigoplus_{n\in N} \Q_2^{[n], 1}\\ 
 \bigoplus_{n\in N} \alpha_n \Q_1^{[n], 1} \\
  \oplus\  \bigoplus_{R \in \Rep (G/N)} \dim (R) \Q_1^{R} 
\end{cases}
& 
\begin{tikzpicture}[baseline]
\begin{scope}[shift={(0,-1)}]
\draw [cyan,  fill=cyan] 
(0,0) -- (0,2.5) -- (2.5,2.5) -- (2.5,0) -- (0,0) ; 
\draw [white] (0,0) -- (0,2.5) -- (2.5,2.5) -- (2.5,0) -- (0,0)  ; 
\draw [very thick] (0,0) -- (0,2.5) ;
\node[above] at (0,2.5) {$\fB_{\TwoVec_{\mathbb{G}^{(2)}}}$}; 
\draw [thick, dashed](0,0.5) -- (2.5,0.5);
\draw [thick](0,1.7) -- (2.5,1.7);
\draw [thick, fill=black] (0,0.5) ellipse (0.05 and 0.05);
\draw [thick, fill=black] (0,1.7) ellipse (0.05 and 0.05);
\node[above] at (1.25,1.7) {$\Q_{2}^{[n]}$};
\node[above] at (1.25,0.5) {$\Q_{1}^{[n],1},\Q_{1}^{R}$};
\node at (0, -0.1) {};
\end{scope}
\end{tikzpicture}
\cr \hline
\Neu (H),t \ H \not\trianglelefteq G:\ \TwoRep (\mathbb{G}^{(2)}) 
&
\begin{cases}
     \bigoplus_{h\in H} \Q_2^{[h], 1}\cr 
  \bigoplus_{h\in H} \Q_1^{[h], 1} \ \oplus \ 
 \bigoplus_{R \in  \ker (\kappa)}  \Q_1^{R}  \cr 
\end{cases}
&
\begin{tikzpicture}[baseline]
\begin{scope}[shift={(0,-1)}]
\draw [cyan,  fill=cyan] 
(0,0) -- (0,2.5) -- (2.5,2.5) -- (2.5,0) -- (0,0) ; 
\draw [white] (0,0) -- (0,2.5) -- (2.5,2.5) -- (2.5,0) -- (0,0)  ; 
\draw [very thick] (0,0) -- (0,2.5) ;
\node[above] at (0,2.5) {$\fB_{\TwoRep (\mathbb{G}^{(2)})}$}; 
\draw [thick, dashed](0,0.5) -- (2.5,0.5);
\draw [thick](0,1.7) -- (2.5,1.7);
\draw [thick, fill=black] (0,0.5) ellipse (0.05 and 0.05);
\draw [thick, fill=black] (0,1.7) ellipse (0.05 and 0.05);
\node[above] at (1.25,1.7) {$\Q_{2}^{[h]}$};
\node[above] at (1.25,0.5) {$\Q_{1}^{[h],1}, \Q_{1}^{R}$};
\node at (0, -0.1) {};
\end{scope}
\end{tikzpicture}
\cr \hline 
\Neu (G),t: \ \TwoRep (G) & 
\begin{cases}
     \bigoplus_{g} \Q_2^{[g], 1} \cr  
\bigoplus_{g} \Q_1^{[g], 1} 
\end{cases}
&
\begin{tikzpicture}[baseline]
\begin{scope}[shift={(0,-1)}]
\draw [cyan,  fill=cyan] 
(0,0) -- (0,2.5) -- (2.5,2.5) -- (2.5,0) -- (0,0) ; 
\draw [white] (0,0) -- (0,2.5) -- (2.5,2.5) -- (2.5,0) -- (0,0)  ; 
\draw [very thick] (0,0) -- (0,2.5) ;
\node[above] at (0,2.5) {$\fB_{\TwoRep (S_3)}$}; 
\draw [thick, dashed](0,0.5) -- (2.5,0.5);
\draw [thick](0,1.7) -- (2.5,1.7);
\draw [thick, fill=black] (0,0.5) ellipse (0.05 and 0.05);
\draw [thick, fill=black] (0,1.7) ellipse (0.05 and 0.05);
\node[above] at (1.25,1.7) {$\Q_{2}^{[g],1}$};
\node[above] at (1.25,0.5) {$\Q_{1}^{[g],1}$};
\node at (0, -0.1) {};
\end{scope}
\end{tikzpicture}
\cr \hline
\end{array}
$$
\caption{Summary of minimal gapped BC for $\cZ(\TwoVec_G)$. Note that $N \trianglelefteq G$ is a normal subgroup whereas $H$ is not normal. The two-group is $\mathbb{G}^{(2)} = N^{(1)} \rtimes (G/N)^{(0)}$. In the case of non-normal subgroup $H$ gauging, the  map $\kappa:\Rep (G) \to \Rep (H)$ corresponds to the projection of any $G$-irrep onto $H$. The lines that are in the kernel can end in this BC. For abelian $H$ this is simply $\Rep (G/H)$. Recall that dashed lines are 1d topological defects, solid lines are 2d. The choices of SPTs are indicated by the parameters $t$. Finally the multiplicities $\alpha_n$ depend on how the conjugacy classes of $G$ split into those of $N$, see beginning of section \ref{sec:Minies}. } \label{tab:MinBCTab}
\end{table}

\subsection{Gapped Boundary Conditions: Non-Minimal}

Non-minimal boundary conditions can be obtained by stacking the minimal Dirichlet boundary condition with a $G$ symmetric 3d TFT and subsequently gauging the diagonal $G$ symmetry.
We refer to this process as generalized gauging.

\subsubsection{Non-minimal Dirichlet Boundary Condition}
A simple class of non-minimal boundary conditions is obtained by simply stacking the Dirichlet boundary condition by a 3d topological order.
Such boundaries are classified by a modular tensor categories (MTCs). 
Given any MTC $\cM$, which describes the topological line defects of a 3d TFT $\fT$, one can stack the Dirichlet boundary condition with $\fT$ to obtain a new topological boundary 
\begin{equation}
    \fB^{\fT}_{\Dir}= \fB_{\Dir}\boxtimes \fT\,.
\end{equation}
The 2-fusion category of topological defects on this boundary are 
\begin{equation}
    \cS= \TwoVec_{G}\boxtimes \Sigma\cM\,,
\end{equation}
where $\Sigma \cM$ is the 2-category obtained by including all condensation defects constructible from the lines in $\cM$, i.e., the Karoubi completion of $\cM$ into a 2-category.

 In relation to the bulk SymTFT defects, this boundary condition looks almost like $\fB_{\Dir}$, except that the bulk identity surface ends on $\fB^{\fT}_{\Dir}$ not only on the identity line but on any line in $\cM$.
Likewise any other line or surface defect on the boundary can be stacked with an object in $\Sigma\cM$ without affecting its bulk lift. 

\subsubsection{Other Non-minimal Boundary Conditions}
Now we describe the non-minimal generalization of the $\fB_{\Neu(H),\tau}$ boundary condition.
%
%
Such a boundary condition is defined by stacking $\fB_{\Dir}$ with a $G$ symmetric $\fT_{G}$ whose lines defects form the modular tensor category $\cM$.
The $G^{(0)}$ symmetry is implemented on $\fT_{G}$ via condensation defects \cite{Carqueville:2017ono}, more precisely by prescribing a monoidal 2-functor \cite{Etingof:2009yvg, Barkeshli:2014cna}
\begin{equation}
    \cF:\TwoVec_{G}\longrightarrow {\rm BrPic}(\cM)\,,
\end{equation}
where ${\rm BrPic}(\cM)$ is the 2-category whose objects are invertible condensation surfaces, 1-morphisms are invertible lines between invertible surfaces and 2-morphisms are invertible local operators operators.
Concretely, prescribing such a functor, involves picking various levels of data.
At the top level, a group homomorphism 
\begin{equation}
    \phi: G\to {\rm Aut}(\cM)\,,
\end{equation}
where $\rm Aut(\cM)$ are braided auto-equivalences of $\cM$.
The choice of $\phi$ maps objects in $\TwoVec_{G}$ to condensation surfaces that implement the $G$ symmetry.
At the next level, one picks an an invertible morphisms in $\cM$ between the surface $\phi(g)\otimes \phi(h)$ and $\phi(gh)$. 
This choice is constrained by associativity of fusion of surfaces.
This amounts to picking an element of $\eta\in H^{2}(G,{\rm Inv}(\cM))$, where ${\rm Inv}(\cM)$ are invertible lines in $\cM$.
This datum captures the symmetry fractionalization properties of the $G$ symmetry.
Finally there is a choice of isomorphism 
\begin{equation}
    D_{0}^{g,h,k}:\quad D_1^{g,h}\otimes D_1^{gh,k}\longrightarrow D_1^{g,hk}\otimes D_1^{h,k}\,,
\end{equation}
which needs to be invertible, i.e., a $U(1)$ number. 
This number is further required to satisfy higher associativity constraints which fixes it to be a 3-cocycle.
This is precisely the choice of discrete torsion $\tau$ we have encountered before.

 Now given $\fT_{G}$, i.e., an MTC $\cM$ with a choice of 2-functor $\cF$, the non-minimal generalization of the $\fB_{\Neu(H),\tau}$ boundary condition is
\begin{equation}
\fB_{\Neu(H),\tau}^{\fT}=\frac{\fB_{\Dir}\boxtimes \fT_{G}}{(H^{\rm diag}\,, \tau)}\,.
\end{equation}
The choice of $\cF$ defines a $G$-crossed braided extension of $\cM$, denoted as
\begin{equation}
    \cM^{\times}_{G}=\bigoplus_{g\in G}\cM^{g}\,.
\end{equation}
Here $\cM^{g}$ denotes the category of lines living at the end of the $g$ condensation defect defined via $\cF$.
From $\cM^{\times}_{G}$, there is a well defined construction known as equivariantization \cite{Etingof:2009yvg, Barkeshli:2014cna} to gauge the $H\subseteq G$ symmetry to obtain a larger MTC $\wt{\cM}$.
The lines in $\wt{\cM}$ have the structure
\begin{equation}
    \wt{\cM}=\bigoplus_{[h]}\wt{\cM}^{[h]}\,,
\end{equation}
where the sum is over $H$-conjugacy classes.
Here the grade $\wt\cM^{[h]}$ is obtained from $\cM^{\times}_H$ as
\begin{equation}
    \wt\cM^{[h]}=[\bigoplus_{h\in [h]}\cM^h]/H\,.
\end{equation}
These fusion of lines in $\wt{\cM}$ is compatible with the product structure on $H$-conjugacy classes.
In particular $\wt{\cM}^{[\id]}$ contains lines $D_{1}^{R_H}$ that generate the dual $\Rep(H)$ 1-form symmetry. 
As in the case of the minimal boundary condition, these lines arise as the boundary projection of bulk line $\Q_1^{R}$ via the projection (see \eqref{L1Shaped})
\begin{equation}
    \kappa:\quad\Rep(G)\longrightarrow \Rep(H)\,.
\end{equation}
Additionally, there are lines in $\cM^{[\id]}$ that do not lift to any bulk SymTFT lines.
All such lines are remotely detectable in the sense of \cite{Lan:2018vjb}, i.e., they braid non-trivially with at least one other line in $\wt{\cM}^{[\id]}$.
Lines in all other grades $\wt{\cM}^{[h]}$ ($[h]\neq [\id]$) are attached to some bulk SymTFT surface.

 A ends of a bulk surface $\Q_2^{[g]}$ can be understood from the splitting of $[g]$ into $[h]$ conjugacy classes.
As before, let us assume
\begin{equation}
    [g]=\oplus_i[g_i]_{H}\,,
\end{equation}
where $[g_i]_{H}$ are $H$-conjugacy classes.
Correspondingly for every $g_i\in H$, the surface $\Q_2^{[g]}$ can end on lines in $\wt{\cM}^{[g_i]_H}$.
For $g_i\notin H$, the surface ends on a twisted sector line $\cE_1^{[g_i]_H}$ attached to the boundary surface defect $D_2^{[g_i]_H}$.
The properties of $\cE_1^{[g_i]_H}$ will be determined explicitly for concrete examples in the later sections of this paper.
The general strategy is to consider the direct sum of twisted sector lines
\begin{equation}
    \bigoplus_{g\in [g_i]_H}\left(D_2^{g}\,, \cE_{1}^{g}\right)\,,
\end{equation}
on $\fB_{\Dir}^{\fT}$
and consider all $H$-symmetric stackings with lines in $\cM$.
The $H$ gauging of these stacked twisted sector lines gives $(D_2^{[g_i]_H}\,, \cE_1^{[g_i]_H})$ on $\fB^{\fT}_{\Neu(H),\tau}$.

\subsection{Gapped Phases}

To construct gapped phases in the SymTFT we fix a symmetry boundary $\Bsym$, which fixes the symmetry fusion 2-category $\cS$.
Any $\cS$-symmetric gapped phase is then obtained by closing the quiche to a sandwich with a gapped physical boundary condition. 
More precisely, choosing a gapped physical boundary condition and compactifying the SymTFT produces an $\cS$ symmetric 3d TFT that describes the IR fixed point of an $\cS$ symmetric gapped phase.

\subsubsection{Generalities}
\label{MumboJumbo}

A gapped phase is specified by two boundary conditions.
Firstly the bulk SymTFT operators that have untwisted or genuine ends on the physical boundary become topological defects of the 3d TFT describing the IR gapped phase.
These topological defects are nothing but order parameters of the $\cS$ symmetric gapped phase.
Data characterizing this gapped phase such as number of vacua, structure of order parameters and the symmetry action on the vacua and order parameters can be systematically and conveniently extracted from the SymTFT sandwich.
The first important property that can be obtained from the SymTFT is the number of vacua: it is given by 
\be
\# \text{vacua} = \sum_{R\in \Rep(G)} {\mathsf n}^R_{\rm sym} {\mathsf n}^R_{\rm phys} \,, 
\ee
where ${\mathsf n}^R_{\rm sym}$ and ${\mathsf n}^R_{\rm phys}$ are the number of linearly independent untwisted sector local operator that $\Q_1^{R}$ has on the symmetry and physical boundaries, respectively. 
More precisely we will get local operators from the interval compactification of $\Q_1^R$ with ends on the LHS and RHS given by $\cE_0^{R,i}$ and $\wt{\cal E}_0^{R,j}$, respectively, as shown in (\ref{QuicheSandwich}), that we denote by 
\be
    \cO^{R\,,i,j}=(\cE_0^{R,i}\,, \Q_1^{R}\,, \wt{\cE}_0^{R,j})\,.
\ee
The action of the symmetry on these order parameters is given by the linking of defects in the bulk. 
Furthermore we have non-genuine local operators, that can be order parameters.
In fact the non-genuine local operators can be in the same symmetry multiplets as local OPs for non-invertible symmetries. 
These are obtained from L-shaped configurations, i.e., bulk SymTFT lines that have genuine ends on $\Bphys$ but non-genuine ends on $\Bsym$ 
\be
\begin{split}
\begin{tikzpicture}
 \begin{scope}[shift={(0,0)},scale=0.8] 
\draw [cyan, fill=cyan!80!red, opacity =0.5]
(0,0) -- (0,4) -- (2,5) -- (5,5) -- (5,1) -- (3,0)--(0,0);
\draw [black, thick, fill=white,opacity=1]
(0,0) -- (0, 4) -- (2, 5) -- (2,1) -- (0,0);
\draw [cyan, thick, fill=cyan, opacity=0.2]
(0,0) -- (0, 4) -- (2, 5) -- (2,1) -- (0,0);
\draw[line width=1pt] (1,2.5) -- (3,2.5);
\draw[line width=1pt,dashed] (3,2.5) -- (4,2.5);
\fill[red!80!black] (1,2.5) circle (3pt);
\draw [fill=blue!40!red!60,opacity=0.2]
(3,0) -- (3, 4) -- (5, 5) -- (5,1) -- (3,0);
\fill[red!80!black] (1,2.5) circle (3pt);
\fill[red!80!black] (4,2.5) circle (3pt);
\draw [black, thick, opacity=1]
(3,0) -- (3, 4) -- (5, 5) -- (5,1) -- (3,0);
\node at (2,5.4) {$\Bsym$};
\node at (5,5.4) {$\Bphys$};
\draw[dashed] (0,0) -- (3,0);
\draw[dashed] (0,4) -- (3,4);
\draw[dashed] (2,5) -- (5,5);
\draw[dashed] (2,1) -- (5,1);
\draw[line width=1pt] (1,2.5) -- (1,4.5);
\draw [cyan, thick, fill=cyan, opacity=0.1]
(0,4) -- (3, 4) -- (5, 5) -- (2,5) -- (0,4);
\node[below, red!80!black] at (.8, 2.4) {$\cE_0^{R,i}$};
\node[below, red!80!black] at (3.8, 2.4) {$\wt{\cE}_0^{R,j}$};
\node  at (2.5, 3) {$\Q_{1}^{R}$};
\node  at (0.5, 3.3) {$D_{1}^{R'}$};
\node  at (6.7, 2.5) {$=$};
\draw [fill=blue!60!green!30,opacity=0.7]
(8,0) -- (8, 4) -- (10, 5) -- (10,1) -- (8,0);
\draw [black, thick, opacity=1]
(8,0) -- (8, 4) -- (10, 5) -- (10,1) -- (8,0);
\draw[line width=1pt] (9,2.5) -- (9,4.5);
\fill[red!80!black] (9,2.5) circle (3pt);
\node[below, red!80!black] at (8.8, 2.4) {${\cO}^{i,j}$};
\node  at (8.5, 3.3) {$D_{1}^{R'}$};
\node[below, black] at (10.8, 3) {$,$};
\end{scope}
\end{tikzpicture}    
\end{split}
\label{eq:local OP}
\ee
which are denoted as
\begin{equation}
(D_1^{R'}\,, \cO^{i,j})=\left((D_1^{R'}\,, \cE_0^{i})\,, \Q_1^R\,, \wt{\cE}_0^{R,j}\right)\,.
\end{equation}
Next, compactifying the SymTFT with bulk surface operators stretched between both boundaries produces line operators in the 3d TFT obtained after compactification. 
Consider a surface $\Q_2^{[g]}$ that ends on both the physical and symmetry boundary.
This produces a genuine line operator in the 3d TFT
\be
\begin{split}
\begin{tikzpicture}
 \begin{scope}[shift={(0,0)},scale=0.8] 
\draw [cyan, fill=cyan!80!red, opacity =0.5]
(0,0) -- (0,4) -- (2,5) -- (5,5) -- (5,1) -- (3,0)--(0,0);
\draw [black, thick, fill=white,opacity=1]
(0,0) -- (0, 4) -- (2, 5) -- (2,1) -- (0,0);
\draw [cyan, thick, fill=cyan, opacity=0.2]
(0,0) -- (0, 4) -- (2, 5) -- (2,1) -- (0,0);
\draw [fill=blue!40!red!60,opacity=0.2]
(3,0) -- (3, 4) -- (5, 5) -- (5,1) -- (3,0);
\draw [black, thick, opacity=1]
(3,0) -- (3, 4) -- (5, 5) -- (5,1) -- (3,0);
\node at (2,5.4) {$\Bsym$};
\node at (5,5.4) {$\Bphys$};
\draw[dashed] (0,0) -- (3,0);
\draw[dashed] (0,4) -- (3,4);
\draw[dashed] (2,5) -- (5,5);
\draw[dashed] (2,1) -- (5,1);
\draw [cyan, thick, fill=cyan, opacity=0.1]
(0,4) -- (3, 4) -- (5, 5) -- (2,5) -- (0,4);
\draw [black, thick, fill=black, opacity=0.3]
(0.,2) -- (3, 2) -- (5., 3) -- (2, 3) -- (0.,2);
\node  at (2.5, 2.5) {$\Q_{2}^{[g]}$};
\node[below, red!80!black] at (-0.5, 2.4) {$\cE_1^{i}$};
\node[below, red!80!black] at (4, 2.4) {$\wt{\cE}_1^{j}$};
\draw [line width=1pt, red!80!black] (0.0, 2.) -- (2, 3);
\draw [line width=1pt, red!80!black] (3, 2.) -- (5, 3);
\node  at (6.7, 2.5) {$=$};
\draw [fill=blue!60!green!30,opacity=0.7]
(8,0) -- (8, 4) -- (10, 5) -- (10,1) -- (8,0);
\draw [black, thick, opacity=1]
(8,0) -- (8, 4) -- (10, 5) -- (10,1) -- (8,0);
\draw [line width=1pt, red!80!black] (8, 2.) -- (10, 3);
\node[below, red!80!black] at (9, 2.4) {${\cL}^{i,j}$};
\node[below, red!80!black] at (10, 2.4) {$,$};
\end{scope}
\end{tikzpicture}    
\end{split}
\ee
which are denoted as
\begin{equation}
    \cL^{i,j}=\left(\cE_1^{i}\,, \Q_2^{[a]}\,, \wt{\cE}_1^{j}\right)\,.
\end{equation}
In the 3d TFT, $\cL^{i,j}$ braid with lines obtained from the boundary projection of SymTFT lines $\Q_1^{R}$, with a braiding phase determined by the character $\chi_{R}([g])$.
The lines $\cL^{i,j}$ serve as order parameters for 1-form symmetry breaking.
Finally, consider a surface $\Q_2^{[g]}$ that has a genuine end on  the physical boundary and a non-genuine end on the symmetry boundary.
This produces a non-genuine line operator in the 3d TFT, i.e., a line defect that is attached to a symmetry generator
\be
\begin{split}
\begin{tikzpicture}
 \begin{scope}[shift={(0,0)},scale=0.8] 
\draw [cyan, fill=cyan!80!red, opacity =0.5]
(0,0) -- (0,4) -- (2,5) -- (5,5) -- (5,1) -- (3,0)--(0,0);
\draw [black, thick, fill=white,opacity=1]
(0,0) -- (0, 4) -- (2, 5) -- (2,1) -- (0,0);
\draw [cyan, thick, fill=cyan, opacity=0.2]
(0,0) -- (0, 4) -- (2, 5) -- (2,1) -- (0,0);
\draw [fill=blue!40!red!60,opacity=0.2]
(3,0) -- (3, 4) -- (5, 5) -- (5,1) -- (3,0);
\draw [black, thick, opacity=1]
(3,0) -- (3, 4) -- (5, 5) -- (5,1) -- (3,0);
\node at (2,5.4) {$\Bsym$};
\node at (5,5.4) {$\Bphys$};
\draw[dashed] (0,0) -- (3,0);
\draw[dashed] (0,4) -- (3,4);
\draw[dashed] (2,5) -- (5,5);
\draw[dashed] (2,1) -- (5,1);
\draw [cyan, thick, fill=cyan, opacity=0.1]
(0,4) -- (3, 4) -- (5, 5) -- (2,5) -- (0,4);
\draw [black, thick, fill=black, opacity=0.3]
(0.,2) -- (3, 2) -- (5., 3) -- (2, 3) -- (0.,2);
\draw [black, thick, fill=black, opacity=0.3]
(0.,2) -- (0., 4) -- (2, 5) -- (2,3) -- (0,2);
\node  at (2.5, 2.5) {$\Q_{2}^{[g]}$};
\node[below, red!80!black] at (-0.5, 2.4) {$\cE_1^{i}$};
\node[below, red!80!black] at (4, 2.4) {$\wt{\cE}_1^{j}$};
\draw [line width=1pt, red!80!black] (0.0, 2.) -- (2, 3);
\draw [line width=1pt, red!80!black] (3, 2.) -- (5, 3);
\node  at (6.7, 2.5) {$=$};
\draw [fill=blue!60!green!30,opacity=0.7]
(8,0) -- (8, 4) -- (10, 5) -- (10,1) -- (8,0);
\draw [black, thick, opacity=1]
(8,0) -- (8, 4) -- (10, 5) -- (10,1) -- (8,0);
\draw [black, thick, fill=black, opacity=0.3]
(8,2) -- (8, 4) -- (10, 5) -- (10,3) -- (8,2);
\draw [line width=1pt, red!80!black] (8, 2.) -- (10, 3);
\node[below, black] at (9, 3.9) {$D_2^{g}$};
\node[below, black] at (1., 3.9) {$D_2^{g}$};
\node[below, red!80!black] at (9, 2.4) {${\cL}^{i,j}$};
\end{scope}
\end{tikzpicture}    
\end{split}
\ee
which are denoted as 
\begin{equation}
    (D_2^g\,, \cL^{i,j})=\left((D_2^{g}\,,\cE_1^{i})\,, \Q_2^{[a]}\,, \wt{\cE}_1^{j}\right)\,.
\end{equation}
In particular the resulting lines can have non-trivial associators, usually labeled by $t\in H^3 (H, U(1))$, where $H$ is the subgroup that labels the surfaces that end on the physical boundary. These do not affect the total number of vacua, but correspond to non-trivial SPT phases.


\subsubsection{Gapped Phases for $\TwoVec_{G}$}
\label{sec:VecGGP}

In practice it is useful to start a systematic analysis by considering all gapped phases for the choice $\Bsym=\mathfrak{B}_{\text{Dir}}$, and then to perform a generalized gauging on the symmetry boundary, that maps $\mathfrak{B}_{\text{Dir}}$ to any other gapped BCs, and thus symmetry boundary (minimal or non-minimal). Of course these phases include (extensions by SPTs of) standard Landau gapped phases, but we will see that the generalized gauging vastly extends the possibilities. 

Starting with the minimal gapped BCs as summarized in table \ref{tab:MinBCTab} we get four types of gapped phases: denoting these by $\Bsym-\Bphys$ we find 
\begin{itemize}
\item Dir-Dir: $G$ SSB with $|G|$ vacua 
\item Dir-Neu$(G),t$: $G$-SPT labeled by $t$ with single vacua
\item Dir-Neu$(N),t$: $G/N$ SSB where $N$ is a normal subgroup, times an $N$-SPT labeled by $t$
\item Dir-Neu$(H),t$: $G/H$ SSB, where $H$ is non-normal, times an $H$-SPT labeled by $t$.
\end{itemize}
Let us discuss some of their salient properties here -- we will consider several in depth examples in the following. 

\paragraph{$G$ SSB phase.} 
Choosing Dir-Dir means that all lines $\Q_1^{R}$ with $R$ an irreducible representation of $G$ can end on both boundaries. 
As discussed in section \ref{MumboJumbo}, the number of vacua for each line $\Q_1^R$ is then $(\dim (R))^2$, as there are that many local operators, so that the total number of vacua is 
\be
\# \text{vacua}= \sum_{R= G-\text{irrep}} \dim (R)^2 = |G| \,.
\ee
This is the $G$-SSB phase.

\paragraph{$G$ SPT phases.} The other extreme is obtained by considering the pairing Dir-Neu$(G, t)$ where $t \in H^3 (G, U(1))$. There is only one local operataor, coming from the identify line, and this is a single-vacuum theory. Whether or not it is the trivial phase or an SPT is determined by the fusion of lines on the physical boundary, which give rise to non-genuine OPs: the defects $\Q_2^{[g]}$ for $g\in G$ can end on the Neu-BC, and have associators given by $t$. These surfaces can however not end on the symmetry boundary and thus give rise to non-genuine lines defects. 

\paragraph{($G/N$-SSB) $\boxtimes (N$-SPTs).} 
For a normal subgroup, we can consider Dir-Neu$(N),t$. In this case the SymTFT lines $\Q_1$ that can end on both boundaries are in $\Rep (G/N)$. There are thus 
\be
\# \text{vacua}= \sum_{R= G/N-\text{irrep} } \dim (R)^2 = |G/N|\,.
\ee
In each vacuum we can have a, possibly non-trivial, SPT for $N$. 
Again there is a choice of associator, which is part of the definition of the physical boundary Neu$(N),t$, which specifies the SPT. 

\paragraph{($G/H$-SSB) $\boxtimes (H$-SPTs)}
For $H$ non-normal Dir-Neu$(H),t$ gives again a mixed SSB and SPT phase. Again the lines determine the number of vacua which in this case is 
\be
\# \text{vacua} = |G/H|
\ee
and there can be a non-trivial unbroken symmetry $H$-SPT. 
Again there is a choice of SPTs for the unbroken symmetry $H$ that is encoded in the Neumann BC. 

\paragraph{SymTFT derivation of vacua for $2\Vec_G$ phases.}
The SymTFT lines $\Q_1^{R}$ for (a subset of) $R\in \Rep(G)$ that can end on both boundaries provide topological local operators after compactification
\be
\begin{split}
\begin{tikzpicture}
 \begin{scope}[shift={(0,0)},scale=0.8] 
\draw [cyan, fill=cyan!80!red, opacity =0.5]
(0,0) -- (0,4) -- (2,5) -- (5,5) -- (5,1) -- (3,0)--(0,0);
\draw [black, thick, fill=white,opacity=1]
(0,0) -- (0, 4) -- (2, 5) -- (2,1) -- (0,0);
\draw [cyan, thick, fill=cyan, opacity=0.2]
(0,0) -- (0, 4) -- (2, 5) -- (2,1) -- (0,0);
\draw[line width=1pt] (1,2.5) -- (3,2.5);
\draw[line width=1pt,dashed] (3,2.5) -- (4,2.5);
\fill[red!80!black] (1,2.5) circle (3pt);
\draw [fill=blue!40!red!60,opacity=0.2]
(3,0) -- (3, 4) -- (5, 5) -- (5,1) -- (3,0);
\fill[red!80!black] (1,2.5) circle (3pt);
\fill[red!80!black] (4,2.5) circle (3pt);
\draw [black, thick, opacity=1]
(3,0) -- (3, 4) -- (5, 5) -- (5,1) -- (3,0);
\node at (2,5.4) {$\Bsym=\fB_{\Dir}$};
\node at (5,5.4) {$\Bphys$};
\draw[dashed] (0,0) -- (3,0);
\draw[dashed] (0,4) -- (3,4);
\draw[dashed] (2,5) -- (5,5);
\draw[dashed] (2,1) -- (5,1);
\draw [cyan, thick, fill=cyan, opacity=0.1]
(0,4) -- (3, 4) -- (5, 5) -- (2,5) -- (0,4);
\node[below, red!80!black] at (.8, 2.4) {$\cE_0^{R,i}$};
\node[below, red!80!black] at (3.8, 2.4) {$\wt{\cE}_0^{R,j}$};
\node  at (2.5, 3) {$\Q_{1}^{R}$};
\node  at (6.7, 2.5) {$=$};
\draw [fill=blue!60!green!30,opacity=0.7]
(8,0) -- (8, 4) -- (10, 5) -- (10,1) -- (8,0);
\draw [black, thick, opacity=1]
(8,0) -- (8, 4) -- (10, 5) -- (10,1) -- (8,0);
\fill[red!80!black] (9,2.5) circle (3pt);
\node[below, red!80!black] at (8.8, 2.4) {${\cO}^{R\,,i,j}$};
\end{scope}
\end{tikzpicture}    
\end{split}
\label{eq:local OP}
\ee
where $i,j=1,...\dim(R)$ and
\begin{equation}
    \cO^{R\,,i,j}=(\cE_0^{R,i}\,, \Q_1^{R}\,, \wt{\cE}_0^{R,j})\,.
\end{equation}
{Note that the number of independent ends of $\Q_1^R$ on $\Bphys=\fB_{\Neu(H),\tau}$ for $H\subseteq G$ is less than or equal to $\dim(R)$ and given by linear combinations uncharged under the $H$ subgroup being gauged.} The $G$ 0-form symmetry acts on these local operators as
\begin{equation}
    g:\;\cO^R\mapsto \cD^R(g)\cO^{R}
\end{equation}
where $\cD^R(g)$ is the matrix representation of $g$ in the irreducible representation $R$ and $\cO^R$ comes from the SymTFT compactfication. 
By introducing a basis of group elements $\ket{g}$, we can decompose
\be
    \cO^R=\sum_g\cD^R(g)\ket{g}\bra{g}\,.
\ee
Using the great orthogonality theorem for irreducibile representations, we can write the projector onto $\bar{g}\in G$ as:
\be \label{eq:projector}
    \ket{\bar{g}}\bra{\bar{g}}=\sum_R\frac{\dim R}{|G|}\,\cD^R(\bar{g})^{-1}\,\cO^R\,.
\ee
Each pysical boundary is specified by a subgroup $H\subseteq G$, which is (up to conjugation) the preserved symmetry in each vacuum. The vacua can therefore be labeled by (left) cosets $G/H$, whose group elements we can write in terms of the local operators $\cO^R$ as sums of projectors \eqref{eq:projector}
\be \label{eq:Gsym_vacua}
     v_{kH}=\sum_{\bar{g}\in kH}\ket{\bar{g}}\bra{\bar{g}}=\sum_{\bar{g}\in kH}\sum_R\frac{\dim R}{|G|}\,\cD^R(\bar{g})^{-1}\,\cO^R\,.
\ee
which are idempotents: $v_iv_j=\delta_{i,j}\,v_i$. The action of $g\in G$ on $v_k$ is $g\cdot v_k=v_{gk}$.

\subsubsection{Gapped Phases for $\TwoRep(G)$}
Choosing $\Bsym=\fB_\Neu$ and $\Bphys=\fB_{\Neu(H),\tau}$ we obtain the various minimal gapped phases for $\TwoRep(G)$, or in other words for $\Rep(G)$ 1-form symmetry.
Such a phase has a single vacuum described by an $H$ Dijkgraaf-Witten theory with twist $\tau$. This 3d TFT has bosonic lines forming a $\Rep(H)$ braided fusion category, which can be identified with Wilson lines for $H$. The 1-form symmetry is then realized by the forgetful functor
\be\label{func2}
    \kappa:\quad \Rep(G)\to\Rep(H)\,.
\ee
This result is easy to derive. We know that the bulk lines $\Q_1^R$ project onto the physical boundary according to the functor \eqref{func2}. The physical boundary lines $D_1^R$ for $R\in\Rep(H)$ provide one set of lines (the electric lines) for the DW theory, and the surfaces $\Q_2^{[g]}$ ending on both boundaries provide the magnetic lines labeled by conjugacy classes of the DW theory.

\subsubsection{Gapped Phases for any $\Bsym$}

To obtain the gapped phases for any other symmetry,
there are two ways: 
\begin{itemize}
\item We can either change the symmetry boundary from Dir to 
any other gapped BC, where the topological defects form a fusion 2-category $\cS$. And then consider in turn again all the gapped BC for the physical boundary. 
\item Alternatively, we can take the gapped phases for $\TwoVec_G$ from section \ref{sec:VecGGP} and gauge the symmetry boundary -- by stacking and gauging. 
\end{itemize}
Both paths are equally applicable and straightforward to implement. We will consider them in turn in various examples.

\section{Non-Invertible Symmetries and Phases from $\cZ(2\Vec_{S_3})$}
\label{sec:S3}

The class of examples we will consider in the present paper generalize those in \cite{Bhardwaj:2024qiv} by considering SymTFTs that are DW theories for non-abelian finite groups $G$. The main new feature is the presence non-invertible lines forming e.g. $\Rep (G)$, and thereby generalizations of two-group symmetries and their 2-representation categories. 

The simplest example is obtained by considering the 4d DW theory for $S_3$, in which case, the symmetries correspond to generalized gauging of  $\TwoVec_{S_3}$, which we discuss now.

\subsection{The SymTFT}
We present the group $S_3$ as 
\begin{equation}
    S_{3}=\big\langle a\,, b \ \big|  \ a^3=b^2=1\,, bab=a^2\,\big\rangle\,.
\end{equation}
The SymTFT which plays a central role in our analysis of symmetries and phases is the  4d $S_3$ Dijkgraaf-Witten theory \cite{Dijkgraaf:1989pz}.
In general, the (co-dimension-2 and higher) topological defects of the 4d $G$ Dijkgraaf-Witten theory are organized as \cite{Kong:2019brm}
\be\label{2Center}
\cZ(\TwoVec_{G}) = \boxplus_{[g]} \TwoRep (H_g)\,,
\ee
where $[g]$ are conjugacy classes in $G$ and $H_{g}$ is the centralizer of a representative element in $[g]$.
Since, $S_3$ has three conjugacy classes $[\id]\,, [a]=\{a\,,a^2\}$ and $[b]=\{b\,,ab\,,a^2b\}$ with $H_{\id}=S_3\,, H_{a}=\Z_3$ and $H_{b}=\Z_2$, we obtain
\be
\cZ(\TwoVec_{S_3}) = \TwoRep(S_3) \boxplus \TwoRep (\Z_3)  \boxplus \TwoRep (\Z_2) \,. 
\ee
Up to condensations, each 2-category $\TwoRep(H_g)$ has a single simple (non-condensation) object, i.e. topological 2d surface defect. We denote these simple objects by
\begin{equation}\label{Q2s}
    \Q_{2}^{[\id]}\,, \ \Q_2^{[a]}\,, \  \Q_2^{[b]}\,,
\end{equation}
labelled by the conjugacy classes of $S_3$.
The topological surface $\Q_2^{[\id]}$ is the transparent or identity surface. The geniune topological lines of the SymTFT form the category $\Rep(S_3)$
\begin{equation}
    {1\rm End}(\Q_2^{[\id]})=\left\{\Q_1^{1}\,, 
    \Q_1^{P}\,, \Q_1^{E}\right\}\cong \Rep(S_3)\,.
\end{equation}
The proper notation for these lines is $\Q_1^{[\id], R}$, but we drop the $[\id]$ for brevity.
Here $P$ denotes the 1-dimensional irreducible representation of $S_3$ that transforms with the sign $(-1)^j$ under $a^ib^j$ and $E$ denotes the 2-dimensional irreducible representation whose representation space is spanned by vectors $v_1$ and $v_2$ that transform as
\begin{equation}
    a: v_i \longmapsto \omega^i v_i\,, \qquad b: v_1\longleftrightarrow v_2\,, 
\end{equation}
with $\omega=\exp(2\pi i/3)$.
Meanwhile the lines on $\Q_2^{[a]}$ and $\Q_2^{[b]}$ form
\begin{equation}
\begin{split}
    {1\rm End}(\Q_2^{[a]})&=\left\{\Q_1^{[a],1}\,, 
    \Q_1^{[a],\omega}\,, \Q_1^{[a],\omega^2}\right\}\cong \Rep(\Z_3)\,, \\
    {1\rm End}(\Q_2^{[b]})&=\left\{\Q_1^{[b],+}\,, 
    \Q_1^{[b],-}\right\}\cong \Rep(\Z_2)\,.
\end{split}
\end{equation}
There are additional condensation surface defects which correspond to condensing an algebra $\cA\in \Rep(H_g)$ on $\Q_2^{[g]}$ \cite{Gaiotto:2019xmp, Roumpedakis:2022aik}, which we denote by $\Q_2^{[g]}/\cA$.
%
%
There are four condensation surfaces in the identity component in  $\cZ(\TwoRep(S_3))$
\begin{equation} \label{eq:RepS3 condensation on id surface}
\left\{{\Q_2^{[\id]}}=\frac{\Q_2^{[\id]}}{\cA_1} \,,   \quad 
\frac{\Q_2^{[\id]}}{\cA_{1\oplus P}}\,,\quad    
\frac{\Q_2^{[\id]}}{\cA_{1\oplus E}}\,,\quad 
\frac{\Q_2^{[\id]}}{\cA_{1\oplus P \oplus 2E}}\right\}\,,    
\end{equation}
corresponding to the four algebras in $\Rep(S_3)$ which are 
\begin{equation}\label{eq:algebras in Rep(S_3)}
\begin{split}
    \cA_1&= \Q_1^{1}\,, \\
    \cA_{1\oplus P}&= \Q_1^{1}\oplus \Q_1^{P}\,, \\
    \cA_{1\oplus E}&= \Q_1^{1}\oplus \Q_1^{E}\,, \\
    \cA_{1\oplus P\oplus 2E}&= \Q_1^{1}\oplus \Q_1^{P}\oplus 2\Q_1^E\,.
\end{split}    
\end{equation}
Similarly there are two defects each in the $[a]$ and $[b]$ connected components 
\begin{equation}
    \left\{{\Q_2^{[a]}} \,,   \quad 
\frac{\Q_2^{[a]}}{\cA_{\Z_3}}\right\}
\, \quad \text{and}\quad
  \left\{{\Q_2^{[b]}} \,,   \quad 
\frac{\Q_2^{[b]}}{\cA_{\Z_2}}\right\}\,
\end{equation}
where 
\begin{equation}
\begin{split}
\cA_{\Z_3}&=\Q^{[a],1}_1\oplus\Q^{[a],\omega}_1\oplus \Q^{[a],\omega^2}_1\,, \\
\cA_{\Z_2}&=\Q^{[b],+}_1\oplus\Q^{[b],-}_1\,.
\end{split}
\end{equation}
The fusion of lines within the surface $\Q_{2}^{[g]}$ is simply given by the fusion of lines in $\Rep(H_g)$. 
Fusions among lines on different surfaces are 
\begin{equation}\label{eq:Fusions of lines in Z2VecS3}
\begin{split}
\Q_1^{P}\otimes \Q_1^{[a],\omega^p}&=\Q_1^{[a],\omega^p}\,, \\
\Q_1^{E}\otimes \Q_1^{[a],\omega^p}&=\Q_1^{[a],\omega^{p
+1}} \oplus 
\Q_1^{[a],\omega^{p+2}}\,, \\
\Q_1^{P}\otimes \Q_1^{[b],\pm}&=\Q_1^{[b],\mp}\,, \\
\Q_1^{E}\otimes \Q_1^{[b],\pm}&=\Q_1^{[b],+} \oplus \Q_1^{[b],-}\,, \\
\Q_1^{[a],\omega^p}\otimes \Q_1^{[a],\omega^p} &=\Q_1^1\oplus \Q_1^P\oplus \Q_1^{[a],\omega^p}\,, \\
\Q_1^{[a],\omega^p}\otimes \Q_1^{[a],\omega^{p+1}} &=\Q_1^E\oplus \Q_1^{[a],\omega^{p+2}}\,, \\
\Q_1^{[a],\omega^p}\otimes \Q_1^{[b],\pm} &=\Q_1^{[b],+}\oplus \Q_1^{[b],-}\,, \\
\Q_1^{[b],\pm}\otimes \Q_1^{[b],\pm} &=\bigoplus_{p=0}^2\Q_1^{[a],\omega^p}\oplus \Q_1^{1}\oplus \Q_1^E\,, \\
\Q_1^{[b],\pm}\otimes \Q_1^{[b],\mp} &=\bigoplus_{p=0}^2\Q_1^{[a],\omega^p}\oplus \Q_1^{P}\oplus \Q_1^E\,. 
\end{split}
\end{equation}
The fusion of surfaces can be read off from the fusion of identity lines on surfaces.
For instance, since $\Q_1^{1}\oplus \Q_1^{ P} \in \Q_1^{[a],1} \otimes \Q_1^{[a],1}$, the fusion outcome of $\Q_2^{[a]}\otimes \Q_2^{[a]}$ contains $\Q_2^{[\id]}$ with $\Q_1^{1}\oplus \Q_1^{P}$ condensed on it. 
The fusion of non-condensation bulk surfaces of the SymTFT is 
\begin{equation}\label{eq:surface fusion Z2VecS3}
\begin{split}
    \Q_2^{[a]}\otimes \Q_2^{[a]}&= {\Q_2^{[\id]} \over \cA_{1\oplus P}}\oplus \Q_2^{[a]} \\     
    \Q_2^{[a]}\otimes \Q_2^{[b]}&= {\Q_2^{[b]} \over \cA_{\Z_2}} , \\     
    \Q_2^{[b]}\otimes \Q_2^{[b]}&= {\Q_2^{[\id]} \over \cA_{1\oplus E}} \oplus 
    { \Q_2^{[a]} \over \cA_{\Z_3}}\,.     
\end{split}
\end{equation}
The fusion of all other condensation surfaces can be derived similarly. 
For instance, since $\Q_1^{[a],1}\otimes (\Q_1^1 \oplus \Q_1^{E})=\cA_{\Z_3}$, we obtain
\begin{equation}
    \Q_2^{[a]}\otimes \frac{\Q_2^{[\id]}}{\cA_{1\oplus E}}=\frac{\Q_2^{[a]}}{\cA_{\Z_3}}\,.
\end{equation}
For conciseness, we avoid listing the fusions between the condensation defects.
The braiding among the topological defects in $\cZ(\TwoVec_G)$ has been discussed in \cite{Kong:2019brm, decoppet2022drinfeld}.
The simplest kind of braiding is that between a topological surface $\Q_2^{[g]}/\cA$ on $S^2$ with a genuine line $\Q_1^{R}$ on $S_1$ which links with $S_2$.
Since the condensation of $\cA$ plays no role on $S^2$, the braiding phase is independent of $\cA$ and is proportional to the character $\chi_R([g])$.
Additionally, there can also be more intricate braiding between multiple surfaces in 4d Dijkgraaf-Witten theory \cite{Wang:2014xba, Chen:2015gma, Tiwari:2016zru, Putrov:2016qdo}. 

\paragraph{SymTFT from gauging.} The $S_3$ 4d Dijkgraaf-Witten theory can be obtained from gauging the 0-form $S_3$ symmetry in the trivial $S_3$ symmetric 4d TQFT (i.e., the IR fixed point of the $S_3$ disordered gapped phase).
From this perspective the various surface defects $\Q_2^{[g]}$ arise by combining uncharged operators in the different $g\in [g]$ twisted sectors.
The topological defects $\Q_2^{[\id]}/\cA$ in \eqref{eq:RepS3 condensation on id surface} are nothing but the theta defects discussed in \cite{Bhardwaj:2022kot} obtained by embedding a 2d $S_3$ symmetric gapped phase and gauging the diagonal $S_3$ symmetry. 
Specifically, the defects corresponding to the algebras $\cA_1$, $\cA_{1\oplus P}$, $\cA_{1\oplus E}$ and $\cA_{1\oplus P \oplus 2E}$ correspond to the stacking of the Trivial, $\Z_2$ SSB, $\Z_3$ SSB and $S_3$ SSB 2d $S_3$ symmetric gapped phases respectively.
Similarly, the condensation defects in the $[a]$ and $[b]$ Schur components can be obtained by appropriately stacking $\Z_3$ and $\Z_2$ symmetric 2d TQFTs to the $g\in [a]$ and $g\in [b]$ twisted sector operators and gauging a diagonal $S_3$.


\subsection{Minimal Gapped Boundary Conditions}
We now describe the minimal topological boundaries of $\cZ(\TwoVec(S_3))$.
We start from the canonical Dirichlet boundary condition on which the topological defects form the 0-form symmetry $S_3$, i.e. categorically $\TwoVec_{S_3}$. 
We subsequently gauge various subcategories of defects on this boundary to obtain the other minimal boundaries.
We describe all the minimal boundaries both in terms of gaugings of the canonical boundary condition and in terms of condensation of bulk SymTFT defects.  
The detailed description of gauging subcategories of $\TwoVec_{S_3}$ can be found in \cite{Bhardwaj:2022maz}. 


Let us start with a brief summary of all minimal gapped boundary conditions, as shown in table \ref{tab:SumBC}.

\begin{table}
$$
\begin{array}{|c|c|c|}\hline
\text{BC, Symmetry} & \Q_p\text{ with Dirichlet BCs} & \text{SymTFT Quiche} \cr \hline \hline 
{\rm Dir}: \TwoVec_{S_3} &
\begin{cases}
     \Q_2^{[\id]}\,,\\ 
 \Q_1^{1} \oplus \Q_1^{P}\oplus 2\,\Q_1^{E}
\end{cases}
&
\begin{tikzpicture}[baseline]
\begin{scope}[shift={(0,-1)}]
\draw [cyan,  fill=cyan] 
(0,0) -- (0,2) -- (2,2) -- (2,0) -- (0,0) ; 
\draw [white] (0,0) -- (0,2) -- (2,2) -- (2,0) -- (0,0)  ; 
\draw [very thick] (0,0) -- (0,2) ;
\node[above] at (0,2) {$\fB_{\rm Dir}$}; 
\draw [thick, dashed](0,1) -- (2,1);
\draw [thick, fill=black] (0,1) ellipse (0.05 and 0.05);
\node[] at (1,1.3) {$\Q_{1}^{\bm{R}}$};
\node at (0, -0.1) {};
\end{scope}
\end{tikzpicture}
\cr\hline
\Neu(\Z_3),p:\  \TwoVec_{\mathbb{G}^{(2)}} & 
\begin{cases}
     \Q_2^{[\id]}\oplus \Q_2^{[a]}\\ 
 \Q_1^{1}\oplus \Q_1^{P}\oplus 2 \,\Q_1^{[a],1}   
\end{cases}
& 
\begin{tikzpicture}[baseline]
\begin{scope}[shift={(0,-1)}]
\draw [cyan,  fill=cyan] 
(0,0) -- (0,2.5) -- (2.5,2.5) -- (2.5,0) -- (0,0) ; 
\draw [white] (0,0) -- (0,2.5) -- (2.5,2.5) -- (2.5,0) -- (0,0)  ; 
\draw [very thick] (0,0) -- (0,2.5) ;
\node[above] at (0,2.5) {$\fB_{\TwoVec_{\mathbb{G}^{(2)}}}$}; 
\draw [thick, dashed](0,0.5) -- (2.5,0.5);
\draw [thick](0,1.7) -- (2.5,1.7);
\draw [thick, fill=black] (0,0.5) ellipse (0.05 and 0.05);
\draw [thick, fill=black] (0,1.7) ellipse (0.05 and 0.05);
\node[above] at (1.25,1.7) {$\Q_{2}^{[a]}$};
\node[above] at (1.25,0.5) {$2\Q_{1}^{[a],1}, \Q_{1}^{P}$};
\node at (0, -0.1) {};
\end{scope}
\end{tikzpicture}
\cr \hline
\Neu (\Z_2),t: \ \TwoRep (\mathbb{G}^{(2)}) 
&
\begin{cases}
     \Q_2^{[\id]}\oplus \Q_2^{[b]}\cr 
 \Q_1^{1}\oplus \Q_1^{E}\oplus \Q_1^{[b],+}     
\end{cases}
&
\begin{tikzpicture}[baseline]
\begin{scope}[shift={(0,-1)}]
\draw [cyan,  fill=cyan] 
(0,0) -- (0,2.5) -- (2.5,2.5) -- (2.5,0) -- (0,0) ; 
\draw [white] (0,0) -- (0,2.5) -- (2.5,2.5) -- (2.5,0) -- (0,0)  ; 
\draw [very thick] (0,0) -- (0,2.5) ;
\node[above] at (0,2.5) {$\fB_{\TwoRep (\mathbb{G}^{(2)})}$}; 
\draw [thick, dashed](0,0.5) -- (2.5,0.5);
\draw [thick](0,1.7) -- (2.5,1.7);
\draw [thick, fill=black] (0,0.5) ellipse (0.05 and 0.05);
\draw [thick, fill=black] (0,1.7) ellipse (0.05 and 0.05);
\node[above] at (1.25,1.7) {$\Q_{2}^{[b]}$};
\node[above] at (1.25,0.5) {$\Q_{1}^{[b],1}, \Q_{1}^{E}$};
\node at (0, -0.1) {};
\end{scope}
\end{tikzpicture}
\cr \hline 
\Neu (S_3),(p,t): \ \TwoRep (S_3) & 
\begin{cases}
     \Q_2^{[\id]}\oplus \Q_2^{[a]}\oplus \Q_2^{[b]}\cr  
 \Q_1^{1} \oplus \Q_1^{[a],1}\oplus \Q_1^{[b],+}     
\end{cases}
&
\begin{tikzpicture}[baseline]
\begin{scope}[shift={(0,-1)}]
\draw [cyan,  fill=cyan] 
(0,0) -- (0,2.5) -- (2.5,2.5) -- (2.5,0) -- (0,0) ; 
\draw [white] (0,0) -- (0,2.5) -- (2.5,2.5) -- (2.5,0) -- (0,0)  ; 
\draw [very thick] (0,0) -- (0,2.5) ;
\node[above] at (0,2.5) {$\fB_{\TwoRep (S_3)}$}; 
\draw [thick, dashed](0,0.5) -- (2.5,0.5);
\draw [thick](0,1.7) -- (2.5,1.7);
\draw [thick, fill=black] (0,0.5) ellipse (0.05 and 0.05);
\draw [thick, fill=black] (0,1.7) ellipse (0.05 and 0.05);
\node[above] at (1.25,1.7) {$\Q_{2}^{[a]}, \Q_{2}^{[b]}$};
\node[above] at (1.25,0.5) {$\Q_{1}^{[a],1}, \Q_{1}^{[b], +}$};
\node at (0, -0.1) {};
\end{scope}
\end{tikzpicture}
\cr \hline
\end{array}
$$
\caption{Summary of minimal gapped BC for $\cZ(\TwoVec_{S_3})$, where the two-group is \\$\mathbb{G}^{(2)}= \mathbb{Z}_3^{(1)}\rtimes \Z_2^{(0)}$.  \label{tab:SumBC}}
\end{table}

\subsubsection{Dirichlet Boundary Condition}

The  canonical Dirichlet boundary condition, was discussed in general in section \ref{sec:Dir} and is denoted by $\fB_{\rm Dir}$. The defects on this boundary condition form the category $\TwoVec(S_3)$.
Each of the SymTFT genuine lines are condensed on this boundary.
More precisely, a SymTFT line carrying a label $R\in \Rep(G)$ has $\dim(R)$ number of topological ends on this boundary which transform in the $R$ representation of $S_3$:
\be
\begin{split}
\begin{tikzpicture}
 \begin{scope}[shift={(0,0)},scale=0.8] 
\draw [cyan, fill=cyan!80!red, opacity =0.5]
(0,0) -- (0,4) -- (2,5) -- (5,5) -- (5,1) -- (3,0)--(0,0);
\draw [black, thick, fill=white,opacity=1]
(0,0) -- (0, 4) -- (2, 5) -- (2,1) -- (0,0);
\draw [cyan, thick, fill=cyan, opacity=0.2]
(0,0) -- (0, 4) -- (2, 5) -- (2,1) -- (0,0);
\draw[line width=1pt] (1,2.5) -- (3,2.5);
\draw[line width=1pt] (3,2.5) -- (4.2,2.5);
\fill[red!80!black] (1,2.5) circle (3pt);
\node at (2,5.4) {$\fB_\Dir$};
\draw[dashed] (0,0) -- (3,0);
\draw[dashed] (0,4) -- (3,4);
\draw[dashed] (2,5) -- (5,5);
\draw[dashed] (2,1) -- (5,1);
\draw [cyan, thick, fill=cyan, opacity=0.1]
(0,4) -- (3, 4) -- (5, 5) -- (2,5) -- (0,4);
\draw [black, thick, dashed]
(3,0) -- (3, 4) -- (5, 5) -- (5,1) -- (3,0);
\node[below, red!80!black] at (.5, 3.2) {$\cE_0^{R,i}$};
\node  at (2.5, 3) {$\Q_{1}^{R}$};
\node  at (11, 2.5) {$i=1\,,\dots\,, {\rm dim}(R)\,.$};
\end{scope}
\end{tikzpicture}    
\end{split}
\ee
We denote these as
\begin{equation}
\begin{split}
    \Q_{1}^{P}\Bigg|_{{\Dir}}=\cE_0^{P}\,, \qquad         \Q_{1}^{E}\Bigg|_{{\Dir}}=\left\{\cE_0^{E,1}\,, \cE_0^{E,2}\right\}\,.
\end{split}
\end{equation}
The surface operators $\Q_{2}^{[g]}$ splits on $\fB_{\Dir}$ as a direct sum of indecomposable objects labeled by $g\in[g]$ as in  (\ref{VecSym}) and the resulting surface operators denoted by $D_2^{g}$ generate the 0-form symmetry group $S_3$, and are the objects of $\TwoVec_{S_3}$.
The symmetry operators have the following linking action on the local operators
\begin{equation}
    D_{2}^{g}:\quad \cE_0^{R,i}\longmapsto \sum_j\cD_{R}(g)_{ij}\cE_0^{R,j}\,.
\end{equation}
Furthermore $\Q_2^{[g]}$ cannot end on $\fB_{\Dir}$, but form  L-shaped configurations as follows: 
\be\label{LShapey}
\begin{split}
\begin{tikzpicture}
 \begin{scope}[shift={(0,0)},scale=0.8] 
\draw [cyan, fill=cyan!80!red, opacity =0.5]
(0,0) -- (0,4) -- (2,5) -- (5,5) -- (5,1) -- (3,0)--(0,0);
\draw [black, thick, fill=white,opacity=1]
(0,0) -- (0, 4) -- (2, 5) -- (2,1) -- (0,0);
\draw [cyan, thick, fill=cyan, opacity=0.2]
(0,0) -- (0, 4) -- (2, 5) -- (2,1) -- (0,0);
\draw [black, thick, fill=black, opacity=0.3]
(0.,2) -- (0., 4) -- (2, 5) -- (2,3) -- (0,2);
\draw [black, thick, fill=black, opacity=0.3]
(0.,2) -- (3.1, 2) -- (5., 3) -- (2, 3) -- (0.,2);
\draw [black, thick, dashed]
(3,0) -- (3, 4) -- (5, 5) -- (5,1) -- (3,0);
\node at (2,5.4) {$\fB_\Dir$};
\draw[dashed] (0,0) -- (3,0);
\draw[dashed] (0,4) -- (3,4);
\draw[dashed] (2,5) -- (5,5);
\draw[dashed] (2,1) -- (5,1);
\draw [cyan, thick, fill=cyan, opacity=0.1]
(0,4) -- (3, 4) -- (5, 5) -- (2,5) -- (0,4);
\draw [thick, red] (0.0, 2.) -- (2, 3) ;
\node[below, red!80!black] at (-0.5, 2.4) {$\cE_1^g$};
\node  at (2.5, 2.55) {$\Q_{2}^{[g]}$};
\node[above] at (1.1, 3) {$D_2^{g}$};
\end{scope}
\end{tikzpicture}    
\end{split}
\ee
corresponding to non-genuine lines $\cE_1^{g}$, attached to surfaces $D^{g}_2$ for all $g\in [g]$
\begin{equation}
\begin{split}
    \Q_{2}^{[a]}\Bigg|_{{\Dir}}&=\left\{(D_2^{a},\cE_1^{a})\,,(D_2^{a^2},\cE_1^{a^2})\right\}\,, \\ 
    \Q_{2}^{[b]}\Bigg|_{{\Dir}}&=\left\{
    (D_2^{b},\cE_1^{b})\,, (D_2^{ab},\cE_1^{ab})\,, (D_2^{a^2b},\cE_1^{a^2b})
    \right\}\,.
\end{split}
\end{equation}
The $S_3$ acts on the twisted sector lines as
\begin{equation}
\label{eq:conj action S3 Dir}
    D_2^{g}:\qquad (D_2^{h},\cE_1^{h})\longmapsto (D_2^{ghg^{-1}},\cE_1^{ghg^{-1}})\,.
\end{equation}

\subsubsection{Neumann$(\Z_3)$ Boundary Condition} \label{sec:S3_NeuZ3}

Now we describe the boundary condition obtained by gauging the $\Z_3$ subsymmetry on $\fB_{\Dir}$.
While gauging  the $\Z_3$ symmetry, one has the freedom of including discrete torsion valued in $H^{3}(\Z_3,U(1))\cong \Z_3$.
Physically, this may be understood as first stacking a $\Z_3$ symmetry protected topological (SPT) phase and then gauging.
We label the different choices of discrete torsion by $p\in \Z_3$.
The resulting boundary condition will be denoted as
\begin{equation}\label{eq:2group BC}
\fB_{\Neu(\Z_3),p}= \frac{\fB_{\Dir}\boxtimes {\rm SPT}_p }{\Z_3} \,.
\end{equation}
The symmetry on this boundary is $\TwoVec_{\mathbb{G}^{(2)}}$ with the two-group\cite{Bhardwaj:2022maz} 
\be
\mathbb{G}^{(2)} = \mathbb{Z}_3^{(1)} \rtimes \mathbb{Z}_2^{(0)} \,.
\ee
Let us describe the properties of ends of various bulk SymTFT defects which we derive by performing the gauging $\fB_{\Dir}$.

 Since the operator $\cE_0^P$ at the end of $\Q_1^{P}$ was untwisted and uncharged under the $\Z_3$ symmetry, it remains unaltered under such a gauging 
\be
\begin{split}
\begin{tikzpicture}
 \begin{scope}[shift={(0,0)},scale=0.8] 
\draw [cyan, fill=cyan!80!red, opacity =0.5]
(0,0) -- (0,4) -- (2,5) -- (5,5) -- (5,1) -- (3,0)--(0,0);
\draw [black, thick, fill=white,opacity=1]
(0,0) -- (0, 4) -- (2, 5) -- (2,1) -- (0,0);
\draw [cyan, thick, fill=cyan, opacity=0.2]
(0,0) -- (0, 4) -- (2, 5) -- (2,1) -- (0,0);
\draw[line width=1pt] (1,2.5) -- (3,2.5);
\draw[line width=1pt] (3,2.5) -- (4.2,2.5);
\fill[red!80!black] (1,2.5) circle (3pt);
\node at (2,5.4) {$\fB_{\Neu(\Z_3),p}$};
\draw [black, thick, dashed]
(3,0) -- (3, 4) -- (5, 5) -- (5,1) -- (3,0);
\draw[dashed] (0,0) -- (3,0);
\draw[dashed] (0,4) -- (3,4);
\draw[dashed] (2,5) -- (5,5);
\draw[dashed] (2,1) -- (5,1);
\draw [cyan, thick, fill=cyan, opacity=0.1]
(0,4) -- (3, 4) -- (5, 5) -- (2,5) -- (0,4);
\node[below, red!80!black] at (.8, 2.4) {$\cE_0^{P}$};
\node  at (2.5, 3) {$\Q_{1}^{P}$};
\node  at (5.5, 2.5) {$,$};
\node  at (11, 2.5) {$\Q_{1}^{P}\Bigg|_{{\Neu}(\Z_3),p}=\cE_0^{P}\,,$};
\end{scope}
\end{tikzpicture}    
\end{split}
\ee
where $\cE_0^P$ is charged under $\Z_2^{(0)}\in \mathbb G^{(2)}$.
On $\fB_{\Dir}$, the operator $\cE_0^{E,j}$ transforms in the $\omega^{j}$ representation of $\Z_3$ 0-form symmetry generated by $D_2^{a}$.
Therefore upon gauging $\Z_3$, being charged these operators get attached to the dual $\Z_3^{(1)}$ symmetry generators, which we denote as $D_1^{\omega}$ and $D_1^{\omega^2}$.
\be
\begin{split}
\begin{tikzpicture}
 \begin{scope}[shift={(0,0)},scale=0.8] 
\draw [cyan, fill=cyan!80!red, opacity =0.5]
(0,0) -- (0,4) -- (2,5) -- (5,5) -- (5,1) -- (3,0)--(0,0);
\draw [black, thick, fill=white,opacity=1]
(0,0) -- (0, 4) -- (2, 5) -- (2,1) -- (0,0);
\draw [cyan, thick, fill=cyan, opacity=0.2]
(0,0) -- (0, 4) -- (2, 5) -- (2,1) -- (0,0);
\draw[line width=1pt] (1,2.5) -- (3,2.5);
\draw[line width=1pt] (1,2.5) -- (1,4.5);
\draw[line width=1pt] (3,2.5) -- (4.2,2.5);
\fill[red!80!black] (1,2.5) circle (3pt);
\node at (2,5.4) {$\fB_{\Neu(\Z_3),p}$};
\draw[dashed] (0,0) -- (3,0);
\draw[dashed] (0,4) -- (3,4);
\draw[dashed] (2,5) -- (5,5);
\draw[dashed] (2,1) -- (5,1);
\draw [cyan, thick, fill=cyan, opacity=0.1]
(0,4) -- (3, 4) -- (5, 5) -- (2,5) -- (0,4);
\node[below, red!80!black] at (.8, 2.4) {$\cE_0^{E,j}$};
\draw [black, thick, dashed]
(3,0) -- (3, 4) -- (5, 5) -- (5,1) -- (3,0);
\node  at (2.5 , 3) {$\Q_{1}^{E}$};
\node  at (0.5, 3.3) {$D_{1}^{\omega^j}$};
\node  at (5.5, 2.5) {$,$};
\node  at (11, 2.5) {$\Q_{1}^{E}\Bigg|_{{\Neu}(\Z_3),p}=\left\{(D_1^{\omega},\cE_0^{E,1})\,, (D_1^{\omega^2}, \cE_0^{E,2})\right\}\,.$};
\end{scope}
\end{tikzpicture}    
\end{split}
\ee
Furthermore, since $\cD_{E}(b): \cE_0^{E,1}\leftrightarrow \cE_0^{E,2}$ on $\fB_{\Dir}$, the corresponding twisted sector operators in $\fB_{\Neu(\Z_3),p}$ are also acted upon by the $\Z_2^{(0)}$ symmetry
\begin{equation}
    D_2^{b}: \quad (D_1^{\omega},\cE_0^{E,1}) \longleftrightarrow 
    (D_1^{\omega^2}, \cE_0^{E,2})\,.
\end{equation}
From which we read off the 2-group action of $\Z_2^{(0)}$ on $\Z_{3}^{(1)}$ 
\begin{equation}
    D_2^{b}:\quad  D_1^{\omega} \longleftrightarrow 
    D_1^{\omega^2}\,.
\end{equation}
The $\Z_3^{(0)}$ twisted sector lines $\cE_1^{a}$ and $\cE_1^{a^2}$ on $\fB_{\Dir}$ become untwisted sector lines that are charged under the dual $\Z_3^{(1)}$ symmetry on $\fB_{\Neu(\Z_3),p}$, i.e.,
\be
\begin{split}
\begin{tikzpicture}
 \begin{scope}[shift={(0,0)},scale=0.8] 
\draw [cyan, fill=cyan!80!red, opacity =0.5]
(0,0) -- (0,4) -- (2,5) -- (5,5) -- (5,1) -- (3,0)--(0,0);
\draw [black, thick, fill=white,opacity=1]
(0,0) -- (0, 4) -- (2, 5) -- (2,1) -- (0,0);
\draw [cyan, thick, fill=cyan, opacity=0.2]
(0,0) -- (0, 4) -- (2, 5) -- (2,1) -- (0,0);
\draw [black, thick, fill=black, opacity=0.3]
(0.,2) -- (3.1, 2) -- (5., 3) -- (2, 3) -- (0.,2);
\node at (2,5.4) {$\fB_{\Neu(\Z_3),p}$};
\draw[dashed] (0,0) -- (3,0);
\draw[dashed] (0,4) -- (3,4);
\draw[dashed] (2,5) -- (5,5);
\draw[dashed] (2,1) -- (5,1);
\draw [black, thick, dashed]
(3,0) -- (3, 4) -- (5, 5) -- (5,1) -- (3,0);
\draw [cyan, thick, fill=cyan, opacity=0.1]
(0,4) -- (3, 4) -- (5, 5) -- (2,5) -- (0,4);
\draw [line width=1pt, red!80!black] (0.0, 2.) -- (2, 3) ;
\node[below, red!80!black] at (-0.5, 2.4) {$\cE_1^{a^j}$};
\node  at (2.5, 2.55) {$\Q_{2}^{[a]}$};
\node  at (5.5, 2.55) {$,$};
\node  at (10.5, 2.55) {$\Q_2^{[a]}\Bigg|_{{\Neu(\Z_3),p}}= \left\{\cE_1^{a}\,, \cE_1^{a^2}\right\}\,.$};
\end{scope}
\end{tikzpicture}    
\end{split}
\ee
%
The Hopf-linking phase between $\cE_1^{a^n}$ and $D_1^{\omega^m}$ is $\omega^{nm}$ where $\omega=\exp\{2\pi i/3\}$.
Meanwhile the F-symbols and fusion properties of $\cE_1^{a^{n}}$ are controlled by the choice of $p\in \Z_3$.
Specifically these satisfy
\begin{equation}
    [\cE_1^{a}]^{3}= D_1^{\omega^{2p}}\,, \quad \quad 
    F(\cE_1^{a^l}\,, \cE_1^{a^m}\,, \cE_1^{a^n}) = \exp\left\{\frac{2\pi i p}{9}p(q+r-[q+r]_3)\right\}\,,
    \label{eq:Z3 1fs charge line algebra}
\end{equation}
where $[q+r]_n\equiv q+r \text{ mod }n$.

 Finally, the three L-shaped ends of $\Q_2^{[b]}$ on $\fB_{\Dir}$, mediated by the lines $\cE_1^{b}, \cE_1^{ab}$ and $\cE_1^{a^2b}$ combine into a single line $\cE_1^{[b]}$ on $\fB_{\Neu(\Z_3),p}$, i.e.,
\be
\begin{split}
\begin{tikzpicture}
 \begin{scope}[shift={(0,0)},scale=0.8] 
\draw [cyan, fill=cyan!80!red, opacity =0.5]
(0,0) -- (0,4) -- (2,5) -- (5,5) -- (5,1) -- (3,0)--(0,0);
\draw [black, thick, fill=white,opacity=1]
(0,0) -- (0, 4) -- (2, 5) -- (2,1) -- (0,0);
\draw [cyan, thick, fill=cyan, opacity=0.2]
(0,0) -- (0, 4) -- (2, 5) -- (2,1) -- (0,0);
\draw [black, thick, fill=black, opacity=0.3]
(0.,2) -- (0., 4) -- (2, 5) -- (2,3) -- (0,2);
\draw [black, thick, fill=black, opacity=0.3]
(0.,2) -- (3.1, 2) -- (5., 3) -- (2, 3) -- (0.,2);
\node at (2,5.4) {$\fB_{\Neu(\Z_3),p}$};
\draw[dashed] (0,0) -- (3,0);
\draw[dashed] (0,4) -- (3,4);
\draw[dashed] (2,5) -- (5,5);
\draw[dashed] (2,1) -- (5,1);
\draw [black, thick, dashed]
(3,0) -- (3, 4) -- (5, 5) -- (5,1) -- (3,0);
\draw [cyan, thick, fill=cyan, opacity=0.1]
(0,4) -- (3, 4) -- (5, 5) -- (2,5) -- (0,4);
\draw [thick, red] (0.0, 2.) -- (2, 3) ;
\node[below, red!80!black] at (-0.5, 2.4) {$\cE_1^{[b]}$};
\node  at (2.5, 2.55) {$\Q_{2}^{[b]}$};
\node[above] at (1.1, 3) {$D_2^{b}/\cA_{\Z_3}$};
\node[above] at (5.5, 2.55) {$,$};
\node[above] at (10.1, 2) {$\Q^{[b]}_{2}\Bigg|_{{\Neu(\Z_3),p}}=(D_2^{b}/\cA_{\Z_3}\,,\cE_1^{[b]})\,,$};
\end{scope}
\end{tikzpicture}    
\end{split}
\ee
where $\cA_{\Z_3}=D_1^{\id}\oplus D_{1}^{\omega}\oplus D_{1}^{\omega^{2}}$.
The fact that the (parallel) boundary projection of $\Q_2^{[b]}$ is the condensation surface defect $D_2^{b}/\cA_{\Z_3}$ can be understood as follows.
The pre-image of this defect in $\fB_{\Dir}$ was the non-simple surface $D_2^{b}\oplus D_{2}^{ab}\oplus D_{2}^{a^2b}$ which naturally had a three dimensional space of topological local operators spanned by $D_{0}^{b}\,, D_{0}^{ab}$ and $D_{0}^{a^{2}b}$.
One can construct operators carrying any representation $\omega^s \in \Rep(\Z_3)$ as
\begin{equation}
    D_0^{[b],\omega^{s}}= D_{0}^{b} + \omega^{-s}D_{0}^{ab}+\omega^{-2s}D_{0}^{a^2b}\,.
\end{equation}
After gauging $\Z_3^{(0)}$, this operator is in the twisted sector of $D_1^{\omega^{s}}$.
This implies that all the $\Z_3^{(1)}$ lines can be absorbed by the boundary projection of $\Q_2^{[b]}$ which therefore is a condensation of $\cA_{\Z_3}$ on the bare $\Z_2^{(0)}$ defect $D_2^{b}$.

\subsubsection{Neumann$(\Z_2)$ Boundary Condition}
We now describe the boundary condition obtained by gauging the 0-form $\Z_2^{b}$ symmetry on $\fB_{\Dir}$ generated by $D_2^{b}$.
As with the previous case of $\Z_3$ gauging, there is a choice of discrete torsion (or equivalently SPT pasting) which is valued in $H^{3}(\Z_2,U(1))\cong \Z_2$.
We parametrize this by $t=\pm 1$ and denote the corresponding boundary condition as 
\begin{equation}
\fB_{\Neu(\Z_2),t}=\frac{\fB_{\Dir}\boxtimes {\rm SPT}_{t}}{\Z_2}\,.
\end{equation}
\paragraph{$\TwoRep(\Z_3^{(1)}\rtimes \Z_2^{(0)})$ Symmetry.} The symmetry obtained upon gauging $\Z_2^{b}\subset S_3$ is known to be $\TwoRep(\mathbb G^{(2)})$ and was studied in detail in \cite{Bhardwaj:2022maz}.
Let us briefly recall this symmetry category.
Firstly, there is a $\Z_2^{(1)}$ subsymmetry whose generator we denote as $D_1^{\wh{b}}$.
More precisely, this generates the subcategory $\TwoRep(\Z_2)$.
Next, there is a non-condensation non-invertible surface defect which we denote as $D_2^{A}$.
This surface defect descends from $D_2^{a}\oplus D_2^{a^2}$ on $\fB_{\Dir}$ which form an indecomposable orbit under the $\Z_2^b$ action and therefore combine into an indecomposable topological defect in $\fB_{\Neu(\Z_2),t}$.
The fusion rules of this defect are
\begin{equation}
    D_2^{A}\otimes D_2^{A}= D_2^{A}\oplus \frac{D_{2}^{\id}}{D_1^{\id}\oplus D_1^{\wh{b}}}\,. 
\end{equation}
\paragraph{Generalized Charges.} We now describe the ends of the bulk SymTFT defects on $\fB_{\Neu(\Z_2),t}$. 
The operator $\cE_0^{P}$ at the end of $\Q_1^P$ on $\fB_{\Dir}$ being charged under $\Z_2^{b}$, goes to the twisted sector of the dual $\Z_2^{(1)}$ symmetry generated by $D_1^{\widehat{b}}$.
\be
\begin{split}
\begin{tikzpicture}
 \begin{scope}[shift={(0,0)},scale=0.8] 
\draw [cyan, fill=cyan!80!red, opacity =0.5]
(0,0) -- (0,4) -- (2,5) -- (5,5) -- (5,1) -- (3,0)--(0,0);
\draw [black, thick, fill=white,opacity=1]
(0,0) -- (0, 4) -- (2, 5) -- (2,1) -- (0,0);
\draw [cyan, thick, fill=cyan, opacity=0.2]
(0,0) -- (0, 4) -- (2, 5) -- (2,1) -- (0,0);
\draw[line width=1pt] (1,2.5) -- (3,2.5);
\draw[line width=1pt] (1,2.5) -- (1,4.5);
\draw[line width=1pt] (3,2.5) -- (4.2,2.5);
\fill[red!80!black] (1,2.5) circle (3pt);
\node at (2,5.4) {$\fB_{\Neu(\Z_2),t}$};
\draw[dashed] (0,0) -- (3,0);
\draw[dashed] (0,4) -- (3,4);
\draw[dashed] (2,5) -- (5,5);
\draw[dashed] (2,1) -- (5,1);
\draw [black, thick, dashed]
(3,0) -- (3, 4) -- (5, 5) -- (5,1) -- (3,0);
\draw [cyan, thick, fill=cyan, opacity=0.1]
(0,4) -- (3, 4) -- (5, 5) -- (2,5) -- (0,4);
\node[below, red!80!black] at (.8, 2.4) {$\cE_0^{P}$};
\node  at (2.5, 3) {$\Q_{1}^{P}$};
\node  at (0.5, 3.3) {$D_{1}^{\wh{b}}$};
\node  at (5.5, 2.5) {$,$};
\node  at (11, 2.5) {$\Q_1^{P}\Bigg|_{\Neu(\Z_2),t}=(D_1^{\wh{b}}\,, \cE_0^P)\,.$};
\end{scope}
\end{tikzpicture}    
\end{split}
\ee
%
On $\fB_{\Dir}$, the doublet of operators $\cE_0^{E,I}$ with $I=1,2$ at the end on $\Q_1^{E}$ can be written in a rotated basis as
\begin{equation}
    \cE_0^{E\,, \pm}:= \cE_0^{E,1}\pm \cE_0^{E,2}\,.
\end{equation}
Among these $\cE_0^{E,+}$ is uncharged while $\cE_0^{E,-}$ transforms in the non-trivial sign representation of $\Z_2^b$.
Therefore upon gauging $\Z_2^{b}$, we find
\begin{equation}
    \Q_1^{E}=(\cE_0^{E,+}\,, (D_1^{\wh{b}}\,, \cE_0^{E,-}))\,.
\end{equation}
\be
\begin{split}
\begin{tikzpicture}
 \begin{scope}[shift={(0,0)},scale=0.8] 
\draw [cyan, fill=cyan!80!red, opacity =0.5]
(0,0) -- (0,4) -- (2,5) -- (5,5) -- (5,1) -- (3,0)--(0,0);
\draw [black, thick, fill=white,opacity=1]
(0,0) -- (0, 4) -- (2, 5) -- (2,1) -- (0,0);
\draw [cyan, thick, fill=cyan, opacity=0.2]
(0,0) -- (0, 4) -- (2, 5) -- (2,1) -- (0,0);
\draw[line width=1pt] (1,2.5) -- (3,2.5);
\draw[line width=1pt] (3,2.5) -- (4.2,2.5);
\fill[red!80!black] (1,2.5) circle (3pt);
\node at (2,5.4) {$\fB_{\Neu(\Z_2),t}$};
\draw[dashed] (0,0) -- (3,0);
\draw[dashed] (0,4) -- (3,4);
\draw[dashed] (2,5) -- (5,5);
\draw[dashed] (2,1) -- (5,1);
\draw [black, thick, dashed]
(3,0) -- (3, 4) -- (5, 5) -- (5,1) -- (3,0);
\draw [cyan, thick, fill=cyan, opacity=0.1]
(0,4) -- (3, 4) -- (5, 5) -- (2,5) -- (0,4);
\node[below, red!80!black] at (.8, 2.4) {$\cE_0^{E,+}$};
\node  at (2.5, 3) {$\Q_{1}^{E}$};
\node  at (5.5, 2.5) {$,$};
\end{scope}
\begin{scope}[shift={(7,0)},scale=0.8] 
\draw [cyan, fill=cyan!80!red, opacity =0.5]
(0,0) -- (0,4) -- (2,5) -- (5,5) -- (5,1) -- (3,0)--(0,0);
\draw [black, thick, fill=white,opacity=1]
(0,0) -- (0, 4) -- (2, 5) -- (2,1) -- (0,0);
\draw [cyan, thick, fill=cyan, opacity=0.2]
(0,0) -- (0, 4) -- (2, 5) -- (2,1) -- (0,0);
\draw[line width=1pt] (1,2.5) -- (3,2.5);
\draw[line width=1pt] (1,2.5) -- (1,4.5);
\draw[line width=1pt] (3,2.5) -- (4.2,2.5);
\fill[red!80!black] (1,2.5) circle (3pt);
\node at (2,5.4) {$\fB_{\Neu(\Z_2),t}$};
\draw[dashed] (0,0) -- (3,0);
\draw[dashed] (0,4) -- (3,4);
\draw[dashed] (2,5) -- (5,5);
\draw[dashed] (2,1) -- (5,1);
\draw [black, thick, dashed]
(3,0) -- (3, 4) -- (5, 5) -- (5,1) -- (3,0);
\draw [cyan, thick, fill=cyan, opacity=0.1]
(0,4) -- (3, 4) -- (5, 5) -- (2,5) -- (0,4);
\node[below, red!80!black] at (.8, 2.4) {$\cE_0^{E,-}$};
\node  at (2.5, 3) {$\Q_{1}^{E}$};
\node  at (0.5, 3.3) {$D_{1}^{\wh{b}}$};
\node  at (5.5, 2.5) {$,$};
\end{scope}
\end{tikzpicture}    
\end{split}
\ee
Now we move onto the generalized charges corresponding to the bulk surface defects.
The surface $\Q_2^{[a]}$ had two ends $\cE_1^{a}$ and $\cE_1^{a^2}$ attached to $D_2^a$ and $D_2^{a^2}$ on $\fB_{\Dir}$.
These were exchanged under $\Z_2^{b}$ and therefore upon gauging $\Z_2^{b}$, they combine into a single indecomposable line attached to the  indecomposable non-invertible surface $D_2^{A}$, i.e., 
\be
\begin{split}
\begin{tikzpicture}
 \begin{scope}[shift={(0,0)},scale=0.8] 
\draw [cyan, fill=cyan!80!red, opacity =0.5]
(0,0) -- (0,4) -- (2,5) -- (5,5) -- (5,1) -- (3,0)--(0,0);
\draw [black, thick, fill=white,opacity=1]
(0,0) -- (0, 4) -- (2, 5) -- (2,1) -- (0,0);
\draw [cyan, thick, fill=cyan, opacity=0.2]
(0,0) -- (0, 4) -- (2, 5) -- (2,1) -- (0,0);
\draw [black, thick, fill=black, opacity=0.3]
(0.,2) -- (0., 4) -- (2, 5) -- (2,3) -- (0,2);
\draw [black, thick, fill=black, opacity=0.3]
(0.,2) -- (3.1, 2) -- (5., 3) -- (2, 3) -- (0.,2);
\node at (2,5.4) {$\fB_{\Neu(\Z_2),t}$};
\draw[dashed] (0,0) -- (3,0);
\draw[dashed] (0,4) -- (3,4);
\draw[dashed] (2,5) -- (5,5);
\draw[dashed] (2,1) -- (5,1);
\draw [black, thick, dashed]
(3,0) -- (3, 4) -- (5, 5) -- (5,1) -- (3,0);
\draw [cyan, thick, fill=cyan, opacity=0.1]
(0,4) -- (3, 4) -- (5, 5) -- (2,5) -- (0,4);
\draw [thick, red] (0.0, 2.) -- (2, 3) ;
\node[below, red!80!black] at (-0.5, 2.4) {$\cE_1^{A}$};
\node  at (2.5, 2.55) {$\Q_{2}^{[a]}$};
\node[above] at (1.1, 3) {$D_2^{A}$};
\node[above] at (5.5, 2.55) {$,$};
\node[above] at (10.1, 2) {$\Q_2^{[a]}=(D_2^{A}\,, \cE_1^{[a]})\,.$};
\end{scope}
\end{tikzpicture}    
\end{split}
\ee
Finally, the surface $\Q_{2}^{[b]}$ had three ends on $\fB_{\Dir}$ labeled as $\cE_1^{a^qb}$  attached to the $D_2^{a^qb}$ defects respectively for $q=0,1,2$.
Among these, $\cE_1^{b}$ is invariant while $\cE_1^{ab}$ and $\cE_1^{a^2b}$ are interchanged under the $\Z_2^b$ action.
Consequently, upon gauging $\Z_2^{b}$, $\cE_1^{b}$ becomes an untwisted sector line operator while $\cE_1^{ab}$ and $\cE_1^{a^2b}$ combine into a indecomposable line denoted as $\cE_1^{Ab}$ in the $D_2^{A}$ twisted sector
\begin{equation}
\Q_2^{[b]}\Bigg|_{{\Neu(\Z_2),t}}=\left\{\cE_1^{b}\,,(D_2^{A}\,, \cE_1^{Ab})\right\}\,.
\end{equation}
\be
\begin{split}
\begin{tikzpicture}
 \begin{scope}[shift={(0,0)},scale=0.8] 
\draw [cyan, fill=cyan!80!red, opacity =0.5]
(0,0) -- (0,4) -- (2,5) -- (5,5) -- (5,1) -- (3,0)--(0,0);
\draw [black, thick, fill=white,opacity=1]
(0,0) -- (0, 4) -- (2, 5) -- (2,1) -- (0,0);
\draw [cyan, thick, fill=cyan, opacity=0.2]
(0,0) -- (0, 4) -- (2, 5) -- (2,1) -- (0,0);
\draw [black, thick, fill=black, opacity=0.3]
(0.,2) -- (3.1, 2) -- (5., 3) -- (2, 3) -- (0.,2);
\node at (2,5.4) {$\fB_{\Neu(\Z_2),t}$};
\draw[dashed] (0,0) -- (3,0);
\draw[dashed] (0,4) -- (3,4);
\draw[dashed] (2,5) -- (5,5);
\draw[dashed] (2,1) -- (5,1);
\draw [black, thick, dashed]
(3,0) -- (3, 4) -- (5, 5) -- (5,1) -- (3,0);
\draw [cyan, thick, fill=cyan, opacity=0.1]
(0,4) -- (3, 4) -- (5, 5) -- (2,5) -- (0,4);
\draw [thick, red] (0.0, 2.) -- (2, 3) ;
\node[below, red!80!black] at (-0.5, 2.4) {$\cE_1^{b}$};
\node  at (2.5, 2.55) {$\Q_{2}^{[b]}$};
\node[above] at (5.5, 2.55) {$,$};
\end{scope}
\begin{scope}[shift={(7,0)},scale=0.8] 
\draw [cyan, fill=cyan!80!red, opacity =0.5]
(0,0) -- (0,4) -- (2,5) -- (5,5) -- (5,1) -- (3,0)--(0,0);
\draw [black, thick, fill=white,opacity=1]
(0,0) -- (0, 4) -- (2, 5) -- (2,1) -- (0,0);
\draw [cyan, thick, fill=cyan, opacity=0.2]
(0,0) -- (0, 4) -- (2, 5) -- (2,1) -- (0,0);
\draw [black, thick, fill=black, opacity=0.3]
(0.,2) -- (0., 4) -- (2, 5) -- (2,3) -- (0,2);
\draw [black, thick, fill=black, opacity=0.3]
(0.,2) -- (3.1, 2) -- (5., 3) -- (2, 3) -- (0.,2);
\node at (2,5.4) {$\fB_{\Neu(\Z_2),t}$};
\draw[dashed] (0,0) -- (3,0);
\draw[dashed] (0,4) -- (3,4);
\draw[dashed] (2,5) -- (5,5);
\draw[dashed] (2,1) -- (5,1);
\draw [black, thick, dashed]
(3,0) -- (3, 4) -- (5, 5) -- (5,1) -- (3,0);
\draw [cyan, thick, fill=cyan, opacity=0.1]
(0,4) -- (3, 4) -- (5, 5) -- (2,5) -- (0,4);
\draw [thick, red] (0.0, 2.) -- (2, 3) ;
\node[below, red!80!black] at (-0.5, 2.4) {$\cE_1^{Ab}$};
\node  at (2.5, 2.55) {$\Q_{2}^{[b]}$};
\node[above] at (1.1, 3) {$D_2^{A}$};
\node[above] at (5.5, 2.55) {$,$};
\end{scope}
\end{tikzpicture}    
\end{split}
\ee
The line $\cE_1^{b}$ is charged under the $\Z_2^{(1)}$ symmetry generated by $D_1^{\wh{b}}$ and has F-symbols given by $t\in H^{3}(\Z_2,U(1))$.
The twisted sector line $\cE_1^{Ab}$ satisfies $\cE_1^{Ab}=\cE_1^{[a]}\otimes \cE_1^{b}$.
The line $D_1^{\wh{b}}$ can end on $\cE_1^{Ab}$, which follows from the fact that it had a $\Z_2^{b}$ odd local operator on $\fB_{\Dir}$.

\subsubsection{Neumann$(S_3)$ Boundary Condition}
We finally discuss the boundary condition obtained by gauging the full $S_3$ 0-form symmetry on $\fB_{\Dir}$.
We will practically study this via sequential gauging, specifically by gauging the $\Z_2^{(0)}$ symmetry on $\fB_{\Neu(\Z_3),p}$.
As usual, there is a choice of discrete torsion
given by $H^{3}(S_3,U(1))\cong \Z_6 =\Z_3\times \Z_2$.
We parametrize this discrete torsion by the tuple $(p,t)\in \Z_3\times \Z_2$.
The corresponding boundary condition is denoted as
\begin{equation}
    \fB_{\Neu(S_3),(p,t)}= \frac{\fB_{\Neu(\Z_3),p}\boxtimes {\rm SPT}_{t}}{\Z_2}\,,
\end{equation}
where ${\rm SPT}_t$ here denotes a $\Z_2$ SPT in the class $t\in H^{3}(\Z_2,U(1))$.

\paragraph{$\TwoRep(S_3)$ Symmetry.} Recall that the symmetry category on $\fB_{\Neu(\Z_3),p}$ is the 2-group $\mathbb G^{(2)}=\Z_3^{(1)}\rtimes \Z_2^{(0)}$.
Gauging the $\Z_2^{0}$ sub-symmetry in $\mathbb G^{(2)}$ furnishes the symmetry category $\TwoRep(S_3)$ \cite{Bhardwaj:2022maz}.
To see this, note that after such a gauging, there is a $\Z_2^{(1)}$ sub-symmetry which is dual to the $\Z_2^{(0)}$ being gauged.
We denote its generator as $D_1^{P}$
Furthermore, since in $\mathbb G^{(2)}$, the $\Z_2^{(0)}$ symmetry acts on $\Z_{3}^{(1)}$ by exchanging the symmetry generators $D_1^{\omega}$ and $D_1^{\omega^2}$, after gauging, these combine into a single non-invertible line we denote as $D_1^{E}$.
It can be easily confirmed that the fusion rules of $D_1^{P}$ and $D_1^{E}$ correspond to the category $\Rep(S_3)$.
There are additionally four condensation defects which correspond to gauging the algebras \eqref{eq:algebras in Rep(S_3)} on the identity surface defects.
By including these condensation defects, one obtains the 2-fusion category $\TwoRep(S_3)$.

\paragraph{Generalized Charges.} 
We now describe how the bulk SymTFT defects end on $\fB_{\Neu(S_3),(p,t)}$.
Again we focus on non-condensation defects.
Our approach will be to perform a $\Z_2^{(0)}$ gauging on the ends of the SymTFT defects on $\fB_{\Neu(\Z_3),p}$.

 On the boundary $\fB_{\Neu(\Z_3),p}$, the line $\Q_1^P$ could end on an untwisted sector operator $\cE_0^{P}$ that transformed in the sign representation of the $\Z_2^{(0)}$ symmetry.
Upon gauging the $\Z_2^{(0)}$, $\cE_0^P$ goes to the twisted sector of the dual $\Z_2^{(1)}$ symmetry whose generator we denote as $D_1^{P}$
\be
\begin{split}
\begin{tikzpicture}
 \begin{scope}[shift={(0,0)},scale=0.8] 
\draw [cyan, fill=cyan!80!red, opacity =0.5]
(0,0) -- (0,4) -- (2,5) -- (5,5) -- (5,1) -- (3,0)--(0,0);
\draw [black, thick, fill=white,opacity=1]
(0,0) -- (0, 4) -- (2, 5) -- (2,1) -- (0,0);
\draw [cyan, thick, fill=cyan, opacity=0.2]
(0,0) -- (0, 4) -- (2, 5) -- (2,1) -- (0,0);
\draw[line width=1pt] (1,2.5) -- (3,2.5);
\draw[line width=1pt] (1,2.5) -- (1,4.5);
\draw[line width=1pt] (3,2.5) -- (4.2,2.5);
\fill[red!80!black] (1,2.5) circle (3pt);
\node at (2,5.4) {$\fB_{\Neu(S_3),(p,t)}$};
\draw[dashed] (0,0) -- (3,0);
\draw[dashed] (0,4) -- (3,4);
\draw[dashed] (2,5) -- (5,5);
\draw[dashed] (2,1) -- (5,1);
\draw [black, thick, dashed]
(3,0) -- (3, 4) -- (5, 5) -- (5,1) -- (3,0);
\draw [cyan, thick, fill=cyan, opacity=0.1]
(0,4) -- (3, 4) -- (5, 5) -- (2,5) -- (0,4);
\node[below, red!80!black] at (.8, 2.4) {$\cE_0^{P}$};
\node  at (2.5, 3) {$\Q_{1}^{P}$};
\node  at (0.5, 3.3) {$D_{1}^{P}$};
\node  at (5.5, 2.5) {$,$};
\node  at (11, 2.5) {$\Q_1^P\Bigg|_{{\Neu(S_3),(p,t)}}=(D_1^P,\cE_0^P)\,.$};
\end{scope}
\end{tikzpicture}    
\end{split}
\ee
Next, on $\fB_{\Neu(\Z_3),p}$, the line $\Q_1^{E}$, had two ends $\cE_0^{E,1}$ and $\cE_0^{E,2}$ in the twisted sectors of $D_1^{\omega}$ and $D_1^{\omega^2}$ respectively.
These two operators were mapped into one another under $\Z_2^{(0)}$ and therefore combine to a single twisted sector operator in the $\Z_2^{(0)}$ gauged boundary
\be
\begin{split}
\begin{tikzpicture}
 \begin{scope}[shift={(0,0)},scale=0.8] 
\draw [cyan, fill=cyan!80!red, opacity =0.5]
(0,0) -- (0,4) -- (2,5) -- (5,5) -- (5,1) -- (3,0)--(0,0);
\draw [black, thick, fill=white,opacity=1]
(0,0) -- (0, 4) -- (2, 5) -- (2,1) -- (0,0);
\draw [cyan, thick, fill=cyan, opacity=0.2]
(0,0) -- (0, 4) -- (2, 5) -- (2,1) -- (0,0);
\draw[line width=1pt] (1,2.5) -- (3,2.5);
\draw[line width=1pt] (1,2.5) -- (1,4.5);
\draw[line width=1pt] (3,2.5) -- (4.2,2.5);
\fill[red!80!black] (1,2.5) circle (3pt);
\node at (2,5.4) {$\fB_{\Neu(S_3),(p,t)}$};
\draw[dashed] (0,0) -- (3,0);
\draw[dashed] (0,4) -- (3,4);
\draw[dashed] (2,5) -- (5,5);
\draw[dashed] (2,1) -- (5,1);
\draw [black, thick, dashed]
(3,0) -- (3, 4) -- (5, 5) -- (5,1) -- (3,0);
\draw [cyan, thick, fill=cyan, opacity=0.1]
(0,4) -- (3, 4) -- (5, 5) -- (2,5) -- (0,4);
\node[below, red!80!black] at (.8, 2.4) {$\cE_0^{E}$};
\node  at (2.5, 3) {$\Q_{1}^{E}$};
\node  at (0.5, 3.3) {$D_{1}^{E}$};
\node  at (5.5, 2.5) {$,$};
\node  at (11, 2.5) {$\Q_1^E\Bigg|_{{\Neu(S_3),(p,t)}}=(D_1^E,\cE_0^E)\,.$};
\end{scope}
\end{tikzpicture}    
\end{split}
\ee
For similar reasons, the two lines at the end of the $\Q_2^{[a]}$ surface $\cE_1^{a}$ and $\cE_1^{a^2}$ combine to a single line $\cE_1^{[a]}$.
\be
\begin{split}
\begin{tikzpicture}
 \begin{scope}[shift={(0,0)},scale=0.8] 
\draw [cyan, fill=cyan!80!red, opacity =0.5]
(0,0) -- (0,4) -- (2,5) -- (5,5) -- (5,1) -- (3,0)--(0,0);
\draw [black, thick, fill=white,opacity=1]
(0,0) -- (0, 4) -- (2, 5) -- (2,1) -- (0,0);
\draw [cyan, thick, fill=cyan, opacity=0.2]
(0,0) -- (0, 4) -- (2, 5) -- (2,1) -- (0,0);
\draw [black, thick, fill=black, opacity=0.3]
(0.,2) -- (3.1, 2) -- (5., 3) -- (2, 3) -- (0.,2);
\node at (2,5.4) {$\fB_{\Neu(S_3),(p,t)}$};
\draw[dashed] (0,0) -- (3,0);
\draw[dashed] (0,4) -- (3,4);
\draw[dashed] (2,5) -- (5,5);
\draw[dashed] (2,1) -- (5,1);
\draw [black, thick, dashed]
(3,0) -- (3, 4) -- (5, 5) -- (5,1) -- (3,0);
\draw [cyan, thick, fill=cyan, opacity=0.1]
(0,4) -- (3, 4) -- (5, 5) -- (2,5) -- (0,4);
\draw [line width=1pt, red!80!black] (0.0, 2.) -- (2, 3) ;
\node[below, red!80!black] at (-0.5, 2.4) {$\cE_1^{[a]}$};
\node  at (2.5, 2.55) {$\Q_{2}^{[a]}$};
\node  at (5.5, 2.55) {$,$};
\node  at (10.5, 2.55) {$\Q_2^{[a]}\Bigg|_{{\Neu(S_3),(p,t)}}=\cE_1^{[a]}\,.$};
\end{scope}
\end{tikzpicture}    
\end{split}
\ee
The lines $\cE_1^{a}$ and $\cE_1^{a^2}$ were charged under the $\Z_3^{(1)}$ symmetry on $\fB_{\Neu(\Z_3),p}$, therefore the corresponding line $\cE_1^{[a]}$ is charged under $D_1^{E}$.
Since $D_1^{E}$ descends from $D_1^{\omega}\oplus D_1^{\omega^2}$, the linking phase is 
\begin{equation}
    {\rm Link}(D_1^{E}\,, L^{[a]})= \omega +\omega^2=-1=\chi_E([a])\,.
\end{equation}
Finally, we now describe the ends of $\Q_2^{[b]}$.
This surface had a single L-shaped end on $\fB_{\Neu(Z_3),p}$ given by the line $\cE_1^{[b]}$ attached to a condensate of $\Z_3^{(1)}$ lines on the $\Z_2^{(0)}$ surface.
Once we gauge $\Z_2^{(0)}$, $\cE_1^{[b]}$ is attached to the identity surface with a condensate of $D_1^{\id}\oplus D_1^{E}$.
Since this condensate can be shrunk to a line we obtain
\be
\begin{split}
\begin{tikzpicture}
 \begin{scope}[shift={(0,0)},scale=0.8] 
\draw [cyan, fill=cyan!80!red, opacity =0.5]
(0,0) -- (0,4) -- (2,5) -- (5,5) -- (5,1) -- (3,0)--(0,0);
\draw [black, thick, fill=white,opacity=1]
(0,0) -- (0, 4) -- (2, 5) -- (2,1) -- (0,0);
\draw [cyan, thick, fill=cyan, opacity=0.2]
(0,0) -- (0, 4) -- (2, 5) -- (2,1) -- (0,0);
\draw [black, thick, fill=black, opacity=0.3]
(0.,2) -- (3.1, 2) -- (5., 3) -- (2, 3) -- (0.,2);
\node at (2,5.4) {$\fB_{\Neu(S_3),(p,t)}$};
\draw[dashed] (0,0) -- (3,0);
\draw[dashed] (0,4) -- (3,4);
\draw[dashed] (2,5) -- (5,5);
\draw[dashed] (2,1) -- (5,1);
\draw [black, thick, dashed]
(3,0) -- (3, 4) -- (5, 5) -- (5,1) -- (3,0);
\draw [cyan, thick, fill=cyan, opacity=0.1]
(0,4) -- (3, 4) -- (5, 5) -- (2,5) -- (0,4);
\draw [line width=1pt, red!80!black] (0.0, 2.) -- (2, 3) ;
\node[below, red!80!black] at (-0.5, 2.4) {$\cE_1^{[b]}$};
\node  at (2.5, 2.55) {$\Q_{2}^{[b]}$};
\node  at (5.5, 2.55) {$,$};
\node  at (10.5, 2.55) {$\Q_2^{[b]}\Bigg|_{{\Neu(S_3),(p,t)}}=\cE_1^{[b]}\,.$};
\end{scope}
\end{tikzpicture}    
\end{split}
\ee
This line is charged under $D_1^{P}$.
The F-symbols of $L^{[a]}$ and $L^{[b]}$ are determined by $(p,t)$.
%


\subsection{$\TwoVec(S_3)$ Gapped Phases}
We now describe the minimal phases with $\TwoVec(S_3)$ symmetry.
For this we study the SymTFT with the symmetry boundary chosen to be 
\begin{equation}
    \Bsym=\fB_{\Dir}
\end{equation}
and the physical boundary to be one of the minimal topological boundaries for $\cZ(\TwoVec(S_3))$.
The order parameters for the gapped phase obtained after compactifying the SymTFT are given by the SymTFT defects that have untwisted sector ends on the physical boundary.
Of course these are well-understood phases, that follow up to SPTs, the Landau paradigm. It is still useful to spell these out in the SymTFT framework as it is the starting point for the generalized gauging and the non-Landau phases. 

\subsubsection{$S_3$ SSB Phase}
\label{sec:papaya1}
Consider the physical boundary to be
\begin{equation}
    \Bphys=\fB_{\Dir}\,.
\end{equation}
The SymTFT lines $\Q_1^{R}$ for any $R\in \Rep(S_3)$ can end on both boundaries and have ${\rm dim}(R)$ ends on either boundary.
Therefore the SymTFT line $\Q_1^{R}$ extended between the two boundaries furnishes ${\rm dim}(R)^2$ topological local operators after compactification
\be
\begin{split}
\begin{tikzpicture}
 \begin{scope}[shift={(0,0)},scale=0.8] 
\draw [cyan, fill=cyan!80!red, opacity =0.5]
(0,0) -- (0,4) -- (2,5) -- (5,5) -- (5,1) -- (3,0)--(0,0);
\draw [black, thick, fill=white,opacity=1]
(0,0) -- (0, 4) -- (2, 5) -- (2,1) -- (0,0);
\draw [cyan, thick, fill=cyan, opacity=0.2]
(0,0) -- (0, 4) -- (2, 5) -- (2,1) -- (0,0);
\draw[line width=1pt] (1,2.5) -- (3,2.5);
\draw[line width=1pt,dashed] (3,2.5) -- (4,2.5);
\fill[red!80!black] (1,2.5) circle (3pt);
\draw [fill=blue!40!red!60,opacity=0.2]
(3,0) -- (3, 4) -- (5, 5) -- (5,1) -- (3,0);
\fill[red!80!black] (1,2.5) circle (3pt);
\fill[red!80!black] (4,2.5) circle (3pt);
\draw [black, thick, opacity=1]
(3,0) -- (3, 4) -- (5, 5) -- (5,1) -- (3,0);
\node at (2,5.4) {$\fB_{\Dir}$};
\node at (5,5.4) {$\fB_{\Dir}$};
\draw[dashed] (0,0) -- (3,0);
\draw[dashed] (0,4) -- (3,4);
\draw[dashed] (2,5) -- (5,5);
\draw[dashed] (2,1) -- (5,1);
\draw [cyan, thick, fill=cyan, opacity=0.1]
(0,4) -- (3, 4) -- (5, 5) -- (2,5) -- (0,4);
\node[below, red!80!black] at (.8, 2.4) {$\cE_0^{R,i}$};
\node[below, red!80!black] at (3.8, 2.4) {$\wt{\cE}_0^{R,j}$};
\node  at (2.5, 3) {$\Q_{1}^{R}$};
\node  at (6.7, 2.5) {$=$};
\draw [fill=blue!60!green!30,opacity=0.7]
(8,0) -- (8, 4) -- (10, 5) -- (10,1) -- (8,0);
\draw [black, thick, opacity=1]
(8,0) -- (8, 4) -- (10, 5) -- (10,1) -- (8,0);
\fill[red!80!black] (9,2.5) circle (3pt);
\node[below, red!80!black] at (8.8, 2.4) {${\cO}^{R\,,i,j}$};
\end{scope}
\end{tikzpicture}    
\end{split}
\label{eq:local OP}
\ee
%
We denote the SymTFT sandwich with the configuration \eqref{eq:local OP} as
\begin{equation}
    \cO^{R\,,i,j}=(\cE_0^{R,i}\,, \Q_1^{R}\,, \wt{\cE}_0^{R,j})\,.
\end{equation}
After compactifying the SymTFT one obtains a 2+1 dimensional TFT with 
\begin{equation}
\#(\text{local operators}) = \#(\text{vacua}) =\sum_{R}{\rm dim}(R)^2=6\,,    
\end{equation}
where $\#$ denotes number of. 
%
We denote these operators with the following shorthand notation
\begin{equation}
    \cO^{\id}:=1\,, \qquad \cO^{P,1,1}:=\cO^P\,, \qquad \cO^{E,i,j}=\cO^{i,j}\,.
\end{equation}
The $S_3$ symmetry acts on these operators as
\begin{equation}
    g:(\cO^{P}\,, \cO^{i,j})\longmapsto (\cD_{P}(g)\cO^{P}\,, \sum_{i'}\cD_{E}(g)_{ii'}\cO^{i',j}) \,.
    \label{eq:S3 action S3 SSB phase}
\end{equation}
The algebra of these operators can be worked out using precisely the same computations as in Sec.~4.6.1 of \cite{Bhardwaj:2023idu}. 
%
There are six vacua which correspond to linear combinations of the local operators which are  idempotent i.e., satisfy
\begin{equation}
    v_iv_{j}=\delta_{ij}v_i\,,
\end{equation}
for $i=0,\dots, 5$.
These serve as projectors onto the six vacua in this gapped phase.
The $S_3$ action can be computed using \eqref{eq:S3 action S3 SSB phase} and takes the form 
\begin{equation}
\begin{split}
    D_2^{a}&:(v_0\,,v_1\,,v_2\,,v_3\,,v_4\,,v_5)\longmapsto (v_1\,,v_2\,,v_0\,,v_4\,,v_5\,,v_3)\,, \\
    D_2^{b}&:(v_0\,,v_1\,,v_2\,,v_3\,,v_4\,,v_5)\longmapsto (v_3\,,v_5\,,v_4\,,v_0\,,v_2\,,v_1)\,.
\end{split}
\end{equation}
Thus the $S_3$ symmetry is broken spontaneously in all the 6 vacua.
Moreover there are no twisted sector operators since none of the surfaces $\Q_2^{[g]}$ end on the physical boundary.
The underlying 3d TFT is a direct sum of 6 trivial theories on which the $S_3$ symmetry is represented as:
\begin{tcolorbox}[
colback=white,
coltitle= black,
colbacktitle=ourcolorforheader,
colframe=black,
title= $\TwoVec_{S_3}$: $S_3$ SSB Phase, 
sharp corners]
\be
\bigoplus_{g\in S_3}{\rm Triv}_{g}\,, \qquad D_2^{g}: {\text{Triv}_{g'}\mapsto {\text{Triv}_{gg'}}}
\ee
\end{tcolorbox}

\subsubsection{$\Z_2$ SSB $\boxtimes$ $\Z_3$ SPT Phase}
\label{sec:papaya3}
Now consider the physical boundary to be
\begin{equation}
    \Bphys=\fB_{\Neu(\Z_3),p}\,.
\end{equation}
The order parameters for this phase carry generalized charges $\Q_1^{P}$ and $\Q_2^{[a]}$ since these SymTFT defects have untwisted ends on $\Bphys$.
The line $\Q_1^{P}$ has a single end each on the physical and symmetry boundaries and therefore provides an additional local operator $\cO^P$ upon compactification. 
\be
\begin{split}
\begin{tikzpicture}
 \begin{scope}[shift={(0,0)},scale=0.8] 
\draw [cyan, fill=cyan!80!red, opacity =0.5]
(0,0) -- (0,4) -- (2,5) -- (5,5) -- (5,1) -- (3,0)--(0,0);
\draw [black, thick, fill=white,opacity=1]
(0,0) -- (0, 4) -- (2, 5) -- (2,1) -- (0,0);
\draw [cyan, thick, fill=cyan, opacity=0.2]
(0,0) -- (0, 4) -- (2, 5) -- (2,1) -- (0,0);
\draw[line width=1pt] (1,2.5) -- (3,2.5);
\draw[line width=1pt,dashed] (3,2.5) -- (4,2.5);
\fill[red!80!black] (1,2.5) circle (3pt);
\draw [fill=blue!40!red!60,opacity=0.2]
(3,0) -- (3, 4) -- (5, 5) -- (5,1) -- (3,0);
\fill[red!80!black] (1,2.5) circle (3pt);
\fill[red!80!black] (4,2.5) circle (3pt);
\draw [black, thick, opacity=1]
(3,0) -- (3, 4) -- (5, 5) -- (5,1) -- (3,0);
\node at (2,5.4) {$\fB_{\Dir}$};
\node at (5,5.4) {$\fB_{\Neu(\Z_3),p}$};
\draw[dashed] (0,0) -- (3,0);
\draw[dashed] (0,4) -- (3,4);
\draw[dashed] (2,5) -- (5,5);
\draw[dashed] (2,1) -- (5,1);
\draw [cyan, thick, fill=cyan, opacity=0.1]
(0,4) -- (3, 4) -- (5, 5) -- (2,5) -- (0,4);
\node[below, red!80!black] at (.8, 2.4) {$\cE_0^{P}$};
\node[below, red!80!black] at (3.8, 2.4) {$\wt{\cE}_0^{P}$};
\node  at (2.5, 3) {$\Q_{1}^{P}$};
\node  at (6.7, 2.5) {$=$};
\draw [fill=blue!60!green!30,opacity=0.7]
(8,0) -- (8, 4) -- (10, 5) -- (10,1) -- (8,0);
\draw [black, thick, opacity=1]
(8,0) -- (8, 4) -- (10, 5) -- (10,1) -- (8,0);
\fill[red!80!black] (9,2.5) circle (3pt);
\node[below, red!80!black] at (8.8, 2.4) {${\cO}^{P}$};
\end{scope}
\end{tikzpicture}    
\end{split}
\label{eq:local OP}
\ee
Since $\Q_1^{P}$ has $\Z_2$ fusion rules, we may rescale this operator such that 
\begin{equation}
    (\cO^P)^2=1\,.
\end{equation}
Then the vacua (idempotent operators) are determined to be
\begin{equation}
    v_0=\frac{1+\cO^P}{2}\,, \qquad v_1=\frac{1-\cO^P}{2}\,.
\end{equation}
The operator $\cO^P$ is charged under the $\Z_2$ symmetry, i.e., 
\begin{equation}
    D_2^{a^ib^j}: \cO_P \longmapsto (-1)^{j}\cO_P\,,
\end{equation}
which implies that the vacua transform as
\begin{equation}
    D_2^{a^ib^j}: (v_0\,, v_1) \longmapsto
    \begin{cases}
        (v_0\,, v_1)\quad j=0\,,     \\       (v_1\,, v_0)\quad j=1\,,
    \end{cases}
\end{equation}
Therefore this phase spontaneously breaks the $\Z_2$ symmetry.

Let us now describe the surface $\Q_2^{[a]}$ which also provides order parameters in this phase.
Specifically, the surface $\Q_2^{[a]}$ ends on two untwisted sector lines $\wt{\cE}_1^{a}$ and $\wt{\cE}_1^{a^2}$ on the physical boundary and on two lines $\cE_1^{a}$ and $\cE_1^{a^2}$ in the $D_2^{a}$ and $D_2^{a^2}$ twisted sectors on the symmetry boundary.
After compactifying the SymTFT sandwich with the $\Q_2^{[a]}$ surface extending between the two boundaries, one obtains four lines in $g\in [a]$ twisted sectors
\be
\begin{split}
\begin{tikzpicture}
 \begin{scope}[shift={(0,0)},scale=0.8] 
\draw [cyan, fill=cyan!80!red, opacity =0.5]
(0,0) -- (0,4) -- (2,5) -- (5,5) -- (5,1) -- (3,0)--(0,0);
\draw [black, thick, fill=white,opacity=1]
(0,0) -- (0, 4) -- (2, 5) -- (2,1) -- (0,0);
\draw [cyan, thick, fill=cyan, opacity=0.2]
(0,0) -- (0, 4) -- (2, 5) -- (2,1) -- (0,0);
\draw [fill=blue!40!red!60,opacity=0.2]
(3,0) -- (3, 4) -- (5, 5) -- (5,1) -- (3,0);
\draw [black, thick, opacity=1]
(3,0) -- (3, 4) -- (5, 5) -- (5,1) -- (3,0);
\node at (2,5.4) {$\fB_{\Dir}$};
\node at (5,5.4) {$\fB_{\Neu(\Z_3),p}$};
\draw[dashed] (0,0) -- (3,0);
\draw[dashed] (0,4) -- (3,4);
\draw[dashed] (2,5) -- (5,5);
\draw[dashed] (2,1) -- (5,1);
\draw [cyan, thick, fill=cyan, opacity=0.1]
(0,4) -- (3, 4) -- (5, 5) -- (2,5) -- (0,4);
\draw [black, thick, fill=black, opacity=0.3]
(0.,2) -- (3, 2) -- (5., 3) -- (2, 3) -- (0.,2);
\draw [black, thick, fill=black, opacity=0.3]
(0.,2) -- (0., 4) -- (2, 5) -- (2,3) -- (0,2);
\node  at (2.5, 2.5) {$\Q_{2}^{[a]}$};
\node[below, red!80!black] at (-0.5, 2.4) {$\cE_1^{a^j}$};
\node[below, red!80!black] at (4, 2.4) {$\wt{\cE}_1^{a^{k}}$};
\draw [line width=1pt, red!80!black] (0.0, 2.) -- (2, 3);
\draw [line width=1pt, red!80!black] (3, 2.) -- (5, 3);
\node  at (6.7, 2.5) {$=$};
\draw [fill=blue!60!green!30,opacity=0.7]
(8,0) -- (8, 4) -- (10, 5) -- (10,1) -- (8,0);
\draw [black, thick, opacity=1]
(8,0) -- (8, 4) -- (10, 5) -- (10,1) -- (8,0);
\draw [black, thick, fill=black, opacity=0.3]
(8,2) -- (8, 4) -- (10, 5) -- (10,3) -- (8,2);
\draw [line width=1pt, red!80!black] (8, 2.) -- (10, 3);
\node[below, black] at (9, 3.9) {$D_2^{a^j}$};
\node[below, black] at (1., 3.9) {$D_2^{a^j}$};
\node[below, red!80!black] at (9, 2.4) {${\cL}^{j,k}$};
\end{scope}
\end{tikzpicture}    
\end{split}
\ee
The set of lines is labeled 
\begin{equation}
    \left\{(D_2^a\,, \cL^{1,1})\,, (D_2^a\,, \cL^{1,2})\,, (D_2^{a^2}\,,\cL^{2,1})\,,
    (D_2^{a^2}\,, \cL^{2,2})\right\}\,.
\end{equation}
These lines are all invariant under the $\Z^{(0)}_3\subset S_3$ symmetry and transform under the $\Z^{(0)}_2$ action as
\begin{equation}
    D_2^{b}: (D_2^{a^i}\,, \cL^{i,j})\longmapsto (D_2^{a^{-i}}\,, \cL^{-i,j})\,,
    \label{eq:Z2 action on Z3 lines}
\end{equation}
where $-i\equiv -i \text{ mod 2}$.
Let us construct the fusion algebra of these twisted sector lines.
Firstly, the lines $\left\{\wt{\cE}_1^{a}\,,\wt{\cE}_1^{a^{2}}\right\}$ and  $\left\{\cE_1^{a}\,, \cE_1^{a^{2}}\right\}$ generate a $\Z_3$ fusion algbera on the physical and symmetry boundary respectively.
Therefore by consistency, we find that the above lines generate a $\Z_3\oplus \Z_3$ algebra.
Specifically the sets of lines $\{D_1^{\id}\,,\cL^{1,1}\,, \cL^{2,2})\}$ and $\{D_1^{\id}\,,L^{1,2}\,, \cL^{2,1})\}$ each independently satisfy $\Z_3$ fusion rules.
Meanwhile any fusion outcome among non-identity lines in these two sets vanishes
\begin{equation}
    \cL^{i,i}\otimes \cL^{j,-j}=0\,.
\end{equation}
This implies that these two sets of lines act within different vacua.
Since the two vacua are indistinguishable, without loss of generality, we can assume that $\cL^{1,2}$ and $L^{2,1}$ are the $\Z_3$ twisted sector lines in $v_0$.
Then by the $\Z_2$ action \eqref{eq:Z2 action on Z3 lines} it follows that $\cL^{1,1}$ and $\cL^{2,2}$ are the $\Z_3$ twisted sector lines in $v_1$.
The F-symbols for these lines are determined by the ends on $\Bphys$ before compactification.
Indeed if $p\neq [0] \in H^3(\Z_3,U(1))$, both the vacua realize non-trivial $\Z_3^{(0)}$ SPTs and are mapped into one another under $\Z_2^{(0)}$.

\begin{tcolorbox}[
colback=white,
coltitle= black,
colbacktitle=ourcolorforheader,
colframe=black,
title= $\TwoVec_{S_3}$: $\Z_2$ SSB $\boxtimes$ $\Z_3$ SPT Phase, 
sharp corners]
\be
\begin{split}
\begin{tikzpicture}
\begin{scope}[shift={(0,0)},scale=1] 
 \pgfmathsetmacro{\del}{-0.5}
\node at  (0,0)  {${\rm Triv}_0 \quad \boxplus \quad {\rm Triv}_1$} ;
\node at  (-1,-0.6)  {\footnotesize$(\Z_3^{a}\text{-}{\rm SPT}_p)$} ;
\node at  (1,-0.6)  {\footnotesize$(\Z_3^{a^2}\text{-}{\rm SPT}_p)$} ;
\end{scope}
\end{tikzpicture}    
\end{split}
\ee
$\TwoVec_{S_3}$ realization on vacua
\begin{equation}
    D_2^{a}\simeq 1_{00}\oplus 1_{11}\,, \qquad
    D_2^{b}= 1_{01}\oplus 1_{10}\,.
\end{equation}
\end{tcolorbox}

\subsubsection{$\Z_3$ SSB $\boxtimes$ $\Z_2$ SPT Phase}
\label{sec:papaya2}

Now consider the physical boundary to be
\begin{equation}
    \Bphys=\fB_{\Neu(\Z_2),t}\,.
\end{equation}
The order parameters in the gapped phase obtained from this SymTFT sandwich carry the generalized charges in the $\Q_1^E$ and $\Q_2^{[b]}$ representations.
Recall that the bulk line $\Q_1^{E}$ has a single untwisted sector end given by $\wt{\cE}_0^{E,+}$ on this physical boundary and two ends $\cE^{E,i}_0$ ($i=1,2$) on the symmetry boundary.
One therefore obtains a total of 3 topological local operators after compactifying the SymTFT.
1 from the $\Q_1^{\id}$ and 2 from $\Q_1^{E}$ which we denote as $\cO_1$ and $\cO_2$.
\be
\begin{split}
\begin{tikzpicture}
 \begin{scope}[shift={(0,0)},scale=0.8] 
\draw [cyan, fill=cyan!80!red, opacity =0.5]
(0,0) -- (0,4) -- (2,5) -- (5,5) -- (5,1) -- (3,0)--(0,0);
\draw [black, thick, fill=white,opacity=1]
(0,0) -- (0, 4) -- (2, 5) -- (2,1) -- (0,0);
\draw [cyan, thick, fill=cyan, opacity=0.2]
(0,0) -- (0, 4) -- (2, 5) -- (2,1) -- (0,0);
\draw[line width=1pt] (1,2.5) -- (3,2.5);
\draw[line width=1pt,dashed] (3,2.5) -- (4,2.5);
\fill[red!80!black] (1,2.5) circle (3pt);
\draw [fill=blue!40!red!60,opacity=0.2]
(3,0) -- (3, 4) -- (5, 5) -- (5,1) -- (3,0);
\fill[red!80!black] (1,2.5) circle (3pt);
\fill[red!80!black] (4,2.5) circle (3pt);
\draw [black, thick, opacity=1]
(3,0) -- (3, 4) -- (5, 5) -- (5,1) -- (3,0);
\node at (2,5.4) {$\fB_{\Dir}$};
\node at (5,5.4) {$\fB_{\Neu(\Z_2),t}$};
\draw[dashed] (0,0) -- (3,0);
\draw[dashed] (0,4) -- (3,4);
\draw[dashed] (2,5) -- (5,5);
\draw[dashed] (2,1) -- (5,1);
\draw [cyan, thick, fill=cyan, opacity=0.1]
(0,4) -- (3, 4) -- (5, 5) -- (2,5) -- (0,4);
\node[below, red!80!black] at (.8, 2.4) {$\cE_0^{E,i}$};
\node[below, red!80!black] at (3.8, 2.4) {$\wt{\cE}_0^{E,+}$};
\node  at (2.5, 3) {$\Q_{1}^{E}$};
\node  at (6.7, 2.5) {$=$};
\draw [fill=blue!60!green!30,opacity=0.7]
(8,0) -- (8, 4) -- (10, 5) -- (10,1) -- (8,0);
\draw [black, thick, opacity=1]
(8,0) -- (8, 4) -- (10, 5) -- (10,1) -- (8,0);
\fill[red!80!black] (9,2.5) circle (3pt);
\node[below, red!80!black] at (8.8, 2.4) {${\cO}_{i}$};
\end{scope}
\end{tikzpicture}    
\end{split}
\label{eq:local OP}
\ee
Since $\cO_1$ and $\cO_2$ transform under the $E$-representation of $S_3$, we have
\begin{equation}
\begin{split}
\label{eq:S3 action}
    D_{2}^{a}&:(\cO_1\,, \cO_2)\longmapsto    (\omega\cO_1\,, \omega^2\cO_2)\,, \\
    D_{2}^{b}&:(\cO_1\,, \cO_2)\longmapsto    (\cO_2\,, \cO_1)\,,
\end{split}
\end{equation}
which implies that 
\begin{equation}
    \cO_1 \cO_2 \propto 1\,, \quad \cO_1\cO_1\propto \cO_2\,, \quad \cO_2\cO_2\propto \cO_1\,.
\end{equation}
These operators can be rescaled such that all the proportionality constants are 1.
The local operators then satisfy the $\Z_3$ algebra 
\begin{equation}
\label{eq:Z3:alg}
    \cO_1 \cO_2 = 1\,, \quad \cO_1\cO_1= \cO_2\,, \quad \cO_2\cO_2= \cO_1\,.
\end{equation}
From which we determine the vacua (idempotent operators) to take the form
\begin{equation}
\label{eq:Z3_vac}
\begin{split}
    v_{j}&= \frac{1+\omega^{j}\cO_1+\omega^{2j}\cO_2}{3}
\end{split}
\end{equation}
 Using \eqref{eq:S3 action}, the $S_3$ action on the vacua is determined to be
 \begin{equation}
\begin{split}
D_2^a&:(v_0\,, v_1\,, v_2)\longmapsto (v_1\,, v_2\,, v_0)\,, \\
D_2^b&:(v_0\,, v_1\,, v_2)\longmapsto (v_0\,, v_2\,, v_1)\,.
\label{eq:S3 action on Z3 SSB vacua}
\end{split}
\end{equation}
Therefore this is an $\Z_3$ spontaneous symmetry broken phase.
In addition to $\Q_1^{E}$, $\Q_2^{[b]}$ also plays the role of an order parameter as it ends on a single (untwisted sector) line $\wt{\cE}_1^{b}$ on the physical boundary and has three lines, one each in the $D_2^{b}$, $D_2^{ab}$ and $D_2^{a^2b}$ twisted sectors.
After compactifying the SymTFT, one obtains a topological line $\cL^{a^{j}b}$ in the $D_2^{a^jb}$ twisted sector for $j=0,1,2$.
\be
\begin{split}
\begin{tikzpicture}
 \begin{scope}[shift={(0,0)},scale=0.8] 
\draw [cyan, fill=cyan!80!red, opacity =0.5]
(0,0) -- (0,4) -- (2,5) -- (5,5) -- (5,1) -- (3,0)--(0,0);
\draw [black, thick, fill=white,opacity=1]
(0,0) -- (0, 4) -- (2, 5) -- (2,1) -- (0,0);
\draw [cyan, thick, fill=cyan, opacity=0.2]
(0,0) -- (0, 4) -- (2, 5) -- (2,1) -- (0,0);
\draw [fill=blue!40!red!60,opacity=0.2]
(3,0) -- (3, 4) -- (5, 5) -- (5,1) -- (3,0);
\draw [black, thick, opacity=1]
(3,0) -- (3, 4) -- (5, 5) -- (5,1) -- (3,0);
\node at (2,5.4) {$\fB_{\Dir}$};
\node at (5,5.4) {$\fB_{\Neu(\Z_2),t}$};
\draw[dashed] (0,0) -- (3,0);
\draw[dashed] (0,4) -- (3,4);
\draw[dashed] (2,5) -- (5,5);
\draw[dashed] (2,1) -- (5,1);
\draw [cyan, thick, fill=cyan, opacity=0.1]
(0,4) -- (3, 4) -- (5, 5) -- (2,5) -- (0,4);
\draw [black, thick, fill=black, opacity=0.3]
(0.,2) -- (3, 2) -- (5., 3) -- (2, 3) -- (0.,2);
\draw [black, thick, fill=black, opacity=0.3]
(0.,2) -- (0., 4) -- (2, 5) -- (2,3) -- (0,2);
\node  at (2.5, 2.5) {$\Q_{2}^{[b]}$};
\node[below, red!80!black] at (-0.5, 2.4) {$\cE_1^{a^jb}$};
\node[below, red!80!black] at (4, 2.4) {$\wt{\cE}_1^{b}$};
\draw [line width=1pt, red!80!black] (0.0, 2.) -- (2, 3);
\draw [line width=1pt, red!80!black] (3, 2.) -- (5, 3);
\node  at (6.7, 2.5) {$=$};
\draw [fill=blue!60!green!30,opacity=0.7]
(8,0) -- (8, 4) -- (10, 5) -- (10,1) -- (8,0);
\draw [black, thick, opacity=1]
(8,0) -- (8, 4) -- (10, 5) -- (10,1) -- (8,0);
\draw [black, thick, fill=black, opacity=0.3]
(8,2) -- (8, 4) -- (10, 5) -- (10,3) -- (8,2);
\draw [line width=1pt, red!80!black] (8, 2.) -- (10, 3);
\node[below, black] at (9, 3.9) {$D_2^{a^jb}$};
\node[below, black] at (1., 3.9) {$D_2^{a^jb}$};
\node[below, red!80!black] at (9, 2.4) {${\cL}_1^{a^jb}$};
\end{scope}
\end{tikzpicture}    
\end{split}
\ee
These lines are all isomorphic and their F-symbols are given by $t\in H^{3}(\Z_2,U(1))$.
This is defining property of a $\Z_2$ SPT.
To summarize, the $S_3$ action in this gapped phase is realized on the three vacua as \eqref{eq:S3 action on Z3 SSB vacua}.
The vacua $v_j$ is symmetric under the $\Z_2$ symmetry generated by $D_2^{a^jb}$ and realizes a $t$ SPT for the preserved $\Z_2$. 
\begin{tcolorbox}[
colback=white,
coltitle= black,
colbacktitle=ourcolorforheader,
colframe=black,
title= $\TwoVec_{S_3}$: $\Z_3$ SSB $\boxtimes$ $\Z_2$ SPT Phase, 
sharp corners]
\be
\begin{split}
\begin{tikzpicture}
\begin{scope}[shift={(0,0)},scale=1] 
 \pgfmathsetmacro{\del}{-0.5}
\node at  (0,0)  {${\rm Triv}_0 \quad \boxplus \quad {\rm Triv}_1 \quad \boxplus \quad {\rm Triv}_2$} ;
\node at  (-2,-0.6)  {\footnotesize$(\Z_2^{b}\text{-}{\rm SPT}_t)$} ;
\node at  (0,-0.6)  {\footnotesize$(\Z_2^{ab}\text{-}{\rm SPT}_t)$} ;
\node at  (2,-0.6)  {\footnotesize$(\Z_2^{a^2b}\text{-}{\rm SPT}_t)$} ;
\end{scope}
\end{tikzpicture}    
\end{split}
\ee
$\TwoVec_{S_3}$ realization on vacua
\begin{equation}
\begin{split}
    D_2^{a}= 1_{01}\oplus 1_{12} \oplus 1_{20}\,, \\
    D_2^{b}\simeq 1_{00}\oplus 1_{12} \oplus 1_{21}\,,
\end{split}
\end{equation}
where $1_{ij}$ is the surface operator that maps ${\rm Triv}_i$ to ${\rm Triv}_j$. 
\end{tcolorbox}

\subsubsection{$S_3$ SPT Phase}
\label{sec:papaya4}
Now we consider the physical boundary to be
\begin{equation}
    \Bphys=\mathfrak{B}_{\Neu(S_3),(p,t)}\,.
\end{equation}
None of the SymTFT lines end on this physical boundary and therefore after compactification, there is a single local operator and consequently a single vacua in this phase.
Instead each SymTFT surface $\Q_2^{[g]}$ has a single end on the physical boundary and $|[g]|$ ends on the symmetry boundary, one in each twisted sector $D_2^{g}$ for $g\in [g]$.
After compactifying the SymTFT, this provides twisted sector lines $(D_2^g,\cL^{(g)})$ for all $g\in D_2^g$ which are the order parameters for this gapped phase.
\be
\begin{split}
\begin{tikzpicture}
 \begin{scope}[shift={(0,0)},scale=0.8] 
\draw [cyan, fill=cyan!80!red, opacity =0.5]
(0,0) -- (0,4) -- (2,5) -- (5,5) -- (5,1) -- (3,0)--(0,0);
\draw [black, thick, fill=white,opacity=1]
(0,0) -- (0, 4) -- (2, 5) -- (2,1) -- (0,0);
\draw [cyan, thick, fill=cyan, opacity=0.2]
(0,0) -- (0, 4) -- (2, 5) -- (2,1) -- (0,0);
\draw [fill=blue!40!red!60,opacity=0.2]
(3,0) -- (3, 4) -- (5, 5) -- (5,1) -- (3,0);
\draw [black, thick, opacity=1]
(3,0) -- (3, 4) -- (5, 5) -- (5,1) -- (3,0);
\node at (2,5.4) {$\fB_{\Dir}$};
\node at (5,5.4) {$\fB_{\Neu(S_3),(p,t)}$};
\draw[dashed] (0,0) -- (3,0);
\draw[dashed] (0,4) -- (3,4);
\draw[dashed] (2,5) -- (5,5);
\draw[dashed] (2,1) -- (5,1);
\draw [cyan, thick, fill=cyan, opacity=0.1]
(0,4) -- (3, 4) -- (5, 5) -- (2,5) -- (0,4);
\draw [black, thick, fill=black, opacity=0.3]
(0.,2) -- (3, 2) -- (5., 3) -- (2, 3) -- (0.,2);
\draw [black, thick, fill=black, opacity=0.3]
(0.,2) -- (0., 4) -- (2, 5) -- (2,3) -- (0,2);
\node  at (2.5, 2.5) {$\Q_{2}^{[g]}$};
\node[below, red!80!black] at (-0.5, 2.4) {$\cE_1^{g}$};
\node[below, red!80!black] at (4, 2.4) {$\wt{\cE}_1^{[g]}$};
\draw [line width=1pt, red!80!black] (0.0, 2.) -- (2, 3);
\draw [line width=1pt, red!80!black] (3, 2.) -- (5, 3);
\node  at (6.7, 2.5) {$=$};
\draw [fill=blue!60!green!30,opacity=0.7]
(8,0) -- (8, 4) -- (10, 5) -- (10,1) -- (8,0);
\draw [black, thick, opacity=1]
(8,0) -- (8, 4) -- (10, 5) -- (10,1) -- (8,0);
\draw [black, thick, fill=black, opacity=0.3]
(8,2) -- (8, 4) -- (10, 5) -- (10,3) -- (8,2);
\draw [line width=1pt, red!80!black] (8, 2.) -- (10, 3);
\node[below, black] at (9, 3.9) {$D_2^{g}$};
\node[below, black] at (1., 3.9) {$D_2^{g}$};
\node[below, red!80!black] at (9, 2.4) {${\cL}^{g}$};
\end{scope}
\end{tikzpicture}    
\end{split}
\ee
The F-symbols of these twisted sector lines are determined by the discrete torsion $(p,t)\in H^{3}(S_3,U(1))$.

To concretely compute the F-symbols, we take an alternate route. 
We study the $S_3$ SPT phase as the gauging of the $\mathbb G^{(2)}$ phase studied in Sec.~\ref{sec:minimal 2grp phases} where the $\Z_3^{(1)}\subset \mathbb G_2$ is spontaneously broken and $\Z_2^{(0)}$ is spontaneously preserved.
More precisely, the $\mathbb G^{(2)}$ gapped phase realizes a $\Z_2$ SPT labelled by $t\in \Z_2$ such that the F-symbols of three lines in the twisted sector of $D_2^{b^{n_1}}$, $D_2^{b^{n_2}}$ and $D_2^{b^{n_3}}$ is
\begin{equation}
   F_{\Z_2}\left((D_2^{b^{n_1}},\cL^{b^{n_1}}),(D_2^{b^{n_2}},\cL^{b^{n_2}}),(D_2^{b^{n_3}},\cL^{b^{n_3}})\right)=(-1)^{tn_1n_2n_3}\,. 
\end{equation}
The $\Z_3^{(1)}$ symmetry breaking pattern is realized by a $\Z_3$ Dijkgraaf-Witten theory with the topological action given by the 3-cocycle $p\in H^3(\Z_3,U(1))$.
The lines charged under the $\Z_3^{(1)}$ symmetry are $\cL^{a}$ and $\cL^{a^2}$ whose F-symbols are 
\begin{equation}
    F_{\Z_3}(\cL^{a^{m_1}}\,, \cL^{a^{m_2}}\,, \cL^{a^{m_3}})=\exp\left\{\frac{2\pi ip}{9}m_1(m_2+m_3-[m_2+m_3]_3)\right\}\,.
\end{equation}
where $[m_2+m_3]_3=m_2+m_3 \text{ mod 3}$.
Upon gauging $\Z_3^{(0)}$, the lines $cL^{a^m}$ become attached to the symmetry defect $D_2^{a^m}$.
The $\Z_2^{(0)}$ action on $\Z_3^{(1)}$ implies the $S_3$ group composition rules
\begin{equation}
    (m_1,n_1)\cdot (m_2,n_2)= ([m_1+m_2]_2\,, [(-1)^{m_2}n_1+n_2]_3)\,.
\end{equation}
The $\Z_3$ 3-cocycle is further twisted by this outer-automorphism action, such that the resulting $S_3$ 3-cocycle is
\begin{equation}
\begin{split}
&F_{S_3}\left((D_2^{a^{m_1}b^{n_1}},\cL^{a^{m_1}b^{n_1}}),(D_2^{a^{m_2}b^{n_2}},\cL^{a^{m_2}b^{n_2}}),(D_2^{a^{m_3}b^{n_3}},\cL^{a^{m_3}b^{n_3}})\right) \\
    &=\exp\left\{\frac{2\pi ip}{9}(-1)^{n_2+n_3}m_1((-1)^{n_3}m_2+m_3-[(-1)^{n_3}m_2+m_3]_3)\right\}(-1)^{tn_1n_2n_3}\,.     
\end{split}
\end{equation}
Which is the explicit form of a 3-cocycle $(p,t)\in H^3(S_3,U(1))$ \cite{propitius1995, Coste:2000tq}.
This therefore realizes an $S_3$ SPT phase.
\begin{tcolorbox}[
colback=white,
coltitle= black,
colbacktitle=ourcolorforheader,
colframe=black,
title= $\TwoVec_{S_3}$: $S_3$ SPT Phase, 
sharp corners]
\be
\begin{split}
\begin{tikzpicture}
\begin{scope}[shift={(0,0)},scale=1] 
 \pgfmathsetmacro{\del}{-0.5}
\node at  (0,0)  {${\rm Triv}_0$} ;
\node at  (0.4,-0.6)  {\footnotesize$(S_3\text{-}{\rm SPT}_{(p,t)})$} ;
\end{scope}
\end{tikzpicture}    
\end{split}
\ee
$\TwoVec_{S_3}$ symmetry action: $D_2^{g} \simeq  1_{00}$.
\end{tcolorbox}

\subsection{$\TwoVec(\Z_3^{(1)}\rtimes \Z_2^{(0)})$ Gapped Phases}\label{sec:minimal 2grp phases}
We now describe the minimal gapped phases with the 2-group symmetry $\mathbb{G}^{(2)}= \Z_3^{(1)}\rtimes \Z_2^{(0)}$.
This has 
\begin{equation}
 \Bsym = \fB_{\Neu(\Z_3),p}\,,   
\end{equation}
and the physical boundary chosen to be one of the minimal boundaries of $\cZ(\TwoVec(S_3))$.

\subsubsection{$\Z^{(0)}_2$ SSB $\boxtimes$  $\Z_3^{(1)}$ Trivial Phase}
\label{sec:pineapple1}
Consider the physical boundary to be
\begin{equation}
    \Bphys=\fB_{\Dir}\,.
\end{equation}
The order parameters for this phase correspond to the SymTFT defects which can end on $\Bphys$.
These are the lines $\Q_1^P$ and $\Q_1^{E}$.
The line $\Q_1^{P}$ can end on both symmetry and physical boundary and therefore furnishes a topological local operator (denoted as $\cO^P$) in the IR gapped phase.
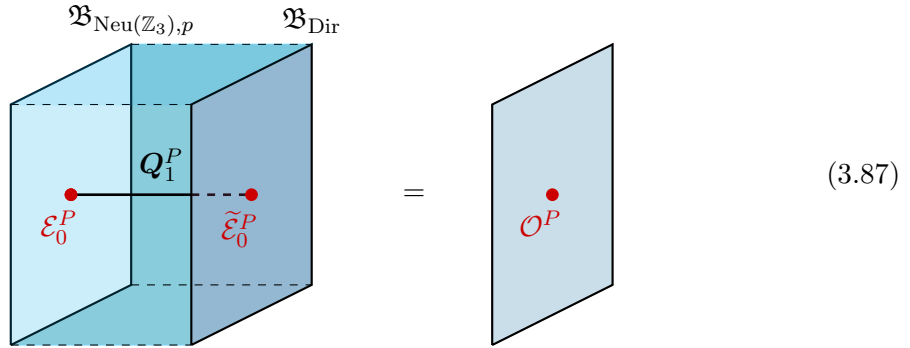
\be
\begin{split}
\begin{tikzpicture}
 \begin{scope}[shift={(0,0)},scale=0.8] 
\draw [cyan, fill=cyan!80!red, opacity =0.5]
(0,0) -- (0,4) -- (2,5) -- (5,5) -- (5,1) -- (3,0)--(0,0);
\draw [black, thick, fill=white,opacity=1]
(0,0) -- (0, 4) -- (2, 5) -- (2,1) -- (0,0);
\draw [cyan, thick, fill=cyan, opacity=0.2]
(0,0) -- (0, 4) -- (2, 5) -- (2,1) -- (0,0);
\draw[line width=1pt] (1,2.5) -- (3,2.5);
\draw[line width=1pt,dashed] (3,2.5) -- (4,2.5);
\fill[red!80!black] (1,2.5) circle (3pt);
\draw [fill=blue!40!red!60,opacity=0.2]
(3,0) -- (3, 4) -- (5, 5) -- (5,1) -- (3,0);
\fill[red!80!black] (1,2.5) circle (3pt);
\fill[red!80!black] (4,2.5) circle (3pt);
\draw [black, thick, opacity=1]
(3,0) -- (3, 4) -- (5, 5) -- (5,1) -- (3,0);
\node at (2,5.4) {$\fB_{\Neu(\Z_3),p}$};
\node at (5,5.4) {$\fB_{\Dir}$};
\draw[dashed] (0,0) -- (3,0);
\draw[dashed] (0,4) -- (3,4);
\draw[dashed] (2,5) -- (5,5);
\draw[dashed] (2,1) -- (5,1);
\draw [cyan, thick, fill=cyan, opacity=0.1]
(0,4) -- (3, 4) -- (5, 5) -- (2,5) -- (0,4);
\node[below, red!80!black] at (.8, 2.4) {$\cE_0^{P}$};
\node[below, red!80!black] at (3.8, 2.4) {$\wt{\cE}_0^{P}$};
\node  at (2.5, 3) {$\Q_{1}^{P}$};
\node  at (6.7, 2.5) {$=$};
\draw [fill=blue!60!green!30,opacity=0.7]
(8,0) -- (8, 4) -- (10, 5) -- (10,1) -- (8,0);
\draw [black, thick, opacity=1]
(8,0) -- (8, 4) -- (10, 5) -- (10,1) -- (8,0);
\fill[red!80!black] (9,2.5) circle (3pt);
\node[below, red!80!black] at (8.8, 2.4) {${\cO}^{P}$};
\end{scope}
\end{tikzpicture}    
\end{split}
\label{eq:local OP}
\ee
The algebra of this operator is inherited from the fusion rules of $\Q_1^P$ and it can be normalized such that $(\cO^P)^2=1$.
Then there are two vacua which correspond to the idempotent operators
\begin{equation}\label{eq:Z2 SSB vacua}
    v_0=\frac{1+\cO^P}{2}\,, \quad     v_1=\frac{1-\cO^P}{2}\,.
\end{equation}
Since the $\cO^P$ is charged under the $\Z_2^{(0)}$ generated by $D_2^{b}$, we have the following action on the vacua
\begin{equation}
    D_2^{b}:\quad v_0 \longleftrightarrow v_1\,.
\end{equation}
Next, the line $\Q_1^{E}$ has two ends on $\Bphys$ denoted as $\wt{\cE}_1^{E,1}$ and $\wt{\cE}_1^{E,2}$.
Similarly, it has two ends on $\Bsym$ which we denote as ${\cE}^{E,1}_1$ and ${\cE}_1^{E,2}$ which are in the twisted sectors of $D_1^{\omega}$ and $D_1^{\omega^2}$.
After compactifying the SymTFT, we obtain four twisted sector local operators 
\be
\begin{split}
\begin{tikzpicture}
 \begin{scope}[shift={(0,0)},scale=0.8] 
\draw [cyan, fill=cyan!80!red, opacity =0.5]
(0,0) -- (0,4) -- (2,5) -- (5,5) -- (5,1) -- (3,0)--(0,0);
\draw [black, thick, fill=white,opacity=1]
(0,0) -- (0, 4) -- (2, 5) -- (2,1) -- (0,0);
\draw [cyan, thick, fill=cyan, opacity=0.2]
(0,0) -- (0, 4) -- (2, 5) -- (2,1) -- (0,0);
\draw[line width=1pt] (1,2.5) -- (3,2.5);
\draw[line width=1pt,dashed] (3,2.5) -- (4,2.5);
\fill[red!80!black] (1,2.5) circle (3pt);
\draw [fill=blue!40!red!60,opacity=0.2]
(3,0) -- (3, 4) -- (5, 5) -- (5,1) -- (3,0);
\fill[red!80!black] (1,2.5) circle (3pt);
\fill[red!80!black] (4,2.5) circle (3pt);
\draw [black, thick, opacity=1]
(3,0) -- (3, 4) -- (5, 5) -- (5,1) -- (3,0);
\node at (2,5.4) {$\fB_{\Neu(\Z_3),p}$};
\node at (5,5.4) {$\fB_{\Dir}$};
\draw[dashed] (0,0) -- (3,0);
\draw[dashed] (0,4) -- (3,4);
\draw[dashed] (2,5) -- (5,5);
\draw[dashed] (2,1) -- (5,1);
\draw[line width=1pt] (1,2.5) -- (1,4.5);
\draw [cyan, thick, fill=cyan, opacity=0.1]
(0,4) -- (3, 4) -- (5, 5) -- (2,5) -- (0,4);
\node[below, red!80!black] at (.8, 2.4) {$\cE_0^{E,i}$};
\node[below, red!80!black] at (3.8, 2.4) {$\wt{\cE}_0^{E,j}$};
\node  at (2.5, 3) {$\Q_{1}^{E}$};
\node  at (0.5, 3.3) {$D_{1}^{\omega^{i}}$};
\node  at (6.7, 2.5) {$=$};
\draw [fill=blue!60!green!30,opacity=0.7]
(8,0) -- (8, 4) -- (10, 5) -- (10,1) -- (8,0);
\draw [black, thick, opacity=1]
(8,0) -- (8, 4) -- (10, 5) -- (10,1) -- (8,0);
\draw[line width=1pt] (9,2.5) -- (9,4.5);
\fill[red!80!black] (9,2.5) circle (3pt);
\node[below, red!80!black] at (8.8, 2.4) {${\cO}^{i,j}$};
\node  at (8.5, 3.3) {$D_{1}^{\omega^{i}}$};
\node[below, black] at (10.8, 3) {$,$};
\end{scope}
\end{tikzpicture}    
\end{split}
\label{eq:local OP}
\ee
which are denoted as
\begin{equation}
(D_1^{\omega^i}\,, \cO^{i,j})=\left((D_1^{\omega^i}\,, \cE_0^{i})\,, \Q_1^E\,, \wt{\cE}_0^{E,j}\right)\,, \qquad i,j=1,2\,.
\end{equation}
The $\Z_2^{(0)}$ acts on end of $\Q_1^{E}$ on $\Bsym$ as $\cE_0^{E,1}\leftrightarrow \cE_0^{E,2}$ and therefore acts on the twisted sector operators as
\begin{equation}
    D_2^{b}:\quad \cO^{i,j}\longleftrightarrow \cO^{[i+1]_2,j}\,.
\end{equation}
These operators decompose into two disjoint sets that are $\cO^{i,i}$ and $\cO^{i,[j+1]_2}$.
Operators in these sets are mutually orthogonal i.e.
\begin{equation}
   \cO^{i,i}\cO^{i,[j+1]_2}=0\,,
\end{equation}
and can normalized to satisfy the algebra
\begin{equation}
\begin{split}
    \cO^{1,1}\cO^{2,2}=v_0\,, \qquad \cO^{1,1}\cO^{1,1}=\cO^{2,2}\,, \qquad \cO^{2,2}\cO^{2,2}=\cO^{1,1}\,, \\
    \cO^{1,2}\cO^{2,1}=v_1\,, \qquad \cO^{1,2}\cO^{1,2}=\cO^{2,1}\,, \qquad \cO^{2,1}\cO^{2,1}=\cO^{1,2}\,. \\    
\end{split}
\end{equation}
From this it is clear that $\cO^{1,1}$ and $\cO^{2,2}$ trivialize the $\Z_3^{(1)}$ symmetry within the vacuum $v_0$, meanwhile $\cO^{1,2}$ and $\cO^{2,1}$ trivialize the $\Z_3^{(1)}$ symmetry within $v_1$.
As required by consistency with the $\Z_2^{(0)}$ action, we have 
\begin{equation}
    D_2^{b}:(v_0,\cO^{1,1}\,, \cO^{2,2}) \longleftrightarrow (v^1,\cO^{2,1}\,, \cO^{1,2})\,.
\end{equation}
Therefore the resulting gapped phase can be summarized as as a $\Z_2$-SSB with 1-form symmetry $\Z_3$ preserved in each vacuum, i.e. a confining phase: 
\begin{tcolorbox}[
colback=white,
coltitle= black,
colbacktitle=ourcolorforheader,
colframe=black,
title= $\TwoVec(\Z_3^{(1)}\rtimes \Z_2^{(0)})$: $\Z_2^{(0)}$ SSB $\boxtimes$ $\Z_3^{(1)}$ Trivial (Confining Phase), 
sharp corners]
\be
\begin{split}
\begin{tikzpicture}
\begin{scope}[shift={(0,0)},scale=1] 
 \pgfmathsetmacro{\del}{-0.5}
 \draw[<->, thick, rounded corners = 10pt] (1,0)--(4,0);
\node at  (2.5,0.3)  {$D_2^b $} ;
\node at  (0.3,0)  {${\rm Triv}_0 $} ;
\node at  (4.7,0)  {${\rm Triv}_1 $} ;
\node at  (0.3,-0.6)  {\footnotesize$(\Z_3^{(1)}\text{-}{\rm Triv})$} ;
\node at  (4.7,-0.6)  {\footnotesize$(\Z_3^{(1)}\text{-}{\rm Triv})$} ;
\end{scope}
\end{tikzpicture}    
\end{split}
\ee 
\end{tcolorbox}

\subsubsection{$\Z_3^{(1)}\rtimes \Z_2^{(0)}$ SSB Phase}
\label{sec:pineapple2}
Consider the physical boundary to be 
\begin{equation}
    \Bphys=\fB_{\Neu(\Z_3),p'}\,.
\end{equation}
The order parameters carry the generalized charges $\Q_1^{P}$ and $\Q_2^{[a]}$ in the corresponding gapped phase.
As in the previous section, $\Q_1^{P}$ can end on both the physical and symmetry boundary and therefore becomes a topological 
local operator $\cO^P$ after compactifying the SymTFT.
\be
\begin{split}
\begin{tikzpicture}
 \begin{scope}[shift={(0,0)},scale=0.8] 
\draw [cyan, fill=cyan!80!red, opacity =0.5]
(0,0) -- (0,4) -- (2,5) -- (5,5) -- (5,1) -- (3,0)--(0,0);
\draw [black, thick, fill=white,opacity=1]
(0,0) -- (0, 4) -- (2, 5) -- (2,1) -- (0,0);
\draw [cyan, thick, fill=cyan, opacity=0.2]
(0,0) -- (0, 4) -- (2, 5) -- (2,1) -- (0,0);
\draw[line width=1pt] (1,2.5) -- (3,2.5);
\draw[line width=1pt,dashed] (3,2.5) -- (4,2.5);
\fill[red!80!black] (1,2.5) circle (3pt);
\draw [fill=blue!40!red!60,opacity=0.2]
(3,0) -- (3, 4) -- (5, 5) -- (5,1) -- (3,0);
\fill[red!80!black] (1,2.5) circle (3pt);
\fill[red!80!black] (4,2.5) circle (3pt);
\draw [black, thick, opacity=1]
(3,0) -- (3, 4) -- (5, 5) -- (5,1) -- (3,0);
\node at (2,5.4) {$\fB_{\Neu(\Z_3),p}$};
\node at (5,5.4) {$\fB_{\Neu(\Z_3),p'}$};
\draw[dashed] (0,0) -- (3,0);
\draw[dashed] (0,4) -- (3,4);
\draw[dashed] (2,5) -- (5,5);
\draw[dashed] (2,1) -- (5,1);
\draw [cyan, thick, fill=cyan, opacity=0.1]
(0,4) -- (3, 4) -- (5, 5) -- (2,5) -- (0,4);
\node[below, red!80!black] at (.8, 2.4) {$\cE_0^{P}$};
\node[below, red!80!black] at (3.8, 2.4) {$\wt{\cE}_0^{P}$};
\node  at (2.5, 3) {$\Q_{1}^{P}$};
\node  at (6.7, 2.5) {$=$};
\draw [fill=blue!60!green!30,opacity=0.7]
(8,0) -- (8, 4) -- (10, 5) -- (10,1) -- (8,0);
\draw [black, thick, opacity=1]
(8,0) -- (8, 4) -- (10, 5) -- (10,1) -- (8,0);
\fill[red!80!black] (9,2.5) circle (3pt);
\node[below, red!80!black] at (8.8, 2.4) {${\cO}^{P}$};
\end{scope}
\end{tikzpicture}    
\end{split}
\label{eq:local OP}
\ee
There are no additional local operators and therefore we obtain two vacua $v_0$ and $v_1$ as in \eqref{eq:Z2 SSB vacua} which are exchanged under the $\Z_2^{(0)}$ symmetry.

 Next, the surface $\Q_2^{[a]}$ can also end on untwisted sector lines both the physical and symmetry boundary.
We denote its ends on the physical boundary as $\wt{\cE}_1^{a}$ and $\wt{\cE}_1^{a^2}$ and on the symmetry boundary as ${\cE}_1^{a}$ and ${\cE}_1^{a^2}$.
After compactifying the SymTFT, with the $\Q_2^{[a]}$ surface extended between both boundaries, one gets four lines as
\be
\begin{split}
\begin{tikzpicture}
 \begin{scope}[shift={(0,0)},scale=0.8] 
\draw [cyan, fill=cyan!80!red, opacity =0.5]
(0,0) -- (0,4) -- (2,5) -- (5,5) -- (5,1) -- (3,0)--(0,0);
\draw [black, thick, fill=white,opacity=1]
(0,0) -- (0, 4) -- (2, 5) -- (2,1) -- (0,0);
\draw [cyan, thick, fill=cyan, opacity=0.2]
(0,0) -- (0, 4) -- (2, 5) -- (2,1) -- (0,0);
\draw [fill=blue!40!red!60,opacity=0.2]
(3,0) -- (3, 4) -- (5, 5) -- (5,1) -- (3,0);
\draw [black, thick, opacity=1]
(3,0) -- (3, 4) -- (5, 5) -- (5,1) -- (3,0);
\node at (2,5.4) {$\fB_{\Neu(\Z_3),p}$};
\node at (5,5.4) {$\fB_{\Neu(\Z_3),p'}$};
\draw[dashed] (0,0) -- (3,0);
\draw[dashed] (0,4) -- (3,4);
\draw[dashed] (2,5) -- (5,5);
\draw[dashed] (2,1) -- (5,1);
\draw [cyan, thick, fill=cyan, opacity=0.1]
(0,4) -- (3, 4) -- (5, 5) -- (2,5) -- (0,4);
\draw [black, thick, fill=black, opacity=0.3]
(0.,2) -- (3, 2) -- (5., 3) -- (2, 3) -- (0.,2);
\node  at (2.5, 2.5) {$\Q_{2}^{[a]}$};
\node[below, red!80!black] at (-0.5, 2.4) {$\cE_1^{a^j}$};
\node[below, red!80!black] at (4, 2.4) {$\wt{\cE}_1^{a^{k}}$};
\draw [line width=1pt, red!80!black] (0.0, 2.) -- (2, 3);
\draw [line width=1pt, red!80!black] (3, 2.) -- (5, 3);
\node  at (6.7, 2.5) {$=$};
\draw [fill=blue!60!green!30,opacity=0.7]
(8,0) -- (8, 4) -- (10, 5) -- (10,1) -- (8,0);
\draw [black, thick, opacity=1]
(8,0) -- (8, 4) -- (10, 5) -- (10,1) -- (8,0);
\draw [line width=1pt, red!80!black] (8, 2.) -- (10, 3);
\node[below, red!80!black] at (9, 2.4) {${\cL}^{j,k}$};
\node[below, red!80!black] at (10, 2.4) {$,$};
\end{scope}
\end{tikzpicture}    
\end{split}
\ee
which are denoted as
\begin{equation}
    \cL^{j,k}=({\cE}^{a^j}_1\,, \Q_2^{[a]}\,, \wt{\cE}^{a^k}_1)\,, \qquad j,k=1,2\,.
\end{equation}
The lines $\cE_1^{a^j}$ and $\wt{\cE}_1^{a^j}$ at the ends of $\Q_2^{[a]}$ on the two boundaries satisfy the fusion rules and F-symbols in \eqref{eq:Z3 1fs charge line algebra}\footnote{The relations satisfied by $\cL_{i}$ are obtained by replacing $p$ with $-p'$ in \eqref{eq:Z3 1fs charge line algebra}.}
\begin{equation}
\begin{split}
    \cE_1^j\otimes \cE_1^j&=\cE_1^{[j+1]_2}\,, \qquad 
    \cE_1^j\otimes \cE_1^{[j+1]_2}=D_1^{\omega^{p}}\,, 
    \\
    \wt{\cE}_1^j\otimes \wt{\cE}_1^j&=\wt{\cE}_1^{[j+1]_2}\,, \qquad 
    \wt{\cE}_1^j\otimes \wt{\cE}_1^{[j+1]_2}=D_1^{\omega^{p'}}\,. 
\end{split}
\end{equation}
Using these and consistency with the bulk fusion of 
$\Q_2^{[a]}$, we obtain the following fusion rules for $\cL^{j,k}$.
\begin{equation}
    \cL^{j,j}\cL^{[j+1]_2,j}=0\,, \qquad 
         \cL^{j,j}\cL^{j,[j+1]_2}=0\,.
\end{equation}
The lines therefore decompose into two sets $\{\cL^{1,1},\cL^{2,2}\}$ and $\{\cL^{1,2},\cL^{2,1}\}$ that have vanishing fusion product among them.
These sets of lines are exchanged under the $\Z_2^{(0)}$ symmetry and act on distinct vacua of the $\Z_2^{(0)}$ SSB phase.
Without loss of generality, we may assume that $\{\cL^{1,1}\,,\cL^{2,2}\}$ act within the vacuum $v_0$ while $\{\cL^{1,2}\,,\cL^{2,1}\}$ act within the vacuum $v_1$.
The fusion rules of these lines can be determined to be
\begin{equation}
\begin{split}   
    \cL^{1,1}\cL^{1,1}&=\cL^{2,2}\,, \\ 
    \cL^{2,2}\cL^{1,1}&=D_1^{\omega^{p-p'}} \\
    \cL^{1,2}\cL^{1,2}&=D_1^{\omega^{-p'}}\cL^{2,1}\,, \\
    \cL^{2,1}\cL^{1,2}&=D_1^{\omega^{p-p'}}\,. \\
\end{split}
\end{equation}
These lines are charged under the $\Z_3^{(1)}$ symmetry such that the line $\cL^{i,j}$ has a braiding phase of $\omega^{in}$ with $D_1^{\omega^n}$.
Finally, there F-symbols incur contributions form both the symmetry and physical boundaries that are labeled by $p$ and $p'$ in $H^3(\Z_3,U(1))$ respectively.
Concretely these are
\begin{equation}
    F(\cL^{i,i'}\,, \cL^{j,j'}\,, \cL^{k,k'})=\exp\left\{\frac{2\pi ip}{9}i(j+k-[j+k]_3)
    -
    \frac{2\pi ip'}{9}i'(j'+k'-[j'+k']_3)
    \right\}\,.
\end{equation}
As this phase has 2 vacua exchanged by $\Z_2^{(0)}$ and topological lines charged under the $\Z_3^{(1)}$ symmetry, it is the phase where the full 2-group symmetry is spontaneously broken.
The symmetry breaking pattern of $\Z_3^{(1)}$ symmetry is labeled by $p-p'\in H^{3}(\Z_3,U(1))$ and the underlying TFT realizing this phase is a 2-group SSB phase: 
\begin{tcolorbox}[
colback=white,
coltitle= black,
colbacktitle=ourcolorforheader,
colframe=black,
title= $\TwoVec(\Z_3^{(1)}\rtimes \Z_2^{(0)})$: $\Z_3^{(1)}\rtimes \Z_2^{(0)}$ 2-Group SSB Phase, 
sharp corners]
\be
\begin{split}
\begin{tikzpicture}
\begin{scope}[shift={(0,0)},scale=1] 
 \pgfmathsetmacro{\del}{-0.5}
 \draw[<->, thick, rounded corners = 10pt] (1,0)--(4,0);
\node at  (2.5,0.3)  {$D_2^b $} ;
\node at  (-0.3,0)  {${\rm DW(\Z_3)}_{p-p'}$} ;
\node at  (5.3,0)  {${\rm DW(\Z_3)}_{p-p'}$} ;
\node at  (-0.3,-0.6)  {\footnotesize$(\Z_3^{(1)}\text{-}{\rm SSB})$} ;
\node at  (5.3,-0.6)  {\footnotesize$(\Z_3^{(1)}\text{-}{\rm SSB})$} ;
\end{scope}
\end{tikzpicture}    
\end{split}
\ee 
\end{tcolorbox}

\subsubsection{$\Z_2^{(0)}$ SPT Phase}
\label{sec:pineapple3}
Consider the physical boundary to be 
\begin{equation}
    \Bphys=\fB_{\Neu(\Z_2),t}\,.
\end{equation}
The order parameters are given by the defects that can end on $\Bphys$. 
These are the line $\Q_1^{E}$ and the surface $\Q_2^{[b]}$.
The line $\Q_1^{E}$ does not end on $\Bsym$, therefore after compactifying the SymTFT, one obtains a single topological local operator, which is the identity.
Correspondingly, there is a single vacuum on which the $\Z_2^{0}$ acts trivially. 
The $\Q_1^{E}$ line has two ends on $\Bsym$ which are uncharged and in the twisted sectors of the $\Z_3^{(1)}$ generators.
\be
\begin{split}
\begin{tikzpicture}
 \begin{scope}[shift={(0,0)},scale=0.8] 
\draw [cyan, fill=cyan!80!red, opacity =0.5]
(0,0) -- (0,4) -- (2,5) -- (5,5) -- (5,1) -- (3,0)--(0,0);
\draw [black, thick, fill=white,opacity=1]
(0,0) -- (0, 4) -- (2, 5) -- (2,1) -- (0,0);
\draw [cyan, thick, fill=cyan, opacity=0.2]
(0,0) -- (0, 4) -- (2, 5) -- (2,1) -- (0,0);
\draw[line width=1pt] (1,2.5) -- (3,2.5);
\draw[line width=1pt,dashed] (3,2.5) -- (4,2.5);
\fill[red!80!black] (1,2.5) circle (3pt);
\draw [fill=blue!40!red!60,opacity=0.2]
(3,0) -- (3, 4) -- (5, 5) -- (5,1) -- (3,0);
\fill[red!80!black] (1,2.5) circle (3pt);
\fill[red!80!black] (4,2.5) circle (3pt);
\draw [black, thick, opacity=1]
(3,0) -- (3, 4) -- (5, 5) -- (5,1) -- (3,0);
\node at (2,5.4) {$\fB_{\Neu(\Z_3),p}$};
\node at (5,5.4) {$\fB_{\Neu(\Z_2),t}$};
\draw[dashed] (0,0) -- (3,0);
\draw[dashed] (0,4) -- (3,4);
\draw[dashed] (2,5) -- (5,5);
\draw[dashed] (2,1) -- (5,1);
\draw[line width=1pt] (1,2.5) -- (1,4.5);
\draw [cyan, thick, fill=cyan, opacity=0.1]
(0,4) -- (3, 4) -- (5, 5) -- (2,5) -- (0,4);
\node[below, red!80!black] at (.8, 2.4) {$\cE_0^{E,i}$};
\node[below, red!80!black] at (3.8, 2.4) {$\wt{\cE}_0^{E,j}$};
\node  at (2.5, 3) {$\Q_{1}^{E}$};
\node  at (0.5, 3.3) {$D_{1}^{\omega^{i}}$};
\node  at (6.7, 2.5) {$=$};
\draw [fill=blue!60!green!30,opacity=0.7]
(8,0) -- (8, 4) -- (10, 5) -- (10,1) -- (8,0);
\draw [black, thick, opacity=1]
(8,0) -- (8, 4) -- (10, 5) -- (10,1) -- (8,0);
\draw[line width=1pt] (9,2.5) -- (9,4.5);
\fill[red!80!black] (9,2.5) circle (3pt);
\node[below, red!80!black] at (8.8, 2.4) {${\cO}^{i,j}$};
\node  at (8.5, 3.3) {$D_{1}^{\omega^{i}}$};
\node[below, black] at (10.8, 3) {$,$};
\end{scope}
\end{tikzpicture}    
\end{split}
\ee
The resulting gapped phase is therefore trivial from the point of view of 1-form symmetry.
The $\Q_2^{[b]}$ surface ends on a single untwisted line $\wt{\cE}_1^{b}$ on $\Bphys$ which has F-symbols given by $t$.
On $\Bsym$, $\Q_2^{[b]}$ ends on the line $(D_2^{b}\,,\wt{\cE}_1^{[b]})$.
After compactifying the SymTFT, one obtains a line in the $D_2^{b}$ twisted sector with F-symbols given by $t\in H^{3}(\Z_2,U(1))$:
\be
\begin{split}
\begin{tikzpicture}
 \begin{scope}[shift={(0,0)},scale=0.8] 
\draw [cyan, fill=cyan!80!red, opacity =0.5]
(0,0) -- (0,4) -- (2,5) -- (5,5) -- (5,1) -- (3,0)--(0,0);
\draw [black, thick, fill=white,opacity=1]
(0,0) -- (0, 4) -- (2, 5) -- (2,1) -- (0,0);
\draw [cyan, thick, fill=cyan, opacity=0.2]
(0,0) -- (0, 4) -- (2, 5) -- (2,1) -- (0,0);
\draw [fill=blue!40!red!60,opacity=0.2]
(3,0) -- (3, 4) -- (5, 5) -- (5,1) -- (3,0);
\draw [black, thick, opacity=1]
(3,0) -- (3, 4) -- (5, 5) -- (5,1) -- (3,0);
\node at (2,5.4) {$\fB_{\Neu(\Z_3),p}$};
\node at (5,5.4) {$\fB_{\Neu(\Z_2),t}$};
\draw[dashed] (0,0) -- (3,0);
\draw[dashed] (0,4) -- (3,4);
\draw[dashed] (2,5) -- (5,5);
\draw[dashed] (2,1) -- (5,1);
\draw [cyan, thick, fill=cyan, opacity=0.1]
(0,4) -- (3, 4) -- (5, 5) -- (2,5) -- (0,4);
\draw [black, thick, fill=black, opacity=0.3]
(0.,2) -- (3, 2) -- (5., 3) -- (2, 3) -- (0.,2);
\draw [black, thick, fill=black, opacity=0.3]
(0.,2) -- (0., 4) -- (2, 5) -- (2,3) -- (0,2);
\node  at (2.5, 2.5) {$\Q_{2}^{[b]}$};
\node[below, red!80!black] at (-0.5, 2.4) {$\cE_1^{[b]}$};
\node[below, red!80!black] at (4, 2.4) {$\wt{\cE}_1^{b}$};
\draw [line width=1pt, red!80!black] (0.0, 2.) -- (2, 3);
\draw [line width=1pt, red!80!black] (3, 2.) -- (5, 3);
\node  at (6.7, 2.5) {$=$};
\draw [fill=blue!60!green!30,opacity=0.7]
(8,0) -- (8, 4) -- (10, 5) -- (10,1) -- (8,0);
\draw [black, thick, opacity=1]
(8,0) -- (8, 4) -- (10, 5) -- (10,1) -- (8,0);
\draw [black, thick, fill=black, opacity=0.3]
(8,2) -- (8, 4) -- (10, 5) -- (10,3) -- (8,2);
\draw [line width=1pt, red!80!black] (8, 2.) -- (10, 3);
\node[below, black] at (9, 3.9) {$D_2^{b}$};
\node[below, black] at (1., 3.9) {$D_2^{b}$};
\node[below, red!80!black] at (9, 2.4) {${\cL}^{b}$};
\end{scope}
\end{tikzpicture}    
\end{split}
\ee
This gapped phase is therefore a $\Z_2$ SPT with $\Z_3^{(1)}$ realized trivially as the identity line.

\begin{tcolorbox}[
colback=white,
coltitle= black,
colbacktitle=ourcolorforheader,
colframe=black,
title= $\TwoVec(\Z_3^{(1)}\rtimes \Z_2^{(0)})$: $\Z_3^{(1)}\rtimes \Z_2^{(0)}$ SPT Phase and $\Z_3^{(1)}$ Trivial (Confining Phase), 
sharp corners]
\be
\begin{split}
\begin{tikzpicture}
\begin{scope}[shift={(0,0)},scale=1] 
 \pgfmathsetmacro{\del}{-0.5}
\node at  (0,0)  {${\rm Triv}_0$} ;
\node at  (0,-0.6)  {\footnotesize$(\Z_3^{(1)}\text{-}{\rm Triv}\,, \Z_2^{(0)}\text{-SPT}_t)$} ;
\end{scope}
\end{tikzpicture}    
\end{split}
\ee 
\end{tcolorbox}

\subsubsection{$\Z_3^{(1)}$ SSB Phase}
\label{sec:pineapple4}
Consider the physical boundary to be 
\begin{equation}
    \Bphys=\fB_{\Neu(S_3),(p,t)}\,,
\end{equation}
The SymTFT surfaces $\Q_2^{[a]}$ and $\Q_{2}^{[b]}$, end on $\Bphys$ on untwisted sector lines which we denote as $\wt{\cE}_1^{[a]}$ and $\wt{\cE}_1^{[b]}$.
Consequently the order parameters for this gapped phase carry the generalized charges $\Q_2^{[a]}$ and $\Q_{2}^{[b]}$.
The $\Q_2^{[a]}$ surface has two untwisted ends on $\Bsym$ denoted ${\cE}_1^{a}$ and ${\cE}_1^{a^2}$.
These provide two topological lines in the IR TQFT describing this gapped phase as
\be
\begin{split}
\begin{tikzpicture}
 \begin{scope}[shift={(0,0)},scale=0.8] 
\draw [cyan, fill=cyan!80!red, opacity =0.5]
(0,0) -- (0,4) -- (2,5) -- (5,5) -- (5,1) -- (3,0)--(0,0);
\draw [black, thick, fill=white,opacity=1]
(0,0) -- (0, 4) -- (2, 5) -- (2,1) -- (0,0);
\draw [cyan, thick, fill=cyan, opacity=0.2]
(0,0) -- (0, 4) -- (2, 5) -- (2,1) -- (0,0);
\draw [fill=blue!40!red!60,opacity=0.2]
(3,0) -- (3, 4) -- (5, 5) -- (5,1) -- (3,0);
\draw [black, thick, opacity=1]
(3,0) -- (3, 4) -- (5, 5) -- (5,1) -- (3,0);
\node at (2,5.4) {$\fB_{\Neu(\Z_3),p}$};
\node at (5,5.4) {$\fB_{\Neu(S_3),(p',t)}$};
\draw[dashed] (0,0) -- (3,0);
\draw[dashed] (0,4) -- (3,4);
\draw[dashed] (2,5) -- (5,5);
\draw[dashed] (2,1) -- (5,1);
\draw [cyan, thick, fill=cyan, opacity=0.1]
(0,4) -- (3, 4) -- (5, 5) -- (2,5) -- (0,4);
\draw [black, thick, fill=black, opacity=0.3]
(0.,2) -- (3, 2) -- (5., 3) -- (2, 3) -- (0.,2);
\node  at (2.5, 2.5) {$\Q_{2}^{[a]}$};
\node[below, red!80!black] at (-0.5, 2.4) {$\cE_1^{a^j}$};
\node[below, red!80!black] at (4, 2.4) {$\wt{\cE}_1^{[a]}$};
\draw [line width=1pt, red!80!black] (0.0, 2.) -- (2, 3);
\draw [line width=1pt, red!80!black] (3, 2.) -- (5, 3);
\node  at (6.7, 2.5) {$=$};
\draw [fill=blue!60!green!30,opacity=0.7]
(8,0) -- (8, 4) -- (10, 5) -- (10,1) -- (8,0);
\draw [black, thick, opacity=1]
(8,0) -- (8, 4) -- (10, 5) -- (10,1) -- (8,0);
\draw [line width=1pt, red!80!black] (8, 2.) -- (10, 3);
\node[below, red!80!black] at (9, 2.4) {${\cL}^{j}$};
\node[below, red!80!black] at (10, 2.4) {$,$};
\end{scope}
\end{tikzpicture}    
\end{split}
\ee
We denote the two genuine lines obtained after compactifying the SymTFT with $\Q_2^{[a]}$ stretched between both ends as $\cL^{j}$.
Using compatibility between bulk fusions of $\Q_{2}^{[a]}$ and fusion of the ends $\cE_1^{a^j}$ and $\wt{\cE}_1^{[a]}$.
It follows that $\cL^{j}$ have the fusion rules of flux lines in $\Z_3$ DW theory with the topological action given by $p-p'\in H^{3}(\Z_3,U(1))$.
These ends braid with the $\Z_3$ 1-form generators on the symmetry boundary and therefore correspond to the $\Z_3^{(1)}$ SSB phase.
Finally the surface $\Q_2^{[b]}$ has a single untwisted end $\wt{\cE}_1^{[b]}$ on the physical boundary and a single twisted end $(D_2^{b}/\cA_{\Z_3}, \cE^{[b]}_1)$ on the symmetry boundary.
Therefore the generalized charge corresponding to $\Q_2^{[b]}$ is in the $D_2^{b}$ twisted sector implying that the $\Z_2^{(0)}$ symmetry is spontaneously preserved in this phase.

\medskip\noindent To summarize, this gapped phase corresponds to a $\Z_3^{(1)}$ SSB phase and is realized by a 3d DW theory with a 3-cocycle twist determined by $p$ and $p'$.
\begin{tcolorbox}[
colback=white,
coltitle= black,
colbacktitle=ourcolorforheader,
colframe=black,
title= $\TwoVec(\Z_3^{(1)}\rtimes \Z_2^{(0)})$: $\Z_3^{(1)}$ SSB Phase (Deconfined Phase), 
sharp corners]
\be
\begin{split}
\begin{tikzpicture}
\begin{scope}[shift={(0,0)},scale=1] 
 \pgfmathsetmacro{\del}{-0.5}
\node at  (0,0)  {${\rm DW}(\Z_3)_{p-p'}$} ;
\node at  (0,-0.6)  {\footnotesize$(\Z_3^{(1)}\text{-SSB})$} ;
\end{scope}
\end{tikzpicture}    
\end{split}
\ee 
\end{tcolorbox}

\subsection{$\TwoRep(\Z_3^{(1)}\rtimes \Z_2^{(0)})$ Gapped Phases}
We now consider the symmetry boundary to be 
\begin{equation}
    \Bsym=\fB_{\Neu(\Z_2),t}\,,
\end{equation}
which carries the $\TwoRep(\Z_3^{(1)}\rtimes \Z_2^{(0)})$ symmetry on it, which is a non-invertible symmetry of 2-representations of the 2-group.

\subsubsection{$\TwoRep(\Z_3^{(1)}\rtimes \Z_2^{(0)})/\Z_2^{(1)}$ SSB Phase}
\label{sec:apple1}
Consider the physical boundary to be 
\begin{equation}
    \Bphys=\fB_{\Dir}\,.
\end{equation}
The order parameters carry the generalized charges $\Q_1^P$ and $\Q_1^{E}$ as these are the bulk lines that have untwisted ends on $\Bphys$.
First, let us consider the bulk line $\Q_1^{E}$.
This has two untwisted sector ends on $\Bphys$ denoted as $\wt\cE_{0}^{E,1}$ and $\wt\cE_0^{E,2}$.
On $\Bsym$, $\Q_1^{E}$ has one untwisted end  
${\cE}_{0}^{E,+}={\cE}_{0}^{E,1}+{\cE}_{0}^{E,2}$ and one $D_{1}^{\wh{b}}$ twisted end ${\cE}_{0}^{E,-}={\cE}_{0}^{E,1}-{\cE}_{0}^{E,2}$.
Compactifying the SymTFT, one obtains a three dimensional space of untwisted local operators spanned by
\begin{equation}
    \left\{1\,, \cO^{1}\,, \cO^{2}\right\}\,,
\end{equation}
where $\cO^{i}$ denotes the local operator obtained by compactifying $\Q_1^{E}$ with the ends ${\cE}_0^{E,+}$ and $\wt\cE_0^{E,i}$ on the symmetry and physical boundary respectively.
\be
\begin{split}
\begin{tikzpicture}
 \begin{scope}[shift={(0,0)},scale=0.8] 
\draw [cyan, fill=cyan!80!red, opacity =0.5]
(0,0) -- (0,4) -- (2,5) -- (5,5) -- (5,1) -- (3,0)--(0,0);
\draw [black, thick, fill=white,opacity=1]
(0,0) -- (0, 4) -- (2, 5) -- (2,1) -- (0,0);
\draw [cyan, thick, fill=cyan, opacity=0.2]
(0,0) -- (0, 4) -- (2, 5) -- (2,1) -- (0,0);
\draw[line width=1pt] (1,2.5) -- (3,2.5);
\draw[line width=1pt,dashed] (3,2.5) -- (4,2.5);
\fill[red!80!black] (1,2.5) circle (3pt);
\draw [fill=blue!40!red!60,opacity=0.2]
(3,0) -- (3, 4) -- (5, 5) -- (5,1) -- (3,0);
\fill[red!80!black] (1,2.5) circle (3pt);
\fill[red!80!black] (4,2.5) circle (3pt);
\draw [black, thick, opacity=1]
(3,0) -- (3, 4) -- (5, 5) -- (5,1) -- (3,0);
\node at (2,5.4) {$\fB_{\Neu(\Z_2),t}$};
\node at (5,5.4) {$\fB_{\Dir}$};
\draw[dashed] (0,0) -- (3,0);
\draw[dashed] (0,4) -- (3,4);
\draw[dashed] (2,5) -- (5,5);
\draw[dashed] (2,1) -- (5,1);
\draw [cyan, thick, fill=cyan, opacity=0.1]
(0,4) -- (3, 4) -- (5, 5) -- (2,5) -- (0,4);
\node[below, red!80!black] at (.8, 2.4) {$\cE_0^{E,+}$};
\node[below, red!80!black] at (3.8, 2.4) {$\wt{\cE}_0^{E,i}$};
\node  at (2.5, 3) {$\Q_{1}^{E}$};
\node  at (6.7, 2.5) {$=$};
\draw [fill=blue!60!green!30,opacity=0.7]
(8,0) -- (8, 4) -- (10, 5) -- (10,1) -- (8,0);
\draw [black, thick, opacity=1]
(8,0) -- (8, 4) -- (10, 5) -- (10,1) -- (8,0);
\fill[red!80!black] (9,2.5) circle (3pt);
\node[below, red!80!black] at (8.8, 2.4) {${\cO}^{i}$};
\end{scope}
\end{tikzpicture}    
\end{split}
\label{eq:local OP}
\ee
This SymTFT sandwich is denoted as
\begin{equation}
   \cO^{i}= ({\cE}_0^{E,+}\,,\Q_1^E\,,\wt\cE_0^{E,i})\,. 
\end{equation}
The algebra of these operators can be obtained from the algebra of ends of $\Q_1^{E}$. 
Recall that these ends transformed as a doublet under the $S_3$ on $\fB_{\Dir}$.
Then by using the intertwiners of $\Rep(S_3)$, one obtains
\begin{equation}
\begin{split}
\cE_0^{E,1}\cE_0^{E,1}&=\cE_0^{E,2}\,, \\
\cE_0^{E,2}\cE_0^{E,2}&=\cE_0^{E,1}\,, \\ 
\cE_0^{E,1}\cE_0^{E,2}&=\lambda(1+\cE_0^{P})\,, \\
\cE_0^{E,2}\cE_0^{E,1}&=\lambda(1-\cE_0^{P})\,.
\end{split}
\end{equation}
with exactly the same relations for $\wt\cE^{E,i}_0$.
These rules imply that 
\begin{equation}
\begin{split}
\label{eq:fusion of E end ops}
\cE_0^{E,+}\cE_0^{E,+}&=\cE_0^{E,+}+2\lambda\,, \\   
\cE_0^{E,-}\cE_0^{E,-}&=\cE_0^{E,+}-2\lambda\,, \\   
\cE_0^{E,-}\cE_0^{E,+}&=-\cE_0^{E,-}+2\lambda\cE_0^P\,, \\   
\end{split}
\end{equation}
Using these one can compute the OPE of operators local operators $\cO^1$ and $\cO^2$ as
\begin{equation}
\begin{split}
    \cO^1 \cO^1&= ({\cE}_0^{E,+}{\cE}_0^{E,+}\,,\Q_1^E\otimes \Q_1^E\,,\wt\cE_0^{E,1}\wt\cE_0^{E,1}) \\ 
    &= ({\cE}_0^{E,+}\,,\Q_1^E\,,\wt\cE_0^{E,2})\,, \\
    &=\cO^2\,, \\
    \cO^1 \cO^2&=({\cE}_0^{E,+}{\cE}_0^{E,+}\,,\Q_1^E\otimes \Q_1^E\,,\wt\cE_0^{E,1}\wt\cE_0^{E,2}) \\ 
    &= (2\lambda\,,\Q_1^{\id}\,,\lambda) \\
    &=2\lambda^2\,.
\end{split}
\end{equation}
We choose the normalization $\lambda=1/\sqrt{2}$, such that the fusion rules take the convenient form
Similarly, we get
\begin{equation}
   \cO^1 \cO^2=\cO^2\cO^1=1\,,\qquad \cO^1\cO^1=\cO^2\,, \qquad \cO^2\cO^2=\cO^1\,.
\end{equation}
These are $\Z_3$ fusion rules and the idempotents (vacua) are constructed by a $\Z_3$ fourier transform as
\begin{equation}
    v_j= \frac{1+\omega^j \cO^1+\omega^{2j}\cO^2}{3}\,.
\end{equation}
To compute the action of the non-invertible symmetry $D_2^A$ on these vacua, we recall that $D_2^{A}$ descends from $D_2^{a}\oplus D_2^{a^2}$ on upon gauging $\Z_2^{b}$ on $\Bsym$.
The sphere linking action of $D_2^{a}\oplus D_2^{a^2}$ on $\cE^{E,j}$ is given by $\omega^j+\omega^{2j}=-1$.
Therefore we have 
\begin{equation}
    D_2^{A}:(1,\cO^{1}\,, \cO^{2}) \longmapsto 
    (2,-\cO^{1}\,, -\cO^{2})\,.
\end{equation}
The action on the vacua can be immediately read off
\begin{equation}
    D_2^{A}:(v_0\,, v_1\,, v_2)\longmapsto (v_1+ v_2\,, v_2+v_0\,, v_0+v_1)\,.
\end{equation}
This is rather unsurprising as this is essentially the $\Z_3^{a}$ broken phase where $D_2^{A}$ is realized as $D_2^a\oplus D_2^{a^2}$.
Next we move on to the $\Z_2$ 1-form symmetry generated by $D_1^{\wh{b}}$.
There are three $D_1^{\widehat{b}}$ twisted order parameters in this gapped phase. 
These are
\begin{equation}
\begin{split}
    (D_1^{\wh{b}}\,,\cO^P)&=\left((D_1^{\wh{b}}\,,{\cE}_0^P)\,, \Q_1^P\,, \wt\cE_0^P\right)\,, \\    
(D_1^{\wh{b}}\,,\overline{\cO}^1)&=\left((D_1^{\wh{b}}\,,{\cE}_0^{E,-})\,, \Q_1^E\,, \wt\cE^{E,1}_0\right)\,, \\    
(D_1^{\wh{b}}\,,\overline{\cO}^2)&=\left((D_1^{\wh{b}}\,,{\cE}_0^{E,-})\,, \Q_1^E\,, \wt\cE^{E,2}_0\right)\,.     
\end{split}
\end{equation}
Using similar methods as above, the algebra of these operators can be computed and we can construct their linear combinations that act within the three vacua $v_{0}\,, v_1$ and $v_2$ and are mapped into each other approprately under the $D_2^A$ action.
Since there is an end of $D_1^{\wh{b}}$ line in each vacuum, the $\Z_2^{(1)}$ symmetry acts trivially in this phase.
\begin{tcolorbox}[
colback=white,
coltitle= black,
colbacktitle=ourcolorforheader,
colframe=black,
title= $\TwoRep(\Z_3^{(1)}\rtimes \Z_2^{(0)})$: $\TwoRep(\Z_3^{(1)}\rtimes \Z_2^{(0)})/\Z_2^{(1)}$ SSB with Trivial $\Z_2^{(1)}$ (Confining Phase), 
sharp corners]
\be
\begin{split}
\begin{tikzpicture}
\begin{scope}[shift={(0,0)},scale=1] 
 \pgfmathsetmacro{\del}{-0.5}
\node at  (0,0)  {${\rm Triv}_0 \quad \boxplus \quad {\rm Triv}_1 \quad \boxplus \quad {\rm Triv}_2$} ;
\node at  (-2,-0.6)  {\footnotesize$(\Z_2^{(1)}\text{-}{\rm Triv})$} ;
\node at  (0,-0.6)  {\footnotesize$(\Z_2^{(1)}\text{-}{\rm Triv})$} ;
\node at  (2,-0.6)  {\footnotesize$(\Z_2^{(1)}\text{-}{\rm Triv})$} ;
\end{scope}
\end{tikzpicture}    
\end{split}
\ee
The non-invertible 0-form symmetry is realized on the vacua as
\begin{equation}
    D_2^{A}=1_{01}\oplus 1_{02}\oplus 1_{12}\oplus 1_{10}\oplus 1_{21}\oplus 1_{20}\,. 
\end{equation}
\end{tcolorbox}

\subsubsection{$\TwoRep(\Z_3^{(1)}\rtimes \Z_2^{(0)})$ SPT Phase}
\label{sec:apple2}
Let us consider the physical boundary to be
\begin{equation}
    \Bphys=\fB_{\Neu(\Z_3),p}\,.
\end{equation}
The order parameters correspond to the generalized charges $\Q_1^{P}$ and $\Q_2^{[a]}$.
Firstly there in no bulk SymTFT line that ends on an untwisted operator on both boundaries.
Therefore after compactifying the SymTFT, one obtains a unique topological operator which is the identity.
Correspondingly, this phase has a unique vacuum.
Next, compactifying the SymTFT with the $\Q_1^{P}$ between both boundaries produces a local operator attached to the $D_1^{\wh{b}}$ line.
\begin{equation}
    (D_{1}^{\wh{b}}\,, \cO^{P})=\left((D_{1}^{\wh{b}}\,,{\cE}_0^P)\,, \Q_1^P\,, \wt\cE_0^P\right)\,.
\end{equation}
This configuration is depicted as
\be
\begin{split}
\begin{tikzpicture}
 \begin{scope}[shift={(0,0)},scale=0.8] 
\draw [cyan, fill=cyan!80!red, opacity =0.5]
(0,0) -- (0,4) -- (2,5) -- (5,5) -- (5,1) -- (3,0)--(0,0);
\draw [black, thick, fill=white,opacity=1]
(0,0) -- (0, 4) -- (2, 5) -- (2,1) -- (0,0);
\draw [cyan, thick, fill=cyan, opacity=0.2]
(0,0) -- (0, 4) -- (2, 5) -- (2,1) -- (0,0);
\draw[line width=1pt] (1,2.5) -- (3,2.5);
\draw[line width=1pt,dashed] (3,2.5) -- (4,2.5);
\fill[red!80!black] (1,2.5) circle (3pt);
\draw [fill=blue!40!red!60,opacity=0.2]
(3,0) -- (3, 4) -- (5, 5) -- (5,1) -- (3,0);
\fill[red!80!black] (1,2.5) circle (3pt);
\fill[red!80!black] (4,2.5) circle (3pt);
\draw [black, thick, opacity=1]
(3,0) -- (3, 4) -- (5, 5) -- (5,1) -- (3,0);
\node at (2,5.4) {$\fB_{\Neu(\Z_2),t}$};
\node at (5,5.4) {$\fB_{\Neu(\Z_3),p}$};
\draw[dashed] (0,0) -- (3,0);
\draw[dashed] (0,4) -- (3,4);
\draw[dashed] (2,5) -- (5,5);
\draw[dashed] (2,1) -- (5,1);
\draw[line width=1pt] (1,2.5) -- (1,4.5);
\draw [cyan, thick, fill=cyan, opacity=0.1]
(0,4) -- (3, 4) -- (5, 5) -- (2,5) -- (0,4);
\node[below, red!80!black] at (.8, 2.4) {$\cE_0^{P}$};
\node[below, red!80!black] at (3.8, 2.4) {$\wt{\cE}_0^{P}$};
\node  at (2.5, 3) {$\Q_{1}^{P}$};
\node  at (0.5, 3.3) {$D_{1}^{\wh{b}}$};
\node  at (6.7, 2.5) {$=$};
\draw [fill=blue!60!green!30,opacity=0.7]
(8,0) -- (8, 4) -- (10, 5) -- (10,1) -- (8,0);
\draw [black, thick, opacity=1]
(8,0) -- (8, 4) -- (10, 5) -- (10,1) -- (8,0);
\draw[line width=1pt] (9,2.5) -- (9,4.5);
\fill[red!80!black] (9,2.5) circle (3pt);
\node[below, red!80!black] at (8.8, 2.4) {${\cO}^{P}$};
\node  at (8.5, 3.3) {$D_{1}^{\wh{b}}$};
\node[below, black] at (10.8, 3) {$,$};
\end{scope}
\end{tikzpicture}    
\end{split}
\ee
Therefore the $D_1^{\wh{b}}$ line is isomorphic to the identity in this phase.
Finally, compactifying the SymTFT with the $\Q_2^{[a]}$ surface extended between both boundaries.
There are two $D_2^{A}$ twisted sector lines one obtains upon such a compactification, which are labeled by the ends of the $\Q_2^{[a]}$ surface on $\Bphys$
\be
\begin{split}
\begin{tikzpicture}
 \begin{scope}[shift={(0,0)},scale=0.8] 
\draw [cyan, fill=cyan!80!red, opacity =0.5]
(0,0) -- (0,4) -- (2,5) -- (5,5) -- (5,1) -- (3,0)--(0,0);
\draw [black, thick, fill=white,opacity=1]
(0,0) -- (0, 4) -- (2, 5) -- (2,1) -- (0,0);
\draw [cyan, thick, fill=cyan, opacity=0.2]
(0,0) -- (0, 4) -- (2, 5) -- (2,1) -- (0,0);
\draw [fill=blue!40!red!60,opacity=0.2]
(3,0) -- (3, 4) -- (5, 5) -- (5,1) -- (3,0);
\draw [black, thick, opacity=1]
(3,0) -- (3, 4) -- (5, 5) -- (5,1) -- (3,0);
\node at (2,5.4) {$\fB_{\Neu(\Z_2),t}$};
\node at (5,5.4) {$\fB_{\Neu(\Z_3),p}$};
\draw[dashed] (0,0) -- (3,0);
\draw[dashed] (0,4) -- (3,4);
\draw[dashed] (2,5) -- (5,5);
\draw[dashed] (2,1) -- (5,1);
\draw [cyan, thick, fill=cyan, opacity=0.1]
(0,4) -- (3, 4) -- (5, 5) -- (2,5) -- (0,4);
\draw [black, thick, fill=black, opacity=0.3]
(0.,2) -- (3, 2) -- (5., 3) -- (2, 3) -- (0.,2);
\draw [black, thick, fill=black, opacity=0.3]
(0.,2) -- (0., 4) -- (2, 5) -- (2,3) -- (0,2);
\node  at (2.5, 2.5) {$\Q_{2}^{[a]}$};
\node[below, red!80!black] at (-0.5, 2.4) {$\cE_1^{A}$};
\node[below, red!80!black] at (4, 2.4) {$\wt{\cE}_1^{a^j}$};
\draw [line width=1pt, red!80!black] (0.0, 2.) -- (2, 3);
\draw [line width=1pt, red!80!black] (3, 2.) -- (5, 3);
\node  at (6.7, 2.5) {$=$};
\draw [fill=blue!60!green!30,opacity=0.7]
(8,0) -- (8, 4) -- (10, 5) -- (10,1) -- (8,0);
\draw [black, thick, opacity=1]
(8,0) -- (8, 4) -- (10, 5) -- (10,1) -- (8,0);
\draw [black, thick, fill=black, opacity=0.3]
(8,2) -- (8, 4) -- (10, 5) -- (10,3) -- (8,2);
\draw [line width=1pt, red!80!black] (8, 2.) -- (10, 3);
\node[below, black] at (9, 3.9) {$D_2^{A}$};
\node[below, black] at (1., 3.9) {$D_2^{A}$};
\node[below, red!80!black] at (9, 2.4) {${\cL}^{j}$};
\node[below, black] at (11, 2.4) {$,$};
\end{scope}
\end{tikzpicture}    
\end{split}
\ee
denoted as
\begin{equation}
   (D_{2}^{A}\,, \cL^{i})=(\wt{\cE}_1^b\,, \Q_2^{[a]}\,, \cE^{a^i}_1)\,, \quad i=1,2\,. 
\end{equation}
The F-symbols of these lines are controlled by $p\in H^3(\Z_3\,,U(1))$.
We leave the detailed computation of these F-symbols for the future.

\begin{tcolorbox}[
colback=white,
coltitle= black,
colbacktitle=ourcolorforheader,
colframe=black,
title= $\TwoRep(\Z_3^{(1)}\rtimes \Z_2^{(0)})$: $\TwoRep(\Z_3^{(1)}\rtimes \Z_2^{(0)})$ SPT Phase and Trivial $\Z_2^{(1)}$ (Confining Phase), 
sharp corners]
\be
\begin{split}
\begin{tikzpicture}
\begin{scope}[shift={(0,0)},scale=1] 
 \pgfmathsetmacro{\del}{-0.5}
\node at  (0,0)  {${\rm Triv}_0$} ;
\node at  (0,-0.6)  {\footnotesize$(\Z_2^{(1)}\text{-}{\rm Triv})$} ;
\end{scope}
\end{tikzpicture}    
\end{split}
\ee
The non-invertible 0-form symmetry is realized on the single vacuum as
\begin{equation}
    D_2^{A}=1_{00}\oplus 1_{00}\,. 
\end{equation}
\end{tcolorbox}

\subsubsection{$\TwoRep(\Z_3^{(1)}\rtimes \Z_2^{(0)})$ SSB Phase}
\label{sec:apple3}
We now consider the physical boundary to be 
\begin{equation}
    \Bphys=\fB_{\Neu(\Z_2)\,, t}\,.
\end{equation}
In this case, the bulk SymTFT defects $\Q_2^{[b]}$ and $\Q_1^{E}$ play the role of order parameters.
Let us first compute the vacua in this phase.
The SymTFT configuration with the $\Q_1^{E}$ line extending between the two boundaries produces a unique local operator after compactification, since this line has a unique untwisted end on each boundary. 
\be
\begin{split}
\begin{tikzpicture}
 \begin{scope}[shift={(0,0)},scale=0.8] 
\draw [cyan, fill=cyan!80!red, opacity =0.5]
(0,0) -- (0,4) -- (2,5) -- (5,5) -- (5,1) -- (3,0)--(0,0);
\draw [black, thick, fill=white,opacity=1]
(0,0) -- (0, 4) -- (2, 5) -- (2,1) -- (0,0);
\draw [cyan, thick, fill=cyan, opacity=0.2]
(0,0) -- (0, 4) -- (2, 5) -- (2,1) -- (0,0);
\draw[line width=1pt] (1,2.5) -- (3,2.5);
\draw[line width=1pt,dashed] (3,2.5) -- (4,2.5);
\fill[red!80!black] (1,2.5) circle (3pt);
\draw [fill=blue!40!red!60,opacity=0.2]
(3,0) -- (3, 4) -- (5, 5) -- (5,1) -- (3,0);
\fill[red!80!black] (1,2.5) circle (3pt);
\fill[red!80!black] (4,2.5) circle (3pt);
\draw [black, thick, opacity=1]
(3,0) -- (3, 4) -- (5, 5) -- (5,1) -- (3,0);
\node at (2,5.4) {$\fB_{\Neu(\Z_2),t}$};
\node at (5,5.4) {$\fB_{\Neu(\Z_2),t'}$};
\draw[dashed] (0,0) -- (3,0);
\draw[dashed] (0,4) -- (3,4);
\draw[dashed] (2,5) -- (5,5);
\draw[dashed] (2,1) -- (5,1);
\draw [cyan, thick, fill=cyan, opacity=0.1]
(0,4) -- (3, 4) -- (5, 5) -- (2,5) -- (0,4);
\node[below, red!80!black] at (.8, 2.4) {$\cE_0^{E,+}$};
\node[below, red!80!black] at (3.8, 2.4) {$\wt{\cE}_0^{E,+}$};
\node  at (2.5, 3) {$\Q_{1}^{E}$};
\node  at (6.7, 2.5) {$=$};
\draw [fill=blue!60!green!30,opacity=0.7]
(8,0) -- (8, 4) -- (10, 5) -- (10,1) -- (8,0);
\draw [black, thick, opacity=1]
(8,0) -- (8, 4) -- (10, 5) -- (10,1) -- (8,0);
\fill[red!80!black] (9,2.5) circle (3pt);
\node[below, red!80!black] at (8.8, 2.4) {${\cO}$};
\node[below, black] at (10.8, 3) {$,$};
\end{scope}
\end{tikzpicture}    
\end{split}
\ee
We denote this operator as $\cO$ and concretely, it is obtained as
\begin{equation}
    \cO= \left(\cE_0^{E,+}\,, \Q_1^{E}\,, \wt\cE_0^{E,+}\right)\,.
\end{equation}
Recall from \eqref{eq:fusion of E end ops} that $(\cE_0^{E,+})^2=\cE_0^{E,+}+ \sqrt{2} $ (we again fix $\lambda=1/\sqrt{2}$).
Then the OPE of the $\cO$ operator is
\begin{equation}
\begin{split}
    \cO^2 &= ({\cE}_0^{E,+}{\cE}_0^{E,+}\,,\Q_1^E\otimes \Q_1^E\,,\wt\cE_0^{E,+}\wt\cE_0^{E,+}) \\ 
    &= ({\cE}_0^{E,+}\,,\Q_1^E\,,\wt\cE_0^{E,+}) + 2({\cE}_{\id}\,,\Q_1^{\id}\,,\wt\cE_0^{\id}) \,, \\
    &=\cO +2\,.
\end{split}
\end{equation}
Using the OPE of local operators, the idempotent operators are determined to be
\begin{equation}
v_0=\frac{2-\cO}{3}\,, \qquad v_1=\frac{1+\cO}{3}\,. 
\end{equation} 
The $D_2^{A}$ symmetry action can be computed using the fact that the sphere linking of $D_2^A$ is 
\begin{equation}
    D_2^{A}:(1,\cO) \longmapsto 
    (2,-\cO)\,,
\end{equation}
which implies that the action on the vacua is 
\begin{equation}
\label{eq:vac action superstar}
    D_2^A:(v_0\,, v_1) \longmapsto (v_0+2v_1\,, v_0)\,. 
\end{equation}
In addition to the topological local operator $\cO$, compactifying the line $\Q_1^{E}$ also furnishes an operator $\overline{\cO}$ in the $D_1^{\wh{b}}$ twisted sector 
\be
\begin{split}
\begin{tikzpicture}
 \begin{scope}[shift={(0,0)},scale=0.8] 
\draw [cyan, fill=cyan!80!red, opacity =0.5]
(0,0) -- (0,4) -- (2,5) -- (5,5) -- (5,1) -- (3,0)--(0,0);
\draw [black, thick, fill=white,opacity=1]
(0,0) -- (0, 4) -- (2, 5) -- (2,1) -- (0,0);
\draw [cyan, thick, fill=cyan, opacity=0.2]
(0,0) -- (0, 4) -- (2, 5) -- (2,1) -- (0,0);
\draw[line width=1pt] (1,2.5) -- (3,2.5);
\draw[line width=1pt,dashed] (3,2.5) -- (4,2.5);
\fill[red!80!black] (1,2.5) circle (3pt);
\draw [fill=blue!40!red!60,opacity=0.2]
(3,0) -- (3, 4) -- (5, 5) -- (5,1) -- (3,0);
\fill[red!80!black] (1,2.5) circle (3pt);
\fill[red!80!black] (4,2.5) circle (3pt);
\draw [black, thick, opacity=1]
(3,0) -- (3, 4) -- (5, 5) -- (5,1) -- (3,0);
\node at (2,5.4) {$\fB_{\Neu(\Z_2),t}$};
\node at (5,5.4) {$\fB_{\Neu(\Z_2),t'}$};
\draw[dashed] (0,0) -- (3,0);
\draw[dashed] (0,4) -- (3,4);
\draw[dashed] (2,5) -- (5,5);
\draw[dashed] (2,1) -- (5,1);
\draw[line width=1pt] (1,2.5) -- (1,4.5);
\draw [cyan, thick, fill=cyan, opacity=0.1]
(0,4) -- (3, 4) -- (5, 5) -- (2,5) -- (0,4);
\node[below, red!80!black] at (.8, 2.4) {$\cE_0^{E,-}$};
\node[below, red!80!black] at (3.8, 2.4) {$\wt{\cE}_0^{E,+}$};
\node  at (2.5, 3) {$\Q_{1}^{E}$};
\node  at (0.5, 3.3) {$D_{1}^{\wh{b}}$};
\node  at (6.7, 2.5) {$=$};
\draw [fill=blue!60!green!30,opacity=0.7]
(8,0) -- (8, 4) -- (10, 5) -- (10,1) -- (8,0);
\draw [black, thick, opacity=1]
(8,0) -- (8, 4) -- (10, 5) -- (10,1) -- (8,0);
\draw[line width=1pt] (9,2.5) -- (9,4.5);
\fill[red!80!black] (9,2.5) circle (3pt);
\node[below, red!80!black] at (8.8, 2.4) {$\overline{\cO}$};
\node  at (8.5, 3.3) {$D_{1}^{\wh{b}}$};
\node[below, black] at (10.8, 3) {$,$};
\end{scope}
\end{tikzpicture}    
\end{split}
\ee
denoted as
\begin{equation}
(D_1^{\wh{b}}\,,\overline{\cO})=\left((D_1^{\wh{b}}\,,\cE_0^{E,-})\,, \Q_1^E\,, \wt\cE^{E,+}_0\right)\,.
\end{equation}
In fact, $\overline{\cO}$ is the only $D_1^{\wh{b}}$-twisted operator in the IR TQFT describing this gapped phase.
The OPE between the twisted and untwisted sector local operators is determined to be 
\begin{equation}
   (D_1^{\wh{b}}\,,\overline{\cO})\cdot \cO=-(D_1^{\wh{b}}\,,\overline{\cO}) 
\end{equation}
where we have used the fact $\cE_0^{E,-}\cE_0^{E,+}=-\cE_0^{E,-}$. 
Now, the fate of the $\Z_2^{(1)}$ symmetry in the two vacua can be understood from
\begin{equation}
\begin{split}
    (D_1^{\wh{b}}\,,\overline{\cO})\cdot v_0&= (D_1^{\wh{b}}\,,\overline{\cO}) \,, \\
(D_1^{\wh{b}}\,,\overline{\cO})\cdot v_1&= 0 \,.
\end{split}
\end{equation}
Which implies that the 1-form symmetry generator is isomorphic to the identity in the vacua $v_0$, i.e., $v_0$ realizes a $\Z_2^{(1)}$ trivial state.
Meanwhile there is no way to make the $\Z_2^{(1)}$ line end on $v_1$.
This implies that the $\Z_2^{(1)}$ symmetry is spontaneously broken in $v_1$.

Let us now move onto the order parameters carrying the generalized charge $\Q_2^{[b]}$.
For the order parameters, we only consider the untwisted sector ends of SymTFT operators on $\Bphys$.
Therefore we obtain two lines from $\Q_2^{[b]}$, one genuine line and one line in the $D_2^{A}$ twisted sector via the following configurations
\begin{equation}
\begin{split}
\cL^b&= \left(\cE_1^b\,, \Q_2^{[b]}\,, \cE_1^{b}\right)\,, \qquad (D_2^A\,,\cL^{Ab})= \left((D_2^{A}\,,\cE_1^{Ab})\,, \Q_2^{[b]}\,, \cE_1^{b}\right)\,.
\end{split}
\end{equation}
These are depicted as
\be
\begin{split}
\hspace{-20pt}
\begin{tikzpicture}
 \begin{scope}[shift={(0,0)},scale=0.8] 
 \pgfmathsetmacro{\del}{-0.5}
\draw [cyan, fill=cyan!80!red, opacity =0.5]
(0,0) -- (0,4) -- (2,5) -- (5,5) -- (5,1) -- (3,0)--(0,0);
\draw [black, thick, fill=white,opacity=1]
(0,0) -- (0, 4) -- (2, 5) -- (2,1) -- (0,0);
\draw [cyan, thick, fill=cyan, opacity=0.2]
(0,0) -- (0, 4) -- (2, 5) -- (2,1) -- (0,0);
\draw [fill=blue!40!red!60,opacity=0.2]
(3,0) -- (3, 4) -- (5, 5) -- (5,1) -- (3,0);
\draw [black, thick, opacity=1]
(3,0) -- (3, 4) -- (5, 5) -- (5,1) -- (3,0);
\node at (2,5.4) {$\fB_{\Neu(\Z_2),t}$};
\node at (5,5.4) {$\fB_{\Neu(\Z_3),p}$};
\draw[dashed] (0,0) -- (3,0);
\draw[dashed] (0,4) -- (3,4);
\draw[dashed] (2,5) -- (5,5);
\draw[dashed] (2,1) -- (5,1);
\draw [cyan, thick, fill=cyan, opacity=0.1]
(0,4) -- (3, 4) -- (5, 5) -- (2,5) -- (0,4);
\draw [black, thick, fill=black, opacity=0.3]
(0.,2) -- (3, 2) -- (5., 3) -- (2, 3) -- (0.,2);
\node  at (2.5, 2.5) {$\Q_{2}^{[b]}$};
\node[below, red!80!black] at (0.5, 2.) {$\cE_1^{b}$};
\node[below, red!80!black] at (4, 2.4) {$\wt{\cE}_1^{A}$};
\draw [line width=1pt, red!80!black] (0.0, 2.) -- (2, 3);
\draw [line width=1pt, red!80!black] (3, 2.) -- (5, 3);
\node  at (5.4, 2.5) {$=$};
\draw [fill=blue!60!green!30,opacity=0.7]
(6.5+\del,0) -- (6.5+\del, 4) -- (8.5+\del, 5) -- (8.5+\del,1) -- (6.5+\del,0);
\draw [black, thick, opacity=1]
(6.5+\del,0) -- (6.5+\del, 4) -- (8.5+\del, 5) -- (8.5+\del,1) -- (6.5+\del,0);
\draw [black, thick, fill=black, opacity=0.3]
(6.5+\del,2) -- (6.5+\del, 4) -- (8.5+\del, 5) -- (8.5+\del,3) -- (6.5+\del,2);
\draw [line width=1pt, red!80!black] (6.5+\del, 2.) -- (8.5+\del, 3);
\node[below, red!80!black] at (7.5+\del, 2.4) {${\cL}^{b}$};
\node[below, black] at (9+\del, 2.4) {$,$};
\end{scope}
\begin{scope}[shift={(7.5,0)},scale=0.8] 
 \pgfmathsetmacro{\del}{-0.5}
\draw [cyan, fill=cyan!80!red, opacity =0.5]
(0,0) -- (0,4) -- (2,5) -- (5,5) -- (5,1) -- (3,0)--(0,0);
\draw [black, thick, fill=white,opacity=1]
(0,0) -- (0, 4) -- (2, 5) -- (2,1) -- (0,0);
\draw [cyan, thick, fill=cyan, opacity=0.2]
(0,0) -- (0, 4) -- (2, 5) -- (2,1) -- (0,0);
\draw [fill=blue!40!red!60,opacity=0.2]
(3,0) -- (3, 4) -- (5, 5) -- (5,1) -- (3,0);
\draw [black, thick, opacity=1]
(3,0) -- (3, 4) -- (5, 5) -- (5,1) -- (3,0);
\node at (2,5.4) {$\fB_{\Neu(\Z_2),t}$};
\node at (5,5.4) {$\fB_{\Neu(\Z_3),p}$};
\draw[dashed] (0,0) -- (3,0);
\draw[dashed] (0,4) -- (3,4);
\draw[dashed] (2,5) -- (5,5);
\draw[dashed] (2,1) -- (5,1);
\draw [cyan, thick, fill=cyan, opacity=0.1]
(0,4) -- (3, 4) -- (5, 5) -- (2,5) -- (0,4);
\draw [black, thick, fill=black, opacity=0.3]
(0.,2) -- (3, 2) -- (5., 3) -- (2, 3) -- (0.,2);
\draw [black, thick, fill=black, opacity=0.3]
(0.,2) -- (0., 4) -- (2, 5) -- (2,3) -- (0,2);
\node  at (2.5, 2.5) {$\Q_{2}^{[b]}$};
\node[below, red!80!black] at (0.5, 2) {$\cE_1^{Ab}$};
\node[below, red!80!black] at (4, 2.4) {$\wt{\cE}_1^{A}$};
\draw [line width=1pt, red!80!black] (0.0, 2.) -- (2, 3);
\draw [line width=1pt, red!80!black] (3, 2.) -- (5, 3);
\node  at (5.4, 2.5) {$=$};
\draw [fill=blue!60!green!30,opacity=0.7]
(6.5+\del,0) -- (6.5+\del, 4) -- (8.5+\del, 5) -- (8.5+\del,1) -- (6.5+\del,0);
\draw [black, thick, opacity=1]
(6.5+\del,0) -- (6.5+\del, 4) -- (8.5+\del, 5) -- (8.5+\del,1) -- (6.5+\del,0);
\draw [black, thick, fill=black, opacity=0.3]
(6.5+\del,2) -- (6.5+\del, 4) -- (8.5+\del, 5) -- (8.5+\del,3) -- (6.5+\del,2);
\draw [line width=1pt, red!80!black] (6.5+\del, 2.) -- (8.5+\del, 3);
\node[below, black] at (7.5+\del, 3.9) {$D_2^{A}$};
\node[below, black] at (1., 3.9) {$D_2^{A}$};
\node[below, red!80!black] at (7.5+\del, 2.4) {${\cL}^{Ab}$};
\end{scope}
\end{tikzpicture}    
\end{split}
\ee
The fusion rule $\cL^b\otimes \cL^b$ follows from the fact that $\cE_1^b\otimes \cE_1^b= \cE^{\id}_1$ as
\begin{equation}
    \cL^b\otimes \cL^b= \left(\cE_1^b\otimes \cE_1^b\,, \Q_2^{[b]}\otimes \Q_2^{[b]} \,, \wt\cE_1^{b}\otimes \wt\cE_1^{b}\right)=D_1^{\id}\,, \\
\end{equation}
Next, to compute the fusion rules $\cL^{Ab}\otimes \cL^{Ab}$, we require $\cE_1^{Ab}\otimes \cE_1^{Ab}$.
Recall that $\cE^{Ab}_1$ descends from the direct sum of lines $\cE_1^{ab}\oplus \cE_1^{a^2b}$ in the twisted sectors of $D_2^{ab}$ and $D_{2}^{a^b}$ respectively on $\fB_{\Dir}$.
The fusion of these lines is given by the group composition in $S_3$, i.e.
\begin{equation}
    \left(\cE_1^{ab}\oplus \cE_1^{a^2b}\right)\otimes 
    \left(\cE_1^{ab}\oplus \cE_1^{a^2b}\right)=
    \left(\cE_1^{ab}\oplus \cE_1^{a^2b}\right) \oplus \left(\cE_1^{ab}\otimes \cE_1^{ab}\oplus \cE_1^{a^2b}\otimes \cE_1^{a^2b}\right)\,.
\end{equation}
The decomposable line in the first parenthesis on the right hand side becomes $\cE^{Ab}$ upon gauging $\Z_2^b$.
Meanwhile the second parenthesis $\cE_1^{\id}\oplus \cE_1^{\id}$ such that the two lines in the direct sum are exchanged under $\Z_2^{b}$.
Therefore the 2-dimensional space of endomorphsims of $\cE_1^{\id}\oplus \cE_1^{\id}$ decomposes into a $\Z_2^b$ even and a $\Z_2^b$ odd space.
As a result after gauging $\Z_2^{b}$, we find 
\begin{equation}
    \cE_1^{Ab}\otimes \cE_1^{Ab}= \cE_1^{Ab} \oplus \cE_1^{\id}\oplus D_1^{\wh{b}}\,. 
\end{equation}
Similarly, one may compute the fusion rule $\cE_1^{Ab}\otimes \cE_1^{b}$. By lifting to the ungauged boundary $\fB_{\Dir}$ and then subsequently gauging $\Z_2^b$\,, one finds
\begin{equation}
    \cE_1^{Ab}\otimes \cE_1^{b}=\cE_1^{A}\,.
\end{equation}
We are now ready to compute the remaining fusions among $\cL^{Ab}$ and $\cL^{b}$
\begin{equation}
\begin{split}
    \cL^{Ab}\otimes \cL^{Ab}&= \left(\cE_1^{Ab}\otimes \cE_1^{Ab}\,, \Q_2^{[b]}\otimes \Q_2^{[b]} \,, \wt\cE_1^{b}\otimes \wt\cE_1^{b}\right) \\
    &= \left(\cE_1^{Ab}\otimes \cE_1^{Ab}\,, \Q_2^{\id} \,, \wt\cE_1^{\id}\right) \\
    &= \left(\cE_1^{\id}\oplus D_1^{\wh{b}}\,, \Q_2^{\id} \,, \wt\cE_1^{\id}\right) \\
    &= D_1^{\id}\oplus D_1^{\wh{b}}\,, \\
    \cL^{Ab}\otimes \cL^{b}&= \left(\cE_1^{Ab}\otimes \cE_1^{b}\,, \Q_2^{[b]}\otimes \Q_2^{[b]} \,, \wt\cE_1^{b}\otimes \wt\cE_1^{b}\right) \\
    &= \left(\cE_1^{A}\,, \Q_2^{\id} \,, \wt\cE_1^{\id}\right) =0\,.
\end{split}
\end{equation}
The fact that $\cL^{Ab}\otimes \cL^{b}=0$ indicates that these lines act within distinct vacua.
Since $\cL^{Ab}$ is a $D_2^A$ twisted line and only the vacuum $v_0$ is symmetric with respect to $D_2^{A}$ (from \eqref{eq:vac action superstar}), it follows that $\cL^{Ab}$ acts within $v_0$.
Meanwhile since $\cL^{b}$ is the order parameter for $\Z_2^{(1)}$ SSB, it only acts within $v_1$ where this symmetry is broken.
Finally the lines $\cL^{b}$ and $\cL^{Ab}$ inherit the F-symbols $t-t'\in H^3(\Z_2,U(1))$.
%
%
The $D_2^{A}$ symmetry in this phase is realized concretely as 
\begin{equation}
    D_2^{A}= 1_{00}\oplus B_{01}\oplus \overline{B}_{10}\,,
\end{equation}
where $1_{00}$ is the identity surface within $v_0$, $B_{01}$ is the interface between ${\rm Triv}_{0}$ and  ${\rm DW(\Z_2)}_{t-t'}$ and $\overline{B}_{10}$ is the converse interface.
As a consistency check, we compute the fusion rules 
\begin{equation}
\begin{split}
 D_2^{A}\otimes D_2^{A}&=    \left(1_{00}\oplus B_{01}\oplus \overline{B}_{10}\right)\otimes \left(1_{00}\oplus B_{01}\oplus \overline{B}_{10}\right)\\
 &= \left(1_{00}\oplus B_{01}\oplus \overline{B}_{10}\right) \oplus (B\overline{B})_{00}\oplus (\overline{B}B)_{11} \\
 &= D_2^{A} \oplus \frac{D_2^{\id}}{D_1^{\id}\oplus D_1^{\widehat{b}}}\,. 
\end{split}
\end{equation}
Where in the line we have used the fact that $(B\overline{B})_{00}=1_{00}\oplus1_{00}$, which is the condensation defect in $v_0$ where $D_1^{\widehat{h}}\sim D_1^{\id}$ and $(\overline{B}B)_{11}$ is indeed the $\Z_2$ condensation defect in $v_1$.

\begin{tcolorbox}[
colback=white,
coltitle= black,
colbacktitle=ourcolorforheader,
colframe=black,
title= $\TwoRep(\Z_3^{(1)}\rtimes \Z_2^{(0)})$: $\TwoRep(\Z_3^{(1)}\rtimes \Z_2^{(0)})$ SSB Phase (``Superstar"), 
sharp corners]
\be
\begin{split}
\begin{tikzpicture}
\begin{scope}[shift={(0,0)},scale=1] 
 \pgfmathsetmacro{\del}{-0.5}
\node at  (0.5,0)  {${\rm Triv}_0 \quad \boxplus \quad \left({\rm DW(\Z_2)_{t-t'}}\right)_1$} ;
\node at  (-1,-0.6)  {\footnotesize$(\Z_2^{(1)}\text{-}{\rm Triv})$} ;
\node at  (1,-0.6)  {\footnotesize$(\Z_2^{(1)}\text{-}{\rm SSB})$} ;
\end{scope}
\end{tikzpicture}    
\end{split}
\ee
The non-invertible 0-form symmetry is realized on the vacua as
\begin{equation}
    D_2^{A}=1_{00}\oplus B_{01}\oplus \overline{B}_{10}\,. 
\end{equation}
\end{tcolorbox}
Note that this phase is very interesting -- we will refer to it as the superstar phase. It has a confining 1-form symmetry in one vaccum, and a deconfined 1-form symmetry in the other, which is thus described by a DW theory. The symmetry acts non-trivially on these two vacua. This is the phase we referred to in the introduction as having two vacua, with different TQFTs in each. this is a new physical feature in this class of models. We will see similar phases for other $\TwoRep (\mathbb{G}^{(2)})$ symmetries.

\subsubsection{$\Z_2^{(1)}$ SSB Phase}
\label{sec:apple4}
Let us now consider the physical boundary to be 
\begin{equation}
    \Bphys=\fB_{\Neu(S_3),(p,t')}\,.
\end{equation}
The order parameters carry generalized charges $\Q_2^{[a]}$ and $\Q_2^{[b]}$.
Since neither of these $\TwoRep(\Z_3^{(1)}\rtimes \Z_2^{(0)})$ symmetry mulitplets contain any local operators, there is only the single identity local operator in the IR TQFT describing this gapped phase.
Correspondingly this phase has a single vacuum $v_0=1$, on which the $D_2^{A}$ symmetry acts as 
\begin{equation}
    D_2^A=\frac{D_2^{\id}}{D_1^{\id}\oplus D_1^{\wh{b}}}\,.
\end{equation}
The SymTFT surface $\Q_2^{[a]}$ provides a single line $\cL^A$ in the $D_2^{A}$ twisted sector 
\begin{equation}
    (D_2^A\,,\cL^A)=\left((D_2^A\,, \cE_1^{A})\,, \Q_2^{[a]}\,, \wt\cE_1^{[a]}\right)\,.
\end{equation}
This configuration is depicted as
\be
\begin{split}
\begin{tikzpicture}
 \begin{scope}[shift={(0,0)},scale=0.8] 
\draw [cyan, fill=cyan!80!red, opacity =0.5]
(0,0) -- (0,4) -- (2,5) -- (5,5) -- (5,1) -- (3,0)--(0,0);
\draw [black, thick, fill=white,opacity=1]
(0,0) -- (0, 4) -- (2, 5) -- (2,1) -- (0,0);
\draw [cyan, thick, fill=cyan, opacity=0.2]
(0,0) -- (0, 4) -- (2, 5) -- (2,1) -- (0,0);
\draw [fill=blue!40!red!60,opacity=0.2]
(3,0) -- (3, 4) -- (5, 5) -- (5,1) -- (3,0);
\draw [black, thick, opacity=1]
(3,0) -- (3, 4) -- (5, 5) -- (5,1) -- (3,0);
\node at (2,5.4) {$\fB_{\Neu(\Z_2),t}$};
\node at (5,5.4) {$\fB_{\Neu(S_3),(p,t')}$};
\draw[dashed] (0,0) -- (3,0);
\draw[dashed] (0,4) -- (3,4);
\draw[dashed] (2,5) -- (5,5);
\draw[dashed] (2,1) -- (5,1);
\draw [cyan, thick, fill=cyan, opacity=0.1]
(0,4) -- (3, 4) -- (5, 5) -- (2,5) -- (0,4);
\draw [black, thick, fill=black, opacity=0.3]
(0.,2) -- (3, 2) -- (5., 3) -- (2, 3) -- (0.,2);
\draw [black, thick, fill=black, opacity=0.3]
(0.,2) -- (0., 4) -- (2, 5) -- (2,3) -- (0,2);
\node  at (2.5, 2.5) {$\Q_{2}^{[a]}$};
\node[below, red!80!black] at (-0.5, 2.4) {$\cE_1^{A}$};
\node[below, red!80!black] at (4, 2.4) {$\wt{\cE}_1^{[a]}$};
\draw [line width=1pt, red!80!black] (0.0, 2.) -- (2, 3);
\draw [line width=1pt, red!80!black] (3, 2.) -- (5, 3);
\node  at (6.7, 2.5) {$=$};
\draw [fill=blue!60!green!30,opacity=0.7]
(8,0) -- (8, 4) -- (10, 5) -- (10,1) -- (8,0);
\draw [black, thick, opacity=1]
(8,0) -- (8, 4) -- (10, 5) -- (10,1) -- (8,0);
\draw [black, thick, fill=black, opacity=0.3]
(8,2) -- (8, 4) -- (10, 5) -- (10,3) -- (8,2);
\draw [line width=1pt, red!80!black] (8, 2.) -- (10, 3);
\node[below, black] at (9, 3.9) {$D_2^{A}$};
\node[below, black] at (1., 3.9) {$D_2^{A}$};
\node[below, red!80!black] at (9, 2.4) {${\cL}^{A}$};
\node[below, black] at (11, 2.4) {$,$};
\end{scope}
\end{tikzpicture}    
\end{split}
\ee
To compute the fusion among $\cL^A$, we need the fusions of the ends of $\Q_2^{[b]}$ on the two boundaries.
These are
\begin{equation}
\begin{split}
    (D_2^A\,, \cE_1^{A})\otimes (D_2^A\,, \cE_1^{A})&= (D_2^A\,, \cE_1^{A}) \oplus \left(D_2^{C}\,, D_1^{C,\id} \right)\,,  \\
    \cE_1^{[a]}\otimes \cE_1^{[a]}&= D_1^{\id}\oplus D_1^{P}\oplus  \cE_1^{[a]}\,.
\end{split}
\end{equation}
where we denote the $D_2^{C}:=D_2^{\id}/\cA_{\Z_2}$, with $\cA_{\Z_2}=D_1^{\id}\oplus D_1^{\wh{b}}$ and the identity on the condensation surface as $D_1^{C,\id}$.
Using these, the fusion rules of $\cL^A$ are determined to be
\begin{equation}
\begin{split}
        \cL^A\otimes \cL^{A}&=D_1^{\id}\oplus D_1^{\wh{b}}\oplus \cL^A\,. 
\end{split}
\end{equation}
The F-symbols of the $\cL^A$ line are controlled by the data $(p,t')\in H^{3}(S^3,U(1))$ on $\Bphys$ and we leave their explicit computation for the future.
Next, compactifying the configuration with the bulk surface $\Q_2^{[b]}$ stretched between 
both boundaries, furnishes both twisted sector and untwisted sector lines which we denote as $\cL^b$ and $\cL^{Ab}$.
Let us focus on the untwisted sector line
\begin{equation}
    \cL^b=\left(\cE_1^{b}\,, \Q_2^{[b]}\,, \wt\cE_1^{[b]}\right)\,,
\end{equation}
depicted as
\be
\begin{split}
\begin{tikzpicture}
 \begin{scope}[shift={(0,0)},scale=0.8] 
\draw [cyan, fill=cyan!80!red, opacity =0.5]
(0,0) -- (0,4) -- (2,5) -- (5,5) -- (5,1) -- (3,0)--(0,0);
\draw [black, thick, fill=white,opacity=1]
(0,0) -- (0, 4) -- (2, 5) -- (2,1) -- (0,0);
\draw [cyan, thick, fill=cyan, opacity=0.2]
(0,0) -- (0, 4) -- (2, 5) -- (2,1) -- (0,0);
\draw [fill=blue!40!red!60,opacity=0.2]
(3,0) -- (3, 4) -- (5, 5) -- (5,1) -- (3,0);
\draw [black, thick, opacity=1]
(3,0) -- (3, 4) -- (5, 5) -- (5,1) -- (3,0);
\node at (2,5.4) {$\fB_{\Neu(\Z_2),t}$};
\node at (5,5.4) {$\fB_{\Neu(S_3),(p,t')}$};
\draw[dashed] (0,0) -- (3,0);
\draw[dashed] (0,4) -- (3,4);
\draw[dashed] (2,5) -- (5,5);
\draw[dashed] (2,1) -- (5,1);
\draw [cyan, thick, fill=cyan, opacity=0.1]
(0,4) -- (3, 4) -- (5, 5) -- (2,5) -- (0,4);
\draw [black, thick, fill=black, opacity=0.3]
(0.,2) -- (3, 2) -- (5., 3) -- (2, 3) -- (0.,2);
\node  at (2.5, 2.5) {$\Q_{2}^{[b]}$};
\node[below, red!80!black] at (-0.5, 2.4) {$\cE_1^{b}$};
\node[below, red!80!black] at (4, 2.4) {$\wt{\cE}_1^{[b]}$};
\draw [line width=1pt, red!80!black] (0.0, 2.) -- (2, 3);
\draw [line width=1pt, red!80!black] (3, 2.) -- (5, 3);
\node  at (6.7, 2.5) {$=$};
\draw [fill=blue!60!green!30,opacity=0.7]
(8,0) -- (8, 4) -- (10, 5) -- (10,1) -- (8,0);
\draw [black, thick, opacity=1]
(8,0) -- (8, 4) -- (10, 5) -- (10,1) -- (8,0);
\draw [line width=1pt, red!80!black] (8, 2.) -- (10, 3);
\node[below, red!80!black] at (9, 2.4) {${\cL}^{b}$};
\node[below, black] at (11, 2.4) {$,$};
\end{scope}
\end{tikzpicture}    
\end{split}
\ee
To compute its fusion rules, we require the fusions of the ends which are
\begin{equation}
\begin{split}
    \cE_1^{b}\otimes \cE_1^{b}&= \cE_1^{\id}\,, \\
    \cE_1^{[b]}\otimes \cE_1^{[b]}&= (D_1^{\id}\oplus D_1^{E})\otimes (D_{1}^{\id}\oplus \cE_1^{[a]})\,, \\
\end{split}  
\end{equation}
Using these, the fusion rules are determined to be
\begin{equation}
    \cL^b\otimes \cL^b=D_1^{\id}\oplus D_1^{C,\id}\,.
\end{equation}
In particular, these lines are charged under the $\Z_2^{(1)}$ symmetry and their F-symbols rules are determined by ${[t-t']}\in H^{3}(\Z_2\,, U(1))$.
Therefore this is the $\Z_2^{(1)}$ symmetry broken phase, realizing the $\Z_2$ Dijkgraaf-Witten theory with 3-cocycle $t-t'$.

\begin{tcolorbox}[
colback=white,
coltitle= black,
colbacktitle=ourcolorforheader,
colframe=black,
title= $\TwoRep(\Z_3^{(1)}\rtimes \Z_2^{(0)})$: $\Z_2^{(1)}$ SSB Phase, 
sharp corners]
\be
\begin{split}
\begin{tikzpicture}
\begin{scope}[shift={(0,0)},scale=1] 
 \pgfmathsetmacro{\del}{-0.5}
\node at  (0,0)  {$ {\rm DW(\Z_2)}_{t-t'}$} ;
\node at  (0,-0.6)  {\footnotesize$(\Z_2^{(1)}\text{-}{\rm SSB})$} ;
\end{scope}
\end{tikzpicture}    
\end{split}
\ee
The non-invertible 0-form symmetry is realized on the vacua as
\begin{equation}
    D_2^{A}=\frac{D_2^{\id}}{D_1^{\id}\oplus D_1^{\wh{b}}}\,. 
\end{equation}
\end{tcolorbox}

\subsection{$\TwoRep(S_3)$ Gapped Phases}
We now describe the minimal gapped phases obtained by choosing the symmetry boundary as 
\begin{equation}
    \Bsym= \fB_{\Neu(S_3)}\,,
\end{equation}
i.e., we choose the boundary with trivial discrete torsion in $H^3(S_3\,,U(1))$. 
In what follows, we allow there to be non-trivial torsion on the physical boundary and construct all minimal gapped phases in this manner.
\subsubsection{$\TwoRep(S_3)$ Trivial Phase}
\label{sec:orange1}

Let us consider the physical boundary to be
\begin{equation}
    \fB_{\phys}=\fB_{\Dir}\,.
\end{equation}
The order parameters carry the generalized charges $\Q_1^{R}$ for $R\in \Rep(S_3)$.
Specifically any line $\Q_1^{R}$ extending accross the SymTFT between the two boundaries, provides ${\rm dim}(R)$ twisted sector order parameters after compactification as
\be
\begin{split}
\begin{tikzpicture}
 \begin{scope}[shift={(0,0)},scale=0.8] 
\draw [cyan, fill=cyan!80!red, opacity =0.5]
(0,0) -- (0,4) -- (2,5) -- (5,5) -- (5,1) -- (3,0)--(0,0);
\draw [black, thick, fill=white,opacity=1]
(0,0) -- (0, 4) -- (2, 5) -- (2,1) -- (0,0);
\draw [cyan, thick, fill=cyan, opacity=0.2]
(0,0) -- (0, 4) -- (2, 5) -- (2,1) -- (0,0);
\draw[line width=1pt] (1,2.5) -- (3,2.5);
\draw[line width=1pt,dashed] (3,2.5) -- (4,2.5);
\fill[red!80!black] (1,2.5) circle (3pt);
\draw [fill=blue!40!red!60,opacity=0.2]
(3,0) -- (3, 4) -- (5, 5) -- (5,1) -- (3,0);
\fill[red!80!black] (1,2.5) circle (3pt);
\fill[red!80!black] (4,2.5) circle (3pt);
\draw [black, thick, opacity=1]
(3,0) -- (3, 4) -- (5, 5) -- (5,1) -- (3,0);
\node at (2,5.4) {$\fB_{\Neu(S_3)}$};
\node at (5,5.4) {$\fB_{\Dir}$};
\draw[dashed] (0,0) -- (3,0);
\draw[dashed] (0,4) -- (3,4);
\draw[dashed] (2,5) -- (5,5);
\draw[dashed] (2,1) -- (5,1);
\draw[line width=1pt] (1,2.5) -- (1,4.5);
\draw [cyan, thick, fill=cyan, opacity=0.1]
(0,4) -- (3, 4) -- (5, 5) -- (2,5) -- (0,4);
\node[below, red!80!black] at (.8, 2.4) {$\cE_0^{R}$};
\node[below, red!80!black] at (3.8, 2.4) {$\wt{\cE}_0^{R,i}$};
\node  at (2.5, 3) {$\Q_{1}^{R}$};
\node  at (0.5, 3.3) {$D_{1}^{R}$};
\node  at (6.7, 2.5) {$=$};
\draw [fill=blue!60!green!30,opacity=0.7]
(8,0) -- (8, 4) -- (10, 5) -- (10,1) -- (8,0);
\draw [black, thick, opacity=1]
(8,0) -- (8, 4) -- (10, 5) -- (10,1) -- (8,0);
\draw[line width=1pt] (9,2.5) -- (9,4.5);
\fill[red!80!black] (9,2.5) circle (3pt);
\node[below, red!80!black] at (8.8, 2.4) {${\cO^{R,i}}$};
\node  at (8.5, 3.3) {$D_{1}^{R}$};
\node[below, black] at (10.8, 3) {$,$};
\end{scope}
\end{tikzpicture}    
\end{split}
\ee
\begin{equation}
(D_1^{R}\,,\cO^{R,i})=\left((D_1^{R}\,,\cE_0^{R})\,, \Q_1^{R}\,, \cE_0^{R,i}\right)\,.
\end{equation}
All these ends transform trivially under the $\TwoRep(S_3)$ symmetry and therefore this corresponds to the single vacuum phase where the entire $\TwoRep(S_3)$ acts trivially.
Specifically any representation $R$ just acts as ${\dim}(R)$ copies of the identity line.

This phase has a non-invertible 1-form symmetry $\Rep (S_3)$ which is preserved. The associated line operators (charges) are confined.

\begin{tcolorbox}[
colback=white,
coltitle= black,
colbacktitle=ourcolorforheader,
colframe=black,
title= $\TwoRep(S_3)$: Trivial Phase with trivial $\Rep (S_3)$ 1-form symmetry (Confining Phase),
sharp corners]
\be
\begin{split}
\begin{tikzpicture}
\begin{scope}[shift={(0,0)},scale=1] 
 \pgfmathsetmacro{\del}{-0.5}
\node at  (0,0)  {$ {\rm Triv}_0$} ;
\end{scope}
\end{tikzpicture}    
\end{split}
\ee
The non-invertible 1-form symmetry is realized as
\begin{equation}
    D_1^{R}=\dim(R) D_1^{\id}\,. 
\end{equation}
\end{tcolorbox}

\subsubsection{$\TwoRep(S_3)/\TwoRep(\Z_2)$ SSB Phase}
\label{sec:orange2}
Let us consider the physical boundary to be 
\begin{equation}
   \Bphys=\fB_{\Neu(\Z_3),p}\,. 
\end{equation}
The order parameters carry generalized charges  $\Q_1^{P}$ and $\Q_2^{[a]}$.
The $\Q_1^{P}$ line stretched between the symmetry and physical boundaries, provides an operator in the twisted sector of $D_1^{P}\in \TwoRep(S_3)$ as
\be
\begin{split}
\begin{tikzpicture}
 \begin{scope}[shift={(0,0)},scale=0.8] 
\draw [cyan, fill=cyan!80!red, opacity =0.5]
(0,0) -- (0,4) -- (2,5) -- (5,5) -- (5,1) -- (3,0)--(0,0);
\draw [black, thick, fill=white,opacity=1]
(0,0) -- (0, 4) -- (2, 5) -- (2,1) -- (0,0);
\draw [cyan, thick, fill=cyan, opacity=0.2]
(0,0) -- (0, 4) -- (2, 5) -- (2,1) -- (0,0);
\draw[line width=1pt] (1,2.5) -- (3,2.5);
\draw[line width=1pt,dashed] (3,2.5) -- (4,2.5);
\fill[red!80!black] (1,2.5) circle (3pt);
\draw [fill=blue!40!red!60,opacity=0.2]
(3,0) -- (3, 4) -- (5, 5) -- (5,1) -- (3,0);
\fill[red!80!black] (1,2.5) circle (3pt);
\fill[red!80!black] (4,2.5) circle (3pt);
\draw [black, thick, opacity=1]
(3,0) -- (3, 4) -- (5, 5) -- (5,1) -- (3,0);
\node at (2,5.4) {$\fB_{\Neu(S_3)}$};
\node at (5,5.4) {$\fB_{\Neu(\Z_3),p}$};
\draw[dashed] (0,0) -- (3,0);
\draw[dashed] (0,4) -- (3,4);
\draw[dashed] (2,5) -- (5,5);
\draw[dashed] (2,1) -- (5,1);
\draw[line width=1pt] (1,2.5) -- (1,4.5);
\draw [cyan, thick, fill=cyan, opacity=0.1]
(0,4) -- (3, 4) -- (5, 5) -- (2,5) -- (0,4);
\node[below, red!80!black] at (.8, 2.4) {$\cE_0^{P}$};
\node[below, red!80!black] at (3.8, 2.4) {$\wt{\cE}_0^{P}$};
\node  at (2.5, 3) {$\Q_{1}^{P}$};
\node  at (0.5, 3.3) {$D_{1}^{P}$};
\node  at (6.7, 2.5) {$=$};
\draw [fill=blue!60!green!30,opacity=0.7]
(8,0) -- (8, 4) -- (10, 5) -- (10,1) -- (8,0);
\draw [black, thick, opacity=1]
(8,0) -- (8, 4) -- (10, 5) -- (10,1) -- (8,0);
\draw[line width=1pt] (9,2.5) -- (9,4.5);
\fill[red!80!black] (9,2.5) circle (3pt);
\node[below, red!80!black] at (8.8, 2.4) {${\cO^{P}}$};
\node  at (8.5, 3.3) {$D_{1}^{P}$};
\node[below, black] at (10.8, 3) {$,$};
\end{scope}
\end{tikzpicture}    
\end{split}
\ee
denoted by 
\begin{equation}
    (D_1^{P}\,, \cO^P)=\left((D_1^{P}\,, \cE_0^P)\,,\Q_1^P\,, \wt{\cE}_0^P \right)\,.
\end{equation}
This implies that the $\Z_2^{(1)}\subset \Rep(S_3)$ is preserved as $D_{1}^P$ is isomorphic to the identity line in the IR TQFT 
describing this gapped phase.
Since there are no non-identity lines that end on both the physical and symmetry boundary, we obtain a unique local operator after compactification and correspondingly, there is a unique vacuum in this phase.
Next, we consider the $\Q_2^{[a]}$ generalized charge.
The surface $\Q_2^{[a]}$ has two untwisted ends on $\Bphys$, which we denote as $\wt{\cE}_1^{a}$ and $\wt{\cE}_1^{a^2}$.
There is a single end on $\Bsym$ which is also untwisted. We denote this as $\cE_1^{[a]}$.
Therefore we obtain two lines
\begin{equation}
    \cL^{[a],a^j}=\left( \cE_1^{[a]}\,,\Q_2^{[a]}\,, \wt{\cE}_1^{a^j} \right)\,,
\end{equation}
depicted as
\be
\begin{split}
\begin{tikzpicture}
 \begin{scope}[shift={(0,0)},scale=0.8] 
\draw [cyan, fill=cyan!80!red, opacity =0.5]
(0,0) -- (0,4) -- (2,5) -- (5,5) -- (5,1) -- (3,0)--(0,0);
\draw [black, thick, fill=white,opacity=1]
(0,0) -- (0, 4) -- (2, 5) -- (2,1) -- (0,0);
\draw [cyan, thick, fill=cyan, opacity=0.2]
(0,0) -- (0, 4) -- (2, 5) -- (2,1) -- (0,0);
\draw [fill=blue!40!red!60,opacity=0.2]
(3,0) -- (3, 4) -- (5, 5) -- (5,1) -- (3,0);
\draw [black, thick, opacity=1]
(3,0) -- (3, 4) -- (5, 5) -- (5,1) -- (3,0);
\node at (2,5.4) {$\fB_{\Neu(S_3)}$};
\node at (5,5.4) {$\fB_{\Neu(Z_3),p}$};
\draw[dashed] (0,0) -- (3,0);
\draw[dashed] (0,4) -- (3,4);
\draw[dashed] (2,5) -- (5,5);
\draw[dashed] (2,1) -- (5,1);
\draw [cyan, thick, fill=cyan, opacity=0.1]
(0,4) -- (3, 4) -- (5, 5) -- (2,5) -- (0,4);
\draw [black, thick, fill=black, opacity=0.3]
(0.,2) -- (3, 2) -- (5., 3) -- (2, 3) -- (0.,2);
\node  at (2.5, 2.5) {$\Q_{2}^{[a]}$};
\node[below, red!80!black] at (-0.5, 2.4) {$\cE_1^{[a]}$};
\node[below, red!80!black] at (4, 2.4) {$\wt{\cE}_1^{a^j}$};
\draw [line width=1pt, red!80!black] (0.0, 2.) -- (2, 3);
\draw [line width=1pt, red!80!black] (3, 2.) -- (5, 3);
\node  at (6.7, 2.5) {$=$};
\draw [fill=blue!60!green!30,opacity=0.7]
(8,0) -- (8, 4) -- (10, 5) -- (10,1) -- (8,0);
\draw [black, thick, opacity=1]
(8,0) -- (8, 4) -- (10, 5) -- (10,1) -- (8,0);
\draw [line width=1pt, red!80!black] (8, 2.) -- (10, 3);
\node[below, red!80!black] at (9, 2.4) {${\cL}^{[a],a^j}$};
\node[below, black] at (11, 2.4) {$,$};
\end{scope}
\end{tikzpicture}    
\end{split}
\ee
This line actually splits as $\cL^{[a],a^j}$ can emit the $D_1^{P}$ line which can itself end on $\cO^{P}$. 
By shrinking $\cO^{P}$ onto $\cL^{[a],a^{j}}$, we obtain an additional local operator on $\cE_1^{[a]}$, implying that this line must split.
Similarly, the symmetry operator line $D_1^{E}$ splits since it can emit a $D_1^{P}$ line.
We find that in this phase $D_1^{E}$ acts as
\begin{equation}
\label{eq:D1E}
    D_1^{E}=D_1^{\omega}\oplus D_1^{\omega^2}\,.
\end{equation}
To understand the structure of this gapped phase, we find it illustrative to ungauge the $\Z_2^b$ symmetry on the symmetry boundary.
We then obtain the SymTFT with $\Bsym=\fB_{\Neu(\Z_3)}$ which corresponds to $\TwoVec(\Z_3^{(1)}\rtimes \Z_2^{(0)})$ SSB phase.
Recall that this phase had two vacua $v_0$ and $v_1$ exchanged under $\Z_2^{b}$ and each vacuum realized a $\Z_3$ Dijkgraaf-Witten theory with 3-cocycle $p\in H^{3}(\Z_3,U(1))$.
Upon gauging $\Z_2$, we obtain a single vacuum theory with the dual $\Z_2^{(1)}$ spontaneously preserved.
The single vacuum again realizes the $\Z_3$ Dijkgraaf-Witten theory with 3-cocycle $p\in H^{3}(\Z_3,U(1))$ with the $D_1^{E}$ symmetry realized as \eqref{eq:D1E}.
The gapped phase can be summarized as
\begin{tcolorbox}[
colback=white,
coltitle= black,
colbacktitle=ourcolorforheader,
colframe=black,
title= $\TwoRep(S_3)$: $\TwoRep(S_3)/\TwoRep(\Z_2)$ SSB Phase: $\Z_2^{(1)}$ preserved (Confining) and SSB for $\Rep(\Z_3)$ 1-form symmetry, 
sharp corners]
\be
\begin{split}
\begin{tikzpicture}
\begin{scope}[shift={(0,0)},scale=1] 
 \pgfmathsetmacro{\del}{-0.5}
\node at  (0,0)  {$ {\rm DW(\Z_3)}_{p}$} ;
\end{scope}
\end{tikzpicture}    
\end{split}
\ee
The non-invertible 1-form symmetry is realized as
\begin{equation}
\begin{split}
    D_1^{P}\simeq D_1^{\id}\,, \qquad D_1^{E}\simeq D_1^{\omega}\oplus D_1^{\omega^2}\,, 
\end{split}
\end{equation}
i.e., $\TwoRep(S_3)$ acts on this gapped phase via the projection $\TwoRep(S_3)\to \TwoRep(\Z_3)$.
\end{tcolorbox}

\subsubsection{$\TwoRep(\Z_2)$ SSB Phase}
\label{sec:orange3}

Now we consider the physical boundary to be 
\begin{equation}
    \Bphys=\fB_{\Neu(\Z_2),t}\,.
\end{equation}
The order parameters are provided by the generalized charges $\Q_1^{E}$ and $\Q_2^{[b]}$.
The generalized charge $\Q_1^{E}$ contributes a local operator in the twisted sector of $D_1^{E}$ which is uncharged under the $\TwoRep(S_3)$ symmetry
\begin{equation}
    (D_1^{E}\,, \cO^E)=\left((D_1^{E}\,, \cE_0^E)\,,\Q_1^E\,, \wt{\cE}_0^{E,+} \right)\,.
\end{equation}
This configuration is depicted as
\be
\begin{split}
\begin{tikzpicture}
 \begin{scope}[shift={(0,0)},scale=0.8] 
\draw [cyan, fill=cyan!80!red, opacity =0.5]
(0,0) -- (0,4) -- (2,5) -- (5,5) -- (5,1) -- (3,0)--(0,0);
\draw [black, thick, fill=white,opacity=1]
(0,0) -- (0, 4) -- (2, 5) -- (2,1) -- (0,0);
\draw [cyan, thick, fill=cyan, opacity=0.2]
(0,0) -- (0, 4) -- (2, 5) -- (2,1) -- (0,0);
\draw[line width=1pt] (1,2.5) -- (3,2.5);
\draw[line width=1pt,dashed] (3,2.5) -- (4,2.5);
\fill[red!80!black] (1,2.5) circle (3pt);
\draw [fill=blue!40!red!60,opacity=0.2]
(3,0) -- (3, 4) -- (5, 5) -- (5,1) -- (3,0);
\fill[red!80!black] (1,2.5) circle (3pt);
\fill[red!80!black] (4,2.5) circle (3pt);
\draw [black, thick, opacity=1]
(3,0) -- (3, 4) -- (5, 5) -- (5,1) -- (3,0);
\node at (2,5.4) {$\fB_{\Neu(S_3)}$};
\node at (5,5.4) {$\fB_{\Neu(\Z_2),t}$};
\draw[dashed] (0,0) -- (3,0);
\draw[dashed] (0,4) -- (3,4);
\draw[dashed] (2,5) -- (5,5);
\draw[dashed] (2,1) -- (5,1);
\draw[line width=1pt] (1,2.5) -- (1,4.5);
\draw [cyan, thick, fill=cyan, opacity=0.1]
(0,4) -- (3, 4) -- (5, 5) -- (2,5) -- (0,4);
\node[below, red!80!black] at (.8, 2.4) {$\cE_0^{E}$};
\node[below, red!80!black] at (3.8, 2.4) {$\wt{\cE}_0^{E,+}$};
\node  at (2.5, 3) {$\Q_{1}^{E}$};
\node  at (0.5, 3.3) {$D_{1}^{E}$};
\node  at (6.7, 2.5) {$=$};
\draw [fill=blue!60!green!30,opacity=0.7]
(8,0) -- (8, 4) -- (10, 5) -- (10,1) -- (8,0);
\draw [black, thick, opacity=1]
(8,0) -- (8, 4) -- (10, 5) -- (10,1) -- (8,0);
\draw[line width=1pt] (9,2.5) -- (9,4.5);
\fill[red!80!black] (9,2.5) circle (3pt);
\node[below, red!80!black] at (8.8, 2.4) {${\cO^{E}}$};
\node  at (8.5, 3.3) {$D_{1}^{E}$};
\node[below, black] at (10.8, 3) {$,$};
\end{scope}
\end{tikzpicture}    
\end{split}
\ee
Therefore in this phase the $D_1^{E}$ line is isomorphic to the identity.
The $D_1^{P}$ symmetry remains non-trivial as there is no $D_1^{P}$ twisted sector in this phase.
By consistency with $\Rep(S_3)$ fusion rules, it can be determined that $D_1^{E}$ acts as
\begin{equation}
    D_1^{E}=D_{1}^{\id}\oplus D_{1}^{P}\,,
\end{equation}
in this phase.
Next, moving onto the $\Q_2^{[b]}$ generalized charge---this contributes a untwisted sector line operator
\begin{equation}
    \cL^{b}=(\cE_1^{[b]}\,, \Q_2^{[b]}\,, \wt{\cE}_1^{b})\,,
\end{equation}
depicted as
\be
\begin{split}
\begin{tikzpicture}
 \begin{scope}[shift={(0,0)},scale=0.8] 
\draw [cyan, fill=cyan!80!red, opacity =0.5]
(0,0) -- (0,4) -- (2,5) -- (5,5) -- (5,1) -- (3,0)--(0,0);
\draw [black, thick, fill=white,opacity=1]
(0,0) -- (0, 4) -- (2, 5) -- (2,1) -- (0,0);
\draw [cyan, thick, fill=cyan, opacity=0.2]
(0,0) -- (0, 4) -- (2, 5) -- (2,1) -- (0,0);
\draw [fill=blue!40!red!60,opacity=0.2]
(3,0) -- (3, 4) -- (5, 5) -- (5,1) -- (3,0);
\draw [black, thick, opacity=1]
(3,0) -- (3, 4) -- (5, 5) -- (5,1) -- (3,0);
\node at (2,5.4) {$\fB_{\Neu(S_3)}$};
\node at (5,5.4) {$\fB_{\Neu(\Z_2),t}$};
\draw[dashed] (0,0) -- (3,0);
\draw[dashed] (0,4) -- (3,4);
\draw[dashed] (2,5) -- (5,5);
\draw[dashed] (2,1) -- (5,1);
\draw [cyan, thick, fill=cyan, opacity=0.1]
(0,4) -- (3, 4) -- (5, 5) -- (2,5) -- (0,4);
\draw [black, thick, fill=black, opacity=0.3]
(0.,2) -- (3, 2) -- (5., 3) -- (2, 3) -- (0.,2);
\node  at (2.5, 2.5) {$\Q_{2}^{[b]}$};
\node[below, red!80!black] at (-0.5, 2.4) {$\cE_1^{[b]}$};
\node[below, red!80!black] at (4, 2.4) {$\wt{\cE}_1^{b}$};
\draw [line width=1pt, red!80!black] (0.0, 2.) -- (2, 3);
\draw [line width=1pt, red!80!black] (3, 2.) -- (5, 3);
\node  at (6.7, 2.5) {$=$};
\draw [fill=blue!60!green!30,opacity=0.7]
(8,0) -- (8, 4) -- (10, 5) -- (10,1) -- (8,0);
\draw [black, thick, opacity=1]
(8,0) -- (8, 4) -- (10, 5) -- (10,1) -- (8,0);
\draw [line width=1pt, red!80!black] (8, 2.) -- (10, 3);
\node[below, red!80!black] at (9, 2.4) {${\cL}^{b}$};
\node[below, black] at (11, 2.4) {$,$};
\end{scope}
\end{tikzpicture}    
\end{split}
\ee
This line is charged under $D_1^{P}$ and has $\Z_2$ fusion rules with the F-symbols given by $t\in H^{3}(\Z_2\,, U(1))$.
Therefore we determine this phase to be the $\Z_{2}$ Dijkgraaf-Witten theory with a 3-cocycle twist $t\in H^{3}(\Z_2,U(1))$.
From the perspective of the $\TwoRep(S_3)$, this corresponds to the phase where the $\Z_2^{(1)}$ subsymmetry is spontaneously broken as the IR TQFT contains the line $\cL^b$ charged under it.

\begin{tcolorbox}[
colback=white,
coltitle= black,
colbacktitle=ourcolorforheader,
colframe=black,
title= $\TwoRep(S_3)$: $\TwoRep(\Z_2)$ SSB Phase and $\Rep (\Z_3)$ 1-form symmetry preserved (Confining),
sharp corners]
\be
\begin{split}
\begin{tikzpicture}
\begin{scope}[shift={(0,0)},scale=1] 
 \pgfmathsetmacro{\del}{-0.5}
\node at  (0,0)  {$ {\rm DW(\Z_2)}_{t}$} ;
\end{scope}
\end{tikzpicture}    
\end{split}
\ee
The $D_1^{P}$ 1-form symmetry is generated by the topological charge line of the $\Z_2$ Dijkgraaf-Witten theory while the $D_1^{E}$ symmetry is generated by
\begin{equation}
\begin{split}
    D_1^{E}\simeq D_1^{\id}\oplus D_1^{P}\,. 
\end{split}
\end{equation}
In summary, $\TwoRep(S_3)$ acts on this gapped phase via the projection $\TwoRep(S_3)\to \TwoRep(\Z_2)$. 
Only the $\TwoRep(\Z_2)$ acts faithfully within this phase and is spontaneously broken.
\end{tcolorbox}

\subsubsection{$\TwoRep(S_3)$ SSB Phase}
\label{sec:orange4}
Finally we consider the case where the physical boundary is 
\begin{equation}
    \Bphys=\fB_{\Neu(S_3)\,, (p,t)}\,.
\end{equation}
In this case the order parameters are the generalized charges $\Q_2^{[a]}$ and $\Q_{2}^{[b]}$.
Each of these charges provide a single untwisted line charged under the $\TwoRep(S_3)$ symmetry as
\begin{equation}
\begin{split}
    \cL^{[a]}&=(\cE_1^{[a]}\,, \Q_2^{[a]}\,, \wt{\cE}_1^{[a]})\,, \\
    \cL^{[b]}&=(\cE_1^{[b]}\,, \Q_2^{[b]}\,, \wt{\cE}_1^{[b]})\,,
\end{split}
\end{equation}
depicted as
\be
\begin{split}
\begin{tikzpicture}
 \begin{scope}[shift={(0,0)},scale=0.8] 
\draw [cyan, fill=cyan!80!red, opacity =0.5]
(0,0) -- (0,4) -- (2,5) -- (5,5) -- (5,1) -- (3,0)--(0,0);
\draw [black, thick, fill=white,opacity=1]
(0,0) -- (0, 4) -- (2, 5) -- (2,1) -- (0,0);
\draw [cyan, thick, fill=cyan, opacity=0.2]
(0,0) -- (0, 4) -- (2, 5) -- (2,1) -- (0,0);
\draw [fill=blue!40!red!60,opacity=0.2]
(3,0) -- (3, 4) -- (5, 5) -- (5,1) -- (3,0);
\draw [black, thick, opacity=1]
(3,0) -- (3, 4) -- (5, 5) -- (5,1) -- (3,0);
\node at (2,5.4) {$\fB_{\Neu(S_3)}$};
\node at (5,5.4) {$\fB_{\Neu(S_3),(p,t)}$};
\draw[dashed] (0,0) -- (3,0);
\draw[dashed] (0,4) -- (3,4);
\draw[dashed] (2,5) -- (5,5);
\draw[dashed] (2,1) -- (5,1);
\draw [cyan, thick, fill=cyan, opacity=0.1]
(0,4) -- (3, 4) -- (5, 5) -- (2,5) -- (0,4);
\draw [black, thick, fill=black, opacity=0.3]
(0.,2) -- (3, 2) -- (5., 3) -- (2, 3) -- (0.,2);
\node  at (2.5, 2.5) {$\Q_{2}^{[g]}$};
\node[below, red!80!black] at (-0.5, 2.4) {$\cE_1^{[g]}$};
\node[below, red!80!black] at (4, 2.4) {$\wt{\cE}_1^{[g]}$};
\draw [line width=1pt, red!80!black] (0.0, 2.) -- (2, 3);
\draw [line width=1pt, red!80!black] (3, 2.) -- (5, 3);
\node  at (6.7, 2.5) {$=$};
\draw [fill=blue!60!green!30,opacity=0.7]
(8,0) -- (8, 4) -- (10, 5) -- (10,1) -- (8,0);
\draw [black, thick, opacity=1]
(8,0) -- (8, 4) -- (10, 5) -- (10,1) -- (8,0);
\draw [line width=1pt, red!80!black] (8, 2.) -- (10, 3);
\node[below, red!80!black] at (9, 2.4) {${\cL}^{[g]}$};
\node[below, black] at (11, 2.4) {$,$};
\end{scope}
\end{tikzpicture}    
\end{split}
\ee
Since the IR TQFT contains the lines $\cL^{[a]}$ and $\cL^{[b]}$ which carry charges under $D_1^{E}$ and $D_1^{P}$ respectively, the full $\TwoRep(S_3)$ symmetry is spontaneously broken in this phase.
There is a unique vacuum since the IR TQFT contains only a single topological local operator which is the identity.
Finally the F-symbols of the charged lines $\cL^{[a]}$ and $\cL^{[b]}$ are controlled by the 3-cocycle $(p,t)\in H^3(S_3,U(1))$.
This gapped phase is the $S_3$ Dijkgraaf-Witten theory with a $(p,t)$ 3-cocycle twist.
\begin{tcolorbox}[
colback=white,
coltitle= black,
colbacktitle=ourcolorforheader,
colframe=black,
title= $\TwoRep(S_3)$: $\TwoRep(S_3)$ SSB, 
sharp corners]
\be
\begin{split}
\begin{tikzpicture}
\begin{scope}[shift={(0,0)},scale=1] 
 \pgfmathsetmacro{\del}{-0.5}
\node at  (0,0)  {$ {\rm DW(S_3)}_{(p,t)}$} ;
\end{scope}
\end{tikzpicture}    
\end{split}
\ee
The $\Rep(S_3)$ 1-form symmetry is generated by the Wilson lines of the $S_3$ Dijkgraaf-Witten theory. 
Since the IR theory contains topological lines charged under the full symmetry category, this is the $\TwoRep(S_3)$ SSB phase.
\end{tcolorbox}

\subsection{Non-Minimal Gapped Boundary Conditions}
We now describe non-minimal gapped boundary conditions for the $4d$ $S_3$ Dijkgraaf-Witten.
The defining feature of such a boundary condition is that it contains non-trivial topological line defects that are not boundary projections of bulk line defects.
We follow the general strategy developed in \cite{Bhardwaj:2024qiv} and outlined in Sec.~\ref{Sec:generalframework}.
%

\subsubsection{Non-minimal Dirichlet boundary condition}
Clearly a simple class of boundary conditions is obtained by stacking the Dirichlet boundary condition by a 3d topological order.
Such boundaries are classified by a modular tensor categories (MTCs). 
Given an MTC $\cM$, which describes the topological line defects of a 3d TFT $\fT$, one can stack the Dirichlet boundary condition with $\fT$ to obtain a new topological boundary 
\begin{equation}
    \fB^{\fT}_{\Dir}= \fB_{\Dir}\boxtimes \fT\,.
\end{equation}
The 2-fusion category of topological defects on this boundary are 
\begin{equation}
    \cS= \TwoVec_{S_3}\boxtimes \Sigma\cM\,,
\end{equation}
where $\Sigma \cM$ (sometimes also denotes as ${\rm Mod}(\cM)$) is the 2-category obtained by including all condensation defects constructible from the lines in $\cM$, i.e., the Karoubi completion of $\cM$ into a 2-category.

 In relation to bulk SymTFT defects, this boundary condition looks almost like $\fB_{\Dir}$, except that the bulk identity surface ends on $\fB^{\fT}_{\Dir}$ not only on the identity line but on any line in $\cM$.
Likewise any other line or surface defect on the boundary can be stacked with an object in $\Sigma\cM$ without affecting its bulk lift. 

\subsubsection{Non-minimal $\Z_3$ Neumann boundary condition}
We now describe the non-minimal generalization of the $\fB_{\Neu(\Z_3)}$ boundary condition.
We will restrict ourselves to the simpler case where there is no discrete torsion while gauging.
Such a boundary condition is defined by stacking $\fB_{\Dir}$ with a $\Z_3^{(0)}$ symmetric 3d TFT $\fT_{\Z_3}$ whose lines defects form the modular tensor category $\cM$.
Given $\fT_{\Z_3}$, we define the non-minimal generalization of the $\fB_{\Neu(\Z_3)}$ boundary condition as
\begin{equation}
\fB_{\Neu(\Z_3)}^{\fT}=\frac{\fB_{\Dir}\boxtimes \fT_{\Z_3}}{\Z_3^{(0)}}\,.
\end{equation}
The $\Z_3$ symmetry is defined via  a $\Z_3$-crossed braided extension of $\cM$, denoted as
\begin{equation}
    \cM^{\times}_{\Z_3}=\bigoplus_{g\in \Z_3}\cM^{g}\,.
\end{equation}
Here $\cM^{g}$ denotes the category of lines living at the end of the $g$ condensation defect.
From $\cM^{\times}_{\Z_3}$, there is a well defined construction known as equivariantization \cite{Etingof:2009yvg, Barkeshli:2014cna} to gauge the $\Z_3$ symmetry to obtain a larger MTC $\wt{\cM}$.
The  lines in $\wt{\cM}$ have a $\Z_3$ grading
\begin{equation}
\label{eq:wtcM_Z3}
    \wt{\cM}=\bigoplus_{g\in \Z_3}\wt{\cM}^g\,, \qquad \wt{\cM}^g=\cM^g/\Z_3\,.
\end{equation}
Among these the lines in $\wt{\cM}^1$ are unattached to any bulk surface.
In particular $\wt{\cM}^1$ contains lines $D_{1}^{\omega}$ and $D_{1}^{\omega^2}$ that generate the dual $\Rep(\Z_3)$ symmetry and are obtainable from the boundary projection of $\Q_1^{E}$.
\begin{equation}
\Q_{1}^{E}\Bigg|_{\fB^{\fT}_{\Neu(\Z_3)}}=\left\{(D_1^{\omega}\,, \cE_0^{E,1})\,, (D_{1}^{\omega^2}\,, \cE_0^{E,2})\right\}\,.
\end{equation}
The rest of the lines in $\wt{\cM}^1$ are intrinsic to the boundary.
The M\"{u}ger center of $\wt{\cM}^1$ is $\Rep(\Z_3)$ generated by $D_1^{\omega}$, therefore these remaining lines in $\wt{\cM}^1$ are all remotely detectable.
The bulk line $\Q_1^{P}$ continues to end on a local operator that is charged under the $\Z_2^{(0)}$ symmetry
\begin{equation}
\Q_{1}^{P}\Bigg|_{\fB^{\fT}_{\Neu(\Z_3)}}=\left\{\cO^P\right\}\,.
\end{equation}
The lines in $\wt{\cM}^a$ and $\wt{\cM}^{a^2}$ descend from the $a$ and $a^2$ twisted sectors, i.e., $\cM^{a}$ and $\cM^{a^2}$ and become attached to the bulk surface $\Q_{2}^{[a]}$.
\begin{equation}
\Q_{2}^{[a]}\Bigg|_{\fB^{\fT}_{\Neu(\Z_3)}}=\left\{
\wt{\cM}^a\,, \wt{\cM}^{a^2} 
\right\}\,.
\end{equation}
The surface $\Q_2^{[b]}$ has a single end (upto fusion with genuine lines) in the twisted sector of $D_2^{b}/\cA_{\Z_3}$
\begin{equation}
    \Q_2^{[b]}\Bigg|_{\fB^{\fT}_{\Neu(\Z_3)}}=\left\{
(D_2^{b}/\cA_{\Z_3}\,, \cE_1^b)
    \right\}
\end{equation}
The remaining lines in $\wt{\cM}_1$ appear simply at the end of the identity surface and are unattached to any bulk defect in the SymTFT.
Note that the fusion 2-category of genuine topological defects on this boundary is
\begin{equation}
    \Sigma \wt{\cM}_1\,.
\end{equation}

\subsubsection{Non-minimal $\Z_2$ Neumann boundary condition}
To construct non-minimal $\Z_2$ Neumann boundary conditions, we follow the same approach as before.
First we stack the Dirichlet boundary condition with a 3d TFT whose underlying modular category of lines $\cM$ can be enriched by $\Z_2$ to obtain the $\Z_2$ crossed braided extension of $\cM$ denoted as
\begin{equation}
    \cM^{\times}_{\Z_2}=\cM^{1}\oplus \cM^p\,.
\end{equation}
Here the category of lines $\cM^{1}\equiv \cM$ while the lines in the non-trivial grade $\cM^{p}$ appear at the end of a condensation defect (say $D_2^{\phi}$) with $\Z_2$ fusion rules. 
We then gauge the diagonal $\Z_2$ symmetry on $\fB_{\Dir}\boxtimes \fT_{\Z_2}$ generated by $D_2^{b}\otimes D_2^{\phi}$.
We denote the gapped boundary condition thus obtained as
\begin{equation}
    \fB_{\Neu(\Z_2)}^{\fT}=\frac{\fB_{\Dir}\boxtimes \fT_{\Z_2}}{\Z_2^{\rm diag}}\,.
\end{equation}
Let us describe the end of the different bulk SymTFT defects after such a gauging.
Firstly, since the end $\cO^{P}$ of $\Q_1^{P}$ is charged under the $\Z_2$ symmetry being gauged, this end goes to the twisted sector, i.e., it is attached to the $\Rep(\Z_2)$ symmetry defect $D_{1}^{\wh{b}}$.
Similarly, the line $\Q_1^{E}$ also has the same ends as on the minimal boundary $\fB_{\Neu(\Z_2)}$.
Apart from $D_1^{\wh{b}}$, there are additional non-trivial lines on the boundary that come from the $\Z_2$ equivariantization of $\cM^1$.
These lines (including $D_1^{\wh{b}}$) form the braided fusion category 
\begin{equation}
    \wt{\cM}^1\,,
\end{equation}
whose M\"{u}ger center is $\Rep(\Z_2)$ generated by $D_1^{\wh{b}}$.
Moving on to the bulk surfaces in the SymTFT---recall that on $\fB_{\Dir}$, $\Q_2^{[a]}$ has two twisted sector ends $(D_2^a,\cE_1^{a})$ and $(D_2^{a^2},\cE_1^{a^2})$ which are swapped by $\Z_2^b$.
Im the minimal case, these twisted sector lines combined into a single line upon $\Z_2^b$ gauging.
In the process of non-minimal gauging, one may further stack $\cE_1^{a}$ and $\cE_1^{a^2}$ with any line $D_1^{x}$ in $\cM$ and its partner $D_{1}^{\phi(x)}$ under the $\Z_2$ action respectively, and then gauge. 
Doing so, one obtains a non-minimal end of $\Q_2^{[a]}$ in the $D_2^{A}$ twisted sector which we denote as $\cE_1^{Ax}$.
\begin{equation}
   \Q_2^{[a]}=\left\{(D_2^A\,,\cE^{Ax}_1)\quad \Big| \quad x\in \cM^1\right\}\,.
\end{equation}
Finally, let us consider the bulk surface $\Q_2^{[b]}$.
On $\fB_{\Dir}$, it had three ends. 
The end $\cE_1^{b}$ in the $D_2^{b}$ twisted sector, now goes to the charged sector of the dual 1-form symmetry generated by $D_1^{\wh{b}}$.
Additionally any line in $\wt{\cM}^p=\cM^p/\Z_2$ carries a charge under  $D_1^{\wh{b}}$ and become attached to $\Q_2^{[b]}$.
Furthermore, the ends $(D_2^{ab}\,,\cE_1^{ab})$ and $(D_{2}^{a^2b}\,,\cE_1^{a^2b})$ were exchanged under $\Z_2^{b}$ and have a similar fate as the ends of the $\Q_2^{[a]}$ described above.
Following the same logic, we obtain,
\begin{equation}
    \Q_2^{[b]}=\left\{D_1^{x}\,, (D_2^{A}\,,\cE_1^{Aby}) \quad  \Big | \quad x\in \wt{\cM}^p\,, y\in \cM \right\}\,,
\end{equation}
Here $(D_2^{A}\,,\cE_1^{Aby})$ denotes the image of the following $\Z_2^{b}$ invariant defect under the $\Z_2^b$ gauging  
\begin{equation}
    (D_2^{a}\,,\cE_1^{ab}\otimes D_1^{y})\otimes (D_2^{a}\,,\cE_1^{a^2b}\otimes D_1^{\phi(y)}) \xrightarrow{{\rm gauging }\ \Z_2^{b}} (D_2^{A}\,,\cE_1^{Aby})\,. 
\end{equation}

\subsubsection{Non-minimal $S_3$ Neumann boundary condition}
The non-minimal generalization of $\fB_{\Neu(S_3)}$ is defined as
\begin{equation}
    \fB^{\fT}_{\Neu(S_3)}=\frac{\fB_{\Dir}\boxtimes \fT_{S_3}}{S^{\rm diag}_{3}}\,,
\end{equation}
where $\fT_{S_3}$ is an $S_3$ symmetric TFT whose underlying modular tensor category of lines is $\cM$.
The $S_3$ enrichment of $\fT$ is defined via an $S_3$ crossed braided fusion category 
\begin{equation}
    \cM^{\times}_{S_3}=\bigoplus_{g\in S_3}\cM^{g}\,. 
\end{equation}
Physically $\cM^{g}$ describes the lines at the end of the $g$ symmetry condensation defect. 
Let us now describe the gauging of $S_3$ within the SymTFT setup.
On $\fB_{\Dir}\boxtimes \fT_{S_3}$, the SymTFT line $\Q_1^{R}$, ended on a multiplet of local operators $\cE_0^{R,i}$, transforming in the $R$ representation of $S_3$.
Upon gauging $S_3$, this multiplet combines into a single operator attached to the $D_1^{R}$ symmetry defect
\begin{equation}
    \Q_1^{R}\Bigg|_{\fB^{\fT}_{\Neu(S_3)}}= (D_1^{R}\,, \cE_0^R)\,.
\end{equation}
Next, on $\fB_{\Dir}\boxtimes \fT_{S_3}$ the surface $\Q_2^{[\id]}$ ended on lines in $\cM^1$.
Gauging $S_3$, one obtains lines in $\wt{\cM}^1\equiv \cM^1/S_3$.
Concretely, the lines in $\wt{\cM}^1$ descend from indecomposable orbits under the $S_3$ action in $\cM^1$. 
Furthermore if a representative line in an $S_3$ orbit is stabilized by (i.e. is invariant under) $H\subseteq S_3$, then it can further carry a representation $R_H\in \Rep(H)$. 

\medskip\noindent Similarly, the lines in $\cM^a$ and $\cM^{a^2}$ were in the twisted sector of the $\Z_3$ subgroup of the $S_3$ diagonal symmetry being gauged to get from $\fB_{\Dir}\boxtimes \fT_{S_3}$ to $\fB_{\Neu(S_3)}^{\fT}$.
Therefore these lines become untwisted sector lines that are charged under the dual $\TwoRep(S_3)$ symmetry.
Furthermore since any given line in $\cM^{a}$ is mapped to another line in $\cM^{a^2}$ under the $\Z_2$ symmetry in $S_3$, such pairs of lines become simple (indecomposable) lines on the gauged boundary.
These lines are objects in 
\begin{equation}
    \wt{\cM}^{[a]}\equiv [\cM^a \oplus \cM^{a^2}]/S_3\,,
\end{equation}
and are attached to $\Q_2^{[a]}$ in the bulk. We denote this as
\begin{equation}
Q_2^{[a]}\Bigg|_{\fB^{\fT}_{\Neu(S_3)}}=\wt{\cM}^{[a]}\,.
\end{equation}
For precisely the same reason, one obtains lines in 
\begin{equation}
Q_2^{[b]}\Bigg|_{\fB^{\fT}_{\Neu(S_3)}}=\wt{\cM}^{[b]}\,, \qquad 
    \wt{\cM}^{[b]}\equiv [\cM^b \oplus \cM^{ab}\oplus \cM^{a^2b}]/S_3\,.
\end{equation}

\subsection{Non-Minimal Gapped Phases for Minimal Symmetries}
We now discuss the gapped phases obtained by picking the physical boundary to be non-minimal and the symmetry boundary to be a minimal boundary.
For simplicity, we do not consider any discrete torsion on either physical or symmetry boundary.

\subsubsection{$\Bphys=\fB_{\Dir}^{\fT}$}
Since the physical boundary is obtained by simply stacking the minimal Dirichlet boundary with a 3d TFT $\fT$ whose underlying MTC of lines is $\cM$, the corresponding gapped phase is also just the stacking of the minimal gapped phase with $\fT$.

 Let the minimal gapped phase obtained from compactifying the SymTFT sandwich with some minimal symmetry boundary $\Bsym$ and physical boundary $\fB_{\Dir}$ be denoted as 
\begin{equation}
 \left(\Bsym\,, \fB_{\Dir}\right)  =\fT_{\rm min}\,. 
\end{equation}
Then the non-minimal gapped phase with Dirichlet type non-minimal boundary is 
\begin{equation}
 \left(\Bsym\,, \fB_{\Dir}^{\fT}\right)  =\fT_{\rm min}\boxtimes \fT\,, 
\end{equation}
where the minimal symmetry $\cS$ corresponding to $\Bsym$ is realized on $\fT_{\rm min}$ and trivially on $\fT$.
Here the UV symmetry category is $\cS$.
The system flows in the IR to TFT/gapped phase which contains emergent symmetries which include the non-trivial lines in $\cM$ that are all trivial in terms of their $\cS$ symmetry properties.

\subsubsection{$\Bphys=\fB_{\Neu(\Z_3)}^{\fT}$}
Now we consider the physical boundary as 
\begin{equation}    \Bphys=\fB_{\Neu(\Z_3)}^{\fT}\equiv\frac{\fB_{\Dir}\boxtimes \fT_{\Z_3}}{\Z_3^{(0)}}\,.
\end{equation}
The order parameters for the gapped phase with the physical boundary carry the generalized charges $\Q_1^{P}$, $\Q_2^{[a]}$ and $\Q_2^{[\id]}$ since these can end in untwisted sectors on $\Bphys$.
Now let us describe the gapped phase obtained from the SymTFT by choosing different possible minimal symmetry boundaries.
\begin{itemize}
    \item {\bf Dirichlet Symmetry Boundary:} Let us consider $\Bsym=\fB_{\Dir}$ which carries a $\TwoVec_{S_3}$ symmetry. 
    Firstly, $\Q_1^{P}$ becomes a local operator $\cO^P$, which is charged under $\Z_2\subset S_3$, therefore the global symmetry in this phase is broken down to $\Z_3$.
    There are two vacua 
    \begin{equation}
        v_0=\frac{1+\cO^P}{2}\,,  \qquad     v_1=\frac{1-\cO^P}{2}\,.
    \end{equation}
    These are interchanged under $D_2^b$, while the $\Z_3$ generator $D_2^{a}$ acts within each vacuum.
    Next, the identity surface $\Q_2^{\id}$ upon compactification provides lines in $\wt{\cM}^1$ in each vacuum.
    Crucially though, not all lines thus obtained in $\wt{\cM}^{1}$ are independent.
    Some of the lines are identified and some simple lines may split.
    The reason for this is that there is a $\Rep(Z_3)$ category of lines (denoted $D_1^{\omega}$ and $D_{1}^{\omega^2}$) on this boundary that are obtained from the projection of the $\Q_1^{E}$ line onto the physical boundary.
    Since the $\Q_1^{E}$ line is trivialized on $\Bsym$, the $\Rep(Z_3)$ lines in this gapped phase can actually end.
    As a consequence any two lines $D_1^{x}$ and $D_{1}^{y}$ in $\wt{\cM}^{1}$ that are related by the $\Rep(\Z_3)$ action,
    \begin{equation}
       {\rm Hom}_{\wt{\cM}^1}\left(D_1^{x}\otimes D_{1}^{\omega}\,, D_1^{y}\right) \neq 0\,,  
    \end{equation}
    are identified. Furthermore any simple line in $D_1^{x}$ in $\wt{\cM}^1$ that satisfies
    \begin{equation}
        {\rm Hom}_{\wt{\cM}^1}\left(D_1^{x}\,, D_1^{x}\otimes (D_{1}^{\id}\oplus D_{1}^{\omega}\oplus D_{1}^{\omega^2})\right) = \mathbb C^{N}\,,
    \end{equation}
    splits into the direct sum of $N$ simple lines.
    This gives precisely the de-equivariantization of $\wt{\cM}^1$ by $\Rep(Z_3)$.
    This is also referred to as gauging 1-form symmetry or anyon condensation \cite{Eliens:2013epa, Bais:2008ni, Kong:2013aya} in various literature.
    The de-equivariantization of $\wt{\cM}^1$ gives back the MTC $\cM$.
    Therefore each vacuum realizes a $\fT$ topological order described by $\fT=\cZ(\cC)$.
    Lastly, the $\Q_2^{[a]}$ upon compactification provides two end within each vacuum that are attached to $D_2^{g}$ for $g\in [a]$ and carry a line in $\wt{\cM}^{g}\in \cM^{\times}_{\Z_3}$.
    For exactly the same reasons as described above, $\wt{\cM}^{g}\in \cM^{\times}_{\Z_3}$ is de-equivariantized by $\Rep(\Z_3)$ to give back $\cM^{g}$.
    In summary, we realize a two vacuum theory with each vacuum realizing a SET phase described by $\fT_{\Z_3}$, a $\Z_3$ crossed braided fusion category with underlying MTC $\cM$.
    \begin{tcolorbox}[
colback=white,
coltitle= black,
colbacktitle=ourcolorforheader,
colframe=black,
title= $\TwoVec(S_3)$: $\Z_2$ SSB $\boxtimes$ $\Z_3$ SET Phase, 
sharp corners]
    \be
    \label{eq:nonmin 1}
\begin{split}
\begin{tikzpicture}
 \begin{scope}[shift={(0,0)},scale=0.8] 
\node  at (0, 0) {$\fT_{\Z_3} \quad \boxplus \quad \fT_{\Z_3}$};
\draw[<->, thick, rounded corners = 20pt] (-0.8,-0.5)--(0,-1.5)--(0.8,-0.5);
\node  at (0, -1.5) {${{b}}$};
\draw[->, thick, rounded corners = 10pt] (1.6,-0.4)--(2.2,-.7)--(2.6,0)--(2.2, 0.7) -- (1.5,0.35);
\draw[->, thick, rounded corners = 10pt] (-1.7,-0.4)--(-2.3,-.7)--(-2.7,0)--(-2.3, 0.7) -- (-1.7,0.35);
\node  at (3.45, 0) {$\left\{a\,, a^2\right\}$};
\node  at (-3.5, 0) {$\left\{a^2\,, a\right\}$};
\end{scope}
\end{tikzpicture}    
\end{split}
\ee
\end{tcolorbox}

   \item {\bf Neumann $\Z_3$ Symmetry Boundary:} Take the symmetry boundary to be $\fB_{\Neu(\Z_3)}$.
   The corresponding gapped phase can be obtained by gauging the $\Z_3^{(0)}$ symmetry in the phase \eqref{eq:nonmin 1}.
   Since the $\Z_3$ symmetry acts within each vacuum. 
   We can just focus on the $\Z_3$ gauging of a single vacua realizing $\fT_{\Z_3}$.
   This produces an theory $\fT_{\Z_3}/\Z_3$ whose underlying MTC of lines is $\wt{\cM}$ in \eqref{eq:wtcM_Z3}.
   Since the theory $\fT_{\Z_3}/\Z_3$  contains lines that braid non-trivially with the $\Z_3^{(1)}$ symmetry generated by $D_{1}^{\omega}$, the $\Z_3$ 1-form symmetry is spontaneously broken.
   Both the vacua realize the same phase $\fT_{\Z_3}/\Z_3$ and are exchanged by $\Z_2^{b}$.
   In summary, this gapped phase has the form
    \begin{tcolorbox}[
colback=white,
coltitle= black,
colbacktitle=ourcolorforheader,
colframe=black,
title= $\TwoVec(\Z_3^{(1)}\rtimes \Z_2^{(0)})$: Non-minimal 2-group SSB, 
sharp corners]
 \be
    \label{eq:nonmin 2}
    \begin{split}
    \begin{tikzpicture}
     \begin{scope}[shift={(0,0)},scale=0.8] 
      \pgfmathsetmacro{\del}{1}
    \node  at (0, 0) {$\fT_{\Z_3}/\Z_3 
    \quad \boxplus \quad \fT_{\Z_3}/\Z_3$} ;
    \draw[<->, thick, rounded corners = 20pt] (-1.,-0.5)--(0,-1.5)--(1.,-0.5);
    \node  at (0, -1.5) {$\Z_2^{(0)}$};
    \draw[->, thick, rounded corners = 10pt] (1.6+\del,-0.4)--(2.2+\del,-.7)--(2.6+\del,0)--(2.2+\del, 0.7) -- (1.5+\del,0.35);
    \draw[->, thick, rounded corners = 10pt] (-1.7-\del,-0.4)--(-2.3-\del,-.7)--(-2.7-\del,0)--(-2.3-\del, 0.7) -- (-1.7-\del,0.35);
    \node  at (3.45+\del +0.2 , 0) {$\Z_3^{(1)} \ \rm{SSB}$};
    \node  at (-3.5-\del -0.2, 0) {$\Z_3^{(1)} \ \rm{SSB}$};
    \end{scope}
    \end{tikzpicture}    
    \end{split}
    \ee
\end{tcolorbox}

    This phase can also be directly obtained from the SymTFT compactification.
    Since $\Q_1^{P}$ ends on both boundaries, it provides an additional topological local operator which is charged under $\Z_2^{(0)}$ and therefore there are two vacua exchanged by $\Z_2^{(0)}$.
    The surface $\Q_2^{[a]}$ can also end on both boundaries and provides lines in $\wt{\cM}^{a}$ and $\wt{\cM}^{a^2}$ in each vacuum.
    The surface $\Q_2^{[\id]}$ also ends on both boundaries and provides lines in $\wt{\cM}^{1}$ in each vacuum.
    Since $\Q_1^{E}$ no longer ends on $\Bsym$, we do not need to perform any 1-form gauging.
    \item {\bf Neumann $\Z_2$ Symmetry Boundary:} We now take the symmetry boundary to be $\fB_{\Neu(\Z_2)}$.
   The corresponding gapped phase can be obtained by gauging the $\Z_2^{(0)}$ symmetry in the phase \eqref{eq:nonmin 1}.
   Since the $\Z_2$ symmetry acts exchanged the two vacua. 
    We obtain a theory with a single vacuum where the dual $\Z_2^{(1)}$ symmetry generated by $D_1^{\wh{b}}$ acts trivially.
    The non-invertible symmetry $D_2^{A}$ acts as $D_2^{a}\oplus D_2^{a^2}$.
    Therefore we obtain the fusion rules
    \begin{equation}
        D_2^{A}\otimes D_2^{A}=D_2^{A}\oplus 2D_2^{\id}\,,
    \end{equation}
    which are consistent with $\TwoRep(\Z_3^{(1)}\rtimes \Z_2^{(0)})$ fusion rules. 
    In particular the surface $D_2^{A}$ can end, i.e., it is isomorphic to the identity surface in this phase.
    The lines that live at the end of the $D_2^{A}$ surface are obtained from the $\Z_2$ quotient of $\cM^{a}\oplus \cM^{a^2}$.
    Since these two grades of $\cM^{\times}_{\Z_3}$ are exchanged by $\Z_2$, elements in the quotient are simply order 2 $\Z_2$ orbits in $\cM^{a}\oplus \cM^{a^2}$.
    To summarize, the underlying 3d TFT is 
       \begin{tcolorbox}[
colback=white,
coltitle= black,
colbacktitle=ourcolorforheader,
colframe=black,
title= $\TwoRep(\Z_3^{(1)}\rtimes \Z_2^{(0)})$: Non-minimal $\TwoRep(\Z_3^{(1)}\rtimes \Z_2^{(0)})$ SPT Phase, 
sharp corners]
  \begin{equation}
        \fT_{\Z_3}\,,
    \end{equation}
    described by the $\Z_3$ crossed braided fusion category $\cM^{\times}_{\Z_3}$.
    On this the $\TwoRep(\mathbb G^{(2)})$ symmetry acts as $D_2^{A}\simeq D_2^{a}\oplus D_2^{a^2}$ and $D_{1}^{\wh{b}}\simeq D_1^{\id}$.
\end{tcolorbox}
 We can also recover this phase directly from the SymTFT.
    For this choice of symmetry boundary, both the order parameters i.e., are in twisted sectors.
    Since $\Q_1^{P}$ is in the twisted sector of $\D_1^{\wh{b}}$, the $\Z_2^{(1)}$ sub-symmetry in $\TwoRep(\Z_3^{(1)}\rtimes \Z_2^{(0)})$ is trivial in this phase.
    The surface $\Q_{2}^{[a]}$ after compactification, provide $D_2^{A}$ twisted sector lines that are $\Z_2^{b}$ invariant lines in $\cM^a\oplus \cM^{a^2}$. 
    Finally, the identity surface provides lines in $\cM$ that are uncharged under the $\TwoRep(\Z_3^{(1)}\rtimes \Z_2^{(0)})$ symmetry.

   \item {\bf Neumann $S_3$ Symmetry Boundary:} We now take the symmetry boundary to be $\fB_{\Neu(S_3)}$.
   The corresponding gapped phase is obtained via a $\Z_2^{(0)}$ gauging of \eqref{eq:nonmin 2}.
    The $\Z_2^{(0)}$ symmetry was spontaneously broken in the gapped phase \eqref{eq:nonmin 2}.
    An equivalent statement is that the IR TFT describing this phase has a topological local operator $\cO^P$ charged under $\Z_2^{(0)}$.
    Upon gauging the $\Z_2^{(0)}$ symmetry $\cO^P$ becomes a $\Z_2^{(1)}$ twisted sector operator and therefore the $\Z_2^{(1)}$ is realized trivially in this phase.
    Furthermore, this phase has a single (identity) local operator and correspondingly a unique vacuum which realizes the 3d TFT $\fT_{\Z_3}/\Z_{3}$.
    The structure of this gapped phase can be be summarized as
         \begin{tcolorbox}[
colback=white,
coltitle= black,
colbacktitle=ourcolorforheader,
colframe=black,
title= $\TwoRep(\Z_3^{(1)}\rtimes \Z_2^{(0)})$: Non-minimal $\TwoRep(S_3)/\TwoRep(\Z_2)$ SSB Phase, 
sharp corners]
The underlying 3d TFT is 
  \begin{equation}
        \fT_{\Z_3}\,,
    \end{equation}
    described by the $\Z_3$ crossed braided fusion category $\cM^{\times}_{\Z_3}$.
    On this the $\TwoRep(S_3)$ symmetry acts via its projection to $\TwoRep(\Z_3)$.
    Specifically, the $D_1^{E}$ symmetry on this phase is realized via $D_1^{\omega}\oplus D_1^{\omega^2}$.
    \end{tcolorbox}
\end{itemize}

\subsubsection{$\Bphys=\fB_{\Neu(\Z_2)}^{\fT}$}
We now consider the physical boundary to be 
\begin{equation}    \Bphys=\fB_{\Neu(\Z_2)}^{\fT}\equiv\frac{\fB_{\Dir}\boxtimes \fT_{\Z_2}}{\Z_2^{(0)}}\,.
\end{equation}
The order parameters for the gapped phase with the physical boundary carry the generalized charges $\Q_1^{E}$, $\Q_2^{[b]}$ and $\Q_2^{[\id]}$ since these can end in untwisted sectors on $\Bphys$.
Let us describe the gapped phases obtained from the SymTFT sandwich with $\Bsym$ being different possible minimal gapped boundaries.

\begin{itemize}
 \item {\bf Dirichlet Symmetry Boundary:} Let us pick 
 \begin{equation}
     \Bsym=\fB_{\Dir}\,.
 \end{equation}
In addition to the identity operator, there are two more local operators coming from compactifying the $\Q_1^E$ line with the $\cE_0^{E,i}$ end on the symmetry boundary and $\cE_0^{E,+}$ on the physical boundary.
These operators satisfy the $\Z_3$ algebra which was deduced in \eqref{eq:Z3:alg}.
Correspondingly, there are 3 vacua given by \eqref{eq:Z3_vac}.
The generators of the $S_3$ 0-form symmetry acts on these vacua as 
\begin{equation}
\begin{split}
    D_{a}&: (v_0\,, v_1\,, v_{2}) \longmapsto   (v_1\,, v_2\,, v_0)\,, \\  
    D_{b}&: (v_0\,, v_1\,, v_{2}) \longmapsto   (v_0\,, v_2\,, v_1)\,, \\  
\end{split}
\end{equation}
Therefore none of the vacua are invariant under the $\Z_3$ symmetry, while $v_{j}$ is invariant under $\Z_2$ generated by $D_2^{a^{j}b}$.
The surface $\Q_2^{[\id]}$ compactified between the physical and symmetry boundary provide the category of lines $\wt{\cM}_1$.
The M\"{u}ger center of $\wt{\cM}_1$ is $\Rep(\Z_2)$ generated by the projection of $\Q_1^{P}$ onto the physical boundary.
Now since $\Q_1^{P}$ can end on the symmetry boundary, this $\Rep(\Z_2)$ subcategory of lines is trivialized in this phase.
Equivalently, these anyon lines are condensed or equivalently, the lines at the end of $\Q_2^{[\id]}$ are given by the de-equivariantization of $\wt{\cM}^1$ which gives $\cM^1$.
Similarly, the surface $\Q_2^{[b]}$ has ends on the category of lines $\cM^p$ on the physical boundary.
Meanwhile on the symmetry boundary, $\Q_2^{[b]}$ has three ends denoted as $\cE_1^{a^jb}$ in the twisted sector of $D_2^{a^jb}$ respectively.
Since the vacuum $v_j$ is invariant under $D_2^{a^jb}$, this topological surface can end within this vacuum and it does so on a line in $\cM^p$.
In summary, 
\begin{tcolorbox}[
colback=white,
coltitle= black,
colbacktitle=ourcolorforheader,
colframe=black,
title= $\TwoVec_{S_3}$: $\Z_3$ SSB $\boxtimes$ $\Z_2$ SET Phase, 
sharp corners]
The underlying 3d TFT is 
          \be
    \label{eq:nonmin NeuZ2 sym=S3}
\begin{split}
\begin{tikzpicture}
 \begin{scope}[shift={(0,0)},scale=.8] 
 \pgfmathsetmacro{\del}{1.2}
\node  at (0, 0) {$\fT_{\Z_2^b} \quad \boxplus \quad \fT_{\Z_2^{ab}} \quad \boxplus \quad \fT_{\Z_2^{a^2b}}$};
\draw[->, thick, rounded corners = 20pt, color=blue] (-0.8-\del,-0.5)--(0-\del,-1.5)--(0.8-\del,-0.5);
\draw[->, thick, rounded corners = 20pt, color=blue] (-0.8+\del,-0.5)--(0+\del,-1.5)--(0.8+\del,-0.5);
\draw[<-, thick, rounded corners = 20pt, color=blue] (-1-\del,-0.5)-- (-0.8-0.5*\del,-1.5)-- (0,-2.3)--(0.8+0.5*\del,-1.5)--(1+\del,-0.5);
\node[color=blue]  at (0-\del, -0.7) {${{a}}$};
\node[color=blue]  at (0+\del, -0.7) {${{a}}$};
\node[color=blue]  at (0, -1.5) {${{a}}$};
\draw[->, thick, rounded corners = 10pt, color=red!70!black] (1.6+\del,-0.4)--(2.2+\del,-.7)--(2.6+\del,0)--(2.2+\del, 0.7) -- (1.5+\del,0.35);
\draw[->, thick, rounded corners = 10pt, color=red!70!black] (-1.9-\del,-0.4)--(-2.5-\del,-.7)--(-2.9-\del,0)--(-2.5-\del, 0.7) -- (-1.9-\del,0.35);
\draw[->, thick, rounded corners = 10pt, color=red!70!black] (-0.6,0.5)--(-0.9,1)--(-0.2,1.4)--(0.5, 1) -- (0.2,0.4);
\node[color=red!70!black]  at (4.5, 0) {$\Z_2^{a^2b}$};
\node[color=red!70!black]  at (-4.5, 0) {$\Z_2^{b}$};
\node[color=red!70!black]  at (-0.3, 1.8) {$\Z_2^{ab}$};
\end{scope}
\end{tikzpicture}    
\end{split}
\ee
    Each vacuum preserves a $\Z_2$ symmetry and realizes a $Z_2$ SET $\fT_{\Z_2}$ for the preserved $\Z_2$.
The structure of $\fT_{\Z_2}$ is described by a $\Z_2$ crossed braided fusion category $\cM^{\times}_{\Z_2}=\cM^1\oplus \cM^p$.
     \end{tcolorbox}
   \item {\bf Neumann $\Z_3$ Symmetry Boundary:} Take the symmetry boundary to be $\fB_{\Neu(\Z_3)}$.
   The corresponding gapped phase can be obtained by gauging the $\Z_3^{(0)}$ symmetry in the phase \eqref{eq:nonmin NeuZ2 sym=S3}.
    The $\Z_3$ symmetry is completely broken in \eqref{eq:nonmin NeuZ2 sym=S3}, i.e., there are local order parameter operators transforming in each representation of $\Z_3$.
    After gauging the $\Z_3$ symmetry, these operators become attached to the dual $\Z_3^{(1)}$ generators and therefore the $\Z_3^{(1)}$ symmetry is realized trivially on the 2-group symmetric phase thus obtained.
    Meanwhile the $\Z_2$ symmetry is realized precisely how it was within each vacuum in \eqref{eq:nonmin NeuZ2 sym=S3}.
    Therefore we obtain the $\TwoVec(\Z_3^{(1)}\rtimes \Z_2^{(0)})$ symmetric gapped phase
\begin{tcolorbox}[
colback=white,
coltitle= black,
colbacktitle=ourcolorforheader,
colframe=black,
title= $\TwoVec(\Z_3^{(1)}\rtimes \Z_2^{(0)})$: Non-minimal SPT Phase, 
sharp corners]
The underlying 3d TFT is 
         \begin{equation}
    \label{eq:nonmin Z2 sym = 2 group}
        \fT_{\Z_2}\,.
    \end{equation}
    The $\Z_2$ 0-form symmetry is preserved and realized via the the symmetry enrichment of the MTC $\cM$.
The $\Z_{3}^{(1)}$ symmetry is realized trivially on this phase
\begin{equation}
    D_{1}^{\omega}\simeq D_1^{\id}\,.
\end{equation}
 \end{tcolorbox}

    \item {\bf Neumann $\Z_2$ Symmetry Boundary:} Take the symmetry boundary to be $\fB_{\Neu(\Z_2)}$.
    The phase thus obtained is a non-minimal version of the $\TwoRep(\Z_3^{(1)}\rtimes \Z_2^{(0)})$ SSB phase.
    This phase is obtained by gauging the $\Z_2^{b}$ symmetry in \eqref{eq:nonmin NeuZ2 sym=S3}.
    Note that there is a single vacuum i.e., $v_0$ which is invariant under $\Z_2^{b}$, i.e., the $D_2^{b}$ surface may end on this vacuum. 
    Upon gauging $\Z_2^{b}$, the $\Z_2^{b}$-twisted sector lines become lines that are charged under the dual $\Z_2^{(1)}$ symmetry.
    Therefore $v_0$ maps to a vacuum in the gauged theory where $\Z_2^{(1)}$ is spontaneously broken.
    The 3d TFT realized in this phase is $\fT_{\Z_2}/\Z_2$ whose underlying category of lines of $\wt{\cM}=\wt{\cM}^1\oplus \wt{\cM}^{p}$.
    Meanwhile $v_1$ and $v_2$ which were exchanged by $\Z_2^{b}$ combine to a single vacuum in which the $\Z_2^{(1)}$ symmetry is realized trivially.
    This vacuum realizes the 3d TFT $\fT_{\Z_2}$.
    \begin{tcolorbox}[
colback=white,
coltitle= black,
colbacktitle=ourcolorforheader,
colframe=black,
title= $\TwoRep(\Z_3^{(1)}\rtimes \Z_2^{(0)})$: Non-minimal $\TwoRep(\Z_3^{(1)}\rtimes \Z_2^{(0)})$ SSB Phase, 
sharp corners]
The underlying 3d TFT is 
         \be
    \label{eq:nonminZ2 sym=2Rep2group}
    \begin{split}
    \begin{tikzpicture}
     \begin{scope}[shift={(0,0)},scale=0.8] 
      \pgfmathsetmacro{\del}{1}
    \node  at (0, 0) {$\fT_{\Z_2}/\Z_2 
    \quad \boxplus \quad \fT_{\Z_2}$} ;
    \draw[<->, thick, rounded corners = 20pt, color=blue] (-1.,-0.5)--(0,-1.8)--(1.,-0.5);
     \draw[<->, thick, rounded corners = 10pt, color=blue] (1.2,-0.5)--(0.9,-1.)--(1.5,-1.7)--(2.1,-1)--(1.8,-0.5);
    \node[color=blue]  at (0.5, -1.9) {$D_2^{A}$};
    \draw[->, thick, rounded corners = 10pt, color=red!70!black] (1.3+\del,-0.4)--(1.9+\del,-.7)--(2.3+\del,0)--(1.9+\del, 0.7) -- (1.2+\del,0.35);
       \draw[->, thick, rounded corners = 10pt, color=red!70!black] (-1.7-\del,-0.4)--(-2.3-\del,-.7)--(-2.7-\del,0)--(-2.3-\del, 0.7) -- (-1.7-\del,0.35);
    \node[color=red!70!black]  at (3.45+\del +0.2 , 0) {$\Z_2^{(1)} \ \rm{Triv}$};
    \node[color=red!70!black]  at (-3.5-\del -0.2, 0) {$\Z_2^{(1)} \ \rm{SSB}$};
    \end{scope}
    \end{tikzpicture}    
    \end{split}
    \ee 
 \end{tcolorbox}

    \item {\bf Neumann $S_3$ Symmetry Boundary:}
    Now we consider the case with $\Bsym=\fB_{\Neu(S_3)}$. 
    The corresponding gapped phase can be obtained by gauging $\Z_2$ symmetry in \eqref{eq:nonmin Z2 sym = 2 group}.
    Clearly this is simply 
    \begin{equation}
        \fT_{Z_2}/\Z_2\,.
    \end{equation} 
    Since the $\Z_3$ 1-form symmetry was realized trivially on \eqref{eq:nonmin Z2 sym = 2 group}, correspondingly $D_1^{E}\simeq D_1^{\omega}\oplus D_1^{\omega^2}$ is realized trivially on this $2\Rep(S_3)$ symmetric phase.
    Meanwhile the $\Z_2^{(0)}$ symmetry was preserved in \eqref{eq:nonmin Z2 sym = 2 group}, which in fact realized a $\Z_2$ SET.
    Therefore the dual $\Z_2^{(1)}$ symmetry is spontaneously broken in $\fT_{\Z_2}/\Z_2$.
     \begin{tcolorbox}[
colback=white,
coltitle= black,
colbacktitle=ourcolorforheader,
colframe=black,
title= $\TwoRep(\S_3)$: Non-minimal $\TwoRep(\Z_2)$ SSB Phase, 
sharp corners]
The underlying 3d TFT is 
    \begin{equation}
        \fT_{Z_2}/\Z_2\,.
    \end{equation} 
    The non-invertible 1-form symmetry is realized trivially $D_{1}^{E}\simeq D_1^{\id}$, while the $\Rep(\Z_2)$ 1-form symmetry is realized via the $\Z_2$ Wilson line in $\wt{\cM}=\cM^{\times}_{\Z_2}/\Z_2$. 
 This is a non-minimal deconfined phase for the $\TwoRep(\Z_2)$ symmetry.
 \end{tcolorbox}
 
\end{itemize}

\subsubsection{$\Bphys=\fB_{\Neu(S_3)}^{\fT}$}
We now consider the physical boundary to be 
\begin{equation}    \Bphys=\fB_{\Neu(S_3)}^{\fT}\equiv\frac{\fB_{\Dir}\boxtimes \fT_{S_3}}{S_3}\,.
\end{equation}
The order parameters for the gapped phase with the physical boundary carry the generalized charges $\Q_2^{[\id]}$, $\Q_2^{[a]]}$ and $\Q_2^{[b]}$ since these surfaces can end in untwisted sectors on $\Bphys$.
Let us describe the gapped phases obtained from the SymTFT sandwich with $\Bsym$ being different possible minimal gapped boundaries.

\begin{itemize}
    \item {\bf Dirichlet Symmetry Boundary:} 
    We take $\Bsym$ to be the minimal boundary $\fB_{\Dir}$ which carries the $\TwoVec_{S_3}$ symmetry.
    The $\Q_1^{R}$ lines end on $\Bsym$ while they become non-trivial $\Rep(S_3)$ lines on $\Bphys$.
    Recall that the lines on $\fB_{\Neu(S_3)}^{\fT}$ form 
    \begin{equation}
        \wt{\cM}=\cM^{[\id]}\oplus \wt{\cM}^{[a]} \oplus \wt{\cM}^{[b]}\,,
    \end{equation}
    with $\wt{\cM}^{[g]}$ attached to the bulk surface $\Q_2^{[g]}$.
    Within $\wt{\cM}$, there is a $\Rep(S_3)$ fusion sub-category of lines generated by the boundary projections of $\Q_1^{R}$.
    From the above, it is clear that the $\Rep(S_3)$ subcategory of lines is trivialized after compactifying the SymTFT.
    Therefore we obtain the de-equivariantized braided fusion category with respect to $\wt{\cM}$ which is
    \begin{equation}
        \cM^{\times}_{S_3}=\bigoplus_{g\in S^{3}}\cM^g\,.
    \end{equation}
    Specifically the surface $\Q_2^{[a]}$ has two twisted sector ends on $\Bsym$ which are $(D_2^a\,, \cE_{1}^{a})$ and $(D_2^{a^2}\,, \cE_{1}^{a^2})$ and $\wt{\cM}^{[a]}$ worth of ends on $\Bphys$.
    As described above, upon compactifying the SymTFT with $\Q_2^{[a]}$ extended between both boundaries, we obtain the de-equivariantization by $\Rep(S_3)\subset \wt{\cM}$ and therefore one obtains
    \begin{equation}
        (D_2^{a}\,, \cM^{a})\quad \text{and} \quad (D_2^{a^2}\,, \cM^{a^2})\,.
    \end{equation}
    For precisely the same reason, upon compactifying the SymTFT with the $\Q_2^{[b]}$ surface stretched between the two boundaries, we obtain
    \begin{equation}
        (D_2^{b}\,, \cM^{b})\,, \quad (D_2^{ab}\,, \cM^{ab}) \,,  \quad (D_2^{a^2b}\,, \cM^{a^2b})\,.
    \end{equation}
In summary
\begin{tcolorbox}[
colback=white,
coltitle= black,
colbacktitle=ourcolorforheader,
colframe=black,
title= $\TwoVec_{S_3}$: $S_3$ SET Phase, 
sharp corners]
The underlying 3d TFT is
  \begin{equation}
        \fT\,.
    \end{equation}
    As an $S_3$ symmetric theory, it is recognized to be $\fT_{S_3}$ which is an $S_3$ SET whose underlying $S_3$ crossed braided fusion category is $\cM^{\times}_{S_3}$.  
\end{tcolorbox}

    \item {\bf Neumann $\Z_3$ Symmetry Boundary:} Consider the case $\Bsym=\fB_{\Neu(\Z_3)}$ which carries a 2-group symmetry.
    With $\Bphys=\fB_{\Neu(S_3)}^{\fT}$, this phase is directly obtained via the $\Z_3$ gauging of $\fT_{S_3}$ whose underlying $S_3$-crossed braided fusion category is $\cM^{\times}_{S_3}$. 
    We denote the phase thus obtained as
    \begin{equation}
        \frac{\fT_{S_3}}{\Z_3}\,.
    \end{equation}
    Let us describe this phase from the SymTFT sandwich.
    The $\Q_1^{P}$ line can end on the symmetry boundary, and becomes the $D_1^{P}\in \Rep(S_3)$ line on the physical boundary.
    The $\Q_1^{E}$ line has two twisted sector ends $(D_1^{\omega}\,, \cE_0^{E,1})$ and $(D_1^{\omega^2}\,, \cE_0^{E,2})$, and becomes the $D_1^{E}\in \Rep(S_3)$ line on the physical boundary.
    After compactifying the SymTFT, $\Q_1^{E}$ decomposes as $D_1^{\omega}\oplus D_1^{\omega^2}$ which generate the $\Z_3^{(1)}$ symmetry.
    On the physical boundary, the category of lines has the structure
    \begin{equation}
            \wt{\cM}=\cM^{[\id]}\oplus \wt{\cM}^{[a]} \oplus \wt{\cM}^{[b]}\,,
    \end{equation}
    where the fusion rules are compatible with products of conjugacy classes.
    The lines in $\wt{\cM}^{[g]}$ are attached to the $\Q_2^{[g]}$ surface in the bulk.
    Since the $\Q_1^{P}$ ends on a local operator on $\Bsym$, after compactification, we obtain a de-equivariantization of $\wt{\cM}$ by $\Rep(\Z_2)\subset \Rep(S_3)$ generated by $D_1^{P}$.
    After such a de-equivariantization $\wt{\cM}^{[g]}$ become
\begin{equation}
\begin{split}
\wt{\cM}^1 &\longmapsto       \cM^{1}/\Z_3\,, \\
\wt{\cM}^{[a]} &\longmapsto       \cM^{a}/\Z_3 \oplus \cM^{a^2}/\Z_3\,, \\
\wt{\cM}^{[b]} &\longmapsto       (\cM^{b}\oplus \cM^{ab}\oplus \cM^{a^2b})/\Z_3\,.
\end{split}
\end{equation}
Now let us describe the different order parameters in this phase.
First, by considering the SymTFT configuration with $\Q_2^{[a]}$ stretched between both boundaries, we obtain untwisted sector lines in 
\begin{equation}
    \cM^{a}/\Z_3 \oplus \cM^{a^2}/\Z_3\,,
\end{equation}
that are charged under the $\Z_3$ 1-form symmetry. Therefore the $\Z_3$ 1-form symmetry is spontaneously broken in this phase, whose spontaneous breaking is described by a 3d TFT whose underlying category of lines is 
\begin{equation}
  \cM^{1}/\Z_3\oplus    \cM^{a}/\Z_3 \oplus \cM^{a^2}/\Z_3\,.
\end{equation}
Finally, by considering the SymTFT configuration with $\Q_2^{[b]}$ stretched between both boundaries, we obtain 
 lines attached to the $\Z_2^{(0)}$ symmetry defect 
\begin{equation}
\left (D_2^{b}\,, (\cM^{b}\oplus \cM^{ab}\oplus \cM^{a^2b})/\Z_3\right)\,.
\end{equation}
Thus this gapped phase corresponds to the $\Z_2^{(0)}$ enrichment of a topological order that spontaneously breaks the $\Z_3^{(1)}$ symmetry.
\begin{tcolorbox}[
colback=white,
coltitle= black,
colbacktitle=ourcolorforheader,
colframe=black,
title= $\TwoVec(\Z_3^{(1)}\rtimes \Z_2^{(0)})$: $\Z_2$ Symmetry Enriched $\Z_3^{(1)}$ SSB (deconfined) Phase, 
sharp corners]
The underlying 3d TFT is
  \begin{equation}
        \fT_{S_3}/\Z_3\,,
    \end{equation}
whose underlying category of lines is 
\begin{equation}
  \wt{\cM}=\cM^{1}/\Z_3\oplus    \cM^{a}/\Z_3 \oplus \cM^{a^2}/\Z_3\,.
\end{equation}
The lines in $\cM^{a^p}/\Z_3$ carry a charge $\omega^p$ under the $\Z_3^{(1)}$ symmetry, therefore this symmetry is spontaneously broken.
The vacuum preserves the $\Z_2^{(0)}$ symmetry whose twist defects carry lines
\begin{equation}
    \left(\cM^{b}\oplus \cM^{ab}\oplus \cM^{a^2b})/\Z_3\right)\,.
\end{equation}
This forms a $\Z_2$ crossed braided extension of $\wt{\cM}$. 
\end{tcolorbox}

\item {\bf Neumann $\Z_2$ Symmetry Boundary:} Now let us consider the symmetry boundary to be $\fB_{\Neu(\Z_2)}$ which carries the $\TwoRep(\Z_3^{(1)}\rtimes \Z_2^{(0)})$ symmetry on it.
We obtain the gapped phase with $\Bphys=\fB_{\Neu(S_3)}^{\fT}$ by starting from the phase $\fT_{S_3}$ (which was obtained with $\fB_{\Dir}$ as the symmetry boundary) and gauging the $\Z_2$ symmetry.
Recall that $\fT_{S_3}$ corresponded to an $S_3$ crossed braided fusion category $\cM^{\times}_{S_3}$.
Firstly it contains the geniune lines in $\cM^1$.
Upon gauging $\Z_2^{b}$, we obtain the pre-modular tensor category of lines 
\begin{equation}
    \cM^1/\Z_2\,,
\end{equation}
whose M\"{u}ger center is given by the dual $\Rep(\Z_2)$ to the 0-form $\Z_2$ symmetry being gauged.
Moving on, note that the $\Z_2^{b}$ exchanged the grades
\begin{equation}
    D_{2}^{b}: \cM^a \longleftrightarrow \cM^{a^2}\,, \qquad D_{2}^{b}: D_2^a \longleftrightarrow D_2^{a^2}\,,
\end{equation}
hence upon gauging $\Z_2^{b}$, we obtain topological lines in twisted sector of $D_2^{A}$.
In the pre-gauged theory, these lines come from the direct sum of a line in $\cM^a$ and its partner under the $\Z_2^b$ action in $\cM^{a^2}$.
After gauging $\Z_2^{b}$, this becomes an indecomposable $D_2^{A}$ twisted sector line.
Next, we consider the lines in $\cM^b$ attached to  $D_2^{b}$ surfaces in the pregauged theory.
Upon gauging $\Z_2^{b}$, these being untwisted sector lines
\begin{equation}
    \cM^b/\Z_2\,,
\end{equation}
charged under the dual $\Rep(\Z_2)$ symmetry.
Finally since $\Z_2^{b}$ also exchanged the grades
\begin{equation}
    D_{2}^{b}: \cM^{ab} \longleftrightarrow \cM^{a^2b}\,, \qquad D_{2}^{b}: D_2^{ab} \longleftrightarrow D_2^{a^2b}\,,
\end{equation}
upon gauging $\Z_2^{b}$, these become topological lines in twisted sector of $D_2^{A}$.

\begin{tcolorbox}[
colback=white,
coltitle= black,
colbacktitle=ourcolorforheader,
colframe=black,
title= $\TwoRep(\Z_3^{(1)}\rtimes \Z_2^{(0)})$: Non-invertible symmetry enriched $\Z_2^{(1)}$ SSB (deconfined) Phase, 
sharp corners]
This 3d phase corresponds to a $\Z_2$ 1-form symmetry broken phase whose lines form the MTC which has a $\Z_2$ graded structure
\begin{equation}
    \cM^1/\Z_2\oplus \cM^b/\Z_2\,,
\end{equation}
where $\cM^b/\Z_2$ are the lines that are charged under the 1-form symmetry.
The non-invertible 0-form symmetry acts as a condensation defect whose ends contain non-trivial lines in
\begin{equation}
    (\cM^a\oplus \cM^{a^2})/\Z_2 \oplus (\cM^{ab}\oplus \cM^{ab^2})/\Z_2\,. 
\end{equation}
\end{tcolorbox}

\item {\bf Neumann $S_3$ Symmetry Boundary:} We finally describe the gapped phase obtained by picking $\Bsym=\fB_{\Neu(S_3)}$.
This phase is obtained by the $S_3$ gauging of $\fT_{S_3}$.

\begin{tcolorbox}[
colback=white,
coltitle= black,
colbacktitle=ourcolorforheader,
colframe=black,
title= $\TwoRep(S_3)$: Non-minimal $\TwoRep(S_3)$ SSB (deconfined) Phase, 
sharp corners]
The underlying 3d TFT is
\begin{equation}
    \fT_{S_3}/S_3\,,
\end{equation}
whose lines form the MTC  
\begin{equation}
    \wt{\cM}=\wt{\cM}^{[\id]}\oplus \wt{\cM}^{[a]}\oplus \wt{\cM}^{[b]}\,,
\end{equation}
graded by conjugacy classes in $S_3$.
The non-invertible 1-form symmetry is generated by the M\"{u}ger center of $\wt{\cM}^{[\id]}$
\begin{equation}
    \cZ_{2}(\wt{\cM}^{[\id]})=\Rep(S_3)\,.
\end{equation}
Since this MTC contains lines carrying under the  lines the generate the 1-form $\Rep(S_3)$ symmetry, it corresponds to the non-minimal $\TwoRep(S_3)$ SSB phase.
\end{tcolorbox}

\end{itemize}

\section{Non-Invertible Symmetries and Phases from $\mathcal{Z} (\TwoVec_{D_8})$}

The second class of examples that we will present in detail have as their SymTFT the $D_8$ DW theory in 3+1d. This has several features that go beyond the ones observed in the $S_3$ case, which will in fact also extend to  $D_{2n}$, as will be discussed later. 
For brevity (within the expanse of the present paper), we will focus on a discusion of the SymTFT, the minimal gapped BCs, and highlight the phases for the non-invertible 2-representation categories $\TwoRep (\mathbb{G}^{(2)})$ and $\TwoRep (D_8)$.

\subsection{SymTFT} \label{eq:D8_SymTFT}
The group of symmetries of a square, $D_8$, has various presentations. We will use the following:
\be
\ba
D_8=\Z_4\rtimes\Z_2=\big\langle r\,,\; x \ \big|  \ r^4=x^2=\id\,,\; xrx=r^3\,,\big\rangle\,.
\ea
\ee
The conjugacy classes $[g]$ and their centralizer groups $H_g$ are: 
\be
\begin{tabular}{c||c|c|c|c|c}
{$[g]$} & {$[\id]$} & {$\left[r^2\right]$} & {$[r]=\left\{r, r^3\right\}$} & {$[x]=\left\{x, xr^2\right\}$} & {$[xr]=\left\{xr, xr^3\right\}$} \\
\hline
$H_g$ & $D_8$ & $D_8$ & $\mathbb{Z}_4^r$ & $\mathbb{Z}_2^x \times \mathbb{Z}_2^{r^2}$ & $\mathbb{Z}_2^{xr} \times \mathbb{Z}_2^{r^2}$ 
\end{tabular}
\ee
Each conjugacy class labels a topological surface $\Q_2^{[g]}$ in the SymTFT:
\be
\Q_2^{[\id]}\,,\quad \Q_2^{[r^2]}\,,\quad \Q_2^{[r]}\,,\quad \Q_2^{[x]}\,,\quad \Q_2^{[xr]} \,.
\ee
The topological lines are the 1-endomorphisms of these surfaces and transform in the irreducible representations of the centralizers $H_g$ of $g\in [g]$:
\be
\ba
\Q_2^{[\id]}: &\qquad \text{1End} (\Q_2^{[\id]}) = \left\{\Q_1^{R};\ R=1, 1_{r}, 1_x, 1_{xr}, E\right\} \cong \Rep (D_8) \cr 
\Q_2^{[r^2]}: &\qquad \text{1End} (\Q_2^{[r^2]}) = \left\{\Q_1^{[r^2], R};\ R=1, 1_{r}, 1_x, 1_{xr}, E\right\} \cong \Rep (D_8) \cr 
\Q_2^{[xr]}: &\qquad \text{1End} (\Q_2^{[x r]}) = \left\{\Q_1^{[xr], R};\ R= (\epsilon_1, \epsilon_2),\; \epsilon_i = \pm\right\} \cong \Rep (\Z_2\times \Z_2) \cr 
\Q_2^{[x]}: &\qquad \text{1End} (\Q_2^{[x]}) = \left\{\Q_1^{[x], R};\ R= (\epsilon_1, \epsilon_2),\; \epsilon_i = \pm\right\} \cong \Rep (\Z_2\times \Z_2) \cr
\Q_2^{[r]}: &\qquad \text{1End} (\Q_2^{[ r]}) = \left\{\Q_1^{[r], R};\ R= i^p,\; p=0,1,2,3 \right\} \cong \Rep (\Z_4)  \,.\cr
\ea
\ee
In particular, the topological lines on the identity surface, $\Q_1^R$, are labeled by irreducible representations of $D_8$, whose character table is:
\begin{table}[h]
    \centering
    $\begin{array}{|c|ccccc|} \hline
& {[\id]} & {\left[r^2\right]} & {[r]} & {[x]} & {[xr]} \\
\hline 
1 & 1 & 1 & 1 & 1 & 1 \\
1_{r} & 1 & 1 & +1 & -1 & -1 \\
1_{xr} & 1 & 1 & -1 & -1 & +1 \\
1_x & 1 & 1 & -1 & +1 & -1 \\
E & 2 & -2 & 0 & 0 & 0\\\hline
\end{array}$
    \caption{$D_8$ character table}
    \label{tab:D8_chars}
\end{table}

We denote by $1_k$ for $k\in\{r,x,xr\}$ a 1-dimensional irrep that is 1 on $[\id],[r^2],[k]$ and $-1$ otherwise, whereas for the 2-dimensional $E$ we chose the following matrix representation:
\be
    \cD_E(r)=\begin{pmatrix}
        i & 0\\
        0 & -i
    \end{pmatrix}\,,
    \qquad 
    \cD_E(x)=\begin{pmatrix}
        0 & 1\\
        1 & 0
    \end{pmatrix}\,.
\ee

\subsection{Minimal Gapped Boundary Conditions}
We start from the canonical Dirichlet boundary condition on which the topological defects form the 0-form symmetry $\Vec_{D_8}$ and subsequently gauge various subcategories of defects on this
boundary to obtain the other minimal boundaries, which are summarized in table \ref{tab:SumBC_D8}.

\begin{table}
$$
\begin{array}{|c|c|l|}\hline
\text{BC} &\text{Symmetry} & \Q_p\text{ with Dirichlet BCs} \cr \hline \hline 
{\rm Dir} & \TwoVec_{D_8} &
\begin{cases}
     \Q_2^{[\id]}\,,\\ 
 \Q_1^{1} \oplus \Q_1^{1_r}\oplus \Q_1^{1_x}\oplus \Q_1^{1_{xr}}\oplus 2\Q_1^{E}
\end{cases}
\cr\hline
\Neu(\Z_2^{r^2}) &  \TwoVec^\omega_{\Z_2^{(0),x}\times \Z_2^{(0),r}\times \Z_2^{(1),\wh{r}^2}} & 
\begin{cases}
     \Q_2^{[\id]}\oplus \Q_2^{[r^2]}\,,\\ 
 \Q_1^{1} \oplus \Q_1^{1_r}\oplus \Q_1^{1_x}\oplus \Q_1^{1_{xr}}\oplus \\
 \Q_1^{r^2,1} \oplus \Q_1^{r^2,1_r}\oplus \Q_1^{r^2,1_x}\oplus \Q_1^{r^2,1_{xr}}
\end{cases} 
\cr \hline
\Neu(\Z_4^{r}) & \TwoVec_{\Z_4^{(1)}\rtimes \Z_2^{(0),x}} & 
\begin{cases}
     \Q_2^{[\id]}\oplus \Q_2^{[r^2]}\oplus \Q_2^{[r]}\,,\\ 
    \Q_1^{1} \oplus \Q_1^{1_r} \oplus \Q_1^{[r^2],1}\oplus \Q_1^{[r^2],1_r}\oplus 2\Q_1^{[r],1}
\end{cases} 
\cr \hline
\Neu(\Z_2^{xr}\times\Z_2^{xr^3}) &  \TwoVec_{(\Z_2^{(1),\wh{xr}}\times\Z_2^{(1),\wh{xr}^3})\rtimes \Z_2^{(0),x}} & 
\begin{cases}
     \Q_2^{[\id]}\oplus \Q_2^{[r^2]}\oplus \Q_2^{[xr]}\,,\\ 
     \Q_1^{1} \oplus \Q_1^{1_{xr}}\oplus \Q_1^{r^2,1}\oplus \Q_1^{r^2,1_{xr}}\oplus 2\Q_1^{xr,1_{++}}
\end{cases}
\cr \hline\
\Neu(\Z_2^{x}\times\Z_2^{xr^2}) & 
\TwoVec_{(\Z_2^{(1),\wh{x}}\times\Z_2^{(1),\wh{xr}^2})\rtimes \Z_2^{(0),xr}} & 
\begin{cases}
     \Q_2^{[\id]}\oplus \Q_2^{[r^2]}\oplus \Q_2^{[x]}\,,\\ 
     \Q_1^{1} \oplus \Q_1^{1_x}\oplus \Q_1^{r^2,1}\oplus \Q_1^{r^2,1_x}\oplus 2\Q_1^{x,1_{++}}
\end{cases}
\cr \hline\
\Neu(\Z_2^{x}) &
\TwoRep(\Z_4^{(1)}\rtimes \Z_2^{(0),x}) &
\begin{cases}
     \Q_2^{[\id]}\oplus \Q_2^{[x]}\,,\\ 
     \Q_1^{1} \oplus \Q_1^{1_x}\oplus \Q_1^{E}\oplus \Q_1^{x,1_{++}}\oplus 2\Q_1^{x,1_{+-}}
\end{cases} 
\cr \hline\
\Neu(\Z_2^{xr}) & 
\TwoRep(\Z_4^{(1)}\rtimes \Z_2^{(0),xr}) &
\begin{cases}
     \Q_2^{[\id]}\oplus \Q_2^{[xr]}\,,\\ 
     \Q_1^{1} \oplus \Q_1^{1_{xr}}\oplus \Q_1^{E}\oplus \Q_1^{xr,1_{++}}\oplus 2\Q_1^{xr,1_{+-}}
\end{cases} 
\cr \hline\
\Neu(D_8) & 
\TwoRep(D_8) & 
\begin{cases}
     \Q_2^{[\id]}\oplus \Q_2^{[r^2]}\oplus \Q_2^{[r]}\oplus \Q_2^{[x]}\oplus \Q_2^{[xr]}\,,\\ 
     \Q_1^{[\id]}\oplus \Q_1^{[r^2],1}\oplus \Q_1^{[r],1}\oplus \Q_1^{[x],1_{++}}\oplus \Q_1^{[xr],1_{++}}\
\end{cases} \\
 \hline
\end{array}
$$
\caption{Summary of minimal BC for $\cZ(\TwoVec{D_8})$: we list the gapped boundary conditions (BCs), then the interpretation of these in terms of symmetry boundaries by giving the fusion 2-category symmetry, and in the final column we list all the topological defects that can end on this boundary: $\Q_p$ denotes a $p$-dimensional topological defect in the center $\cZ(\TwoVec_{D_8})$. Note that the Neumann BCs can in general have discrete torsion.} \label{tab:SumBC_D8}
\end{table}

\newpage
\subsubsection{Dirichlet Boundary Condition}
\paragraph{Genuine local operators.}
The Dirichlet boundary condition, which when used as the symmetry boundary in the SymTFT gives rise the symmetry $\TwoVec_{D_8}$, is obtained by requiring all genuine bulk topological lines to end on local operators $\cE^{R,i}$, where $R$ is a $D_8$ irrep and $i=1,...,\dim(R)$. 
\be
\begin{split}
\begin{tikzpicture}
 \begin{scope}[shift={(0,0)},scale=0.8] 
\draw [cyan, fill=cyan!80!red, opacity =0.5]
(0,0) -- (0,4) -- (2,5) -- (5,5) -- (5,1) -- (3,0)--(0,0);
\draw [black, thick, fill=white,opacity=1]
(0,0) -- (0, 4) -- (2, 5) -- (2,1) -- (0,0);
\draw [cyan, thick, fill=cyan, opacity=0.2]
(0,0) -- (0, 4) -- (2, 5) -- (2,1) -- (0,0);
\draw[line width=1pt] (1,2.5) --  (4.2,2.5);
\fill[red!80!black] (1,2.5) circle (3pt);
\node at (2,5.4) {$\fB_\Dir$};
\draw[dashed] (0,0) -- (3,0);
\draw[dashed] (0,4) -- (3,4);
\draw[dashed] (2,5) -- (5,5);
\draw[dashed] (2,1) -- (5,1);
\draw [cyan, thick, fill=cyan, opacity=0.1]
(0,4) -- (3, 4) -- (5, 5) -- (2,5) -- (0,4);
\draw [ black, thick, dashed]
(3,0) -- (3, 4) -- (5, 5) -- (5,1) -- (3,0);
\node[below, red!80!black] at (.5, 3.2) {$\cE_0^{R,i}$};
\node  at (2.5, 3) {$\Q_{1}^{R}$};
\node  at (11, 2.5) {$i=1\,,\dots\,, {\rm dim}(R)\,.$};
\end{scope}
\end{tikzpicture}    
\end{split}
\ee
Specifically,
\be
 \Q_{1}^{1_k}\Bigg|_{{\Dir}}=\cE_0^{{1_k}}\,, \quad 1_k\in\{1,1_r,1_x,1_{xr}\}\,,\qquad         \Q_{1}^{E}\Bigg|_{{\Dir}}=\left\{\cE_0^{E,1}\,, \cE_0^{E,2}\right\}\,.
\ee
\paragraph{Absence of non-genuine local operators and of condensation defects.}
Since all topological lines end on $\fB_\Dir$, there are no non-genuine local operators nor condensation defects on this boundary. Indeed, we shall see that all surface operators, described in the next paragraph, are invertible and generate the $2\Vec_{D_8}$ 0-form symmetry group.
\paragraph{Surface operators.}
All topological surface operators have Neumann boundary conditions: their projections onto the boundary become the symmetry generators $D_2^g$ with linking action 
\begin{equation}
    D_{2}^{g}:\quad \cE_0^{R,i}\longmapsto \sum_j\cD_{R}(g)_{ij}\cE_0^{R,j}\,.
\end{equation} 
where $\cD_R(g)$ is the matrix representation of $g\in G$ in the irrep $R$, since the local operators at the end of the $\Q_1^R$ SymTFT line transform in the $R$ representation of $D_8$.
The 1d irreps coincide with their characters, listed in table \ref{tab:D8_chars}, whereas for the 2d irrep $E$ we choose:
\be
    \cD(r)=\begin{pmatrix}
        i & 0\\
        0 & -i
    \end{pmatrix}\,,
    \qquad 
    \cD(x)=\begin{pmatrix}
        0 & 1\\
        1 & 0
    \end{pmatrix}\,.
\ee
The fusion of surfaces follows the (invertible) $D_8$ group multiplication rule:
\be
    D_2^g\otimes D_2^h=D_2^{gh}\,,\qquad g,h\in D_8\,.
\ee
\paragraph{L-shaped surface configurations.}
The non-identity surfaces cannot end on the Dirichlet boundary, and form L-shaped configurations as in (\ref{LShaped}), 
\be
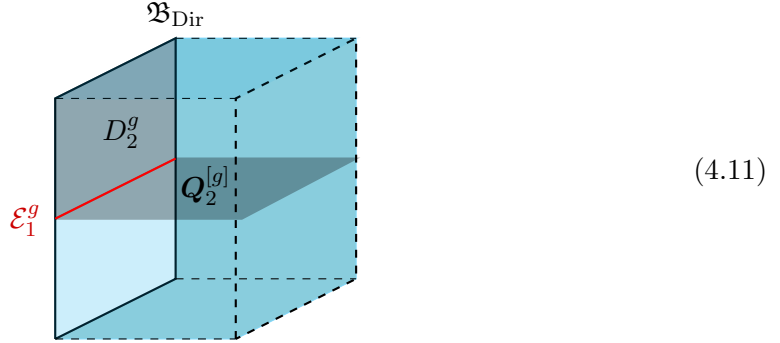
\begin{split}
\begin{tikzpicture}
 \begin{scope}[shift={(0,0)},scale=0.8] 
\draw [cyan, fill=cyan!80!red, opacity =0.5]
(0,0) -- (0,4) -- (2,5) -- (5,5) -- (5,1) -- (3,0)--(0,0);
\draw [black, thick, fill=white,opacity=1]
(0,0) -- (0, 4) -- (2, 5) -- (2,1) -- (0,0);
\draw [cyan, thick, fill=cyan, opacity=0.2]
(0,0) -- (0, 4) -- (2, 5) -- (2,1) -- (0,0);
\draw [black, thick, fill=black, opacity=0.3]
(0.,2) -- (0., 4) -- (2, 5) -- (2,3) -- (0,2);
\draw [black, thick, fill=black, opacity=0.3]
(0.,2) -- (3.1, 2) -- (5., 3) -- (2, 3) -- (0.,2);
\draw [black, thick, dashed]
(3,0) -- (3, 4) -- (5, 5) -- (5,1) -- (3,0);
\node at (2,5.4) {$\fB_\Dir$};
\draw[dashed] (0,0) -- (3,0);
\draw[dashed] (0,4) -- (3,4);
\draw[dashed] (2,5) -- (5,5);
\draw[dashed] (2,1) -- (5,1);
\draw [cyan, thick, fill=cyan, opacity=0.1]
(0,4) -- (3, 4) -- (5, 5) -- (2,5) -- (0,4);
\draw [thick, red] (0.0, 2.) -- (2, 3) ;
\node[below, red!80!black] at (-0.5, 2.4) {$\cE_1^g$};
\node  at (2.5, 2.55) {$\Q_{2}^{[g]}$};
\node[above] at (1.1, 3) {$D_2^{g}$};
\end{scope}
\end{tikzpicture}    
\end{split}
\ee
consisting of non-genuine lines $\cE_1^g$, attached to non-identity surfaces $D_2^g$ for $g\in [g]$
\be
\ba
    \Q_{2}^{[r^2]}\Bigg|_{{\Dir}}&=\left\{(D_2^{r^2},\cE_1^{r^2})\right\}\,, \\
    \Q_{2}^{[xr]}\Bigg|_{{\Dir}}&=\left\{(D_2^{xr},\cE_1^{xr})\,, (D_2^{xr^3},\cE_1^{xr^3})\right\}\,, \\
    \Q_{2}^{[x]}\Bigg|_{{\Dir}}&=\left\{(D_2^{x},\cE_1^{x})\,, (D_2^{xr^2},\cE_1^{xr^2})\right\}\,, \\
    \Q_{2}^{[r]}\Bigg|_{{\Dir}}&=\left\{(D_2^{r},\cE_1^{r})\,,(D_2^{r^3},\cE_1^{r^3})\right\}\,, 
\ea
\ee
with $D_8$ acting on the twisted sector lines as
\begin{equation}
    D_2^{g}:\qquad (D_2^{h},\cE_1^{h})\longmapsto (D_2^{ghg^{-1}},\cE_1^{ghg^{-1}})\,.
\end{equation}

\subsubsection{Neumann($\Z_2^{r^2}$) Boundary Condition} \label{sec:D8_NeuZ2r2}
Let us now gauge the normal $\Z_2^{r^2}$ subsymmetry of $\fB_\Dir$ with discrete torsion 
\be
    s\in H^3(\Z_2, U(1))=\Z_2\,.
\ee
We denote
\be
\fB_{\Neu(\Z_2^{r^2}),s}= \frac{\fB_{\Dir}\boxtimes {\rm SPT}_s }{\Z_2^{r^2}} \,.
\ee
After gauging, one obtains a dual $\Z_2^{\wh{r}^2}$ 1-form symmetry and a residual 0-form symmetry
\be
    D_8/\Z_2^{r^2}=\{1\sim(1,r^2),\,x\sim(x,xr^2),\,r\sim(r,r^3),\,xr\sim(xr,xr^3)\}\cong\Z_2^x\times\Z_2^r\,.
\ee
The symmetry on this boundary is therefore the 2-group 
\be
    \cS=2\Vec^\omega(\Z_2^{(1),\wh{r}^2}\times\Z_2^{(0),r}\times \Z_2^{(0),x})
\ee
with a non-trivial anomaly $\omega$ \cite{Bhardwaj:2022maz}.

\paragraph{Genuine local operators.}
Since the operators $\cE_0^{1_k}$ at the end of $\Q_1^{1_k}$ were untwisted and uncharged under the $\Z_2^{r^2}$ symmetry,
they remain unaltered after gauging 
\be
\begin{split}
\begin{tikzpicture}
 \begin{scope}[shift={(0,0)},scale=0.8] 
\draw [cyan, fill=cyan!80!red, opacity =0.5]
(0,0) -- (0,4) -- (2,5) -- (5,5) -- (5,1) -- (3,0)--(0,0);
\draw [black, thick, fill=white,opacity=1]
(0,0) -- (0, 4) -- (2, 5) -- (2,1) -- (0,0);
\draw [cyan, thick, fill=cyan, opacity=0.2]
(0,0) -- (0, 4) -- (2, 5) -- (2,1) -- (0,0);
\draw[line width=1pt] (1,2.5) -- (4.2,2.5);
\fill[red!80!black] (1,2.5) circle (3pt);
\node at (2,5.4) {$\fB_{{\Neu(\Z_2^{r^2}),s}}$};
\draw [ black, thick, dashed]
(3,0) -- (3, 4) -- (5, 5) -- (5,1) -- (3,0);
\draw[dashed] (0,0) -- (3,0);
\draw[dashed] (0,4) -- (3,4);
\draw[dashed] (2,5) -- (5,5);
\draw[dashed] (2,1) -- (5,1);
\draw [cyan, thick, fill=cyan, opacity=0.1]
(0,4) -- (3, 4) -- (5, 5) -- (2,5) -- (0,4);
\node[below, red!80!black] at (.8, 2.4) {$\cE_0^{1_k}$};
\node  at (2.5, 3) {$\Q_{1}^{1_k}$};
\node  at (5.5, 2.5) {$,$};
\node  at (11, 2.5) {$\Q_{1}^{1_k}\Bigg|_{{\Neu(\Z_2^{r^2}),s}}=\cE_0^{{1_k}}\,, \quad 1_k\in\{1,1_r,1_s,1_{sr}\}\,.$};
\end{scope}
\end{tikzpicture}    
\end{split}
\ee
The charges of these local operators under the $\Z_2^{(0),r}\times\Z_2^{(0),x}$ 0-form symmetry can be read off from the character table \ref{tab:D8_chars}.

\paragraph{Lines generating the dual symmetry after gauging and condensation surfaces.}
After gauging the $\Z_2^{(0),r^2}$ symmetry of $\fB_\Dir$, one obtains a dual $\Z_2^{(1),\wh{r}^2}$ symmetry on $\fB_{\Neu(\Z_2^{r^2}),s}$, whose generator we denote by $D_1^{\wh{r}^2}$. The algebra:
\be
    \cA_{\Z_2^{\wh{r}^2}}=D_1^1\oplus D_1^{\wh{r}^2}\,,
\ee
can be condensed on the identity surface, giving rise to a condensation defect 
\be
   D_2^{\Z_2^{\wh{r}^2}}=\frac{D_2^{[\id]}}{\cA_{\Z_2^{\wh{r}^2}}}\,.
\ee

\paragraph{Non-genuine local operators.}
On $\fB_{\Dir}$, the operators $\cE_0^{E,1}$ and $\cE_0^{E,2}$ transform in the $-1$ representation of $\Z_2^{(0),r^2}$. Therefore after gauging $\Z_2^{r^2}$, they become attached to the dual $\Z_2^{(1),\wh{r}^2}$ symmetry generator $D_1^{\wh{r}^2}$.
\be
\begin{split}
\begin{tikzpicture}
 \begin{scope}[shift={(0,0)},scale=0.8] 
\draw [cyan, fill=cyan!80!red, opacity =0.5]
(0,0) -- (0,4) -- (2,5) -- (5,5) -- (5,1) -- (3,0)--(0,0);
\draw [black, thick, fill=white,opacity=1]
(0,0) -- (0, 4) -- (2, 5) -- (2,1) -- (0,0);
\draw [cyan, thick, fill=cyan, opacity=0.2]
(0,0) -- (0, 4) -- (2, 5) -- (2,1) -- (0,0);
\draw[line width=1pt] (1,2.5) -- (4.2,2.5);
\draw[line width=1pt] (1,2.5) -- (1,4.5);
\fill[red!80!black] (1,2.5) circle (3pt);
\node at (2,5.4) {$\fB_{\Neu(\Z_2^{r^2}),s}$};
\draw[dashed] (0,0) -- (3,0);
\draw[dashed] (0,4) -- (3,4);
\draw[dashed] (2,5) -- (5,5);
\draw[dashed] (2,1) -- (5,1);
\draw [cyan, thick, fill=cyan, opacity=0.1]
(0,4) -- (3, 4) -- (5, 5) -- (2,5) -- (0,4);
\node[below, red!80!black] at (.8, 2.4) {$\cE_0^{E,j}$};
\draw [black, thick, dashed]
(3,0) -- (3, 4) -- (5, 5) -- (5,1) -- (3,0);
\node  at (2.5 , 3) {$\Q_{1}^{E}$};
\node  at (0.5, 3.3) {$D_{1}^{\wh{r}^2}$};
\node  at (5.5, 2.5) {$,$};
\node  at (11, 2.5) {$\Q_{1}^{E}\Bigg|_{\Neu(\Z_2^{r^2}),s}=\left\{(D_1^{\wh{r}^2},\cE_0^{E,1})\,, (D_1^{\wh{r}^2}, \cE_0^{E,2})\right\}\,.$};
\end{scope}
\end{tikzpicture}    
\end{split}
\ee
Since $\cD_{E}(x): \cE_0^{E,1}\leftrightarrow \cE_0^{E,2}$, and $\cD_{E}(r): \cE^{E,1}\mapsto i\cE_0^{E,1},\;\cE_0^{E,2}\mapsto -i\cE_0^{E,2}$ on $\fB_{\Dir}$, the corresponding twisted sector operators on $\fB_{\Neu(\Z_2,s)}$ are also acted upon by the $\Z_2^{(0),x}\times\Z_2^{(0),r}$ 0-form symmetry
\be \label{eq:Neu_r2_frac}
\ba
    D_2^{x}: \quad &(D_1^{\wh{r}^2},\cE_0^{E,1}) \leftrightarrow 
    (D_1^{\wh{r}^2},\cE_0^{E,2})\;, \\
    D_2^{r}: \quad &(D_1^{\wh{r}^2},\cE_0^{E,1}) \mapsto  
    i(D_1^{\wh{r}^2},\cE_0^{E,1})\,, \qquad 
    (D_1^{\wh{r}^2},\cE_0^{E,2}) \mapsto  
    -i(D_1^{\wh{r}^2},\cE_0^{E,2})\,.
\ea
\ee

\paragraph{Genuine topological line.}
The $\Z_2^{(0),r^2}$ 0-form twisted sector line $\cE_1^{r^2}$ on $\fB_{\Dir}$ becomes an untwisted sector line that is charged under the dual $\Z_2^{(1),\wh{r}^2}$ 1-form symmetry on $\fB_{\Neu(\Z_2^{r^2}),s}$, i.e.,
\be
\begin{split}
\begin{tikzpicture}
 \begin{scope}[shift={(0,0)},scale=0.8] 
\draw [cyan, fill=cyan!80!red, opacity =0.5]
(0,0) -- (0,4) -- (2,5) -- (5,5) -- (5,1) -- (3,0)--(0,0);
\draw [black, thick, fill=white,opacity=1]
(0,0) -- (0, 4) -- (2, 5) -- (2,1) -- (0,0);
\draw [cyan, thick, fill=cyan, opacity=0.2]
(0,0) -- (0, 4) -- (2, 5) -- (2,1) -- (0,0);
\draw [black, thick, fill=black, opacity=0.3]
(0.,2) -- (3.1, 2) -- (5., 3) -- (2, 3) -- (0.,2);
\node at (2,5.4) {$\fB_{\Neu(\Z_2^{r^2}),s}$};
\draw[dashed] (0,0) -- (3,0);
\draw[dashed] (0,4) -- (3,4);
\draw[dashed] (2,5) -- (5,5);
\draw[dashed] (2,1) -- (5,1);
\draw [black, thick, dashed]
(3,0) -- (3, 4) -- (5, 5) -- (5,1) -- (3,0);
\draw [cyan, thick, fill=cyan, opacity=0.1]
(0,4) -- (3, 4) -- (5, 5) -- (2,5) -- (0,4);
\draw [line width=1pt, red!80!black] (0.0, 2.) -- (2, 3) ;
\node[below, red!80!black] at (-0.5, 2.4) {$\cE_1^{r^2}$};
\node  at (2.5, 2.55) {$\Q_{2}^{[r^2]}$};
\node  at (5.5, 2.55) {$,$};
\node  at (10.5, 2.55) {$\Q_2^{[r^2]}\Bigg|_{{\Neu(\Z_2^{r^2}),s}}= \cE_1^{r^2}\,.$};
\end{scope}
\end{tikzpicture}    
\end{split}
\ee
with $F$-symbol
\be
    F(\cE_1^{r^2},\cE_1^{r^2},\cE_1^{r^2})=s(\cE_1^{r^2},\cE_1^{r^2},\cE_1^{r^2})\,;
\ee
when the discrete torsion $s$ is non-trivial it is equal to $-1$. 
\paragraph{Local operators on non-identity lines.}
Furthermore, the 1d irreps $1_k$ on $\Q_{1}^{[r^2]}$ give rise to local operators $\cE_0^{1_k}$ on $\cE_1^{r^2}$:
\be
\ba
  \Q_{1}^{[r^2], 1_k}\Bigg|_{{\Neu(\Z_2^{r^2}),s}}&=\cE_1^{r^2}\cE_0^{1_k}\,, 
\ea
\ee
where the charge of $\cE_0^{1_k}$ under the $\Z_2^{(0),x}\times\Z_2^{(0),r}$ 0-form symmetry follows from the character table \ref{tab:D8_chars}.

\paragraph{Surfaces that cannot end on $\fB_{\Neu(\Z_2^{r^2}),s}$.}
On $\fB_{\Dir}$, the surface $\Q_2^{[r]}$ was the non-simple combination $D_2^r\oplus D_2^{r^3}$, with a 2-dimensional space of topological local operators spanned by $D_0^r,\,D_0^{r^3}$. These can be combined into local operators transforming under $\Rep(\Z_2)$:
\be
    D_0^{r,\pm}=D_0^{r}\pm D_0^{r^3}
\ee
of which $D_0^{r,-}$ is charged under $\Z_2^{(0),r^2}$. When gauging $\Z_2^{(0),r^2}$, the SymTFT surface $\Q_2^{[r]}$ projects to a single surface on $\fB_{\Neu(\Z_2^{r^2}),s}$, that we denote as $D_2^{r}/\cA_{\Z_2^{\wh{r}^2}}$, on which $\cA_{\Z_2^{\wh{r}^2}}=D_1^1\oplus D_1^{\wh{r}^2}$ is condensed. Indeed, the $D_0^{r,-}$ local operator, which was charged under $\Z_2^{(0),r^2}$, becomes a twisted sector operator for $\Z_2^{(1),\wh{r}^2}$ therefore the surface containing it can absorb the $ D_1^{\wh{r}^2}$ line. Such surface can also arise from the $L$-shaped configuration:
\be
\begin{split}
\begin{tikzpicture}
 \begin{scope}[shift={(0,0)},scale=0.8] 
\draw [cyan, fill=cyan!80!red, opacity =0.5]
(0,0) -- (0,4) -- (2,5) -- (5,5) -- (5,1) -- (3,0)--(0,0);
\draw [black, thick, fill=white,opacity=1]
(0,0) -- (0, 4) -- (2, 5) -- (2,1) -- (0,0);
\draw [cyan, thick, fill=cyan, opacity=0.2]
(0,0) -- (0, 4) -- (2, 5) -- (2,1) -- (0,0);
\draw [black, thick, fill=black, opacity=0.3]
(0.,2) -- (0., 4) -- (2, 5) -- (2,3) -- (0,2);
\draw [black, thick, fill=black, opacity=0.3]
(0.,2) -- (3.1, 2) -- (5., 3) -- (2, 3) -- (0.,2);
\node at (2,5.4) {$\fB_{\Neu(\Z_2^{r^2}),s}$};
\draw[dashed] (0,0) -- (3,0);
\draw[dashed] (0,4) -- (3,4);
\draw[dashed] (2,5) -- (5,5);
\draw[dashed] (2,1) -- (5,1);
\draw [black, thick, dashed]
(3,0) -- (3, 4) -- (5, 5) -- (5,1) -- (3,0);
\draw [cyan, thick, fill=cyan, opacity=0.1]
(0,4) -- (3, 4) -- (5, 5) -- (2,5) -- (0,4);
\draw [thick, red] (0.0, 2.) -- (2, 3) ;
\node[below, red!80!black] at (-0.5, 2.4) {$\cE_1^{[r]}$};
\node  at (2.5, 2.55) {$\Q_{2}^{[r]}$};
\node[above] at (1.1, 3) {$D_2^{r}/\cA_{\Z_2^{\wh{r}^2}}$};
\node[above] at (5.5, 2.55) {$,$};
\node[above] at (10.1, 2) {$\Q^{[r]}_{2}\Bigg|_{\Neu(\Z_2^{r^2}),s}=(D_2^{r}/\cA_{\Z_2^{\wh{r}^2}}\,,\cE_1^{[r]})\,,$};
\end{scope}
\end{tikzpicture}    
\end{split}
\ee
$D_2^{r}/\cA_{\Z_2^{\wh{r}^2}}$ is in the same Schur component as the invertible surface $D_2^r$ that generates the $\Z_2^{(0),r}$ subsymmetry of $\fB_{\Neu(\Z_2,s)}$ \cite{Bhardwaj:2022maz}. 
 
Analogously, on $\fB_{\Dir}$, the surface $\Q_2^{[x]}$ was the non-simple combination $D_2^x\oplus D_2^{xr^2}$, with a 2-dimensional space of topological local operators spanned by $D_0^x,D_0^{xr^2}$ that can be combined into local operators transforming under $\Rep(\Z_2)$:
\be
    D_0^{x,\pm}=D_0^{x}\pm D_0^{xr^2}
\ee
of which $D_0^{x,-}$ is charged under $\Z_2^{(0),r^2}$. When gauging $\Z_2^{(0),r^2}$, the SymTFT surface $\Q_2^{[x]}$ projects to a single surface on $\fB_{\Neu(\Z_2^{r^2}),s}$, denoted as $D_2^{x}/\cA_{\Z_2^{\wh{r}^2}}$ that can also arise from the $L$-shaped configuration:
\be
\begin{split}
\begin{tikzpicture}
 \begin{scope}[shift={(0,0)},scale=0.8] 
\draw [cyan, fill=cyan!80!red, opacity =0.5]
(0,0) -- (0,4) -- (2,5) -- (5,5) -- (5,1) -- (3,0)--(0,0);
\draw [black, thick, fill=white,opacity=1]
(0,0) -- (0, 4) -- (2, 5) -- (2,1) -- (0,0);
\draw [cyan, thick, fill=cyan, opacity=0.2]
(0,0) -- (0, 4) -- (2, 5) -- (2,1) -- (0,0);
\draw [black, thick, fill=black, opacity=0.3]
(0.,2) -- (0., 4) -- (2, 5) -- (2,3) -- (0,2);
\draw [black, thick, fill=black, opacity=0.3]
(0.,2) -- (3.1, 2) -- (5., 3) -- (2, 3) -- (0.,2);
\node at (2,5.4) {$\fB_{\Neu(\Z_2^{r^2}),s}$};
\draw[dashed] (0,0) -- (3,0);
\draw[dashed] (0,4) -- (3,4);
\draw[dashed] (2,5) -- (5,5);
\draw[dashed] (2,1) -- (5,1);
\draw [black, thick, dashed]
(3,0) -- (3, 4) -- (5, 5) -- (5,1) -- (3,0);
\draw [cyan, thick, fill=cyan, opacity=0.1]
(0,4) -- (3, 4) -- (5, 5) -- (2,5) -- (0,4);
\draw [thick, red] (0.0, 2.) -- (2, 3) ;
\node[below, red!80!black] at (-0.5, 2.4) {$\cE_1^{[x]}$};
\node  at (2.5, 2.55) {$\Q_{2}^{[x]}$};
\node[above] at (1.1, 3) {$D_2^{x}/\cA_{\Z_2^{\wh{r}^2}}$};
\node[above] at (5.5, 2.55) {$,$};
\node[above] at (10.1, 2) {$\Q^{[x]}_{2}\Bigg|_{\Neu(\Z_2^{r^2}),s}=(D_2^{x}/\cA_{\Z_2^{\wh{r}^2}}\,,\cE_1^{[x]})\,,$};
\end{scope}
\end{tikzpicture}    
\end{split}
\ee
$D_2^{x}/\cA_{\Z_2^{\wh{r}^2}}$ is in the same Schur component as the invertible surface $D_2^x$ that generates the $\Z_2^{(0),x}$ subsymmetry of $\fB_{\Neu(\Z_2^{r^2}),s}$ \cite{Bhardwaj:2022maz}. 

Similarly, on $\fB_{\Dir}$, the surface $\Q_2^{[xr]}$ was the non-simple combination $D_2^{xr}\oplus D_2^{xr^3}$, with a 2-dimensional space of topological local operators spanned by $D_0^{xr},D_0^{xr^3}$ that be combined into local operators transforming under $\Rep(\Z_2)$:
\be
    D_0^{xr,\pm}=D_0^{xr}\pm D_0^{xr^3}
\ee
of which $D_0^{xr,-}$ is charged under $\Z_2^{(0),r^2}$. When gauging $\Z_2^{(0),r^2}$, the SymTFT surface $\Q_2^{[xr]}$ projects to a single surface on $\fB_{\Neu(\Z_2^{r^2}),s}$, denoted $D_2^{xr}/\cA_{\Z_2^{\wh{r}^2}}$, that can also arise from the $L$-shaped configuration:
\be
\begin{split}
\begin{tikzpicture}
 \begin{scope}[shift={(0,0)},scale=0.8] 
\draw [cyan, fill=cyan!80!red, opacity =0.5]
(0,0) -- (0,4) -- (2,5) -- (5,5) -- (5,1) -- (3,0)--(0,0);
\draw [black, thick, fill=white,opacity=1]
(0,0) -- (0, 4) -- (2, 5) -- (2,1) -- (0,0);
\draw [cyan, thick, fill=cyan, opacity=0.2]
(0,0) -- (0, 4) -- (2, 5) -- (2,1) -- (0,0);
\draw [black, thick, fill=black, opacity=0.3]
(0.,2) -- (0., 4) -- (2, 5) -- (2,3) -- (0,2);
\draw [black, thick, fill=black, opacity=0.3]
(0.,2) -- (3.1, 2) -- (5., 3) -- (2, 3) -- (0.,2);
\node at (2,5.4) {$\fB_{\Neu(\Z_2^{r^2}),s}$};
\draw[dashed] (0,0) -- (3,0);
\draw[dashed] (0,4) -- (3,4);
\draw[dashed] (2,5) -- (5,5);
\draw[dashed] (2,1) -- (5,1);
\draw [black, thick, dashed]
(3,0) -- (3, 4) -- (5, 5) -- (5,1) -- (3,0);
\draw [cyan, thick, fill=cyan, opacity=0.1]
(0,4) -- (3, 4) -- (5, 5) -- (2,5) -- (0,4);
\draw [thick, red] (0.0, 2.) -- (2, 3) ;
\node[below, red!80!black] at (-0.5, 2.4) {$\cE_1^{[xr]}$};
\node  at (2.5, 2.55) {$\Q_{2}^{[xr]}$};
\node[above] at (1.1, 3) {$D_2^{xr}/\cA_{\Z_2^{\wh{r}^2}}$};
\node[above] at (5.5, 2.55) {$,$};
\node[above] at (10.1, 2) {$\Q^{[xr]}_{2}\Bigg|_{\Neu(\Z_2^{r^2}),s}=(D_2^{xr}/\cA_{\Z_2^{\wh{r}^2}}\,,\cE_1^{[xr]})\,,$};
\end{scope}
\end{tikzpicture}    
\end{split}
\ee
$D_2^{x}/\cA_{\Z_2^{\wh{r}^2}}$ is in the same Schur component as the invertible surface $D_2^x$ that generates the $\Z_2^{(0),x}$ subsymmetry of $\fB_{\Neu(\Z_2,s)}$ \cite{Bhardwaj:2022maz}.

\paragraph{2-group anomaly.} Let us recall the consequences of the 2-group anomaly \cite{Bhardwaj:2022maz}  
\begin{itemize}
    \item  The $\Z_2^{(0),x}\times\Z_2^{(0),r}$ 0-form symmetry is fractionalized on the line operator generating the $\Z_2^{(1),\wh{r}^2}$ 1-form symmetry:
\be
\begin{tikzpicture}
 \begin{scope}[shift={(0,0)},scale=0.8] 
 \draw [cyan, thick, fill=cyan, opacity=0.2]
(0,0) -- (0, 4) -- (2, 5) -- (2,1) -- (0,0);
\draw [cyan, thick]
(0,0) -- (0, 4) -- (2, 5) -- (2,1) -- (0,0);
\draw [thick, red] (-1, 2) -- (0,2) ;
\draw [dashed, red] (0, 2) -- (1,2) ;
\node[below, red] at (3, 2) {$\cE_1^{\wh{r}^2}$};
\draw [thick, red] (1, 2) -- (4,2) ;
\node[below] at (1.1, 3) {$D_2^{r}$};
\end{scope}
\begin{scope}[shift={(3.2,0)},scale=0.8] 
\draw [cyan, thick, fill=cyan, opacity=0.2]
(0,0) -- (0, 4) -- (2, 5) -- (2,1) -- (0,0);
\draw [cyan, thick]
(0,0) -- (0, 4) -- (2, 5) -- (2,1) -- (0,0);
\draw [dashed, red] (0, 2) -- (1,2) ;
\draw [thick, red] (1, 2) -- (3,2) ;
\node[below] at (1.1, 3) {$D_2^{r}$};
\node at (4, 2) {$=\quad\;\; -$};
\draw [thick, red] (5.5, 2) -- (9,2) ;
\node[below, red] at (7, 2) {$\cE_1^{\wh{r}^2}$};
\end{scope}
\end{tikzpicture}
\ee
\be 
\begin{tikzpicture}
 \begin{scope}[shift={(0,0)},scale=0.8] 
 \draw [cyan, thick, fill=cyan, opacity=0.2]
(0,0) -- (0, 4) -- (2, 5) -- (2,1) -- (0,0);
\draw [cyan, thick]
(0,0) -- (0, 4) -- (2, 5) -- (2,1) -- (0,0);
\draw [thick, red] (-1, 2) -- (0,2) ;
\draw [dashed, red] (0, 2) -- (1,2) ;
\node[below, red] at (3, 2) {$\cE_1^{\wh{r}^2}$};
\draw [thick, red] (1, 2) -- (4,2) ;
\node[below] at (1.1, 3) {$D_2^{x}$};
\end{scope}
\begin{scope}[shift={(3.2,0)},scale=0.8] 
\draw [cyan, thick, fill=cyan, opacity=0.2]
(0,0) -- (0, 4) -- (2, 5) -- (2,1) -- (0,0);
\draw [cyan, thick]
(0,0) -- (0, 4) -- (2, 5) -- (2,1) -- (0,0);
\draw [dashed, red] (0, 2) -- (1,2) ;
\draw [thick, red] (1, 2) -- (3,2) ;
\node[below] at (1.1, 3) {$D_2^{r}$};
\node at (4, 2) {$=\;\; -$};
\end{scope}
 \begin{scope}[shift={(8,0)},scale=0.8] 
 \draw [cyan, thick, fill=cyan, opacity=0.2]
(0,0) -- (0, 4) -- (2, 5) -- (2,1) -- (0,0);
\draw [cyan, thick]
(0,0) -- (0, 4) -- (2, 5) -- (2,1) -- (0,0);
\draw [thick, red] (-1, 2) -- (0,2) ;
\draw [dashed, red] (0, 2) -- (1,2) ;
\node[below, red] at (3, 2) {$\cE_1^{\wh{r}^2}$};
\draw [thick, red] (1, 2) -- (4,2) ;
\node[below] at (1.1, 3) {$D_2^{x}$};
\end{scope}
\begin{scope}[shift={(11.2,0)},scale=0.8] 
\draw [cyan, thick, fill=cyan, opacity=0.2]
(0,0) -- (0, 4) -- (2, 5) -- (2,1) -- (0,0);
\draw [cyan, thick]
(0,0) -- (0, 4) -- (2, 5) -- (2,1) -- (0,0);
\draw [dashed, red] (0, 2) -- (1,2) ;
\draw [thick, red] (1, 2) -- (3,2) ;
\node[below] at (1.1, 3) {$D_2^{r}$};
\end{scope}
\end{tikzpicture}
\ee
\item As discussed around eq. \eqref{eq:Neu_r2_frac}, the local operators $\cE_0^{E,1}$ and $\cE_0^{E,2}$ (in the twisted sector for $\Z_2^{(1),\wh{r}^2}$) are fractionally charged under the $\Z_2^{(0),r}$ generator $D_2^r$, transforming as $\pm i$ respectively;
\item Additionally, {$\cE_1^r$ is fractionally charged under $D_1^{\wh{r}^2}$.} This follows from the fact that $\cE_1^{r}\otimes \cE_1^{r}= \cE_1^{r^2}$ that has a braiding phase of $-1$ with $D_1^{\wh{r}^2}$.
\end{itemize}

\subsubsection{Neumann$(\Z_4^r)$ Boundary Condition} 
The boundary condition obtained by gauging the $\Z_4^r$ subsymmetry of $\fB_{\Dir}$ with possible discrete torsion $p\in H^{3}(\Z_4,U(1))\cong \Z_4$ will be denoted by:
\begin{equation}\label{eq:D8_Neu_r}
\fB_{\Neu(\Z_4^r),p}= \frac{\fB_{\Dir}\boxtimes {\rm SPT}_p }{\Z_4} \,.
\end{equation}
After gauging one obtains a dual $\Z_4^{\wh{r}}$ symmetry, and a residual 0-form symmetry given by:
\be
    D_8/\Z_4^r=\{1\sim(1,r,r^2,r^3),\;x\sim(x,xr,xr^2,x^3)\}\cong\Z_2^x
\ee
that acts on the 1-form symmetry by exchanging $r\leftrightarrow r^3$, giving the 2-group symmetry \cite{Bhardwaj:2022maz}
\be
    \cS=\TwoVec_{\Z_4^{(1)}\rtimes \Z_2^{(0),x}}\,.
\ee
\paragraph{Genuine local operator.}
Since the operator $\cE_0^{1_r}$ at the end of $\Q_1^{1_r}$ was untwisted and uncharged under the $\Z_4^{r}$ symmetry, it remains unaltered after gauging 
\be
\begin{split}
\begin{tikzpicture}
 \begin{scope}[shift={(0,0)},scale=0.8] 
\draw [cyan, fill=cyan!80!red, opacity =0.5]
(0,0) -- (0,4) -- (2,5) -- (5,5) -- (5,1) -- (3,0)--(0,0);
\draw [black, thick, fill=white,opacity=1]
(0,0) -- (0, 4) -- (2, 5) -- (2,1) -- (0,0);
\draw [cyan, thick, fill=cyan, opacity=0.2]
(0,0) -- (0, 4) -- (2, 5) -- (2,1) -- (0,0);
\draw[line width=1pt] (1,2.5) -- (4.2,2.5);
\fill[red!80!black] (1,2.5) circle (3pt);
\node at (2,5.4) {$\fB_{\Neu(\Z_4^r),p}$};
\draw [ black, thick, dashed]
(3,0) -- (3, 4) -- (5, 5) -- (5,1) -- (3,0);
\draw[dashed] (0,0) -- (3,0);
\draw[dashed] (0,4) -- (3,4);
\draw[dashed] (2,5) -- (5,5);
\draw[dashed] (2,1) -- (5,1);
\draw [cyan, thick, fill=cyan, opacity=0.1]
(0,4) -- (3, 4) -- (5, 5) -- (2,5) -- (0,4);
\node[below, red!80!black] at (.8, 2.4) {$\cE_0^{1_r}$};
\node  at (2.5, 3) {$\Q_{1}^{1_r}$};
\node  at (5.5, 2.5) {$,$};
\node  at (11, 2.5) {$\Q_{1}^{1_r}\Bigg|_{\Neu(\Z_4^r),p}=\cE_0^{1_r}\;.$};
\end{scope}
\end{tikzpicture}    
\end{split}
\ee
The local operator $\cE_0^{{1_r}}$ transforms with a $-1$ under $\Z_2^{(0),x}$.

\paragraph{Lines generating the dual symmetry after gauging and condensation surfaces.}
After gauging the $\Z_4^{(0),r}$ symmetry of $\fB_\Dir$, one obtains a dual $\Z_4^{(1),\wh{r}}$ symmetry on $\fB_{\Neu(\Z_4^{r}),p}$, whose generator we denote by $D_1^{\wh{r}}$. The algebras:
\begin{align}
     \cA_{\Z_2^{\wh{r}^2}}&=D_1^1\oplus D_1^{\wh{r}^2}\,,\\
    \cA_{\Z_4^{\wh{r}}}&=D_1^1\oplus D_1^{\wh{r}}\oplus D_1^{\wh{r}^2}\oplus D_1^{\wh{r}^3} \,, \label{eq:cA_Z4}
\end{align}
can be condensed on the identity surface, giving rise to condensation defects 
\be
   D_2^{\Z_2^{\wh{r}^2}}=\frac{D_2^{[\id]}}{\cA_{\Z_2^{\wh{r}^2}}}\,,\qquad D_2^{\Z_4^{\wh{r}}}=\frac{D_2^{[\id]}}{\cA_{\Z_4^{\wh{r}}}}\,.
\ee

\paragraph{Non-genuine local operators.}
On $\fB_{\Dir}$, the operators $\cE_0^{1_x}$ and $\cE_0^{1_{xr}}$ transform as $-1$ under $\Z_4^{(0),r}$, therefore after gauging $\Z_4^{r}$, they become attached to the dual $\Z_2^{(1),\wh{r}^2}\subset\Z_4^{(1),\wh{r}}$ symmetry generator $D_1^{\wh{r}^2}$.
\be
\begin{split}
\begin{tikzpicture}
 \begin{scope}[shift={(0,0)},scale=0.8] 
\draw [cyan, fill=cyan!80!red, opacity =0.5]
(0,0) -- (0,4) -- (2,5) -- (5,5) -- (5,1) -- (3,0)--(0,0);
\draw [black, thick, fill=white,opacity=1]
(0,0) -- (0, 4) -- (2, 5) -- (2,1) -- (0,0);
\draw [cyan, thick, fill=cyan, opacity=0.2]
(0,0) -- (0, 4) -- (2, 5) -- (2,1) -- (0,0);
\draw[line width=1pt] (1,2.5) -- (3,2.5);
\draw[line width=1pt] (1,2.5) -- (1,4.5);
\draw[line width=1pt] (3,2.5) -- (4.2,2.5);
\fill[red!80!black] (1,2.5) circle (3pt);
\node at (2,5.4) {$\fB_{\Neu(\Z_4^r),p}$};
\draw[dashed] (0,0) -- (3,0);
\draw[dashed] (0,4) -- (3,4);
\draw[dashed] (2,5) -- (5,5);
\draw[dashed] (2,1) -- (5,1);
\draw [cyan, thick, fill=cyan, opacity=0.1]
(0,4) -- (3, 4) -- (5, 5) -- (2,5) -- (0,4);
\node[below, red!80!black] at (.8, 2.4) {$\cE_0^{1_{x(r)}}$};
\draw [black, thick, dashed]
(3,0) -- (3, 4) -- (5, 5) -- (5,1) -- (3,0);
\node  at (2.75 , 3) {$\Q_{1}^{1_{x(r)}}$};
\node  at (0.5, 3.3) {$D_{1}^{\wh{r}^2}$};
\node  at (5.5, 2.5) {$,$};
\node  at (11, 3.5) {$ \Q_{1}^{1_x}\Bigg|_{\Neu(\Z_4^r),p}=(D_1^{\wh{r}^2},\cE_0^{1_x})\,,
$};
\node  at (11, 1.5) {$\Q_{1}^{1_{xr}}\Bigg|_{\Neu(\Z_4^r),p}=(D_1^{\wh{r}^2},\cE_0^{1_{xr}})\,.
$};
\end{scope}
\end{tikzpicture}    
\end{split}
\ee
Under $\Z_2^{(0),x}$, the local operator $\cE_0^{1_x}$ is uncharged while $\cE_0^{1{xr}}$ carries charge $-1$.

On $\fB_\Dir,\; \cE_0^{E,1}$ and $\cE_0^{E,2}$ transform under $\Z_4^{(0),r}$ as $i$ and $-i$ respectively. Therefore, after gauging $\Z_4^{r}$, they become attached to the dual $\Z_4^{(1)}$ symmetry generators $D_1^{\wh{r}}$ and $D_1^{\wh{r}^3}$ respectively. 
\be
\begin{split}
\begin{tikzpicture}
 \begin{scope}[shift={(0,0)},scale=0.8] 
\draw [cyan, fill=cyan!80!red, opacity =0.5]
(0,0) -- (0,4) -- (2,5) -- (5,5) -- (5,1) -- (3,0)--(0,0);
\draw [black, thick, fill=white,opacity=1]
(0,0) -- (0, 4) -- (2, 5) -- (2,1) -- (0,0);
\draw [cyan, thick, fill=cyan, opacity=0.2]
(0,0) -- (0, 4) -- (2, 5) -- (2,1) -- (0,0);
\draw[line width=1pt] (1,2.5) -- (3,2.5);
\draw[line width=1pt] (1,2.5) -- (1,4.5);
\draw[line width=1pt,dashed] (3,2.5) -- (4.2,2.5);
\fill[red!80!black] (1,2.5) circle (3pt);
\node at (2,5.4) {$\fB_{\Neu(\Z_4^r),p}$};
\draw[dashed] (0,0) -- (3,0);
\draw[dashed] (0,4) -- (3,4);
\draw[dashed] (2,5) -- (5,5);
\draw[dashed] (2,1) -- (5,1);
\draw [cyan, thick, fill=cyan, opacity=0.1]
(0,4) -- (3, 4) -- (5, 5) -- (2,5) -- (0,4);
\node[below, red!80!black] at (.8, 2.4) {$\cE_0^{E,j}$};
\draw [black, thick, dashed]
(3,0) -- (3, 4) -- (5, 5) -- (5,1) -- (3,0);
\node  at (2.5 , 3) {$\Q_{1}^{E}$};
\node  at (0.75, 3.3) {$D_{1}^{\wh{r}^{(2j-1)}}$};
\node  at (5.5, 2.5) {$,$};
\node  at (11, 2.5) {$\Q_{1}^{E}\Bigg|_{{\Neu(\Z_4^r),p}}=\left\{(D_1^{\wh{r}},\cE_0^{E,1})\,, (D_1^{\wh{r}^3}, \cE_0^{E,2})\right\}\,.$};
\end{scope}
\end{tikzpicture}    
\end{split}
\ee
Furthermore, since $\cD_{E}(x): \cE_0^{E,1}\leftrightarrow \cE_0^{E,2}$ on $\fB_{\Dir}$, the corresponding twisted sector operators in $\fB_{\Neu(\Z_4^r),p}$ are also acted upon by the $\Z_2^{(0),x}$ symmetry
\begin{equation}
    D_2^{x}: \quad (D_1^{\wh{r}},\cE_0^{E,1}) \longleftrightarrow 
    (D_1^{\wh{r}^3}, \cE_0^{E,2})\,.
\end{equation}
From which we read off the 2-group action of $\Z_2^{(0),x}$ on $\Z_{4}^{(1),r}$ 
\begin{equation}
    D_2^{x}:\quad  D_1^{\wh{r}} \longleftrightarrow 
    D_1^{\wh{r}^3}\,.
\end{equation}

\paragraph{Genuine topological lines.}
The $\Z_4^{(0),r}$ twisted sector lines $\cE_1^{r},\,\cE_1^{r^2}$ and $\cE_1^{r^3}$ on $\fB_{\Dir}$ become untwisted sector lines that are charged under the dual $\Z_4^{(1),\wh{r}}$ symmetry on $\fB_{\Neu(\Z_4^r),p}$:
\be
\begin{split}
\begin{tikzpicture}
 \begin{scope}[shift={(0,0)},scale=0.8] 
\draw [cyan, fill=cyan!80!red, opacity =0.5]
(0,0) -- (0,4) -- (2,5) -- (5,5) -- (5,1) -- (3,0)--(0,0);
\draw [black, thick, fill=white,opacity=1]
(0,0) -- (0, 4) -- (2, 5) -- (2,1) -- (0,0);
\draw [cyan, thick, fill=cyan, opacity=0.2]
(0,0) -- (0, 4) -- (2, 5) -- (2,1) -- (0,0);
\draw [black, thick, fill=black, opacity=0.3]
(0.,2) -- (3.1, 2) -- (5., 3) -- (2, 3) -- (0.,2);
\node at (2,5.4) {$\fB_{\Neu(\Z_4^r),p}$};
\draw[dashed] (0,0) -- (3,0);
\draw[dashed] (0,4) -- (3,4);
\draw[dashed] (2,5) -- (5,5);
\draw[dashed] (2,1) -- (5,1);
\draw [black, thick, dashed]
(3,0) -- (3, 4) -- (5, 5) -- (5,1) -- (3,0);
\draw [cyan, thick, fill=cyan, opacity=0.1]
(0,4) -- (3, 4) -- (5, 5) -- (2,5) -- (0,4);
\draw [line width=1pt, red!80!black] (0.0, 2.) -- (2, 3) ;
\node[below, red!80!black] at (-0.5, 2.4) {$\cE_1^{r^j}$};
\node  at (2.5, 2.55) {$\Q_{2}^{[r^j]}$};
\node  at (5.5, 2.55) {$,$};
\node  at (10.5, 3.5) {$\Q_2^{[r]}\Bigg|_{{\Neu(\Z_4^r),p}}= \left\{\cE_1^{r}\,, \cE_1^{r^3}\right\}\,,$};
\node  at (10.5, 1.5) {$\Q_2^{[r^2]}\Bigg|_{{\Neu(\Z_4^r),p}}= \cE_1^{r^2}\,.\qquad\quad$};
\end{scope}
\end{tikzpicture}    
\end{split}
\ee
The linking phase between $\cE_1^{r^n}$ and $D_1^{\wh{r}^m}$ is $i^{nm}$, while the F-symbols and fusion properties of $\cE_1^{r^n}$ are controlled by the choice of $p\in \Z_4$:
\begin{equation} 
    F(\cE_1^{r^l}\,, \cE_1^{r^m}\,, \cE_1^{r^n}) = \exp\left\{\frac{2\pi i p}{16}\,l(m+n-[m+n]_4)\right\}\,,
    \label{eq:Z3 1fs charge line algebra}
\end{equation}
where $[m+n]_4\equiv m+n \text{ mod }4$.
\paragraph{Local operators on non-identity lines.}
The 1d irrep $1_r$ on $\Q_{1}^{[r^2]}$ gives rise to a local operator $\cE_0^{1_r}$ on $\cE_1^{r^2}$:
\be
\ba
  \Q_{1}^{[r^2], 1_r}\Bigg|_{{\Neu(\Z_4^{r}),p}}&=\cE_1^{r^2}\cE_0^{1_r}\,, 
\ea
\ee
charged under the $\Z_2^{(0),x}$ 0-form symmetry.

\paragraph{Surfaces that cannot end on $\fB_{\Neu(\Z_4^{r}),p}$.}
When gauging $\Z_4^{(0),r}$, the SymTFT surface $\Q_2^{[r]}$ becomes transparent. Since $\Q_2^{[x]}$ and $\Q_2^{[xr]}$ are related by fusing with $\Q_2^{[r]}$, they must be identified on $\fB_{\Neu(\Z_4^r),p}$.
On $\fB_{\Dir}$, the surface $\Q_2^{[x]}\oplus\Q_2^{[xr]}$ was the non-simple combination $D_2^x\oplus D_2^{xr}\oplus D_2^{xr^2}\oplus D_2^{xr^3}$, with a 4-dimensional space of topological local operators that can be combined into local operators transforming under $\Rep(\Z_4)$:
\be
    D_0^{x,i^{n}}=D_0^{x}+i^{-n}D_0^{xr}+i^{-2n}D_0^{x}+i^{-3n}D_0^{xr^3}
\ee
with charge $i^n$ under $\Z_4^{(0),r}$ on $\fB_\Dir$. When gauging $\Z_4^{(0),r}$, the SymTFT surface $\Q_2^{[x]}\oplus\Q_2^{[xr]}$ projects to a single surface on $\fB_{\Neu(\Z_4^{r}),p}$, denoted as $D_2^{x}/\cA_{\Z_4^{\wh{r}}}$ on which $\cA_{\Z_4^{\wh{r}}}$, eq. \eqref{eq:cA_Z4}, is condensed. $D_2^{x}/\cA_{\Z_4^{\wh{r}}}$ can also arise from the $L$-shaped configuration:
\be
\begin{split}
\begin{tikzpicture}
 \begin{scope}[shift={(0,0)},scale=0.8] 
\draw [cyan, fill=cyan!80!red, opacity =0.5]
(0,0) -- (0,4) -- (2,5) -- (5,5) -- (5,1) -- (3,0)--(0,0);
\draw [black, thick, fill=white,opacity=1]
(0,0) -- (0, 4) -- (2, 5) -- (2,1) -- (0,0);
\draw [cyan, thick, fill=cyan, opacity=0.2]
(0,0) -- (0, 4) -- (2, 5) -- (2,1) -- (0,0);
\draw [black, thick, fill=black, opacity=0.3]
(0.,2) -- (0., 4) -- (2, 5) -- (2,3) -- (0,2);
\draw [black, thick, fill=black, opacity=0.3]
(0.,2) -- (3.1, 2) -- (5., 3) -- (2, 3) -- (0.,2);
\node at (2,5.4) {$\fB_{\Neu(\Z_4^r),p}$};
\draw[dashed] (0,0) -- (3,0);
\draw[dashed] (0,4) -- (3,4);
\draw[dashed] (2,5) -- (5,5);
\draw[dashed] (2,1) -- (5,1);
\draw [black, thick, dashed]
(3,0) -- (3, 4) -- (5, 5) -- (5,1) -- (3,0);
\draw [cyan, thick, fill=cyan, opacity=0.1]
(0,4) -- (3, 4) -- (5, 5) -- (2,5) -- (0,4);
\draw [thick, red] (0.0, 2.) -- (2, 3) ;
\node[below, red!80!black] at (-0.5, 2.4) {$\cE_1^{[x]}$};
\node  at (2.5, 2.55) {$\Q_{2}^{[x]}$};
\node[above] at (1.1, 3) {$D_2^{x}/\cA_{\Z_4^{\wh{r}}}$};
\node[above] at (5.5, 2.55) {$,$};
\node[above] at (10.1, 2) {$\Q^{[x]}_{2}\Bigg|_{\Neu(\Z_4^r),p}=(D_2^{x}/\cA_{\Z_4^{\wh{r}}}\,,\cE_1^{[x]})\,,$};
\end{scope}
\end{tikzpicture}    
\end{split}
\ee
$D_2^{x}/\cA_{\Z_4^{\wh{r}}}$ is in the same Schur component as the invertible surface $D_2^x$ that generates the $\Z_2^{(0),x}$ subsymmetry of $\fB_{\Neu(\Z_4^r),p}$ \cite{Bhardwaj:2022maz}.

\subsubsection{Neumann$(\Z_2^{xr}\times\Z_2^{xr^3})$ and Neumann$(\Z_2^x\times\Z_2^{xr^2})$ Boundary Conditions} \label{sec:D8_NeuZ2xrZ2xr3}

The SymTFT boundary condition obtained by gauging the $\Z_2^{(0),{xr}}\times\Z_2^{(0),xr^3}$ subsymmetry of $\fB_{\Dir}$ with possible discrete torsion 
\be
    t\in H^3(\Z_2\times\Z_2,U(1))=\Z_2\times\Z_2\times\Z_2
\ee
will be denoted as:
\begin{equation}\label{eq:D8_Neu_x_xr2}
 \fB_{\Neu(\Z_2^{xr}\times\Z_2^{xr^3}),t}= \frac{\fB_{\Dir}\boxtimes {\rm SPT}_t }{\Z_2\times\Z_2} \,.
\end{equation}
where the gauged $\Z_2\times\Z_2$ is the diagonal of $\Z_2^{xr}\times\Z_2^{xr^3}\subset D_8$ and the $\Z_2\times\Z_2$ symmetry of SPT$_t$.
Since $D_8$ can be presented equivalently as:
\be \label{eq:D8_out_aut}
\ba
    \mathbb{Z}_4^r \rtimes \mathbb{Z}_2^{xr}:&\qquad \{\id,\,r,\,r^2,\,r^3,xr,xr^2,xr^3,x\} \\
    \mathbb{Z}_4^r \rtimes \mathbb{Z}_2^x:&\qquad \{\id,\,r,\,r^2,\,r^3,x,xr,xr^2,xr^3\} \\
\ea
\ee
there is an outer automorphism that relates $x$ and $xr$, so the features of $\fB_{\Neu(\Z_2^x\times\Z_2^{xr^2},t)}$ can be deduced from those of $\fB_{\Neu(\Z_2^{xr}\times\Z_2^{xr^3}),t}$ by mapping from the first to the second presentation of $D_8$ in equation \eqref{eq:D8_out_aut}.

\noindent After gauging the $\Z_2^{xr}\times\Z_2^{xr^3}$ 0-form symmetry on $\fB_\Dir$, one obtains $\fB_{\Neu(\Z_2^{xr}\times\Z_2^{xr^3}),t}$ with a dual $\Z_2^{\wh{xr}}\times\Z_2^{\wh{xr}^3}$ 1-form symmetry and a residual 0-form symmetry given by:
\be \label{eq:D8_NeuZ2Z2_0-form}
    D_8/(\Z_2^x\times\Z_2^{xr^2})=\{1\sim(1,xr,r^2,xr^3),\;x\sim(x,r,xr^2,r^3)\}\cong\Z_2^x
\ee
that acts on the 1-form symmetry by exchanging $xr\leftrightarrow xr^3$, giving the 2-group symmetry 
\be
    \cS=\TwoVec_{(\Z_2^{(1),xr}\times\Z_2^{(1),xr^3})\rtimes \Z_2^{(0),x}}\,.
\ee
\paragraph{Genuine local operator.}
Since the operator $\cE_0^{1_{xr}}$ at the end of $\Q_1^{1_{xr}}$ was untwisted and uncharged under the $\Z_2^{xr}\times\Z_2^{xr^3}$ symmetry, it remains unaltered after gauging 
\be
\begin{split}
\begin{tikzpicture}
 \begin{scope}[shift={(0,0)},scale=0.8] 
\draw [cyan, fill=cyan!80!red, opacity =0.5]
(0,0) -- (0,4) -- (2,5) -- (5,5) -- (5,1) -- (3,0)--(0,0);
\draw [black, thick, fill=white,opacity=1]
(0,0) -- (0, 4) -- (2, 5) -- (2,1) -- (0,0);
\draw [cyan, thick, fill=cyan, opacity=0.2]
(0,0) -- (0, 4) -- (2, 5) -- (2,1) -- (0,0);
\draw[line width=1pt] (1,2.5) -- (4.2,2.5);
\fill[red!80!black] (1,2.5) circle (3pt);
\node at (2,5.4) {$\fB_{\Neu(\Z_2^{xr}\times\Z_2^{xr^3}),t}$};
\draw [ black, thick, dashed]
(3,0) -- (3, 4) -- (5, 5) -- (5,1) -- (3,0);
\draw[dashed] (0,0) -- (3,0);
\draw[dashed] (0,4) -- (3,4);
\draw[dashed] (2,5) -- (5,5);
\draw[dashed] (2,1) -- (5,1);
\draw [cyan, thick, fill=cyan, opacity=0.1]
(0,4) -- (3, 4) -- (5, 5) -- (2,5) -- (0,4);
\node[below, red!80!black] at (.8, 2.4) {$\cE_0^{1_{xr}}$};
\node  at (2.5, 3) {$\Q_{1}^{1_{xr}}$};
\node  at (5.5, 2.5) {$,$};
\node  at (11, 2.5) {$\Q_{1}^{1_{xr}}\Bigg|_{\Neu(\Z_2^{xr}\times\Z_2^{xr^3}),t}=\cE_0^{1_{xr}}\;.$};
\end{scope}
\end{tikzpicture}    
\end{split}
\ee
The local operator $\cE_0^{{1_{xr}}}$ transforms with a $-1$ under $\Z_2^{(0),x}$.

\paragraph{Lines generating the dual symmetry after gauging and condensation surfaces.}
After gauging the $\Z_2^{(0),xr}\times\Z_2^{(0),xr^3}$ symmetry of $\fB_\Dir$, one obtains a dual $\Z_2^{(1),\wh{xr}}\times\Z_2^{(1),\wh{xr}^3}$ symmetry on $\fB_{\Neu(\Z_2^{xr}\times\Z_2^{xr^3}),t}$, whose generators we denote by $D_1^{\wh{xr}}\,,\;D_1^{\wh{xr}^3}$. The full list of condensation surfaces and their fusion is provided in sec. 5.7 of \cite{Bhardwaj:2022maz}.

\paragraph{Non-genuine local operators.}
On $\fB_{\Dir}$, the operators $\cE_0^{1_{x}}$ and $\cE_0^{1_{r}}$ transform as $-1$ under $\Z_2^{(0),xr}$ and $\Z_2^{(0),xr^3}$, therefore after gauging $\Z_2^{(0),xr}\times\Z_2^{(0),xr^3}$, they become attached to the dual $\Z_2^{(1),\wh{xr}}$ symmetry generator $D_1^{\wh{xr}}$.
\be
\begin{split}
\begin{tikzpicture}
 \begin{scope}[shift={(0,0)},scale=0.8] 
\draw [cyan, fill=cyan!80!red, opacity =0.5]
(0,0) -- (0,4) -- (2,5) -- (5,5) -- (5,1) -- (3,0)--(0,0);
\draw [black, thick, fill=white,opacity=1]
(0,0) -- (0, 4) -- (2, 5) -- (2,1) -- (0,0);
\draw [cyan, thick, fill=cyan, opacity=0.2]
(0,0) -- (0, 4) -- (2, 5) -- (2,1) -- (0,0);
\draw[line width=1pt] (1,2.5) -- (3,2.5);
\draw[line width=1pt] (1,2.5) -- (1,4.5);
\draw[line width=1pt] (3,2.5) -- (4.2,2.5);
\fill[red!80!black] (1,2.5) circle (3pt);
\node at (2,5.4) {$\fB_{\Neu(\Z_2^{xr}\times\Z_2^{xr^3}),t}$};
\draw[dashed] (0,0) -- (3,0);
\draw[dashed] (0,4) -- (3,4);
\draw[dashed] (2,5) -- (5,5);
\draw[dashed] (2,1) -- (5,1);
\draw [cyan, thick, fill=cyan, opacity=0.1]
(0,4) -- (3, 4) -- (5, 5) -- (2,5) -- (0,4);
\node[below, red!80!black] at (.8, 2.4) {$\cE_0^{1_{x,r}}$};
\draw [black, thick, dashed]
(3,0) -- (3, 4) -- (5, 5) -- (5,1) -- (3,0);
\node  at (2.75 , 3) {$\Q_{1}^{1_{x,r}}$};
\node  at (0.5, 3.3) {$D_{1}^{\wh{xr}}$};
\node  at (5.5, 2.5) {$,$};
\node  at (11, 3.5) {$ \Q_{1}^{1_x}\Bigg|_{\Neu(\Z_2^{xr}\times\Z_2^{xr^3}),t}=(D_1^{\wh{xr}},\cE_0^{1_{x}})\,,
$};
\node  at (11, 1.5) {$\Q_{1}^{1_{r}}\Bigg|_{\Neu(\Z_2^{xr}\times\Z_2^{xr^3}),t}=(D_1^{\wh{xr}},\cE_0^{1_{r}})\,.
$};
\end{scope}
\end{tikzpicture}    
\end{split}
\ee
Under $\Z_2^{(0),x}$, the local operator $\cE_0^{1_x}$ is uncharged while $\cE_0^{1_r}$ carries charge $-1$. We can unitarily transform $\cD_E(xr)=\sigma^y=\begin{pmatrix}
    0 & -i \\ i & 0
\end{pmatrix}$ to $\sigma^z=\diag(1,-1)$. Let us denote by 
\be
    \cE_0^{E,\pm i}:=\cE_0^{E,1}\pm i \cE_0^{E,2}
\ee
the new basis elements. The charge of $\cE_0^{E,\pm i}$ under $(\Z_2^{xr},\Z_2^{xr^3})$ is $(+,-)$ and $(-,+)$ respectively. Therefore upon gauging $\Z_2^{xr}\times\Z_2^{xr^3}$, we find
\begin{equation}
    \Q_1^{E}=\left\{(D_1^{\wh{xr}^3},\cE_0^{E,+i})\,, \;(D_1^{\wh{xr}},\cE_0^{E,-i})\right\}\,.
\end{equation}
\be
\begin{split}
\begin{tikzpicture}
 \begin{scope}[shift={(0,0)},scale=0.8] 
\draw [cyan, fill=cyan!80!red, opacity =0.5]
(0,0) -- (0,4) -- (2,5) -- (5,5) -- (5,1) -- (3,0)--(0,0);
\draw [black, thick, fill=white,opacity=1]
(0,0) -- (0, 4) -- (2, 5) -- (2,1) -- (0,0);
\draw [cyan, thick, fill=cyan, opacity=0.2]
(0,0) -- (0, 4) -- (2, 5) -- (2,1) -- (0,0);
\draw[line width=1pt] (1,2.5) -- (3,2.5);
\draw[line width=1pt] (1,2.5) -- (1,4.5);
\draw[line width=1pt] (3,2.5) -- (4.2,2.5);
\fill[red!80!black] (1,2.5) circle (3pt);
\node at (2,5.4) {$\fB_{\Neu(\Z_2^{xr}\times\Z_2^{xr^3}),t}$};
\draw[dashed] (0,0) -- (3,0);
\draw[dashed] (0,4) -- (3,4);
\draw[dashed] (2,5) -- (5,5);
\draw[dashed] (2,1) -- (5,1);
\draw [black, thick, dashed]
(3,0) -- (3, 4) -- (5, 5) -- (5,1) -- (3,0);
\draw [cyan, thick, fill=cyan, opacity=0.1]
(0,4) -- (3, 4) -- (5, 5) -- (2,5) -- (0,4);
\node[below, red!80!black] at (.8, 2.4) {$\cE_0^{E,+i}$};
\node  at (2.5, 3) {$\Q_{1}^{E}$};
\node  at (0.5, 3.3) {$D_{1}^{\wh{xr}^3}$};
\node  at (5.5, 2.5) {$,$};
\end{scope}
\begin{scope}[shift={(7,0)},scale=0.8] 
\draw [cyan, fill=cyan!80!red, opacity =0.5]
(0,0) -- (0,4) -- (2,5) -- (5,5) -- (5,1) -- (3,0)--(0,0);
\draw [black, thick, fill=white,opacity=1]
(0,0) -- (0, 4) -- (2, 5) -- (2,1) -- (0,0);
\draw [cyan, thick, fill=cyan, opacity=0.2]
(0,0) -- (0, 4) -- (2, 5) -- (2,1) -- (0,0);
\draw[line width=1pt] (1,2.5) -- (3,2.5);
\draw[line width=1pt] (1,2.5) -- (1,4.5);
\draw[line width=1pt] (3,2.5) -- (4.2,2.5);
\fill[red!80!black] (1,2.5) circle (3pt);
\node at (2,5.4) {$\fB_{\Neu(\Z_2^{xr}\times\Z_2^{xr^3}),t}$};
\draw[dashed] (0,0) -- (3,0);
\draw[dashed] (0,4) -- (3,4);
\draw[dashed] (2,5) -- (5,5);
\draw[dashed] (2,1) -- (5,1);
\draw [black, thick, dashed]
(3,0) -- (3, 4) -- (5, 5) -- (5,1) -- (3,0);
\draw [cyan, thick, fill=cyan, opacity=0.1]
(0,4) -- (3, 4) -- (5, 5) -- (2,5) -- (0,4);
\node[below, red!80!black] at (.8, 2.4) {$\cE_0^{E,-i}$};
\node  at (2.5, 3) {$\Q_{1}^{E}$};
\node  at (0.5, 3.3) {$D_{1}^{\wh{xr}}$};
\node  at (5.5, 2.5) {$,$};
\end{scope}
\end{tikzpicture}    
\end{split}
\ee
Furthermore, since in this basis $\cD_{E}(x)=\sigma^y$, the 2-group action of $\Z_2^{(0),x}$:
\be
\ba
D_2^{x}: \quad &(D_1^{\wh{xr}^3},\cE_0^{E,+i}) \mapsto  
    i(D_1^{\wh{xr}},\cE_0^{E,-i})\,, \qquad 
    (D_1^{\wh{xr}},\cE_0^{E,-i}) \mapsto  
    -i(D_1^{\wh{xr}^3},\cE_0^{E,+i})\,.
\ea
\ee

\paragraph{Genuine topological lines.}
The $\Z_2^{(0),xr}\times\Z_2^{(0),xr^3}$ twisted sector lines $\cE_1^{xr},\,\cE_1^{r^2},\,\cE^{xr^3}$ on $\fB_{\Dir}$ become untwisted sector lines that are charged under the dual $\Z_2^{(1),\wh{xr}}\times\Z_2^{(1),\wh{xr}^3}$ symmetry on $\fB_{\Neu(\Z_2^{xr}\times\Z_2^{xr^3}),t}$:
\begin{align}
&\begin{tikzpicture}
 \begin{scope}[shift={(0,0)},scale=0.8] 
\draw [cyan, fill=cyan!80!red, opacity =0.5]
(0,0) -- (0,4) -- (2,5) -- (5,5) -- (5,1) -- (3,0)--(0,0);
\draw [black, thick, fill=white,opacity=1]
(0,0) -- (0, 4) -- (2, 5) -- (2,1) -- (0,0);
\draw [cyan, thick, fill=cyan, opacity=0.2]
(0,0) -- (0, 4) -- (2, 5) -- (2,1) -- (0,0);
\draw [black, thick, fill=black, opacity=0.3]
(0.,2) -- (3.1, 2) -- (5., 3) -- (2, 3) -- (0.,2);
\node at (2,5.4) {$\fB_{\Neu(\Z_2^{xr}\times\Z_2^{xr^3}),t}$};
\draw[dashed] (0,0) -- (3,0);
\draw[dashed] (0,4) -- (3,4);
\draw[dashed] (2,5) -- (5,5);
\draw[dashed] (2,1) -- (5,1);
\draw [black, thick, dashed]
(3,0) -- (3, 4) -- (5, 5) -- (5,1) -- (3,0);
\draw [cyan, thick, fill=cyan, opacity=0.1]
(0,4) -- (3, 4) -- (5, 5) -- (2,5) -- (0,4);
\draw [line width=1pt, red!80!black] (0.0, 2.) -- (2, 3) ;
\node[below, red!80!black] at (-1.5, 2.4) {$\cE_1^{xr},\,\cE_1^{xr^3}$};
\node  at (2.5, 2.55) {$\Q_{2}^{[xr]}$};
\node  at (5.5, 2.55) {$,$};
\node  at (10.5, 2.55) {$\Q_2^{[xr]}\Bigg|_{\Neu(\Z_2^{xr}\times\Z_2^{xr^3}),t}= \left\{\cE_1^{xr}\,, \cE_1^{xr^3}\right\}\,,$};
\end{scope}
\end{tikzpicture}    
\\
&
\begin{tikzpicture}
 \begin{scope}[shift={(1,0)},scale=0.8] 
\draw [cyan, fill=cyan!80!red, opacity =0.5]
(0,0) -- (0,4) -- (2,5) -- (5,5) -- (5,1) -- (3,0)--(0,0);
\draw [black, thick, fill=white,opacity=1]
(0,0) -- (0, 4) -- (2, 5) -- (2,1) -- (0,0);
\draw [cyan, thick, fill=cyan, opacity=0.2]
(0,0) -- (0, 4) -- (2, 5) -- (2,1) -- (0,0);
\draw [black, thick, fill=black, opacity=0.3]
(0.,2) -- (3.1, 2) -- (5., 3) -- (2, 3) -- (0.,2);
\node at (2,5.4) {$\fB_{\Neu(\Z_2^{xr}\times\Z_2^{xr^3}),t}$};
\draw[dashed] (0,0) -- (3,0);
\draw[dashed] (0,4) -- (3,4);
\draw[dashed] (2,5) -- (5,5);
\draw[dashed] (2,1) -- (5,1);
\draw [black, thick, dashed]
(3,0) -- (3, 4) -- (5, 5) -- (5,1) -- (3,0);
\draw [cyan, thick, fill=cyan, opacity=0.1]
(0,4) -- (3, 4) -- (5, 5) -- (2,5) -- (0,4);
\draw [line width=1pt, red!80!black] (0.0, 2.) -- (2, 3) ;
\node[below, red!80!black] at (-0.5, 2.4) {$\cE_1^{r^2}$};
\node  at (2.5, 2.55) {$\Q_{2}^{[r^2]}$};
\node  at (5.5, 2.55) {$,$};
\node  at (10.5, 2.55) {$\Q_2^{[r^2]}\Bigg|_{\Neu(\Z_2^{xr}\times\Z_2^{xr^3}),t}= \cE_1^{r^2}\,.$};
\end{scope}
\end{tikzpicture}    
\end{align}
The $F$-symbols for the lines depend on the choice of discrete torsion $t\in\Z_2\times\Z_2\times\Z_2$, whose 3 generators can be written as the $F$-symbols for each $\Z_2$ subgroup of $\Z_2\times\Z_2$ \cite{Hu:2012wx}. Writing elements in $\Z_2\times\Z_2$ as $l=(l_\I,l_\II)$, one has:
\be
\ba
    F_\I(\cE_1^l,\cE_1^m,\cE_1^n)&=\exp\left(\frac{2\pi i}{4}\,l_\I(m_\I+n_\I+[m_\I+n_\I])\right)\,,\\
    F_\II(\cE_1^l,\cE_1^m,\cE_1^n)&=\exp\left(\frac{2\pi i}{4}\,l_\II(m_\II+n_\II+[m_\II+n_\II])\right)\,,\\
    F_{\I,\II}(\cE_1^l,\cE_1^m,\cE_1^n)&=\exp\left(\frac{2\pi i}{4}\,l_\I(m_\II+n_\II+[m_\II+n_\II])\right)\,,\\
\ea
\ee
where $[m_\I+n_\I]\equiv (m_\I+m_\II \mod 2)$ etc.

\paragraph{Local operators on non-identity lines.}
The 1d irrep $1_{xr}$ on $\Q_{1}^{[r^2]}$ gives rise to a local operator $\cE_0^{1_{xr}}$ on $\cE_1^{r^2}$:
\be
\ba
  \Q_{1}^{[r^2], 1_x}\Bigg|_{\Neu(\Z_2^{xr}\times\Z_2^{xr^3}),t}&=\cE_1^{r^2}\cE_0^{1_{xr}}\,, 
\ea
\ee
charged under the $\Z_2^{(0),x}$ 0-form symmetry.

\paragraph{Surfaces that cannot end on $\fB_{\Neu(\Z_2^{xr}\times\Z_2^{xr^3}),t}$.} 
When gauging $\Z_2^{xr}\times\Z_2^{xr^3}$, the SymTFT surface $\Q_2^{[xr]}$ becomes transparent. Since $\Q_2^{[x]}$ and $\Q_2^{[r]}$ are related by fusing with $\Q_2^{[xr]}$, they must be identified on $\fB_{\Neu(\Z_2^{xr}\times\Z_2^{xr^3}),t}$.
On $\fB_{\Dir}$, the surface $\Q_2^{[r]}\oplus\Q_2^{[x]}$ was the non-simple combination $D_2^r\oplus D_2^{r^3}\oplus D_2^{x}\oplus D_2^{xr^2}$, with a 4-dimensional space of topological local operators that can be combined into local operators transforming under $\Rep(\Z_2\times\Z_2)$. When gauging $\Z_2^{xr}\times\Z_2^{xr^3}$, the SymTFT surface $\Q_2^{[x]}\oplus\Q_2^{[r]}$ projects to a single surface on $\fB_{\Neu(\Z_2^{xr}\times\Z_2^{xr^3}),t}$, denoted as $D_2^{x}/\cA_{\Z_2^{\wh{xr}}\times\Z_2^{\wh{xr}^3}}$ which can also arise from the $L$-shaped configuration:
\be
\begin{split}
\begin{tikzpicture}
 \begin{scope}[shift={(0,0)},scale=0.8] 
\draw [cyan, fill=cyan!80!red, opacity =0.5]
(0,0) -- (0,4) -- (2,5) -- (5,5) -- (5,1) -- (3,0)--(0,0);
\draw [black, thick, fill=white,opacity=1]
(0,0) -- (0, 4) -- (2, 5) -- (2,1) -- (0,0);
\draw [cyan, thick, fill=cyan, opacity=0.2]
(0,0) -- (0, 4) -- (2, 5) -- (2,1) -- (0,0);
\draw [black, thick, fill=black, opacity=0.3]
(0.,2) -- (0., 4) -- (2, 5) -- (2,3) -- (0,2);
\draw [black, thick, fill=black, opacity=0.3]
(0.,2) -- (3.1, 2) -- (5., 3) -- (2, 3) -- (0.,2);
\node at (2,5.4) {$\fB_{\Neu(\Z_2^{xr}\times\Z_2^{xr^3}),t}$};
\draw[dashed] (0,0) -- (3,0);
\draw[dashed] (0,4) -- (3,4);
\draw[dashed] (2,5) -- (5,5);
\draw[dashed] (2,1) -- (5,1);
\draw [black, thick, dashed]
(3,0) -- (3, 4) -- (5, 5) -- (5,1) -- (3,0);
\draw [cyan, thick, fill=cyan, opacity=0.1]
(0,4) -- (3, 4) -- (5, 5) -- (2,5) -- (0,4);
\draw [thick, red] (0.0, 2.) -- (2, 3) ;
\node[below, red!80!black] at (-0.5, 2.4) {$\cE_1^{[x]}$};
\node  at (2.5, 2.55) {$\Q_{2}^{[x]}$};
\node[above] at (1.1, 3) {$D_2^{x}/\cA_{\Z_2^{\wh{xr}}\times\Z_2^{\wh{xr}^3}}$};
\node[above] at (5.5, 2.55) {$,$};
\node[above] at (10.1, 2) {$\Q^{[x]}_{2}\Bigg|_{\Neu(\Z_4^r),p}=(D_2^{x}/\cA_{\Z_2^{\wh{xr}}\times\Z_2^{\wh{xr}^3}}\,,$};
\end{scope}
\end{tikzpicture}    
\end{split}
\ee
$D_2^{x}/\cA_{\Z_2^{\wh{xr}}\times\Z_2^{\wh{xr}^3}}$ is in the same Schur component as the invertible surface $D_2^x$ that generates the $\Z_2^{(0),x}$ subsymmetry of $\fB_{\Neu(\Z_2^{xr}\times\Z_2^{xr^3}),t}$ \cite{Bhardwaj:2022maz}.

\subsubsection{Neumann$(\Z_2^{x})$ and Neumann$(\Z_2^{xr})$ Boundary Conditions} \label{sec:D8_NeuZ2x}
Since the features of $\fB_{\Neu(\Z_2^{xr}),s}$ can be deduced from those of $\fB_{\Neu(\Z_2^{x}),s}$ by mapping from the first to the second presentation of $D_8$ in equation \eqref{eq:D8_out_aut}, we will focus on $\fB_{\Neu(\Z_2^{x}),s}$ in the following. This boundary is obtained by gauging the non-normal $\Z_2^{x}$ sub-symmetry of $\fB_\Dir$, with possible discrete torsion $s\in H^3(\Z_2, U(1))=\Z_2$. We denote
\be
\fB_{\Neu(\Z_2^{x}),s}= \frac{\fB_{\Dir}\boxtimes {\rm SPT}_s }{\Z_2^{x}} \,.
\ee
The left-cosets are:
\be
    D_8/\Z_2^x=\{(1,x),\;(r,xr^3),\;(r^2,xr^2),\;(r^3,xr)\}\,,
\ee
therefore, by picking either the first or the second representative for each coset, the  symmetry after gauging can be written either as \cite{Bhardwaj:2022maz}:
\be
    \TwoRep((\Z_2^{(1),xr}\times\Z_2^{(1),xr^3})\rtimes \Z_2^{(0),x})\,.
\ee
or equivalently as
\be
    \TwoRep(\Z_4^{(1)}\rtimes \Z_2^{(0),x})\,.
\ee
We will use the latter presentation in the following sections. $\fB_{\Neu(\Z_2^{x}),s}$ has 1-form symmetry generated by $D_1^{\wh{x}}$, an invertible 0-form symmetry generated by $D_2^{r^2}$ and a non-invertible 0-form symmetry generated by a surface we will denote by $D_2^{[r]}$, which on $\fB_\Dir$ was the composite object $D_2^r\oplus D_2^{r^3}$, constituting an orbit under $\Z_2^x$. It obeys the fusion \cite{Bhardwaj:2022maz}:
\be
    D_2^{[r]}\otimes D_2^{[r]}=\frac{D_2^{\id}}{D_1^1\oplus D_1^{\wh{x}}}\oplus\frac{D_2^{r^2}}{D_1^1\oplus D_1^{\wh{x}}}
\ee

\paragraph{Local operators.} 
Since the genuine local operator $\cE_0^{1_x}$ at the end of $\Q_1^{1_x}$ was untwisted and uncharged under the $\Z_2^{x}$ symmetry, it remains unaltered after gauging 
\be
\begin{split}
\begin{tikzpicture}
 \begin{scope}[shift={(0,0)},scale=0.8] 
\draw [cyan, fill=cyan!80!red, opacity =0.5]
(0,0) -- (0,4) -- (2,5) -- (5,5) -- (5,1) -- (3,0)--(0,0);
\draw [black, thick, fill=white,opacity=1]
(0,0) -- (0, 4) -- (2, 5) -- (2,1) -- (0,0);
\draw [cyan, thick, fill=cyan, opacity=0.2]
(0,0) -- (0, 4) -- (2, 5) -- (2,1) -- (0,0);
\draw[line width=1pt] (1,2.5) -- (4.2,2.5);
\fill[red!80!black] (1,2.5) circle (3pt);
\node at (2,5.4) {$\fB_{{\Neu(\Z_2^{x}),s}}$};
\draw [ black, thick, dashed]
(3,0) -- (3, 4) -- (5, 5) -- (5,1) -- (3,0);
\draw[dashed] (0,0) -- (3,0);
\draw[dashed] (0,4) -- (3,4);
\draw[dashed] (2,5) -- (5,5);
\draw[dashed] (2,1) -- (5,1);
\draw [cyan, thick, fill=cyan, opacity=0.1]
(0,4) -- (3, 4) -- (5, 5) -- (2,5) -- (0,4);
\node[below, red!80!black] at (.8, 2.4) {$\cE_0^{1_x}$};
\node  at (2.5, 3) {$\Q_{1}^{1_x}$};
\node  at (5.5, 2.5) {$,$};
\node  at (11, 2.5) {$\Q_{1}^{1_k}\Bigg|_{\Neu(\Z_2^{x}),s}=\cE_0^{{1_x}}\,.$};
\end{scope}
\end{tikzpicture}    
\end{split}
\ee
$\cE_0^{1_x}$ carries charge $-2$ under the 0-form symmetry generated by $D_2^{[r]}$ since on $\fB_\Dir$ it had charge $-1$ under both $D_2^r$ and $D_2^{r^3}$.

On $\fB_{\Dir}$, the doublet of operators $\cE_0^{E,I}$ with $I=1,2$ at the end on $\Q_1^{E}$ can be written in a rotated basis as
\begin{equation}
    \cE_0^{E\,, \pm}:= \cE_0^{E,1}\pm \cE_0^{E,2}\,.
\end{equation}
Among these $\cE_0^{E,+}$ is uncharged while $\cE_0^{E,-}$ transforms in the non-trivial sign representation of $\Z_2^x$. Therefore upon gauging $\Z_2^{x}$, one has
\begin{equation}
    \Q_1^{E}\Bigg|_{{\Neu(\Z_2^{x}),s}}=(\cE_0^{E,+}\,, (D_1^{\wh{x}}\,, \cE_0^{E,-}))\,.
\end{equation}
\be
\begin{split}
\begin{tikzpicture}
 \begin{scope}[shift={(0,0)},scale=0.8] 
\draw [cyan, fill=cyan!80!red, opacity =0.5]
(0,0) -- (0,4) -- (2,5) -- (5,5) -- (5,1) -- (3,0)--(0,0);
\draw [black, thick, fill=white,opacity=1]
(0,0) -- (0, 4) -- (2, 5) -- (2,1) -- (0,0);
\draw [cyan, thick, fill=cyan, opacity=0.2]
(0,0) -- (0, 4) -- (2, 5) -- (2,1) -- (0,0);
\draw[line width=1pt] (1,2.5) -- (3,2.5);
\draw[line width=1pt] (3,2.5) -- (4.2,2.5);
\fill[red!80!black] (1,2.5) circle (3pt);
\node at (2,5.4) {$\fB_{\Neu(\Z_2^x),s}$};
\draw[dashed] (0,0) -- (3,0);
\draw[dashed] (0,4) -- (3,4);
\draw[dashed] (2,5) -- (5,5);
\draw[dashed] (2,1) -- (5,1);
\draw [black, thick, dashed]
(3,0) -- (3, 4) -- (5, 5) -- (5,1) -- (3,0);
\draw [cyan, thick, fill=cyan, opacity=0.1]
(0,4) -- (3, 4) -- (5, 5) -- (2,5) -- (0,4);
\node[below, red!80!black] at (.8, 2.4) {$\cE_0^{E,+}$};
\node  at (2.5, 3) {$\Q_{1}^{E}$};
\node  at (5.5, 2.5) {$,$};
\end{scope}
\begin{scope}[shift={(7,0)},scale=0.8] 
\draw [cyan, fill=cyan!80!red, opacity =0.5]
(0,0) -- (0,4) -- (2,5) -- (5,5) -- (5,1) -- (3,0)--(0,0);
\draw [black, thick, fill=white,opacity=1]
(0,0) -- (0, 4) -- (2, 5) -- (2,1) -- (0,0);
\draw [cyan, thick, fill=cyan, opacity=0.2]
(0,0) -- (0, 4) -- (2, 5) -- (2,1) -- (0,0);
\draw[line width=1pt] (1,2.5) -- (3,2.5);
\draw[line width=1pt] (1,2.5) -- (1,4.5);
\draw[line width=1pt] (3,2.5) -- (4.2,2.5);
\fill[red!80!black] (1,2.5) circle (3pt);
\node at (2,5.4) {$\fB_{\Neu(\Z_2^x),s}$};
\draw[dashed] (0,0) -- (3,0);
\draw[dashed] (0,4) -- (3,4);
\draw[dashed] (2,5) -- (5,5);
\draw[dashed] (2,1) -- (5,1);
\draw [black, thick, dashed]
(3,0) -- (3, 4) -- (5, 5) -- (5,1) -- (3,0);
\draw [cyan, thick, fill=cyan, opacity=0.1]
(0,4) -- (3, 4) -- (5, 5) -- (2,5) -- (0,4);
\node[below, red!80!black] at (.8, 2.4) {$\cE_0^{E,-}$};
\node  at (2.5, 3) {$\Q_{1}^{E}$};
\node  at (0.5, 3.3) {$D_{1}^{\wh{x}}$};
\node  at (5.5, 2.5) {$,$};
\end{scope}
\end{tikzpicture}    
\end{split}
\ee
Both $\cE_0^+$ and $\cE_0^-$ carry charge 0 under $D_2^{[r]}$ and charge $-1$ under $D_2^{r^2}$.

On $\fB_{\Dir}$, the operators $\cE_0^{1_{xr}}$ and $\cE_0^{1_{xr}}$ transform as $-1$ under $\Z_2^{(0),x}$, therefore after gauging $\Z_2^{x}$, they become attached to the dual $\Z_2^{(1),\wh{x}}$ symmetry generator $D_1^{\wh{x}}$.
\be
\begin{split}
\begin{tikzpicture}
 \begin{scope}[shift={(0,0)},scale=0.8] 
\draw [cyan, fill=cyan!80!red, opacity =0.5]
(0,0) -- (0,4) -- (2,5) -- (5,5) -- (5,1) -- (3,0)--(0,0);
\draw [black, thick, fill=white,opacity=1]
(0,0) -- (0, 4) -- (2, 5) -- (2,1) -- (0,0);
\draw [cyan, thick, fill=cyan, opacity=0.2]
(0,0) -- (0, 4) -- (2, 5) -- (2,1) -- (0,0);
\draw[line width=1pt] (1,2.5) -- (3,2.5);
\draw[line width=1pt] (1,2.5) -- (1,4.5);
\draw[line width=1pt] (3,2.5) -- (4.2,2.5);
\fill[red!80!black] (1,2.5) circle (3pt);
\node at (2,5.4) {$\fB_{\Neu(\Z_2^x),s}$};
\draw[dashed] (0,0) -- (3,0);
\draw[dashed] (0,4) -- (3,4);
\draw[dashed] (2,5) -- (5,5);
\draw[dashed] (2,1) -- (5,1);
\draw [cyan, thick, fill=cyan, opacity=0.1]
(0,4) -- (3, 4) -- (5, 5) -- (2,5) -- (0,4);
\node[below, red!80!black] at (.8, 2.4) {$\cE_0^{1_{(x)r}}$};
\draw [black, thick, dashed]
(3,0) -- (3, 4) -- (5, 5) -- (5,1) -- (3,0);
\node  at (2.75 , 3) {$\Q_{1}^{1_{(x)r}}$};
\node  at (0.5, 3.3) {$D_{1}^{\wh{x}}$};
\node  at (5.5, 2.5) {$,$};
\node  at (11, 3.5) {$ \Q_{1}^{1_{xr}}\Bigg|_{\Neu(\Z_2^x),s}=(D_1^{\wh{x}},\cE_0^{1_{xr}})\,,$};
\node  at (11, 1.5) {$\Q_{1}^{1_{r}}\Bigg|_{\Neu(\Z_2^x),s}=(D_1^{\wh{x}},\cE_0^{1_{r}})\,.
$};
\end{scope}
\end{tikzpicture}    
\end{split}
\ee
Under $D_2^{[r]}$, the local operators $\cE_0^{1_{r}}$ and $\cE_0^{1_{xr}}$ carry charge $+2$ and $-2$ respectively.

\paragraph{Genuine topological lines.}
The SymTFT surface $\Q_1^{[x]}$ ends on two lines, $\cE_1^{x}$ and $\cE_1^{xr^2}$, both of which are invariant under the $\Z_2^x$ action: they therefore furnish genuine topological lines charged under the dual $\Z_2^{\wh{x}}$ 1-form symmetry on $\fB_{\Neu(\Z_2^{x}),s}$, i.e.,
\be
\begin{split}
\begin{tikzpicture}
 \begin{scope}[shift={(0,0)},scale=0.8] 
\draw [cyan, fill=cyan!80!red, opacity =0.5]
(0,0) -- (0,4) -- (2,5) -- (5,5) -- (5,1) -- (3,0)--(0,0);
\draw [black, thick, fill=white,opacity=1]
(0,0) -- (0, 4) -- (2, 5) -- (2,1) -- (0,0);
\draw [cyan, thick, fill=cyan, opacity=0.2]
(0,0) -- (0, 4) -- (2, 5) -- (2,1) -- (0,0);
\draw [black, thick, fill=black, opacity=0.3]
(0.,2) -- (3.1, 2) -- (5., 3) -- (2, 3) -- (0.,2);
\node at (2,5.4) {$\fB_{\Neu(\Z_2^{x}),s}$};
\draw[dashed] (0,0) -- (3,0);
\draw[dashed] (0,4) -- (3,4);
\draw[dashed] (2,5) -- (5,5);
\draw[dashed] (2,1) -- (5,1);
\draw [black, thick, dashed]
(3,0) -- (3, 4) -- (5, 5) -- (5,1) -- (3,0);
\draw [cyan, thick, fill=cyan, opacity=0.1]
(0,4) -- (3, 4) -- (5, 5) -- (2,5) -- (0,4);
\draw [line width=1pt, red!80!black] (0.0, 2.) -- (2, 3) ;
\node[below, red!80!black] at (-1.5, 2.4) {$\cE_1^{x},\cE_1^{xr^2}$};
\node  at (2.5, 2.55) {$\Q_{2}^{[x]}$};
\node  at (5.5, 2.55) {$,$};
\node  at (10.5, 2.55) {$\Q_2^{[x]}\Bigg|_{{\Neu(\Z_2^{x}),s}}= \{\cE_1^{x}\,,\cE_1^{xr^2}\}\,.$};
\end{scope}
\end{tikzpicture}    
\end{split}
\ee
with $F$-symbol coming from $s\in H^3(\Z_2,U(1))$.

\paragraph{Local operators on non-identity lines.}
The non-trivial irrep $1_{+-}$ of $H_x=\Z_2^x\times\Z_2^{r^2}$ on the surface $\Q_{1}^{[x]}$ give rise to local a local operator $\cE_0^{1_{+-}}$ on each line $\cE_1^{x}$, $\cE_1^{xr^2}$:
\be
\ba
  \Q_{1}^{[x], 1_{+-}}\Bigg|_{{\Neu(\Z_2^{x}),s}}&=\{\cE_1^{x}\cE_0^{1_{+-}}\,,\;\cE_1^{xr^2}\cE_0^{1_{+-}}\}\,, 
\ea
\ee
where $\cE_0^{1_{+-}}$ carries charge $-1$ under $\Z_2^{(0),r^2}$.

\paragraph{Surfaces that cannot end on $\fB_{\Neu(\Z_2^{x}),s}$.} On $\fB_\Dir$, $\Q_2^{[r^2]}$ projected to a surface generating $\Z_2^{r^2}$: since the action of $\Z_2^x$ on $r^2$ is trivial, this surfce survives the $\Z_2^x$ gauging:
\be
\begin{split}
\begin{tikzpicture}
 \begin{scope}[shift={(0,0)},scale=0.8] 
\draw [cyan, fill=cyan!80!red, opacity =0.5]
(0,0) -- (0,4) -- (2,5) -- (5,5) -- (5,1) -- (3,0)--(0,0);
\draw [black, thick, fill=white,opacity=1]
(0,0) -- (0, 4) -- (2, 5) -- (2,1) -- (0,0);
\draw [cyan, thick, fill=cyan, opacity=0.2]
(0,0) -- (0, 4) -- (2, 5) -- (2,1) -- (0,0);
\draw [black, thick, fill=black, opacity=0.3]
(0.,2) -- (0., 4) -- (2, 5) -- (2,3) -- (0,2);
\draw [black, thick, fill=black, opacity=0.3]
(0.,2) -- (3.1, 2) -- (5., 3) -- (2, 3) -- (0.,2);
\node at (2,5.4) {$\fB_{\Neu(\Z_2^{x}),s}$};
\draw[dashed] (0,0) -- (3,0);
\draw[dashed] (0,4) -- (3,4);
\draw[dashed] (2,5) -- (5,5);
\draw[dashed] (2,1) -- (5,1);
\draw [black, thick, dashed]
(3,0) -- (3, 4) -- (5, 5) -- (5,1) -- (3,0);
\draw [cyan, thick, fill=cyan, opacity=0.1]
(0,4) -- (3, 4) -- (5, 5) -- (2,5) -- (0,4);
\draw [thick, red] (0.0, 2.) -- (2, 3) ;
\node[below, red!80!black] at (-0.5, 2.4) {$\cE_1^{r^2}$};
\node  at (2.5, 2.55) {$\Q_{2}^{[r^2]}$};
\node[above] at (1.1, 3) {$D_2^{r^2}$};
\node[above] at (5.5, 2.55) {$,$};
\node[above] at (10.1, 2) {$\Q^{[r^2]}_{2}\Bigg|_{\Neu(\Z_2^{x}),s}=(D_2^{r^2}\,,\cE_1^{r^2})\,,$};
\end{scope}
\end{tikzpicture}    
\end{split}
\ee
$\Q_2^{[r]}$ instead projects to $D_2^{[r]}$, which comes from the $\Z_2^x$ orbit $D_2^r\oplus D_2^{r^3}$:
\be
\begin{split}
\begin{tikzpicture}
 \begin{scope}[shift={(0,0)},scale=0.8] 
\draw [cyan, fill=cyan!80!red, opacity =0.5]
(0,0) -- (0,4) -- (2,5) -- (5,5) -- (5,1) -- (3,0)--(0,0);
\draw [black, thick, fill=white,opacity=1]
(0,0) -- (0, 4) -- (2, 5) -- (2,1) -- (0,0);
\draw [cyan, thick, fill=cyan, opacity=0.2]
(0,0) -- (0, 4) -- (2, 5) -- (2,1) -- (0,0);
\draw [black, thick, fill=black, opacity=0.3]
(0.,2) -- (0., 4) -- (2, 5) -- (2,3) -- (0,2);
\draw [black, thick, fill=black, opacity=0.3]
(0.,2) -- (3.1, 2) -- (5., 3) -- (2, 3) -- (0.,2);
\node at (2,5.4) {$\fB_{\Neu(\Z_2^{x}),s}$};
\draw[dashed] (0,0) -- (3,0);
\draw[dashed] (0,4) -- (3,4);
\draw[dashed] (2,5) -- (5,5);
\draw[dashed] (2,1) -- (5,1);
\draw [black, thick, dashed]
(3,0) -- (3, 4) -- (5, 5) -- (5,1) -- (3,0);
\draw [cyan, thick, fill=cyan, opacity=0.1]
(0,4) -- (3, 4) -- (5, 5) -- (2,5) -- (0,4);
\draw [thick, red] (0.0, 2.) -- (2, 3) ;
\node[below, red!80!black] at (-0.5, 2.4) {$\cE_1^{[r]}$};
\node  at (2.5, 2.55) {$\Q_{2}^{[r]}$};
\node[above] at (1.1, 3) {$D_2^{[r]}$};
\node[above] at (5.5, 2.55) {$,$};
\node[above] at (10.1, 2) {$\Q^{[r]}_{2}\Bigg|_{\Neu(\Z_2^{x}),s}=(D_2^{[r]}\,,\cE_1^{[r]})\,,$};
\end{scope}
\end{tikzpicture}    
\end{split}
\ee
the line $\cE_1^{[r]}$ is uncharged under $D_1^{\wh{x}}$.

The SymTFT surface $\Q_2^{[xr]}$ can be written as a $\Z_2^x$ orbit $D_2^{xr}\oplus D_2^{xr^3}$, so projects to $D_2^{[r]}$ on $\fB_{\Neu(\Z_2^{x}),s}$. However, the line at its boundary is now given by $\cE_1^{x[r]}$, which come from the bulk fusion $\cE_1^{[x]}\otimes\cE^{[r]}$, therefore $\cE_1^{x[r]}$ is charged under $\Z_2^{(1),\wh{x}}$.
\be
\begin{split}
\begin{tikzpicture}
 \begin{scope}[shift={(0,0)},scale=0.8] 
\draw [cyan, fill=cyan!80!red, opacity =0.5]
(0,0) -- (0,4) -- (2,5) -- (5,5) -- (5,1) -- (3,0)--(0,0);
\draw [black, thick, fill=white,opacity=1]
(0,0) -- (0, 4) -- (2, 5) -- (2,1) -- (0,0);
\draw [cyan, thick, fill=cyan, opacity=0.2]
(0,0) -- (0, 4) -- (2, 5) -- (2,1) -- (0,0);
\draw [black, thick, fill=black, opacity=0.3]
(0.,2) -- (0., 4) -- (2, 5) -- (2,3) -- (0,2);
\draw [black, thick, fill=black, opacity=0.3]
(0.,2) -- (3.1, 2) -- (5., 3) -- (2, 3) -- (0.,2);
\node at (2,5.4) {$\fB_{\Neu(\Z_2^{x}),s}$};
\draw[dashed] (0,0) -- (3,0);
\draw[dashed] (0,4) -- (3,4);
\draw[dashed] (2,5) -- (5,5);
\draw[dashed] (2,1) -- (5,1);
\draw [black, thick, dashed]
(3,0) -- (3, 4) -- (5, 5) -- (5,1) -- (3,0);
\draw [cyan, thick, fill=cyan, opacity=0.1]
(0,4) -- (3, 4) -- (5, 5) -- (2,5) -- (0,4);
\draw [thick, red] (0.0, 2.) -- (2, 3) ;
\node[below, red!80!black] at (-0.5, 2.4) {$\cE_1^{x[r]}$};
\node  at (2.5, 2.55) {$\Q_{2}^{[xr]}$};
\node[above] at (1.1, 3) {$D_2^{[r]}$};
\node[above] at (5.5, 2.55) {$,$};
\node[above] at (10.1, 2) {$\Q^{[xr]}_{2}\Bigg|_{\Neu(\Z_2^{x}),s}=(D_2^{[r]}\,,\cE_1^{x[r]})\,,$};
\end{scope}
\end{tikzpicture}    
\end{split}
\ee

The boundary $\Neu(\Z_2^{xr}),s$ can be obtained from  $\Neu(\Z_2^{x}),s$
we just described by using the $D_8$ automorphism \eqref{eq:D8_out_aut}. Note that we can unitarily transform $\cD_E(xr)=\sigma^y=\begin{pmatrix}
    0 & -i \\ i & 0
\end{pmatrix}$ to $\sigma^z=\diag(1,-1)$, denoting by 
\be
    \cE_0^{E,\pm i}:=\cE_0^{E,1}\pm i \cE_0^{E,2}
\ee
the new basis elements. The charge of $\cE_0^{E,\pm i}$ under $\Z_2^{xr}$ is $\pm1$ respectively. Therefore, upon gauging $\Z_2^{xr}$, we find
\begin{equation} \label{eq:D8_Z2x_Q1E}
    \Q_1^{E}\Bigg|_{{\Neu(\Z_2^{xr}),s}}=\left\{\cE_0^{E,+i}\,, \;(D_1^{\wh{xr}},\cE_0^{E,-i})\right\}\,.
\end{equation}
\be
\begin{split}
\begin{tikzpicture}
 \begin{scope}[shift={(0,0)},scale=0.8] 
\draw [cyan, fill=cyan!80!red, opacity =0.5]
(0,0) -- (0,4) -- (2,5) -- (5,5) -- (5,1) -- (3,0)--(0,0);
\draw [black, thick, fill=white,opacity=1]
(0,0) -- (0, 4) -- (2, 5) -- (2,1) -- (0,0);
\draw [cyan, thick, fill=cyan, opacity=0.2]
(0,0) -- (0, 4) -- (2, 5) -- (2,1) -- (0,0);
\draw[line width=1pt] (1,2.5) -- (3,2.5);
\draw[line width=1pt] (3,2.5) -- (4.2,2.5);
\fill[red!80!black] (1,2.5) circle (3pt);
\node at (2,5.4) {$\fB_{\Neu(\Z_2^{xr}),s}$};
\draw[dashed] (0,0) -- (3,0);
\draw[dashed] (0,4) -- (3,4);
\draw[dashed] (2,5) -- (5,5);
\draw[dashed] (2,1) -- (5,1);
\draw [black, thick, dashed]
(3,0) -- (3, 4) -- (5, 5) -- (5,1) -- (3,0);
\draw [cyan, thick, fill=cyan, opacity=0.1]
(0,4) -- (3, 4) -- (5, 5) -- (2,5) -- (0,4);
\node[below, red!80!black] at (.8, 2.4) {$\cE_0^{E,+i}$};
\node  at (2.5, 3) {$\Q_{1}^{E}$};
\node  at (5.5, 2.5) {$,$};
\end{scope}
\begin{scope}[shift={(7,0)},scale=0.8] 
\draw [cyan, fill=cyan!80!red, opacity =0.5]
(0,0) -- (0,4) -- (2,5) -- (5,5) -- (5,1) -- (3,0)--(0,0);
\draw [black, thick, fill=white,opacity=1]
(0,0) -- (0, 4) -- (2, 5) -- (2,1) -- (0,0);
\draw [cyan, thick, fill=cyan, opacity=0.2]
(0,0) -- (0, 4) -- (2, 5) -- (2,1) -- (0,0);
\draw[line width=1pt] (1,2.5) -- (3,2.5);
\draw[line width=1pt] (1,2.5) -- (1,4.5);
\draw[line width=1pt] (3,2.5) -- (4.2,2.5);
\fill[red!80!black] (1,2.5) circle (3pt);
\node at (2,5.4) {$\fB_{\Neu(\Z_2^{xr}),s}$};
\draw[dashed] (0,0) -- (3,0);
\draw[dashed] (0,4) -- (3,4);
\draw[dashed] (2,5) -- (5,5);
\draw[dashed] (2,1) -- (5,1);
\draw [black, thick, dashed]
(3,0) -- (3, 4) -- (5, 5) -- (5,1) -- (3,0);
\draw [cyan, thick, fill=cyan, opacity=0.1]
(0,4) -- (3, 4) -- (5, 5) -- (2,5) -- (0,4);
\node[below, red!80!black] at (.8, 2.4) {$\cE_0^{E,-i}$};
\node  at (2.5, 3) {$\Q_{1}^{E}$};
\node  at (0.5, 3.3) {$D_{1}^{\wh{xr}}$};
\node  at (5.5, 2.5) {$,$};
\end{scope}
\end{tikzpicture}    
\end{split}
\ee

\subsubsection{Neumann$(D_8)$ Boundary Condition} \label{sec:D8_NeuD8}
We finally discuss the boundary condition obtained by gauging the full $D_8$ 0-form symmetry of $\fB_{\Dir}$ discrete torsion
given by $\tau\in H^{3}(D_8,U(1))=\Z_4\times\Z_2\times\Z_2$.
The corresponding boundary condition is denoted as
\begin{equation}
    \fB_{\Neu(D_8),\tau}= \frac{\fB_\Dir\boxtimes {\rm SPT}_{\tau}}{D_8}\,.
\end{equation}
The symmetry on this boundary is $\cS=\TwoRep(D_8)$, generated by topological lines $D_1^R$, for $R\in \Rep(D_8)$. All surfaces are condensation defects of condensable algebras in $\Rep(D_8)$.

\paragraph{Non-genuine local operators.}
When gauging the full $D_8$ symmetry, since all $\Q_1^R$ SymTFT lines transform non-trivially, they give rise to twisted sector local operators and there are no genuine local operators on this boundary. 
\be
\begin{split}
\begin{tikzpicture}
 \begin{scope}[shift={(0,0)},scale=0.8] 
\draw [cyan, fill=cyan!80!red, opacity =0.5]
(0,0) -- (0,4) -- (2,5) -- (5,5) -- (5,1) -- (3,0)--(0,0);
\draw [black, thick, fill=white,opacity=1]
(0,0) -- (0, 4) -- (2, 5) -- (2,1) -- (0,0);
\draw [cyan, thick, fill=cyan, opacity=0.2]
(0,0) -- (0, 4) -- (2, 5) -- (2,1) -- (0,0);
\draw[line width=1pt] (1,2.5) -- (3,2.5);
\draw[line width=1pt] (1,2.5) -- (1,4.5);
\draw[line width=1pt] (3,2.5) -- (4.2,2.5);
\fill[red!80!black] (1,2.5) circle (3pt);
\node at (2,5.4) {$\fB_{\Neu(D_8),\tau}$};
\draw[dashed] (0,0) -- (3,0);
\draw[dashed] (0,4) -- (3,4);
\draw[dashed] (2,5) -- (5,5);
\draw[dashed] (2,1) -- (5,1);
\draw [cyan, thick, fill=cyan, opacity=0.1]
(0,4) -- (3, 4) -- (5, 5) -- (2,5) -- (0,4);
\node[below, red!80!black] at (.8, 2.4) {$\cE_0^{R}$};
\draw [black, thick, dashed]
(3,0) -- (3, 4) -- (5, 5) -- (5,1) -- (3,0);
\node  at (2.5 , 3) {$\Q_{1}^{R}$};
\node  at (0.5, 3.3) {$D_{1}^{R}$};
\node  at (5.5, 2.5) {$,$};
\node  at (11, 2.5) {$ \Q_{1}^{R}\Bigg|_{\Neu(D_8),\tau}=(D_1^{R},\cE_0^{R})\,.
$};
\end{scope}
\end{tikzpicture}    
\end{split}
\ee

\paragraph{Genuine topological lines.}
All SymTFT surfaces can end on $\fB_{\Neu(D_8),\tau}$, and gives rise to genuine topological lines labeled by $[g]\in D_8$, listed in section \ref{eq:D8_SymTFT}.
\be
\begin{split}
\begin{tikzpicture}
 \begin{scope}[shift={(0,0)},scale=0.8] 
\draw [cyan, fill=cyan!80!red, opacity =0.5]
(0,0) -- (0,4) -- (2,5) -- (5,5) -- (5,1) -- (3,0)--(0,0);
\draw [black, thick, fill=white,opacity=1]
(0,0) -- (0, 4) -- (2, 5) -- (2,1) -- (0,0);
\draw [cyan, thick, fill=cyan, opacity=0.2]
(0,0) -- (0, 4) -- (2, 5) -- (2,1) -- (0,0);
\draw [black, thick, fill=black, opacity=0.3]
(0.,2) -- (3.1, 2) -- (5., 3) -- (2, 3) -- (0.,2);
\node at (2,5.4) {$\fB_{\Neu(D_8),\tau}$};
\draw[dashed] (0,0) -- (3,0);
\draw[dashed] (0,4) -- (3,4);
\draw[dashed] (2,5) -- (5,5);
\draw[dashed] (2,1) -- (5,1);
\draw [black, thick, dashed]
(3,0) -- (3, 4) -- (5, 5) -- (5,1) -- (3,0);
\draw [cyan, thick, fill=cyan, opacity=0.1]
(0,4) -- (3, 4) -- (5, 5) -- (2,5) -- (0,4);
\draw [line width=1pt, red!80!black] (0.0, 2.) -- (2, 3) ;
\node[below, red!80!black] at (-0.5, 2.4) {$\cE_1^{[g]}$};
\node  at (2.5, 2.55) {$\Q_{2}^{[g]}$};
\node  at (5.5, 2.55) {$,$};
\node  at (10.5, 2.5) {$\Q_2^{[g]}\Bigg|_{\Neu(D_8),\tau}= \cE_1^{[g]}\,,\qquad\quad$};
\end{scope}
\end{tikzpicture}    
\end{split}
\ee
whose $F$-symbols follow from $\tau\in H^3(D_8,U(1))$. The charge of these lines under the $\Rep(D_8)$ 1-form symmetry generators is:
\be
    \text{Link}(D_1^R,\cE_1^{[g]})=\chi_R([g])\,,
\ee
where $\chi_R$ is the character of the irrep $R$. 
There are no twisted sector lines since all SymTFT surfaces can end on this boundary.

\subsection{$\TwoVec_{D_8}$  Gapped Phases}
The minimal gapped phases for $D_8$ 0-form symmetry in 2+1d follow of course the standard Landau paradigm, plus SPTs. Nevertheless we will start with this well-known class of gapped phases, as we will use it to then gauge on the symmetry boundary in order to get the remaining phases for the less well-studied, categorical symmetries. 
A summary can be seen in table \ref{tab:D8phases}.

\begin{table}
$$
\begin{array}{|c|l|}\hline
\text{Physical BC $\Bphys$} & \Q_p\text{ with Dirichlet BCs} \cr \hline \hline 
{\rm Dir}  &
\text{$D_8$ SSB: 8 vacua} 
\cr\hline
\Neu(\Z_2^{r^2}),s & 
\Z_2^r \times \Z_2^x \ \text{SSB, 4 vacua}
\cr \hline
\Neu(\Z_4^{r}),p & 
\Z_2^{x}\ \text{SSB, 2 vacua} 
\cr \hline
\Neu(\Z_2^{xr}\times\Z_2^{xr^3}),t & 
\Z_2^{x}\ \text{SSB, 2 vacua} 
\cr \hline\
\Neu(\Z_2^{x}\times\Z_2^{xr^2}),t & 
\Z_2^{xr}\ \text{SSB, 2 vacua}
\cr \hline\
\Neu(\Z_2^{x}),s & 
\Z_4 \ \text{SSB, 4 vacua}
\cr \hline\
\Neu(\Z_2^{xr}),s & 
\Z_4 \ \text{SSB, 4 vacua}
\cr \hline\
\Neu(D_8),\tau& 
\text{SPTs}
     \cr
 \hline
\end{array}
$$
\caption{Summary of minimal gapped phases for $\TwoVec (D_8)$.} \label{tab:D8phases}
\end{table}

\subsubsection{$D_8$ SSB Phase} \label{sec:D8_D8SSB}
Consider the physical boundary to be
\begin{equation}
    \Bphys=\fB_{\Dir}\,.
\end{equation}
In this case all local operators $\cO^R$ can end with multiplicity $\dim R$, yielding $\sum_R(\dim R)^2=|G|$ many vacua, in this case 8. Setting $H=1$ in equation \eqref{eq:Gsym_vacua}, they are:
\be
\ba
    v_1&=\frac{1}{8}\lb 1+\cO^{1_r}+\cO^{1_x}+\cO^{1_{xr}}+2\cO^{E,1,1}+2\cO^{E,2,2}\rb \,,\\
    v_{xr}&=\frac{1}{8}\lb 1-\cO^{1_r}-\cO^{1_x}+\cO^{1_{xr}}+2i\cO^{E,1,2}-2i\cO^{E,2,1}\rb\,,\\
    v_{xr^3}&=\frac{1}{8}\lb 1-\cO^{1_r}-\cO^{1_x}+\cO^{1_{xr}}-2i\cO^{E,1,2}+2i\cO^{E,2,1}\rb \,,\\
    v_{r^2}&=\frac{1}{8}\lb 1+\cO^{1_r}+\cO^{1_x}+\cO^{1_{xr}}-2\cO^{E,1,1}-2\cO^{E,2,2}\rb \,,\\
    v_x&=\frac{1}{8}\lb 1-\cO^{1_r}+\cO^{1_x}-\cO^{1_{xr}}+2\cO^{E,1,2}+2\cO^{E,2,1}\rb \,,\\
    v_r&=\frac{1}{8}\lb 1+\cO^{1_r}-\cO^{1_x}-\cO^{1_{xr}}-2i\cO^{E,1,1}+2i\cO^{E,2,2}\rb \,,\\
    v_{r^3}&=\frac{1}{8}\lb 1+\cO^{1_r}-\cO^{1_x}-\cO^{1_{xr}}+2i\cO^{E,1,1}-2i\cO^{E,2,2}\rb \,,\\
    v_{xr^2}&=\frac{1}{8}\lb 1-\cO^{1_r}+\cO^{1_x}-\cO^{1_{xr}}-2\cO^{E,1,2}-2\cO^{E,2,1}\rb \,.
\ea
\ee
The action of $g\in D_8$ on each vacuum is 
\be
    g\cdot v_{\bar{g}}=v_{g\bar{g}}\,,
\ee
from which we see that no vacuum is invariant an that the full $D_8$ 0-form symmetry is spontaneously broken so this is a $D_8$ SSB phase. We can summarize this as:
\begin{tcolorbox}[
colback=white,
coltitle= black,
colbacktitle=ourcolorforheader,
colframe=black,
title= $\TwoVec_{D_8}$: $D_8$ SSB Phase,
sharp corners]
\be
\bigoplus_{g\in D_8}{\rm Triv}_{g}\,, \qquad D_2^{g}: {\text{Triv}_{g'}\mapsto {\text{Triv}_{gg'}}}
\ee
\end{tcolorbox}

\subsubsection{$\Z_2^r\times\Z_2^x$ SSB Phase} \label{sec:D8_Z2rZ2x_SSB}
Consider the physical boundary to be
\begin{equation}
    \Bphys=\fB_{\Neu(\Z_2^{r^2}),s}\,.
\end{equation}
The vacua are:
\be
\ba
    v_{\Z_2^{r^2}}=v_{1}+v_{r^2}&=\frac{1}{4}\lb 1+\cO^{1_r}+\cO^{1_x}+\cO^{1_{xr}}\rb \,,\\
    v_{xr\Z_2^{r^2}}=v_{xr}+v_{xr^3}&=\frac{1}{4}\lb 1-\cO^{1_r}-\cO^{1_x}+\cO^{1_{xr}}\rb\,,\\
    v_{x\Z_2^{r^2}}=v_x+v_{xr^2}&=\frac{1}{4}\lb 1-\cO^{1_r}+\cO^{1_x}-\cO^{1_{xr}}\rb \,,\\
    v_{r\Z_2^{r^2}}=v_r+v_{r^3}&=\frac{1}{4}\lb 1+\cO^{1_r}-\cO^{1_x}-\cO^{1_{xr}}\rb \,.
\ea
\ee
This is $\Z_2^r\times\Z_2^x$ SSB phase, in which the action of the generators is as follows:
\be
\ba
    x:&& v_{\Z_2^{r^2}}&\longleftrightarrow v_{x\Z_2^{r^2}}\,, & v_{r\Z_2^{r^2}}&\longleftrightarrow v_{xr\Z_2^{r^2}}\,,\\
    r:&& v_{\Z_2^{r^2}}&\longleftrightarrow v_{r\Z_2^{r^2}}\,, & v_{x\Z_2^{r^2}}&\longleftrightarrow v_{xr\Z_2^{r^2}}\,.
\ea
\ee
In summary we obtain the following TQFT: 
\begin{tcolorbox}[
colback=white,
coltitle= black,
colbacktitle=ourcolorforheader,
colframe=black,
title= $\TwoVec_{D_8}$: $\Z_2^r\times\Z_2^x$ SSB phase,
sharp corners]
\be
\Triv_1 \oplus \Triv_2  \oplus \Triv_3 \oplus \Triv_4: \qquad 
r= 1_{13} \oplus 1_{31} \oplus 1_{24} \oplus 1_{42} \,,\ 
x = 1_{12} \oplus 1_{21} \oplus 1_{34} \oplus 1_{43}
 \,.
\ee
\end{tcolorbox}

\subsubsection{$\Z_2^x$ SSB Phase I} \label{sec:D8_Z2x_SSB1} 
Consider the physical boundary to be
\begin{equation}
    \Bphys=\fB_{\Neu(\Z_4^{r}),p}\,.
\end{equation}
The vacua are:
\be
\ba
    v_{\Z_4^{r}}=v_{1}+v_r+v_{r^2}+v_{r^3}&=\frac{1}{2}\lb 1+\cO^{1_r}\rb \,,\\
     v_{x\Z_4^{r}}=v_x+v_{xr}+v_{xr^2}+v_{xr^3}&=\frac{1}{2}\lb 1-\cO^{1_r}\rb\,.
\ea
\ee
This is $\Z_2^x$ SSB phase, in which $x$ exchanges the two vacua:
\begin{tcolorbox}[
colback=white,
coltitle= black,
colbacktitle=ourcolorforheader,
colframe=black,
title= $\TwoVec_{D_8}$: $\Z_2^x$ SSB Phase $\I$,
sharp corners]
\be
\Triv_1 \oplus \Triv_2 : \qquad 
x = 1_{12} \oplus 1_{21} 
 \,.
\ee
\end{tcolorbox}

\subsubsection{$\Z_2^x$ SSB Phase II}  \label{sec:D8_Z2x_SSB2} 
Consider the physical boundary to be
\begin{equation}
    \Bphys=\fB_{\Neu(\Z_2^{xr}\times\Z_2^{xr^3}),t}\,.
\end{equation}
The vacua are:
\be
\ba
    v_{\Z_2^{xr}\times\Z_2^{xr^3}}=v_{1}+v_{xr}+v_{r^2}+v_{xr^3}&=\frac{1}{2}\lb 1+\cO^{1_{xr}}\rb \,,\\
    v_{x\Z_2^{xr}\times\Z_2^{xr^3}}=v_{x}+v_{r}+v_{xr^2}+v_{r^3}&=\frac{1}{2}\lb 1-\cO^{1_{xr}}\rb\,.
\ea
\ee
This is $\Z_2^x$ SSB phase, in which $x$ exchanges the two vacua.

\subsubsection{$\Z_2^{xr}$ SSB Phase } \label{sec:D8_Z2xr_SSB} 
Consider the physical boundary to be
\begin{equation}
    \Bphys=\fB_{\Neu(\Z_2^{x}\times\Z_2^{xr^2}),t}\,.
\end{equation}
The vacua are:
\be
\ba
    v_{\Z_2^{x}\times\Z_2^{xr^2}}=v_{1}+v_{x}+v_{r^2}+v_{xr^2}&=\frac{1}{2}\lb 1+\cO^{1_{x}}\rb \,,\\
    v_{xr\Z_2^{x}\times\Z_2^{xr^2}}=v_{xr}+v_{r^3}+v_{xr^3}+v_{r}&=\frac{1}{2}\lb 1-\cO^{1_{x}}\rb\,.
\ea
\ee
This is $\Z_2^{xr}$ SSB phase, in which $xr$ exchanges the two vacua.

\subsubsection{$\Z_4$ SSB Phase I} \label{sec:D8_Z4SSB1}
Consider the physical boundary to be
\begin{equation}
    \Bphys=\fB_{\Neu(\Z_2^{x}),s}\,.
\end{equation}
The vacua are:
\be
\ba
    v_{\Z_2^{x}}=v_1+v_x&=\frac{1}{4}\lb 1+\cO^{1_x}+2\cO^{E,1,1}+2\cO^{E,1,2}+2\cO^{E,2,1}+2\cO^{E,2,2}\rb \,,\\
    v_{r\Z_2^{x}}=v_{r}+v_{xr^3}&=\frac{1}{4}\lb 1-\cO^{1_x}-2i\cO^{E,1,1}-2i\cO^{E,1,2}+2i\cO^{E,2,1}+2i\cO^{E,2,2}\rb\,,\\
     v_{r^2\Z_2^{x}}=v_{r^2}+v_{xr^2}&=\frac{1}{4}\lb 1+\cO^{1_x}-2\cO^{E,1,1}-2\cO^{E,1,2}-2\cO^{E,2,1}-2\cO^{E,2,2}\rb \,,\\
    v_{r^3\Z_2^{x}}=v_{r^3}+v_{xr}&=\frac{1}{4}\lb 1-\cO^{1_x}+2i\cO^{E,1,1}+2i\cO^{E,1,2}-2i\cO^{E,2,1}-2i\cO^{E,2,2}\rb\,,
\ea
\ee
In this phase $\Z_4^r$ cyclically permutes the vacua:
\be
    r= 1_{12} \oplus 1_{23} \oplus 1_{34} \oplus 1_{41} 
\ee
\begin{tcolorbox}[
colback=white,
coltitle= black,
colbacktitle=ourcolorforheader,
colframe=black,
title= $\TwoVec_{D_8}$: $\Z_4$ SSB Phase I,
sharp corners]
\be\label{boxD8Z2}
\Triv_1 \oplus \Triv_2 \oplus \Triv_3 \oplus \Triv_4 : \qquad 
r=1_{12} \oplus 1_{23} \oplus 1_{34} \oplus 1_{41}  
 \,.
\ee
\end{tcolorbox}

\subsubsection{$\Z_4$ SSB Phase II} \label{sec:D8_Z4SSB2}
Consider the physical boundary to be
\begin{equation}
    \Bphys=\fB_{\Neu(\Z_2^{xr}),s}\,.
\end{equation}
The vacua are:
\be
\ba
    v_{\Z_2^{xr}}=v_1+v_{xr}&=\frac{1}{4}\lb 1+\cO^{1_{xr}}+2\cO^{E,1,1}+2i\cO^{E,1,2}-2i\cO^{E,2,1}+2\cO^{E,2,2}\rb \,,\\
    v_{r\Z_2^{xr}}=v_{r}+v_{x}&=\frac{1}{4}\lb 1-\cO^{1_{xr}}-2i\cO^{E,1,1}+2\cO^{E,1,2}+2\cO^{E,2,1}+2i\cO^{E,2,2}\rb\,,\\
     v_{r^2\Z_2^{xr}}=v_{r^2}+v_{xr^3}&=\frac{1}{4}\lb 1+\cO^{1_{xr}}+2\cO^{E,1,1}-2i\cO^{E,1,2}+2i\cO^{E,2,1}+2\cO^{E,2,2}\rb \,,\\
     v_{r^3\Z_2^{xr}}=v_{r^3}+v_{xr^2}&=\frac{1}{4}\lb 1-\cO^{1_{xr}}+2i\cO^{E,1,1}-2\cO^{E,1,2}-2\cO^{E,2,1}-2i\cO^{E,2,2}\rb\,.\\
\ea
\ee
In this phase $\Z_4^r$ cyclically permutes the vacua in the same way as (\ref{boxD8Z2})

\subsubsection{$D_8$ SPT Phase} \label{sec:D8_D8-SPT}
Choosing
\begin{equation}
    \Bphys=\fB_{\Neu(D_8),\tau}\,,
\end{equation}
there is a single vacuum:
\be
    v_{D_8}=\sum_{\bar{g}\in D_8}v_{\bar{g}}v_{\bar{g}}=1
\ee
in which the full $D_8$ symmetry is preserved. The SPT phase is labeled by
\be
    \tau\in H^3(D_8,U(1))=\Z_4\times\Z_2\times\Z_2\,.
\ee

\subsection{$\TwoRep(\Z_4^{(1)}\rtimes \Z_2^{(0)})$ Gapped Phases}

Let us now fix the symmetry boundary of the SymTFT to be 
\begin{equation}
    \Bsym=\fB_{\Neu(\Z_2^x),s}\,,
\end{equation}
which carries the $\TwoRep(\Z_4^{(1)}\rtimes \Z_2^{(0)})$ symmetry on it. As discussed in section \ref{sec:D8_NeuZ2x}, this boundary condition is obtained by gauging the non-normal $\Z_2^x$ subgroup of $D_8=\Z_4\rtimes\Z_2$ on $\fB_\Dir$.
The symmetry structure of $\TwoRep(\Z_4^{(1)}\rtimes \Z_2^{(0)})$ comprises of a $\Z_2^{(1)}\times\Z_2^{(0)}$ invertible subsymmetry generated by $(D_1^{\wh x},D_2^{r^2})$ along with a non-invertible 0-form symmetry generated by $D_2^{[r]}$ with fusion rule
\be\label{fus8}
D_2^{[r]}\otimes D_2^{[r]}=D_2^C\otimes(D_2^{\id}\oplus D_2^{r^2}) \,,
\ee
where $D_2^{C}:=D_2^{\id}/(D_1^{\id}\oplus D_1^{\wh{x}})$ is the condensation defect for $D_1^{\wh x}$.

\subsubsection{$\TwoRep (\Z_4^{(1)}\rtimes\Z_2^{(0)})/\Z_2^{(1)}$ SSB Phase}
Taking the physical boundary to be 
\begin{equation}
    \Bphys=\fB_{\Dir}\,,
\end{equation}
the SymTFT lines that have untwisted ends on both $\Bsym$ and $\Bphys$ are:
\be
    \{\Q_1^{1}\,,\quad \Q_1^{1_x}\,,\quad \Q_1^E\}\,,
\ee
On $\Bsym$, $\Q_1^{E}$ has one untwisted end  
${\cE}_{0}^{E,+}={\cE}_{0}^{E,1}+{\cE}_{0}^{E,2}$ and one $D_{1}^{\wh{x}}$ twisted end ${\cE}_{0}^{E,-}={\cE}_{0}^{E,1}-{\cE}_{0}^{E,2}$.
Compactifying the SymTFT, one thus obtains a four dimensional space of untwisted local operators spanned by
\begin{equation}
    \left\{1\,,\; \cO^{E,1}\,,\; \cO^{E,2}\,,\;\cO^{1_x}\right\}\,,
\end{equation}
where $\cO^{E,j}$ for $j=1,2$ denotes the local operator obtained by compactifying $\Q_1^{E}$ with the ends ${\cE}_0^{E,+}$ and $\wt\cE_0^{E,j}$ on the symmetry and physical boundary respectively
\be
\begin{split}
\begin{tikzpicture}
 \begin{scope}[shift={(0,0)},scale=0.8] 
\draw [cyan, fill=cyan!80!red, opacity =0.5]
(0,0) -- (0,4) -- (2,5) -- (5,5) -- (5,1) -- (3,0)--(0,0);
\draw [black, thick, fill=white,opacity=1]
(0,0) -- (0, 4) -- (2, 5) -- (2,1) -- (0,0);
\draw [cyan, thick, fill=cyan, opacity=0.2]
(0,0) -- (0, 4) -- (2, 5) -- (2,1) -- (0,0);
\draw[line width=1pt] (1,2.5) -- (3,2.5);
\draw[line width=1pt,dashed] (3,2.5) -- (4,2.5);
\fill[red!80!black] (1,2.5) circle (3pt);
\draw [fill=blue!40!red!60,opacity=0.2]
(3,0) -- (3, 4) -- (5, 5) -- (5,1) -- (3,0);
\fill[red!80!black] (1,2.5) circle (3pt);
\fill[red!80!black] (4,2.5) circle (3pt);
\draw [black, thick, opacity=1]
(3,0) -- (3, 4) -- (5, 5) -- (5,1) -- (3,0);
\node at (2,5.4) {$\fB_{\Neu(\Z_2^x),s}$};
\node at (5,5.4) {$\fB_{\Dir}$};
\draw[dashed] (0,0) -- (3,0);
\draw[dashed] (0,4) -- (3,4);
\draw[dashed] (2,5) -- (5,5);
\draw[dashed] (2,1) -- (5,1);
\draw [cyan, thick, fill=cyan, opacity=0.1]
(0,4) -- (3, 4) -- (5, 5) -- (2,5) -- (0,4);
\node[below, red!80!black] at (.8, 2.4) {$\cE_0^{E,+}$};
\node[below, red!80!black] at (3.8, 2.4) {$\wt{\cE}_0^{E,j}$};
\node  at (2.5, 3) {$\Q_{1}^{E}$};
\node  at (6.7, 2.5) {$=$};
\draw [fill=blue!60!green!30,opacity=0.7]
(8,0) -- (8, 4) -- (10, 5) -- (10,1) -- (8,0);
\draw [black, thick, opacity=1]
(8,0) -- (8, 4) -- (10, 5) -- (10,1) -- (8,0);
\fill[red!80!black] (9,2.5) circle (3pt);
\node[below, red!80!black] at (8.8, 2.4) {${\cO}^{E,j}$};
\end{scope}
\end{tikzpicture}    
\end{split}
\label{eq:local OP}
\ee
while $\cO^{1_x}$ comes from $\Q_1^{1_x}$ ending on both boundaries
\be
\begin{split}
\begin{tikzpicture}
 \begin{scope}[shift={(0,0)},scale=0.8] 
\draw [cyan, fill=cyan!80!red, opacity =0.5]
(0,0) -- (0,4) -- (2,5) -- (5,5) -- (5,1) -- (3,0)--(0,0);
\draw [black, thick, fill=white,opacity=1]
(0,0) -- (0, 4) -- (2, 5) -- (2,1) -- (0,0);
\draw [cyan, thick, fill=cyan, opacity=0.2]
(0,0) -- (0, 4) -- (2, 5) -- (2,1) -- (0,0);
\draw[line width=1pt] (1,2.5) -- (3,2.5);
\draw[line width=1pt,dashed] (3,2.5) -- (4,2.5);
\fill[red!80!black] (1,2.5) circle (3pt);
\draw [fill=blue!40!red!60,opacity=0.2]
(3,0) -- (3, 4) -- (5, 5) -- (5,1) -- (3,0);
\fill[red!80!black] (1,2.5) circle (3pt);
\fill[red!80!black] (4,2.5) circle (3pt);
\draw [black, thick, opacity=1]
(3,0) -- (3, 4) -- (5, 5) -- (5,1) -- (3,0);
\node at (2,5.4) {$\fB_{\Neu(\Z_2^x),s}$};
\node at (5,5.4) {$\fB_{\Dir}$};
\draw[dashed] (0,0) -- (3,0);
\draw[dashed] (0,4) -- (3,4);
\draw[dashed] (2,5) -- (5,5);
\draw[dashed] (2,1) -- (5,1);
\draw [cyan, thick, fill=cyan, opacity=0.1]
(0,4) -- (3, 4) -- (5, 5) -- (2,5) -- (0,4);
\node[below, red!80!black] at (.8, 2.4) {$\cE_0^{1_x}$};
\node[below, red!80!black] at (3.8, 2.4) {$\wt{\cE}_0^{1_x}$};
\node  at (2.5, 3) {$\Q_{1}^{1_x}$};
\node  at (6.7, 2.5) {$=$};
\draw [fill=blue!60!green!30,opacity=0.7]
(8,0) -- (8, 4) -- (10, 5) -- (10,1) -- (8,0);
\draw [black, thick, opacity=1]
(8,0) -- (8, 4) -- (10, 5) -- (10,1) -- (8,0);
\fill[red!80!black] (9,2.5) circle (3pt);
\node[below, red!80!black] at (8.8, 2.4) {${\cO}^{1_x}$};
\end{scope}
\end{tikzpicture}    
\end{split}
\label{eq:local OP}
\ee
The idempotents can be computed by starting with the $D_8$ SSB phase for $2\Vec_{D_8}$ of section \ref{sec:D8_D8SSB} (which has the same $\Bphys$ as our current sandwich) and gauging $\Z_2^{x}$ symmetry on the symmetry boundary for $2\Vec_{D_8}$ to obtain the symmetry $2\Rep(\Z_4^{(1)}\rtimes\Z_2^{(0)})$ we are interested in. They are:
\be
\ba
    v_{0}&=\frac{1}{4}\lb 1+\cO^{1_x}+2\cO^{E,1}+2\cO^{E,2}\rb \,,\\
    v_{1}&=\frac{1}{4}\lb 1-\cO^{1_x}-2i\cO^{E,1,}+2i\cO^{E,2}\rb\,,\\
    v_{2}&=\frac{1}{4}\lb 1+\cO^{1_x}-2\cO^{E,1}-2\cO^{E,2}\rb \,,\\
    v_{3}&=\frac{1}{4}\lb 1-\cO^{1_x}+2i\cO^{E,1}-2i\cO^{E,2}\rb\,,
\ea
\ee

We obtain a $\TwoRep(\Z_2^{(1)}\rtimes\Z_2^{(0)})$ phase with four vacua. 
$D_2^{r^2}$ acts by $-1$ on both $\cO^{E,1}$ and $\cO^{E,2}$ so it permutes the vacua in pairs:
\be
D_2^{r^2}=1_{02}\oplus 1_{20}\oplus 1_{13}\oplus 1_{31}\,.
\ee
To compute the action of the non-invertible symmetry $D_2^{[r]}$, we recall that $D_2^{[r]}$ descends from $D_2^{r}\oplus D_2^{r^3}$ upon gauging $\Z_2^{x}$. The sphere linking action of $D_2^{r}\oplus D_2^{r^3}$ on $\cO^{1_x}$ is $-1-1=-2$ while on $\cO^{E,1}$ and $\cO^{E,2}$ it is given by $i-i=0$. Therefore, the non-invertible $D_2^{[r]}$ sends each vacuum into the sum of two others:
\be \label{eq:D2r_2R2G1}
D_2^{[r]}=1_{01}\oplus 1_{03}\oplus 1_{12}\oplus 1_{10}\oplus 1_{23}\oplus 1_{21}\oplus 1_{30}\oplus 1_{32}\,.
\ee
Since $\Bphys$ has no (non-identity) genuine lines, $D_1^{\wh x}$ is realized trivially in each vacuum. 
It can be easily verified that the fusion rule \eqref{fus8} holds, by using the fact that
\be
D_2^C=2 \times (1_{00} \oplus  1_{11} \oplus 1_{22} \oplus 1_{33})
\ee
since the 1-form symmetry is generated by the identity line  $D_1^{\wh{x}}\simeq D_1^{1}$ in each vacuum. 
In summary, this is a $\TwoRep (\Z_4^{(1)}\rtimes\Z_2^{(0)})/\Z_2^{(1)}$ SSB with 4 vacua: $\Z_2^{(1)}$ is preserved in each so this is a confining phase,  $D_2^{r^2}$ is spontaneously broken and swaps pairs of vacua, while $D_2^{[r]}$ sends each vacuum to the sum of two others, eq. \eqref{eq:D2r_2R2G1}.
\begin{tcolorbox}[
colback=white,
coltitle= black,
colbacktitle=ourcolorforheader,
colframe=black,
title= $\TwoRep (\Z_4^{(1)}\rtimes\Z_2^{(0)})$:  $\TwoRep (\Z_4^{(1)}\rtimes\Z_2^{(0)})/\Z_2^{(1)}$ SSB with Trivial $\Z_2^{(1)}$ (Confining Phase),
sharp corners]
\be
\Triv_0 \oplus \Triv_1\oplus \Triv_2 \oplus \Triv_3 :\qquad 
\left\{\ba
\Z_2^{(0)}: & \quad  1_{02} \oplus 1_{20} \oplus 1_{13} \oplus 1_{31} \cr 
D_2^{[r]} :& \quad 1_{01}\oplus 1_{03}\oplus 1_{12}\oplus 1_{10}\oplus 1_{23}\oplus 1_{21}\oplus 1_{30}\oplus 1_{32} \cr 
D_2^C: & \quad 2 \times (1_{00} \oplus  1_{11} \oplus 1_{22} \oplus 1_{33})\cr 
\Z_2^{(1)}: & \quad \text{trivial}
 \,.
\ea
\right.
\ee
\end{tcolorbox}

\subsubsection{$\TwoRep (\Z_4^{(1)}\rtimes\Z_2^{(0)})/(\Z_2^{(1)}\times\Z_2^{(0)})$ SSB Phase} 
Let us take the physical boundary to be
\begin{equation}
    \Bphys=\fB_{\Neu(\Z_2^{r^2}),s'}\,.
\end{equation}
The SymTFT lines that have untwisted ends on both $\Bsym$ and $\Bphys$ are:
\be
    \{\Q_1^{1}\,,\quad \Q_1^{1_x}\}\,,
\ee
which, after compactifying the SymTFT, give rise to a two dimensional space of untwisted local operators spanned by
\begin{equation}
    \left\{1\,,\; \cO^{1_x}\right\}\,,
\end{equation}
where $\cO^{1_x}$ comes from $\Q_1^{1_x}$ ending on both boundaries
\be
\begin{split}
\begin{tikzpicture}
 \begin{scope}[shift={(0,0)},scale=0.8] 
\draw [cyan, fill=cyan!80!red, opacity =0.5]
(0,0) -- (0,4) -- (2,5) -- (5,5) -- (5,1) -- (3,0)--(0,0);
\draw [black, thick, fill=white,opacity=1]
(0,0) -- (0, 4) -- (2, 5) -- (2,1) -- (0,0);
\draw [cyan, thick, fill=cyan, opacity=0.2]
(0,0) -- (0, 4) -- (2, 5) -- (2,1) -- (0,0);
\draw[line width=1pt] (1,2.5) -- (3,2.5);
\draw[line width=1pt,dashed] (3,2.5) -- (4,2.5);
\fill[red!80!black] (1,2.5) circle (3pt);
\draw [fill=blue!40!red!60,opacity=0.2]
(3,0) -- (3, 4) -- (5, 5) -- (5,1) -- (3,0);
\fill[red!80!black] (1,2.5) circle (3pt);
\fill[red!80!black] (4,2.5) circle (3pt);
\draw [black, thick, opacity=1]
(3,0) -- (3, 4) -- (5, 5) -- (5,1) -- (3,0);
\node at (2,5.4) {$\fB_{\Neu(\Z_2^x),s}$};
\node at (5,5.4) {$\fB_{\Neu(\Z_2^{r^2}),s'}$};
\draw[dashed] (0,0) -- (3,0);
\draw[dashed] (0,4) -- (3,4);
\draw[dashed] (2,5) -- (5,5);
\draw[dashed] (2,1) -- (5,1);
\draw [cyan, thick, fill=cyan, opacity=0.1]
(0,4) -- (3, 4) -- (5, 5) -- (2,5) -- (0,4);
\node[below, red!80!black] at (.8, 2.4) {$\cE_0^{1_x}$};
\node[below, red!80!black] at (3.8, 2.4) {$\wt{\cE}_0^{1_x}$};
\node  at (2.5, 3) {$\Q_{1}^{1_x}$};
\node  at (6.7, 2.5) {$=$};
\draw [fill=blue!60!green!30,opacity=0.7]
(8,0) -- (8, 4) -- (10, 5) -- (10,1) -- (8,0);
\draw [black, thick, opacity=1]
(8,0) -- (8, 4) -- (10, 5) -- (10,1) -- (8,0);
\fill[red!80!black] (9,2.5) circle (3pt);
\node[below, red!80!black] at (8.8, 2.4) {${\cO}^{1_x}$};
\end{scope}
\end{tikzpicture}    
\end{split}
\label{eq:local OP}
\ee
The idempotents can be computed by starting with the $\Z_2^r\times\Z_2^x$ SSB phase for $2\Vec_{D_8}$ of section \ref{sec:D8_Z2rZ2x_SSB} (which has the same $\Bphys$ as our current sandwich) and gauging $\Z_2^{x}$ on the symmetry boundary. They are:
\be
\ba
    v_0&=\frac{1}{2}(1+\cO^{1_x})\,,\\
    v_1&=\frac{1}{2}(1-\cO^{1_x})\,.
\ea
\ee
This is a $\TwoRep(\Z_4^{(1)}\rtimes\Z_2^{(0)})$ phase with two vacua.

The only non-trivial line on $\Bphys$ is $\wt{\cE}_1^{r^2}$ (coming from the SymTFT surface $\Q_2^{[r^2]}$), which gives rise to a $D_2^{r^2}$ twisted sector line after compactification:
\begin{equation} \label{eq:Lr2}
    (D_2^{r^2}\,,\cL^{r^2})=\left((D_2^{r^2}\,, \cE_1^{r^2})\,, \Q_2^{[r^2]}\,, \wt\cE_1^{r^2}\right)\,.
\end{equation}
This configuration is depicted as
\be
\begin{split}
\begin{tikzpicture}
 \begin{scope}[shift={(0,0)},scale=0.8] 
\draw [cyan, fill=cyan!80!red, opacity =0.5]
(0,0) -- (0,4) -- (2,5) -- (5,5) -- (5,1) -- (3,0)--(0,0);
\draw [black, thick, fill=white,opacity=1]
(0,0) -- (0, 4) -- (2, 5) -- (2,1) -- (0,0);
\draw [cyan, thick, fill=cyan, opacity=0.2]
(0,0) -- (0, 4) -- (2, 5) -- (2,1) -- (0,0);
\draw [fill=blue!40!red!60,opacity=0.2]
(3,0) -- (3, 4) -- (5, 5) -- (5,1) -- (3,0);
\draw [black, thick, opacity=1]
(3,0) -- (3, 4) -- (5, 5) -- (5,1) -- (3,0);
\node at (2,5.4) {$\fB_{\Neu(\Z_2^x),s}$};
\node at (5,5.4) {$\fB_{\Neu(\Z_2^{r^2}),s'}$};
\draw[dashed] (0,0) -- (3,0);
\draw[dashed] (0,4) -- (3,4);
\draw[dashed] (2,5) -- (5,5);
\draw[dashed] (2,1) -- (5,1);
\draw [cyan, thick, fill=cyan, opacity=0.1]
(0,4) -- (3, 4) -- (5, 5) -- (2,5) -- (0,4);
\draw [black, thick, fill=black, opacity=0.3]
(0.,2) -- (3, 2) -- (5., 3) -- (2, 3) -- (0.,2);
\draw [black, thick, fill=black, opacity=0.3]
(0.,2) -- (0., 4) -- (2, 5) -- (2,3) -- (0,2);
\node  at (2.5, 2.5) {$\Q_{2}^{r^2}$};
\node[below, red!80!black] at (-0.5, 2.4) {$\cE_1^{r^2}$};
\node[below, red!80!black] at (4, 2.4) {$\wt{\cE}_1^{r^2}$};
\draw [line width=1pt, red!80!black] (0.0, 2.) -- (2, 3);
\draw [line width=1pt, red!80!black] (3, 2.) -- (5, 3);
\node  at (6.7, 2.5) {$=$};
\draw [fill=blue!60!green!30,opacity=0.7]
(8,0) -- (8, 4) -- (10, 5) -- (10,1) -- (8,0);
\draw [black, thick, opacity=1]
(8,0) -- (8, 4) -- (10, 5) -- (10,1) -- (8,0);
\draw [black, thick, fill=black, opacity=0.3]
(8,2) -- (8, 4) -- (10, 5) -- (10,3) -- (8,2);
\draw [line width=1pt, red!80!black] (8, 2.) -- (10, 3);
\node[below, black] at (9, 3.9) {$D_2^{r^2}$};
\node[below, black] at (1., 3.9) {$D_2^{r^2}$};
\node[below, red!80!black] at (9, 2.4) {${\cL}^{r^2}$};
\node[below, black] at (11, 2.4) {$,$};
\end{scope}
\end{tikzpicture}    
\end{split}
\ee
$\cL^{r^2}$ is uncharged under $D_1^{\wh x}$: the 1-form symmetry is therefore realized trivially in each vacuum.

Let us turn to the 0-form symmetry:
$D_2^{r^2}$ is the identity
\be
D_2^{r^2}=1_{00}\oplus 1_{11}\,,
\ee
whereas the non-invertible $D_2^{[r]}$ is non-trivial, since  $\cO^{1_x}$ carries charge $-1-1=-2$ under $D_2^r\oplus D_2^{r^3}$. Under this symmetry action, the two vacua are thus exchanged and multiplied by a factor of 2:
\be
D_2^{[r]}=2 \times (1_{01}\oplus 1_{10})
\ee
One can verify the fusion rule \eqref{fus8} using the fact that
\be
D_2^C=2 \times (1_{00} \oplus  1_{11})
\ee
This is therefore a $\TwoRep (\Z_4^{(1)}\rtimes\Z_2^{(0)})/(\Z_2^{(1)}\times\Z_2^{(0)})$ SSB Phase since the invertible $\Z_2^{(1)}\times\Z_2^{(0)}$ is preserved and only the $D_2^{[r]}$ non-invertible symmetry of $\TwoRep (\Z_4^{(1)}\rtimes\Z_2^{(0)})$ is spontaneously broken.

\begin{tcolorbox}[
colback=white,
coltitle= black,
colbacktitle=ourcolorforheader,
colframe=black,
title= $\TwoRep (\Z_4^{(1)}\rtimes\Z_2^{(0)})$:  $\TwoRep (\Z_4^{(1)}\rtimes\Z_2^{(0)})/(\Z_2^{(1)}\times\Z_2^{(0)})$ SSB (Confining Phase),
sharp corners]
\be
\Triv_0 \oplus \Triv_1:\qquad 
\left\{\ba
\Z_2^{(0)}: & \quad \text{trivial} \cr 
D_2^{[r]} :& \quad 2 \times (1_{01}\oplus 1_{10})\cr 
D_2^C: & \quad 2 \times (1_{00} \oplus  1_{11})\cr 
\Z_2^{(1)}: & \quad \text{trivial}
 \,.
\ea
\right.
\ee
\end{tcolorbox}

\subsubsection{$\TwoRep(\Z_4^{(1)}\rtimes \Z_2^{(0)})$ SPT Phase with trivial $\Z_2^{(1)}$} 
Let us take the physical boundary to be
\begin{equation}
    \Bphys=\fB_{\Neu(\Z_4^{r}),p}\,.
\end{equation}
The SymTFT line that has untwisted ends on both $\Bsym$ and $\Bphys$ is only the identity $\Q_1^1$,
which, after compactifying the SymTFT gives rise to a single local operator and thus one vacuum. The lines on $\Bphys$ are $\cE_1^{r^2}$, eq. \eqref{eq:Lr2}, and $\cE_1^{[r]}$, which gives rise to a $D_2^{[r]}$ twisted sector line after compactification:
\begin{equation} 
    (D_2^{[r]}\,,\cL^{[r]})=\left((D_2^{[r]}\,, \cE_1^{[r]})\,, \Q_2^{[r]}\,, \wt\cE_1^{[r]}\right)\,,
\end{equation}
depicted as
\be
\begin{split}
\begin{tikzpicture}
 \begin{scope}[shift={(0,0)},scale=0.8] 
\draw [cyan, fill=cyan!80!red, opacity =0.5]
(0,0) -- (0,4) -- (2,5) -- (5,5) -- (5,1) -- (3,0)--(0,0);
\draw [black, thick, fill=white,opacity=1]
(0,0) -- (0, 4) -- (2, 5) -- (2,1) -- (0,0);
\draw [cyan, thick, fill=cyan, opacity=0.2]
(0,0) -- (0, 4) -- (2, 5) -- (2,1) -- (0,0);
\draw [fill=blue!40!red!60,opacity=0.2]
(3,0) -- (3, 4) -- (5, 5) -- (5,1) -- (3,0);
\draw [black, thick, opacity=1]
(3,0) -- (3, 4) -- (5, 5) -- (5,1) -- (3,0);
\node at (2,5.4) {$\fB_{\Neu(\Z_2^x),s}$};
\node at (5,5.4) {$\fB_{\Neu(\Z_4^{r}),p}$};
\draw[dashed] (0,0) -- (3,0);
\draw[dashed] (0,4) -- (3,4);
\draw[dashed] (2,5) -- (5,5);
\draw[dashed] (2,1) -- (5,1);
\draw [cyan, thick, fill=cyan, opacity=0.1]
(0,4) -- (3, 4) -- (5, 5) -- (2,5) -- (0,4);
\draw [black, thick, fill=black, opacity=0.3]
(0.,2) -- (3, 2) -- (5., 3) -- (2, 3) -- (0.,2);
\draw [black, thick, fill=black, opacity=0.3]
(0.,2) -- (0., 4) -- (2, 5) -- (2,3) -- (0,2);
\node  at (2.5, 2.5) {$\Q_{2}^{[r]}$};
\node[below, red!80!black] at (-0.5, 2.4) {$\cE_1^{[r]}$};
\node[below, red!80!black] at (4, 2.4) {$\wt{\cE}_1^{[r]}$};
\draw [line width=1pt, red!80!black] (0.0, 2.) -- (2, 3);
\draw [line width=1pt, red!80!black] (3, 2.) -- (5, 3);
\node  at (6.7, 2.5) {$=$};
\draw [fill=blue!60!green!30,opacity=0.7]
(8,0) -- (8, 4) -- (10, 5) -- (10,1) -- (8,0);
\draw [black, thick, opacity=1]
(8,0) -- (8, 4) -- (10, 5) -- (10,1) -- (8,0);
\draw [black, thick, fill=black, opacity=0.3]
(8,2) -- (8, 4) -- (10, 5) -- (10,3) -- (8,2);
\draw [line width=1pt, red!80!black] (8, 2.) -- (10, 3);
\node[below, black] at (9, 3.9) {$D_2^{[r]}$};
\node[below, black] at (1., 3.9) {$D_2^{[r]}$};
\node[below, red!80!black] at (9, 2.4) {${\cL}^{[r]}$};
\node[below, black] at (11, 2.4) {$,$};
\end{scope}
\end{tikzpicture}    
\end{split}
\ee
Both $\cE^{r^2}$ and $\cE^{[r]}$ are uncharged under $D_1^{\wh x}$: the $\Z_2$ 1-form symmetry is therefore realized trivially.

The idempotents can be computed by starting with the $\Z_2^x$ SSB phase I for $2\Vec_{D_8}$ of section \ref{sec:D8_Z2x_SSB1}  (which has the same $\Bphys$ as our current sandwich) and gauging $\Z_2^{x}$ on the symmetry boundary, confirming that in the thus obtained $2\Rep(\Z_4^{(1)}\rtimes\Z_2^{(0)})$ symmetric phase there is a single one vacuum $v_0$ with  $D_2^{r^2}$ realized trivially and
\be
D_2^{[r]}\cong 2\times 1_{00}
\ee
such that a line living at the end of $D_2^{[r]}$ is not charged under $D_1^{\wh x}$.

\begin{tcolorbox}[
colback=white,
coltitle= black,
colbacktitle=ourcolorforheader,
colframe=black,
title= $\TwoRep(\Z_4^{(1)}\rtimes \Z_2^{(0)})$: $\TwoRep(\Z_4^{(1)}\rtimes \Z_2^{(0)})$ SPT Phase and Trivial $\Z_2^{(1)}$ (Confining Phase), 
sharp corners]
\be
\begin{split}
\begin{tikzpicture}
\begin{scope}[shift={(0,0)},scale=1] 
 \pgfmathsetmacro{\del}{-0.5}
\node at  (0,0)  {${\rm Triv}_0$} ;
\node at  (0,-0.6)  {\footnotesize$(\Z_2^{(1)}\text{-}{\rm Triv})$} ;
\end{scope}
\end{tikzpicture}    
\end{split}
\ee
The non-invertible 0-form symmetry is realized on the single vacuum as
\begin{equation}
    D_2^{[r]}=2\times 1_{00}\,. 
\end{equation}
\end{tcolorbox}

\subsubsection{$\TwoRep(\Z_4^{(1)}\rtimes \Z_2^{(0)})$ Mixed SPT Phase} \label{sec:D8_1-form_non-triv}
Taking the physical boundary to be
\begin{equation}
    \Bphys=\fB_{\Neu(\Z_2^{xr}\times\Z_2^{xr^3}),t}\,,
\end{equation}
the SymTFT line that has untwisted ends on both $\Bsym$ and $\Bphys$ is again only the identity $\Q_1^1$,
which, after compactifying the SymTFT gives rise to a single vacuum. However, there are non-trivial twisted sector local operators: specifically, the compactification of  $\Q_1^{1_{xr}}$ produces a local operator attached to the $D_1^{\wh{x}}$ line.
\begin{equation}
    (D_{1}^{\wh{x}}\,, \cO^{1_{xr}})=\left((D_{1}^{\wh{x}}\,,{\cE}_0^{1_{xr}})\,, \Q_1^{1_{xr}}\,, \wt\cE_0^{1_{xr}}\right)\,.
\end{equation}
This configuration is depicted as
\be
\begin{split}
\begin{tikzpicture}
 \begin{scope}[shift={(0,0)},scale=0.8] 
\draw [cyan, fill=cyan!80!red, opacity =0.5]
(0,0) -- (0,4) -- (2,5) -- (5,5) -- (5,1) -- (3,0)--(0,0);
\draw [black, thick, fill=white,opacity=1]
(0,0) -- (0, 4) -- (2, 5) -- (2,1) -- (0,0);
\draw [cyan, thick, fill=cyan, opacity=0.2]
(0,0) -- (0, 4) -- (2, 5) -- (2,1) -- (0,0);
\draw[line width=1pt] (1,2.5) -- (3,2.5);
\draw[line width=1pt,dashed] (3,2.5) -- (4,2.5);
\fill[red!80!black] (1,2.5) circle (3pt);
\draw [fill=blue!40!red!60,opacity=0.2]
(3,0) -- (3, 4) -- (5, 5) -- (5,1) -- (3,0);
\fill[red!80!black] (1,2.5) circle (3pt);
\fill[red!80!black] (4,2.5) circle (3pt);
\draw [black, thick, opacity=1]
(3,0) -- (3, 4) -- (5, 5) -- (5,1) -- (3,0);
\node at (2,5.4) {$\fB_{\Neu(\Z_2^x),s}$};
\node at (5,5.4) {$\fB_{\Neu(\Z_2^{xr}\times\Z_2^{xr^3}),t}$};
\draw[dashed] (0,0) -- (3,0);
\draw[dashed] (0,4) -- (3,4);
\draw[dashed] (2,5) -- (5,5);
\draw[dashed] (2,1) -- (5,1);
\draw[line width=1pt] (1,2.5) -- (1,4.5);
\draw [cyan, thick, fill=cyan, opacity=0.1]
(0,4) -- (3, 4) -- (5, 5) -- (2,5) -- (0,4);
\node[below, red!80!black] at (.8, 2.4) {$\cE_0^{1_{xr}}$};
\node[below, red!80!black] at (3.8, 2.4) {$\wt{\cE}_0^{1_{xr}}$};
\node  at (2.5, 3) {$\Q_{1}^{1_{xt}}$};
\node  at (0.5, 3.3) {$D_{1}^{\wh{x}}$};
\node  at (6.7, 2.5) {$=$};
\draw [fill=blue!60!green!30,opacity=0.7]
(8,0) -- (8, 4) -- (10, 5) -- (10,1) -- (8,0);
\draw [black, thick, opacity=1]
(8,0) -- (8, 4) -- (10, 5) -- (10,1) -- (8,0);
\draw[line width=1pt] (9,2.5) -- (9,4.5);
\fill[red!80!black] (9,2.5) circle (3pt);
\node[below, red!80!black] at (8.8, 2.4) {${\cO}^{1_{xr}}$};
\node  at (8.5, 3.3) {$D_{1}^{\wh{x}}$};
\node[below, black] at (10.8, 3) {$,$};
\end{scope}
\end{tikzpicture}    
\end{split}
\ee
where $\cO^{1_{xr}}$ carries charge $-2$ under $D_2^{[r]}$.

The lines on $\Bphys$ are $\cE^{r^2}$, eq. \eqref{eq:Lr2}, and $\cE^{x[r]}$, which gives rise to a $D_2^{[r]}$ twisted sector line after compactification:
\begin{equation}
    (D_2^{[r]}\,,\cL^{x[r]})=\left((D_2^{[r]}\,, \cE_1^{x[r]})\,, \Q_2^{[xr]}\,, \wt\cE_1^{x[r]}\right)\,,
\end{equation}
depicted as
\be
\begin{split}
\begin{tikzpicture}
 \begin{scope}[shift={(0,0)},scale=0.8] 
\draw [cyan, fill=cyan!80!red, opacity =0.5]
(0,0) -- (0,4) -- (2,5) -- (5,5) -- (5,1) -- (3,0)--(0,0);
\draw [black, thick, fill=white,opacity=1]
(0,0) -- (0, 4) -- (2, 5) -- (2,1) -- (0,0);
\draw [cyan, thick, fill=cyan, opacity=0.2]
(0,0) -- (0, 4) -- (2, 5) -- (2,1) -- (0,0);
\draw [fill=blue!40!red!60,opacity=0.2]
(3,0) -- (3, 4) -- (5, 5) -- (5,1) -- (3,0);
\draw [black, thick, opacity=1]
(3,0) -- (3, 4) -- (5, 5) -- (5,1) -- (3,0);
\node at (2,5.4) {$\fB_{\Neu(\Z_2^x),s}$};
\node at (5,5.4) {$\fB_{\Neu(\Z_4^{r}),p}$};
\draw[dashed] (0,0) -- (3,0);
\draw[dashed] (0,4) -- (3,4);
\draw[dashed] (2,5) -- (5,5);
\draw[dashed] (2,1) -- (5,1);
\draw [cyan, thick, fill=cyan, opacity=0.1]
(0,4) -- (3, 4) -- (5, 5) -- (2,5) -- (0,4);
\draw [black, thick, fill=black, opacity=0.3]
(0.,2) -- (3, 2) -- (5., 3) -- (2, 3) -- (0.,2);
\draw [black, thick, fill=black, opacity=0.3]
(0.,2) -- (0., 4) -- (2, 5) -- (2,3) -- (0,2);
\node  at (2.5, 2.5) {$\Q_{2}^{r^2}$};
\node[below, red!80!black] at (-0.5, 2.4) {$\cE_1^{x[r]}$};
\node[below, red!80!black] at (4, 2.4) {$\wt{\cE}_1^{x[r]}$};
\draw [line width=1pt, red!80!black] (0.0, 2.) -- (2, 3);
\draw [line width=1pt, red!80!black] (3, 2.) -- (5, 3);
\node  at (6.7, 2.5) {$=$};
\draw [fill=blue!60!green!30,opacity=0.7]
(8,0) -- (8, 4) -- (10, 5) -- (10,1) -- (8,0);
\draw [black, thick, opacity=1]
(8,0) -- (8, 4) -- (10, 5) -- (10,1) -- (8,0);
\draw [black, thick, fill=black, opacity=0.3]
(8,2) -- (8, 4) -- (10, 5) -- (10,3) -- (8,2);
\draw [line width=1pt, red!80!black] (8, 2.) -- (10, 3);
\node[below, black] at (9, 3.9) {$D_2^{[r]}$};
\node[below, black] at (1., 3.9) {$D_2^{[r]}$};
\node[below, red!80!black] at (9, 2.4) {${\cL}^{x[r]}$};
\node[below, black] at (11, 2.4) {$,$};
\end{scope}
\end{tikzpicture}    
\end{split}
\ee
Unlike the previous phase, $\cE_1^{x[r]}$ is charged under $D_1^{\wh x}$, therefore the 1-form symmetry is realized non-trivially this is a mixed `non-trivial SPT' phase between the 1-form symmetry and $D_2^{[r]}$ since there is a non-genuine line (in the twisted sector for the non-invertible $D_2^{[r]}$ 0-form symm) which is charged under the 1-form symm generated by $D_1^{\wh{x}}$: it is the higher-dimensional analogue of the string order-parameters for 1+1d SPTs. No SymTFT surfaces end on both boundaries, so there are no genuine line operators in this phase.

The idempotents can be computed by starting with the $\Z_2^x$ SSB phase II for $2\Vec_{D_8}$ of section \ref{sec:D8_Z2x_SSB2}  (which has the same $\Bphys$ as our current sandwich) and gauging $\Z_2^{x}$ on the symmetry boundary, confirming that in the thus obtained $2\Rep(\Z_4^{(1)}\rtimes\Z_2^{(0)})$ symmetric phase there is a single one vacuum $v_0$ with $D_2^{r^2}$ realized trivially and
\be
D_2^{[r]}\cong 2\times 1_{00}
\ee
such that a line living at the end of $D_2^{[r]}$ is charged under $D_1^{\wh x}$.
\begin{tcolorbox}[
colback=white,
coltitle= black,
colbacktitle=ourcolorforheader,
colframe=black,
title= $\TwoRep(\Z_4^{(1)}\rtimes \Z_2^{(0)})$ Mixed SPT Phase {with non-trivial $\Z_2^{(1)}$}, 
sharp corners]
\be
\begin{split}
\begin{tikzpicture}
\begin{scope}[shift={(0,0)},scale=1] 
 \pgfmathsetmacro{\del}{-0.5}
\node at  (0,0)  {${\rm Triv}_0$} ;
\node at  (0,-0.6)  {\footnotesize ($\Z_2^{(1)}$-Mixed SPT)};
\end{scope}
\end{tikzpicture}    
\end{split}
\ee
\end{tcolorbox}

\subsubsection{$\TwoRep(\Z_4^{(1)}\rtimes \Z_2^{(0)})/\Z_2^{(0)}$ SSB Phase} 
Let us take the physical boundary to be
\begin{equation}
    \Bphys=\fB_{\Neu(\Z_2^{x}\times\Z_2^{xr^2}),t}\,.
\end{equation}
The SymTFT lines that have untwisted ends on both $\Bsym$ and $\Bphys$ are:
\be
    \{\Q_1^{1}\,,\quad \Q_1^{1_x}\}\,,
\ee
which, after compactifying the SymTFT, give rise to a two dimensional space of untwisted local operators spanned by
\begin{equation}
    \left\{1\,,\; \cO^{1_x}\right\}\,,
\end{equation}
where $\cO^{1_x}$ comes from $\Q_1^{1_x}$ ending on both boundaries
\be
\begin{split}
\begin{tikzpicture}
 \begin{scope}[shift={(0,0)},scale=0.8] 
\draw [cyan, fill=cyan!80!red, opacity =0.5]
(0,0) -- (0,4) -- (2,5) -- (5,5) -- (5,1) -- (3,0)--(0,0);
\draw [black, thick, fill=white,opacity=1]
(0,0) -- (0, 4) -- (2, 5) -- (2,1) -- (0,0);
\draw [cyan, thick, fill=cyan, opacity=0.2]
(0,0) -- (0, 4) -- (2, 5) -- (2,1) -- (0,0);
\draw[line width=1pt] (1,2.5) -- (3,2.5);
\draw[line width=1pt,dashed] (3,2.5) -- (4,2.5);
\fill[red!80!black] (1,2.5) circle (3pt);
\draw [fill=blue!40!red!60,opacity=0.2]
(3,0) -- (3, 4) -- (5, 5) -- (5,1) -- (3,0);
\fill[red!80!black] (1,2.5) circle (3pt);
\fill[red!80!black] (4,2.5) circle (3pt);
\draw [black, thick, opacity=1]
(3,0) -- (3, 4) -- (5, 5) -- (5,1) -- (3,0);
\node at (2,5.4) {$\fB_{\Neu(\Z_2^x),s}$};
\node at (5,5.4) {$\fB_{\Neu(\Z_2^{x}\times\Z_2^{xr^2}),t}$};
\draw[dashed] (0,0) -- (3,0);
\draw[dashed] (0,4) -- (3,4);
\draw[dashed] (2,5) -- (5,5);
\draw[dashed] (2,1) -- (5,1);
\draw [cyan, thick, fill=cyan, opacity=0.1]
(0,4) -- (3, 4) -- (5, 5) -- (2,5) -- (0,4);
\node[below, red!80!black] at (.8, 2.4) {$\cE_0^{1_x}$};
\node[below, red!80!black] at (3.8, 2.4) {$\wt{\cE}_0^{1_x}$};
\node  at (2.5, 3) {$\Q_{1}^{1_x}$};
\node  at (6.7, 2.5) {$=$};
\draw [fill=blue!60!green!30,opacity=0.7]
(8,0) -- (8, 4) -- (10, 5) -- (10,1) -- (8,0);
\draw [black, thick, opacity=1]
(8,0) -- (8, 4) -- (10, 5) -- (10,1) -- (8,0);
\fill[red!80!black] (9,2.5) circle (3pt);
\node[below, red!80!black] at (8.8, 2.4) {${\cO}^{1_x}$};
\end{scope}
\end{tikzpicture}    
\end{split}
\label{eq:local OP}
\ee
This phase also presents a genuine line operator, coming from the $\Q_2^{[x]}$ generalized charge ending on both $\Bsym$ and $\Bphys$
\begin{equation}
    \cL^{x}=(\cE_1^{[x]}\,, \Q_2^{[x]}\,, \wt{\cE}_1^{x})\,,
\end{equation}
depicted as
\be
\begin{split}
\begin{tikzpicture}
 \begin{scope}[shift={(0,0)},scale=0.8] 
\draw [cyan, fill=cyan!80!red, opacity =0.5]
(0,0) -- (0,4) -- (2,5) -- (5,5) -- (5,1) -- (3,0)--(0,0);
\draw [black, thick, fill=white,opacity=1]
(0,0) -- (0, 4) -- (2, 5) -- (2,1) -- (0,0);
\draw [cyan, thick, fill=cyan, opacity=0.2]
(0,0) -- (0, 4) -- (2, 5) -- (2,1) -- (0,0);
\draw [fill=blue!40!red!60,opacity=0.2]
(3,0) -- (3, 4) -- (5, 5) -- (5,1) -- (3,0);
\draw [black, thick, opacity=1]
(3,0) -- (3, 4) -- (5, 5) -- (5,1) -- (3,0);
\node at (2,5.4) {$\fB_{\Neu(\Z_2^x),s}$};
\node at (5,5.4) {$\fB_{\fB_{\Neu(\Z_2^{x}\times\Z_2^{xr^2}),t}}$};
\draw[dashed] (0,0) -- (3,0);
\draw[dashed] (0,4) -- (3,4);
\draw[dashed] (2,5) -- (5,5);
\draw[dashed] (2,1) -- (5,1);
\draw [cyan, thick, fill=cyan, opacity=0.1]
(0,4) -- (3, 4) -- (5, 5) -- (2,5) -- (0,4);
\draw [black, thick, fill=black, opacity=0.3]
(0.,2) -- (3, 2) -- (5., 3) -- (2, 3) -- (0.,2);
\node  at (2.5, 2.5) {$\Q_{2}^{[b]}$};
\node[below, red!80!black] at (-0.5, 2.4) {$\cE_1^{[x]}$};
\node[below, red!80!black] at (4, 2.4) {$\wt{\cE}_1^{x}$};
\draw [line width=1pt, red!80!black] (0.0, 2.) -- (2, 3);
\draw [line width=1pt, red!80!black] (3, 2.) -- (5, 3);
\node  at (6.7, 2.5) {$=$};
\draw [fill=blue!60!green!30,opacity=0.7]
(8,0) -- (8, 4) -- (10, 5) -- (10,1) -- (8,0);
\draw [black, thick, opacity=1]
(8,0) -- (8, 4) -- (10, 5) -- (10,1) -- (8,0);
\draw [line width=1pt, red!80!black] (8, 2.) -- (10, 3);
\node[below, red!80!black] at (9, 2.4) {${\cL}^{x}$};
\node[below, black] at (11, 2.4) {$.$};
\end{scope}
\end{tikzpicture}    
\end{split}
\ee
$\cL^x$ is charged under $D_1^{\wh{x}}$, obeys $\Z_2$ fusion rules and has F-symbols given by $s-s'\in H^{3}(\Z_2\,, U(1))$. 

The idempotents are the same as those of the  $\Z_2^{xr}$ SSB phase for $2\Vec_{D_8}$ of section \ref{sec:D8_Z2xr_SSB} (which has the same $\Bphys$ as our current sandwich) since both vacua are already invariant under $x$:
\be
\ba
    v_{0}&=\frac{1}{2}\lb 1+\cO^{1_{x}}\rb \,,\\
    v_{1}&=\frac{1}{2}\lb 1-\cO^{1_{x}}\rb\,.
\ea
\ee
However, the effect of gauging $\Z_2^x$ is visible at the level of lines: after gauging each of the two invariant vacua $v_0$, $v_1$ carry $\Z_{2}$ Dijkgraaf-Witten theory with a 3-cocycle twist $s-s'\in H^{3}(\Z_2,U(1))$, i.e. toric code or double semion, so the underlying 3d TFT is
\be
    \fT=\lb\text{DW}_{s-s'}\rb_0\oplus \lb\text{DW}_{s-s'}\rb_1
\ee
with line operators given by $D_1^{\wh{x}}$, which is bosonic, and $\cL^x$ which is a boson if $s-s'$ is trivial or a semion otherwise. 
The invertible 0-form symmetry generator $D_2^{r^2}$ is the identity for $\fT$, while
non-invertible $D_2^{[r]}$ is realized as 
\be   
    D_2^{[r]} = (D_2^{C})_{01}\oplus (D_2^{C})_{10}
\ee
which obeys the fusion \eqref{fus8}.
\begin{tcolorbox}[
colback=white,
coltitle= black,
colbacktitle=ourcolorforheader,
colframe=black,
title= $\TwoRep(\Z_4^{(1)}\rtimes \Z_2^{(0)})$: $\TwoRep(\Z_4^{(1)}\rtimes \Z_2^{(0)})/\Z_2^{(0)}$ SSB Phase, 
sharp corners]
\be
\begin{split}
\begin{tikzpicture}
\begin{scope}[shift={(0,0)},scale=1] 
 \pgfmathsetmacro{\del}{-0.5}
\node at  (0.5,0)  {$\left({\rm DW(\Z_2)_{s-s'}}\right)_0 \quad \boxplus \quad \left({\rm DW(\Z_2)_{s-s'}}\right)_1$} ;
\node at  (-1.5,-0.6)  {\footnotesize$(\Z_2^{(1)}\text{-}{\rm SSB})$} ;
\node at  (2.3,-0.6)  {\footnotesize$(\Z_2^{(1)}\text{-}{\rm SSB})$} ;
\end{scope}
\end{tikzpicture}    
\end{split}
\ee
$D_2^{r^2}$ is the identity, whereas the non-invertible 0-form symmetry is realized as
\begin{equation}
    D_2^{[r]} = D_2^{[r]} = (D_2^{C})_{01}\oplus (D_2^{C})_{10} \,. 
\end{equation}
\end{tcolorbox}

\subsubsection{$\TwoRep(\Z_4^{(1)}\rtimes \Z_2^{(0)})$ SSB Phase} 
\label{sec:carrot3}
Let us now take the physical boundary to be
\begin{equation}
    \Bphys=\fB_{\Neu(\Z_2^{x}),s'}\,.
\end{equation}
which is the same as $\Bphys$ (upto discrete torsion). The SymTFT lines that end are:
\be
    \{\Q_1^{1}\,,\quad \Q_1^{1_x}\,,\quad \Q_1^E\}\,,
\ee
$\Q_1^{E}$ has one untwisted end  
${\cE}_{0}^{E,+}={\cE}_{0}^{E,1}+{\cE}_{0}^{E,2}$ and one $D_{1}^{\wh{x}}$ twisted end ${\cE}_{0}^{E,-}={\cE}_{0}^{E,1}-{\cE}_{0}^{E,2}$.
Compactifying the SymTFT, one thus obtains a three dimensional space of untwisted local operators spanned by
\begin{equation}
    \left\{1\,,\; \cO^{E,+}\,,\;\cO^{1_x}\right\}\,,
\end{equation}
where $\cO^{E,+}$ denotes the local operator obtained by compactifying $\Q_1^{E}$ with end ${\cE}_0^{E,+}$ on both SymTFT boundaries
\be
\begin{split}
\begin{tikzpicture}
 \begin{scope}[shift={(0,0)},scale=0.8] 
\draw [cyan, fill=cyan!80!red, opacity =0.5]
(0,0) -- (0,4) -- (2,5) -- (5,5) -- (5,1) -- (3,0)--(0,0);
\draw [black, thick, fill=white,opacity=1]
(0,0) -- (0, 4) -- (2, 5) -- (2,1) -- (0,0);
\draw [cyan, thick, fill=cyan, opacity=0.2]
(0,0) -- (0, 4) -- (2, 5) -- (2,1) -- (0,0);
\draw[line width=1pt] (1,2.5) -- (3,2.5);
\draw[line width=1pt,dashed] (3,2.5) -- (4,2.5);
\fill[red!80!black] (1,2.5) circle (3pt);
\draw [fill=blue!40!red!60,opacity=0.2]
(3,0) -- (3, 4) -- (5, 5) -- (5,1) -- (3,0);
\fill[red!80!black] (1,2.5) circle (3pt);
\fill[red!80!black] (4,2.5) circle (3pt);
\draw [black, thick, opacity=1]
(3,0) -- (3, 4) -- (5, 5) -- (5,1) -- (3,0);
\node at (2,5.4) {$\fB_{\Neu(\Z_2^x),s}$};
\node at (5,5.4) {$\fB_{\Neu(\Z_2^{x}),s'}$};
\draw[dashed] (0,0) -- (3,0);
\draw[dashed] (0,4) -- (3,4);
\draw[dashed] (2,5) -- (5,5);
\draw[dashed] (2,1) -- (5,1);
\draw [cyan, thick, fill=cyan, opacity=0.1]
(0,4) -- (3, 4) -- (5, 5) -- (2,5) -- (0,4);
\node[below, red!80!black] at (.8, 2.4) {$\cE_0^{E,+}$};
\node[below, red!80!black] at (3.8, 2.4) {$\wt{\cE}_0^{E,+}$};
\node  at (2.5, 3) {$\Q_{1}^{E}$};
\node  at (6.7, 2.5) {$=$};
\draw [fill=blue!60!green!30,opacity=0.7]
(8,0) -- (8, 4) -- (10, 5) -- (10,1) -- (8,0);
\draw [black, thick, opacity=1]
(8,0) -- (8, 4) -- (10, 5) -- (10,1) -- (8,0);
\fill[red!80!black] (9,2.5) circle (3pt);
\node[below, red!80!black] at (8.8, 2.4) {${\cO}^{E,+}$};
\end{scope}
\end{tikzpicture}    
\end{split}
\label{eq:local OP}
\ee
while $\cO^{1_x}$ comes from $\Q_1^{1_x}$ ending on both boundaries
\be
\begin{split}
\begin{tikzpicture}
 \begin{scope}[shift={(0,0)},scale=0.8] 
\draw [cyan, fill=cyan!80!red, opacity =0.5]
(0,0) -- (0,4) -- (2,5) -- (5,5) -- (5,1) -- (3,0)--(0,0);
\draw [black, thick, fill=white,opacity=1]
(0,0) -- (0, 4) -- (2, 5) -- (2,1) -- (0,0);
\draw [cyan, thick, fill=cyan, opacity=0.2]
(0,0) -- (0, 4) -- (2, 5) -- (2,1) -- (0,0);
\draw[line width=1pt] (1,2.5) -- (3,2.5);
\draw[line width=1pt,dashed] (3,2.5) -- (4,2.5);
\fill[red!80!black] (1,2.5) circle (3pt);
\draw [fill=blue!40!red!60,opacity=0.2]
(3,0) -- (3, 4) -- (5, 5) -- (5,1) -- (3,0);
\fill[red!80!black] (1,2.5) circle (3pt);
\fill[red!80!black] (4,2.5) circle (3pt);
\draw [black, thick, opacity=1]
(3,0) -- (3, 4) -- (5, 5) -- (5,1) -- (3,0);
\node at (2,5.4) {$\fB_{\Neu(\Z_2^x),s}$};
\node at (5,5.4) {$\fB_{\Neu(\Z_2^{x}),s'}$};
\draw[dashed] (0,0) -- (3,0);
\draw[dashed] (0,4) -- (3,4);
\draw[dashed] (2,5) -- (5,5);
\draw[dashed] (2,1) -- (5,1);
\draw [cyan, thick, fill=cyan, opacity=0.1]
(0,4) -- (3, 4) -- (5, 5) -- (2,5) -- (0,4);
\node[below, red!80!black] at (.8, 2.4) {$\cE_0^{1_x}$};
\node[below, red!80!black] at (3.8, 2.4) {$\wt{\cE}_0^{1_x}$};
\node  at (2.5, 3) {$\Q_{1}^{1_x}$};
\node  at (6.7, 2.5) {$=$};
\draw [fill=blue!60!green!30,opacity=0.7]
(8,0) -- (8, 4) -- (10, 5) -- (10,1) -- (8,0);
\draw [black, thick, opacity=1]
(8,0) -- (8, 4) -- (10, 5) -- (10,1) -- (8,0);
\fill[red!80!black] (9,2.5) circle (3pt);
\node[below, red!80!black] at (8.8, 2.4) {${\cO}^{1_x}$};
\end{scope}
\end{tikzpicture}    
\end{split}
\label{eq:local OP}
\ee
This phase also presents a genuine line operator, coming from the $\Q_2^{[x]}$ generalized charge ending on both $\Bsym$ and $\Bphys$
\begin{equation}
    \cL^{x}=(\cE_1^{[x]}\,, \Q_2^{[x]}\,, \wt{\cE}_1^{x})\,,
\end{equation}
depicted as
\be
\begin{split}
\begin{tikzpicture}
 \begin{scope}[shift={(0,0)},scale=0.8] 
\draw [cyan, fill=cyan!80!red, opacity =0.5]
(0,0) -- (0,4) -- (2,5) -- (5,5) -- (5,1) -- (3,0)--(0,0);
\draw [black, thick, fill=white,opacity=1]
(0,0) -- (0, 4) -- (2, 5) -- (2,1) -- (0,0);
\draw [cyan, thick, fill=cyan, opacity=0.2]
(0,0) -- (0, 4) -- (2, 5) -- (2,1) -- (0,0);
\draw [fill=blue!40!red!60,opacity=0.2]
(3,0) -- (3, 4) -- (5, 5) -- (5,1) -- (3,0);
\draw [black, thick, opacity=1]
(3,0) -- (3, 4) -- (5, 5) -- (5,1) -- (3,0);
\node at (2,5.4) {$\fB_{\Neu(\Z_2^x),s}$};
\node at (5,5.4) {$\fB_{\Neu(\Z_2^x),s'}$};
\draw[dashed] (0,0) -- (3,0);
\draw[dashed] (0,4) -- (3,4);
\draw[dashed] (2,5) -- (5,5);
\draw[dashed] (2,1) -- (5,1);
\draw [cyan, thick, fill=cyan, opacity=0.1]
(0,4) -- (3, 4) -- (5, 5) -- (2,5) -- (0,4);
\draw [black, thick, fill=black, opacity=0.3]
(0.,2) -- (3, 2) -- (5., 3) -- (2, 3) -- (0.,2);
\node  at (2.5, 2.5) {$\Q_{2}^{[b]}$};
\node[below, red!80!black] at (-0.5, 2.4) {$\cE_1^{[x]}$};
\node[below, red!80!black] at (4, 2.4) {$\wt{\cE}_1^{x}$};
\draw [line width=1pt, red!80!black] (0.0, 2.) -- (2, 3);
\draw [line width=1pt, red!80!black] (3, 2.) -- (5, 3);
\node  at (6.7, 2.5) {$=$};
\draw [fill=blue!60!green!30,opacity=0.7]
(8,0) -- (8, 4) -- (10, 5) -- (10,1) -- (8,0);
\draw [black, thick, opacity=1]
(8,0) -- (8, 4) -- (10, 5) -- (10,1) -- (8,0);
\draw [line width=1pt, red!80!black] (8, 2.) -- (10, 3);
\node[below, red!80!black] at (9, 2.4) {${\cL}^{x}$};
\node[below, black] at (11, 2.4) {$.$};
\end{scope}
\end{tikzpicture}    
\end{split}
\ee
$\cL^x$ is charged under $D_1^{\wh{x}}$, obeys $\Z_2$ fusion rules and has F-symbols given by $s-s'\in H^{3}(\Z_2\,, U(1))$. 
The idempotents can be computed by starting with the $\Z_4^r$ SSB phase I for $2\Vec_{D_8}$ of section \ref{sec:D8_Z4SSB1} (which has the same $\Bphys$ as our current sandwich) and gauging $\Z_2^{x}$ on the symmetry boundary. They are:
\be
\ba
    v_{0}&=\frac{1}{4}\lb 1+\cO^{1_x}+2\cO^{E,+}\rb \,,\\
     v_{1}&=\frac{1}{4}\lb 1+\cO^{1_x}-2\cO^{E,+}\rb \,,\\
    v_{2}&=\frac{1}{2}\lb 1-\cO^{1_x}\rb \,.\\
\ea
\ee
$v_0$ and $v_1$ were already $\Z_2^x$ symmetric so, after gauging it, they carry a $\Z_{2}$ Dijkgraaf-Witten theory with a 3-cocycle twist $s-s'\in H^{3}(\Z_2,U(1))$, i.e. toric code or double semion, whereas $v_2$ comes from the orbit of two vacua which were not $\Z_2^x$ invariant so after gauging its underline TFT is therefore trivial. In total, the underlying 3d TFT for this $\Rep(\Z_4^{(1)}\rtimes\Z_2^{(0)})$ symmetric phase is
\be
  \fT=\lb\text{DW}(\Z_2)_{s-s'}\rb_0\oplus \lb\text{DW}(\Z_2)_{s-s'}\rb_1 \oplus {\rm Triv}_2
\ee
with line operators in the DW theories given by $D_1^{\wh{x}}$, which is bosonic, and $\cL^x$ which is a boson if $s-s'$ is trivial or a semion otherwise. 
Since $\cO^{E,+}$ carries charge $-1$ under $\Z_2^{(0),r^2}$, the invertible 0-form symmetry generator $D_2^{r^2}$ is realized as 
\be
D_2^{r^2}=1_{01}\oplus 1_{10}\oplus 1_{22}\,.
\ee
$\cO^{E,+}$ carries charge 0 under the  non-invertible  $D_2^{[r]}$ therefore its action on the vacua is 
\begin{equation}
    D_2^{[r]}:(v_0\,,\; v_1\,,\; v_2) \longmapsto (v_2\,,\; v_2\,,\; 2v_0+2v_1)\,. 
\end{equation}
and this surface operator is realized concretely as 
\begin{equation}
    D_2^{[r]}= B_{02}\oplus B_{12}\oplus \overline{B}_{20}\oplus \overline{B}_{21}\,,
\end{equation}
where $B_{i2}$ for $i=0,1$ is the interface between $\lb{\rm DW(\Z_2)}_{s-s'}\rb_i$ and ${\rm Triv}_2$ while $\overline{B}_{2i}$ is the converse interface.
One can verify \eqref{fus8} by using the fact that the product of two $B$ interfaces is a condensation.
\begin{tcolorbox}[
colback=white,
coltitle= black,
colbacktitle=ourcolorforheader,
colframe=black,
title= $\TwoRep(\Z_4^{(1)}\rtimes \Z_2^{(0)})$: $\TwoRep(\Z_4^{(1)}\rtimes \Z_2^{(0)})$ SSB Phase (``Superstar"), 
sharp corners]
\be
\begin{split}
\begin{tikzpicture}
\begin{scope}[shift={(0,0)},scale=1] 
 \pgfmathsetmacro{\del}{-0.5}
\node at  (0.5,0)  {$\lb\text{DW}(\Z_2)_{s-s'}\rb_0\oplus \lb\text{DW}(\Z_2)_{s-s'}\rb_1 \oplus {\rm Triv}_2$} ;
\node at  (0.8,-0.6)  {\footnotesize$(\Z_2^{(1)}\text{-}{\rm SSB})\hspace{1.4cm} (\Z_2^{(1)}\text{-}{\rm SSB})\hspace{1.2cm} (\Z_2^{(1)}\text{-}{\rm Triv})$};
\end{scope}
\end{tikzpicture}    
\end{split}
\ee
The 0-form symmetry is realized on the vacua as
\be
\ba
    D_2^{r^2}&=1_{01}\oplus 1_{10}\oplus 1_{22}\,,\\
    D_2^{[r]}&= B_{02}\oplus B_{12}\oplus \overline{B}_{20}\oplus \overline{B}_{21}\,.
\ea
\ee
\end{tcolorbox}

\subsubsection{$\Z_2^{(0)}$ SSB Phase with mixed SPT} 
Taking the physical boundary to be
\begin{equation}
    \Bphys=\fB_{\Neu(\Z_2^{xr}),s'}\,,
\end{equation}
the SymTFT lines that end are:
\be
    \{\Q_1^{1}\,,\quad \Q_1^E\}\,,
\ee
Compactifying the SymTFT, one thus obtains a two dimensional space of untwisted local operators spanned by
\begin{equation}
    \left\{1\,,\; \cO^{E,+}\right\}\,,
\end{equation}
where $\cO^{E,+}$ denotes the local operator obtained by compactifying $\Q_1^{E}$ with end ${\cE}_0^{E,+}$ on $\Bsym$ and $\cE^{+i}$ (where $i$ is the 4$^{\rm th}$ root of unity) on $\Bphys$, recall eq. \eqref{eq:D8_Z2x_Q1E}. 

 There are non-trivial twisted sector local operators: for example, the compactification of  $\Q_1^{1_{xr}}$ produces a local operator attached to the $D_1^{\wh{x}}$ line.
\begin{equation}
    (D_{1}^{\wh{x}}\,, \cO^{1_{xr}})=\left((D_{1}^{\wh{x}}\,,{\cE}_0^{1_{xr}})\,, \Q_1^{1_{xr}}\,, \wt\cE_0^{1_{xr}}\right)\,.
\end{equation}
This configuration is depicted as
\be
\begin{split}
\begin{tikzpicture}
 \begin{scope}[shift={(0,0)},scale=0.8] 
\draw [cyan, fill=cyan!80!red, opacity =0.5]
(0,0) -- (0,4) -- (2,5) -- (5,5) -- (5,1) -- (3,0)--(0,0);
\draw [black, thick, fill=white,opacity=1]
(0,0) -- (0, 4) -- (2, 5) -- (2,1) -- (0,0);
\draw [cyan, thick, fill=cyan, opacity=0.2]
(0,0) -- (0, 4) -- (2, 5) -- (2,1) -- (0,0);
\draw[line width=1pt] (1,2.5) -- (3,2.5);
\draw[line width=1pt,dashed] (3,2.5) -- (4,2.5);
\fill[red!80!black] (1,2.5) circle (3pt);
\draw [fill=blue!40!red!60,opacity=0.2]
(3,0) -- (3, 4) -- (5, 5) -- (5,1) -- (3,0);
\fill[red!80!black] (1,2.5) circle (3pt);
\fill[red!80!black] (4,2.5) circle (3pt);
\draw [black, thick, opacity=1]
(3,0) -- (3, 4) -- (5, 5) -- (5,1) -- (3,0);
\node at (2,5.4) {$\fB_{\Neu(\Z_2^x),s}$};
\node at (5,5.4) {$\fB_{\Neu(\Z_2^{xr}),s'}$};
\draw[dashed] (0,0) -- (3,0);
\draw[dashed] (0,4) -- (3,4);
\draw[dashed] (2,5) -- (5,5);
\draw[dashed] (2,1) -- (5,1);
\draw[line width=1pt] (1,2.5) -- (1,4.5);
\draw [cyan, thick, fill=cyan, opacity=0.1]
(0,4) -- (3, 4) -- (5, 5) -- (2,5) -- (0,4);
\node[below, red!80!black] at (.8, 2.4) {$\cE_0^{1_{xr}}$};
\node[below, red!80!black] at (3.8, 2.4) {$\wt{\cE}_0^{1_{xr}}$};
\node  at (2.5, 3) {$\Q_{1}^{1_{xt}}$};
\node  at (0.5, 3.3) {$D_{1}^{\wh{x}}$};
\node  at (6.7, 2.5) {$=$};
\draw [fill=blue!60!green!30,opacity=0.7]
(8,0) -- (8, 4) -- (10, 5) -- (10,1) -- (8,0);
\draw [black, thick, opacity=1]
(8,0) -- (8, 4) -- (10, 5) -- (10,1) -- (8,0);
\draw[line width=1pt] (9,2.5) -- (9,4.5);
\fill[red!80!black] (9,2.5) circle (3pt);
\node[below, red!80!black] at (8.8, 2.4) {${\cO}^{1_{xr}}$};
\node  at (8.5, 3.3) {$D_{1}^{\wh{x}}$};
\node[below, black] at (10.8, 3) {$,$};
\end{scope}
\end{tikzpicture}    
\end{split}
\ee
where $\cO^{1_{xr}}$ carries charge $-2$ under $D_2^{[r]}$.

The lines on $\Bphys$ are $\cE^{r^2}$, eq. \eqref{eq:Lr2}, and $\cE^{x[r]}$, which gives rise to a $D_2^{[r]}$ twisted sector line after compactification:
\begin{equation}
    (D_2^{[r]}\,,\cL^{x[r]})=\left((D_2^{[r]}\,, \cE_1^{x[r]})\,, \Q_2^{[xr]}\,, \wt\cE_1^{x[r]}\right)\,,
\end{equation}
depicted as
\be
\begin{split}
\begin{tikzpicture}
 \begin{scope}[shift={(0,0)},scale=0.8] 
\draw [cyan, fill=cyan!80!red, opacity =0.5]
(0,0) -- (0,4) -- (2,5) -- (5,5) -- (5,1) -- (3,0)--(0,0);
\draw [black, thick, fill=white,opacity=1]
(0,0) -- (0, 4) -- (2, 5) -- (2,1) -- (0,0);
\draw [cyan, thick, fill=cyan, opacity=0.2]
(0,0) -- (0, 4) -- (2, 5) -- (2,1) -- (0,0);
\draw [fill=blue!40!red!60,opacity=0.2]
(3,0) -- (3, 4) -- (5, 5) -- (5,1) -- (3,0);
\draw [black, thick, opacity=1]
(3,0) -- (3, 4) -- (5, 5) -- (5,1) -- (3,0);
\node at (2,5.4) {$\fB_{\Neu(\Z_2^x),s}$};
\node at (5,5.4) {$\fB_{\Neu(\Z_2^{xr}),s'}$};
\draw[dashed] (0,0) -- (3,0);
\draw[dashed] (0,4) -- (3,4);
\draw[dashed] (2,5) -- (5,5);
\draw[dashed] (2,1) -- (5,1);
\draw [cyan, thick, fill=cyan, opacity=0.1]
(0,4) -- (3, 4) -- (5, 5) -- (2,5) -- (0,4);
\draw [black, thick, fill=black, opacity=0.3]
(0.,2) -- (3, 2) -- (5., 3) -- (2, 3) -- (0.,2);
\draw [black, thick, fill=black, opacity=0.3]
(0.,2) -- (0., 4) -- (2, 5) -- (2,3) -- (0,2);
\node  at (2.5, 2.5) {$\Q_{2}^{r^2}$};
\node[below, red!80!black] at (-0.5, 2.4) {$\cE_1^{x[r]}$};
\node[below, red!80!black] at (4, 2.4) {$\wt{\cE}_1^{x[r]}$};
\draw [line width=1pt, red!80!black] (0.0, 2.) -- (2, 3);
\draw [line width=1pt, red!80!black] (3, 2.) -- (5, 3);
\node  at (6.7, 2.5) {$=$};
\draw [fill=blue!60!green!30,opacity=0.7]
(8,0) -- (8, 4) -- (10, 5) -- (10,1) -- (8,0);
\draw [black, thick, opacity=1]
(8,0) -- (8, 4) -- (10, 5) -- (10,1) -- (8,0);
\draw [black, thick, fill=black, opacity=0.3]
(8,2) -- (8, 4) -- (10, 5) -- (10,3) -- (8,2);
\draw [line width=1pt, red!80!black] (8, 2.) -- (10, 3);
\node[below, black] at (9, 3.9) {$D_2^{[r]}$};
\node[below, black] at (1., 3.9) {$D_2^{[r]}$};
\node[below, red!80!black] at (9, 2.4) {${\cL}^{x[r]}$};
\node[below, black] at (11, 2.4) {$,$};
\end{scope}
\end{tikzpicture}    
\end{split}
\ee
$\cE_1^{x[r]}$ is charged under $D_1^{\wh x}$, therefore the 1-form symmetry is realized non-trivially, like for the phase in sec. \ref{sec:D8_1-form_non-triv}. No SymTFT surfaces end on both boundaries, so there are no genuine line operators.

The idempotents can be computed by starting with the $\Z_4^r$ SSB phase II for $2\Vec_{D_8}$ of section \ref{sec:D8_Z4SSB2} (which has the same $\Bphys$ as our current sandwich) and gauging $\Z_2^{x}$ on the symmetry boundary. They are:
\be
\ba
    v_{0}&=\frac{1}{2}\lb 1+\cO^{E,+}\rb \,,\\
    v_{1}&=\frac{1}{2}\lb 1-\cO^{E,+}\rb \,.\\
\ea
\ee
Both vacua come from the orbit of two vacua which were not $\Z_2^x$ invariant so the underlying 3d TFT in each vacuum is trivial
\be
\fT = \Triv_0 \oplus \Triv_1\,.
\ee
$D_2^{r^2}$ exchanges the two vacua, since $\cO^{E,+}$ carries charge $-1$ under it:
\be
    D_2^{r^2}=1_{01}\oplus 1_{10}\,.
\ee
This is therefore a $\Z_2^{(0)}$ SSB phase in which the non-invertible 0-form symmetry sends ecah vacua to the sum of both:
\be
D_2^{[r]} = 1_{00}\oplus 1_{01}\oplus 1_{10}\oplus 1_{11}\,.
\ee
\begin{tcolorbox}[
colback=white,
coltitle= black,
colbacktitle=ourcolorforheader,
colframe=black,
title= $\TwoRep (\Z_4^{(1)}\rtimes\Z_2^{(0)})$: $\Z_2^{(0)}$ SSB with non-trivial $\Z_2^{(1)}$,
sharp corners]
\be
\Triv_0 \oplus \Triv_1:\qquad 
\left\{\ba
\Z_2^{(0)}: & \quad D_2^{r^2}=1_{01}\oplus 1_{10} \cr 
D_2^{[r]} :& \quad 1_{00}\oplus 1_{01}\oplus 1_{10}\oplus 1_{11}\cr 
D_2^C: & \quad 2 \times (1_{00} \oplus  1_{11})\cr 
\Z_2^{(1)}: & \quad \text{non-trivial}
 \,.
\ea
\right.
\ee
\end{tcolorbox}

\subsubsection{$\Z_2^{(1)}$ SSB Phase} 
Let us finally take the physical boundary to be 
\begin{equation}
    \Bphys=\fB_{\Neu(D_8),\tau}\,.
\end{equation}
Since no SymTFT lines end on both boundaries, there is only the single identity local operator in the IR TQFT describing this gapped phase. 

The SymTFT surface $\Q_2^{[x]}$ ending on both boundaries produces a genuine line operator
\begin{equation}
    \cL^x=\left(\cE_1^{x}\,, \Q_2^{[x]}\,, \wt\cE_1^{[x]}\right)\,,
\end{equation}
depicted as
\be
\begin{split}
\begin{tikzpicture}
 \begin{scope}[shift={(0,0)},scale=0.8] 
\draw [cyan, fill=cyan!80!red, opacity =0.5]
(0,0) -- (0,4) -- (2,5) -- (5,5) -- (5,1) -- (3,0)--(0,0);
\draw [black, thick, fill=white,opacity=1]
(0,0) -- (0, 4) -- (2, 5) -- (2,1) -- (0,0);
\draw [cyan, thick, fill=cyan, opacity=0.2]
(0,0) -- (0, 4) -- (2, 5) -- (2,1) -- (0,0);
\draw [fill=blue!40!red!60,opacity=0.2]
(3,0) -- (3, 4) -- (5, 5) -- (5,1) -- (3,0);
\draw [black, thick, opacity=1]
(3,0) -- (3, 4) -- (5, 5) -- (5,1) -- (3,0);
\node at (2,5.4) {$\fB_{\Neu(\Z_2^x),s}$};
\node at (5,5.4) {$\fB_{\Neu(D_8),\tau}$};
\draw[dashed] (0,0) -- (3,0);
\draw[dashed] (0,4) -- (3,4);
\draw[dashed] (2,5) -- (5,5);
\draw[dashed] (2,1) -- (5,1);
\draw [cyan, thick, fill=cyan, opacity=0.1]
(0,4) -- (3, 4) -- (5, 5) -- (2,5) -- (0,4);
\draw [black, thick, fill=black, opacity=0.3]
(0.,2) -- (3, 2) -- (5., 3) -- (2, 3) -- (0.,2);
\node  at (2.5, 2.5) {$\Q_{2}^{[x]}$};
\node[below, red!80!black] at (-0.5, 2.4) {$\cE_1^{x}$};
\node[below, red!80!black] at (4, 2.4) {$\wt{\cE}_1^{[x]}$};
\draw [line width=1pt, red!80!black] (0.0, 2.) -- (2, 3);
\draw [line width=1pt, red!80!black] (3, 2.) -- (5, 3);
\node  at (6.7, 2.5) {$=$};
\draw [fill=blue!60!green!30,opacity=0.7]
(8,0) -- (8, 4) -- (10, 5) -- (10,1) -- (8,0);
\draw [black, thick, opacity=1]
(8,0) -- (8, 4) -- (10, 5) -- (10,1) -- (8,0);
\draw [line width=1pt, red!80!black] (8, 2.) -- (10, 3);
\node[below, red!80!black] at (9, 2.4) {${\cL}^{x}$};
\node[below, black] at (11, 2.4) {$,$};
\end{scope}
\end{tikzpicture}    
\end{split}
\ee
In particular, this line is charged under the $\Z_2^{(1)}$ symmetry generated by $D_1^{\wh{x}}$. Therefore this is the $\Z_2^{(1)}$ symmetry broken phase, realizing the $\Z_2$ Dijkgraaf-Witten theory with 3-cocycle determined by the doscrete torsion on both boundaries $s,p,t$. We leave the exact relation to this data to the future.

This phase has a single vacuum $v_0=1$, whose underlying TFT is DW$(\Z_2)$, on which $D_2^{r^2}$ is the identity, while $D_2^{[r]}$ symmetry is realized as
\begin{equation}
    D_2^{[r]}=\frac{D_2^{\id}}{D_1^{1}\oplus D_1^{\wh{x}}}\,.
\end{equation}

\begin{tcolorbox}[
colback=white,
coltitle= black,
colbacktitle=ourcolorforheader,
colframe=black,
title= $\TwoRep(\Z_4^{(1)}\rtimes \Z_2^{(0)})$: $\Z_2^{(1)}$ SSB Phase, 
sharp corners]
\be
\begin{split}
\begin{tikzpicture}
\begin{scope}[shift={(0,0)},scale=1] 
 \pgfmathsetmacro{\del}{-0.5}
\node at  (0,0)  {$ {\rm DW(\Z_2)}$} ;
\node at  (0,-0.6)  {\footnotesize$(\Z_2^{(1)}\text{-}{\rm SSB})$} ;
\end{scope}
\end{tikzpicture}    
\end{split}
\ee
The non-invertible 0-form symmetry is realized as
\begin{equation}
    D_2^{[r]}=\frac{D_2^{\id}}{D_1^{1}\oplus D_1^{\wh{x}}}\,. 
\end{equation}
\end{tcolorbox}

\subsection{$\TwoRep (D_8)$ Gapped Phases}

We now describe the minimal gapped phases obtained by choosing the symmetry boundary
\begin{equation}
    \Bsym= \fB_{\Neu(D_8)}\,,
\end{equation}
described in section \ref{sec:D8_NeuD8}, with trivial discrete torsion $\tau\in H^3(D_8,U(1))$. The symmetry is $\cS=\TwoRep (D_8)$, i.e. $\Rep(D_8)$ 1-form symmetry generated by lines $D_1^R$, for $R\in\Rep(D_8)$:
\be
    D_1^1\,,\quad D_1^{1_x}\,,\quad D_1^{1_r}\,,\quad D_1^{1_{xr}}\,,\quad D_1^E\,.
\ee
Their fusions follow from the tensor products of irreps, consistently with the $D_8$ character table \ref{tab:D8_chars} and are (for $1_k=1_x\,,\;1_r\,,\;1_{xr}$):
\be
\ba
    D_1^{1_x}\otimes D_1^{1_r}&=D_1^{1_{xr}}\,,\quad 
    D_1^{1_k}\otimes D_1^{1_k}=D_1^{1}\,,\\
    D_1^E\otimes D_1^E&=D_1^{1}\oplus D_1^{1_x}\oplus D_1^{1_r}\oplus D_1^{1_{xr}}\,,
\ea
\ee
from which we see that the 1-dimensional irreps comprise of an invertible $\Z_2\times\Z_2$ 1-form symmetry, whereas $D_1^E$ is non-invertible. (1+1)d gapped and gapless phases with $\Rep(D_8)$ categorical symmetry were studied from the SymTFT in \cite{Bhardwaj:2024qrf} and from the lattice in \cite{Warman:2024lir}.

We recall that on the $2\Rep(D_8)$ symmetry boundary in (2+1)d all local operators are non-genuine: this implies that all $\TwoRep(D_8)$-symmetric gapped phases will have a unique vacuum. 

$\Bsym$ carries genuine lines $\cE_1^{[g]}$ coming from the end of the bulk SymTFT surface $\Q_2^{[g]}$, with charge
\be
    \text{Link}(D_1^R,\cE_1^{[g]})=\chi_R([g])\,,
\ee
under $D_1^R$ for $R\in\Rep(D_8)$. The non-identity SymTFT surfaces that also end on $\Bphys$ will in general give rise to line order parameters charged under certain sub-symmetries of $\Rep(D_8)$, revealing 1-form symmetry spontaneous symmetry breaking, as will see below.

\subsubsection{$\TwoRep(D_8)$ Trivial Phase}
Consider the physical boundary to be
\begin{equation}
    \Bphys=\fB_{\Dir}\,.
\end{equation}
The idempotents can be computed by starting with the $D_8$ SSB phase for $2\Vec_{D_8}$ of section \ref{sec:D8_D8SSB} (which has the same $\Bphys$ as our current sandwich) and gauging $D_8$ 0-form symmetry on the symmetry boundary. Since the $D_8$ symmetry is completely broken, after gauging we obtain a single vacuum whose underlying TFT is trivial.

The only genuine line in this phase is the identity, so any representation $R$ just acts as ${\dim}(R)$ copies of the identity line.
This phase has a non-invertible 1-form symmetry $\Rep(D_8)$ which is preserved. The associated line operators (charges) are confined. 
\begin{tcolorbox}[
colback=white,
coltitle= black,
colbacktitle=ourcolorforheader,
colframe=black,
title= $\TwoRep(D_8)$: Trivial Phase with trivial $\Rep (D_8)$ 1-form symmetry (Confining Phase),
sharp corners]
\be
\begin{split}
\begin{tikzpicture}
\begin{scope}[shift={(0,0)},scale=1] 
 \pgfmathsetmacro{\del}{-0.5}
\node at  (0,0)  {$ {\rm Triv}_0$} ;
\end{scope}
\end{tikzpicture}    
\end{split}
\ee
The non-invertible 1-form symmetry is realized as
\begin{equation}
    D_1^{R}=\dim(R) D_1^{1}\,. 
\end{equation}
\end{tcolorbox}

\subsubsection{$\TwoRep(D_8)$: $\TwoRep(D_8)/\TwoRep(\Z_2\times\Z_2)$ SSB Phase}
Let us take the physical boundary to be
\begin{equation}
    \Bphys=\fB_{\Neu(\Z_2^{r^2}),s}\,.
\end{equation}
The idempotents can be computed by starting with the $\Z_2^r\times\Z_2^x$ SSB phase for $2\Vec_{D_8}$ of section \ref{sec:D8_Z2rZ2x_SSB} (which has the same $\Bphys$ as our current sandwich) and gauging $D_8$ on the symmetry boundary. Since the $\Z_2^{r^2}$ symmetry was preserved in each vacuum, after gauging $D_8$ we obtain a unique vacuum whose underlying TFT carries $\Z_{2}$ Dijkgraaf-Witten theory with a 3-cocycle twist $s\in H^{3}(\Z_2,U(1))$, denoted as DW$(\Z_2)_s$, i.e. toric code or double semion.

The SymTFT bulk surface $\Q_2^{[r^2]}$ ending on both boundaries produces, after compactification, a genuine line $\cL^{r^2}$ charged under $D_1^E$ which is realized in this phase as:
\be
    D_1^E\simeq 2D_1^{\wh{r}^2}\,,
\ee
where $D_1^{\wh{r}^2}$ is the $\Z_2^{(1)}$ electric line in DW$(\Z_2)_s$ linking non-trivially with $\cL^x$. The remaining $\Z_2\times\Z_2$ sub-symmetry of $\Rep(D_8)$, generated by $D_1^{1_x},\;D_1^{1_r}$ is preserved in this phase.

\begin{tcolorbox}[
colback=white,
coltitle= black,
colbacktitle=ourcolorforheader,
colframe=black,
title= $\TwoRep(D_8)$: $\TwoRep(D_8)/\TwoRep(\Z_2\times\Z_2)$ SSB Phase, 
sharp corners]
\be
\begin{split}
\begin{tikzpicture}
\begin{scope}[shift={(0,0)},scale=1] 
 \pgfmathsetmacro{\del}{-0.5}
\node at  (0,0)  {$ {\rm DW(\Z_2)}_{s}$} ;
\end{scope}
\end{tikzpicture}    
\end{split}
\ee
The $\Rep(D_8)$ 1-form symmetry is realized as
\be
\ba
    D_1^{1_k}&\simeq D_1^{1}\,,\quad (1_k=1_x\,,\;1_r\,,\;1_{xr})\,,\\
    D_1^{E}&\simeq 2D_1^{\wh{r}^2}
\ea
\ee
i.e. $\TwoRep(D_8)$ acts on this gapped phase via a projection $\TwoRep(D_8)\to \TwoRep(\Z_2)$ whose kernel contains all the 1-dimensional $D_8$ irreps. 
\end{tcolorbox}

\subsubsection{$\TwoRep(\Z_2\times\Z_2)$ SSB Phase I}
Let us take the physical boundary to be
\begin{equation}
    \Bphys=\fB_{\Neu(\Z_4^{r}),p}\,.
\end{equation}
The idempotents can be computed by starting with the $\Z_2^x$ SSB phase I for $2\Vec_{D_8}$ of section \ref{sec:D8_Z2x_SSB1}  (which has the same $\Bphys$ as our current sandwich) and gauging $D_8$ on the symmetry boundary.  Since the $\Z_4^{r}$ symmetry was preserved in each vacuum, after gauging $D_8$ we obtain a unique vacuum whose underlying TFT carries $\Z_{4}$ Dijkgraaf-Witten theory with a 3-cocycle twist $p\in H^{3}(\Z_4,U(1))$. We denote the 3d TFT as DW$(\Z_4)_p$ and the line generating its electric $\Z_4^{(1)}$ symmetry as $D_1^{\wh{r}}$.

The SymTFT bulk surfaces $\Q_2^{[r^2]}$ and $\Q_2^{[r]}$ ending on both boundaries produce, after compactification, genuine lines $\cL^{r^2}\,,\;\cL^r\,,\;\cL^{r^3}$. From the SymTFT bulk linking, we know that $\cL^{r}$ and $\cL^{r^3}$ carry charge $-1$ under $D_1^{1_x}$ and $D_1^{1_{xr}}$ and charge $+1$ under $D_1^{1_r}$. The charge of  $\cL^r$ and $\cL^{r^3}$ under $D_1^E$ is $0$ while that of $\cL^{r^2}$ is $-2$. The $2\Rep(D_8)$ symmetry generators are therefore realized in this phase as follows:
\be
\ba
    D_1^{1_r}&\simeq D_1^{1}\,,\quad \\
    D_1^{1_x}&\simeq D_1^{1_{xr}}\simeq D_1^{\wh{r}^2} \,,\\
    D_1^E&\simeq D_1^{\wh{r}}\oplus D_1^{\wh{r}^3}\,,
\ea
\ee
so it is a $2\Rep(\Z_2\times\Z_2)$ SSB phase for the 1-form symmetry generated by $D_1^{1_x}$ and $D_1^{1_{xr}}$.
\begin{tcolorbox}[
colback=white,
coltitle= black,
colbacktitle=ourcolorforheader,
colframe=black,
title= $\TwoRep(D_8)$: $\TwoRep(\Z_2\times\Z_2)$ SSB Phase I, 
sharp corners]
\be
\begin{split}
\begin{tikzpicture}
\begin{scope}[shift={(0,0)},scale=1] 
 \pgfmathsetmacro{\del}{-0.5}
\node at  (0,0)  {$ {\rm DW(\Z_4)}_{p}$} ;
\end{scope}
\end{tikzpicture}    
\end{split}
\ee
The $\Rep(D_8)$ 1-form symmetry is realized as
\be
\ba
    D_1^{1_r}\simeq D_1^{1}\,, \quad D_1^{1_x}\simeq D_1^{1_{xr}}\simeq D_1^{\wh{r}^2} \,,\quad 
    D_1^E\simeq D_1^{\wh{r}}\oplus D_1^{\wh{r}^3}\,,
\ea
\ee
i.e. $\TwoRep(D_8)$ acts on this gapped phase via a projection $\TwoRep(D_8)\to \TwoRep(\Z_4)$.
\end{tcolorbox}

\subsubsection{$\TwoRep(\Z_2\times\Z_2)$ SSB Phase II}
Taking the physical boundary to be
\begin{equation}
    \Bphys=\fB_{\Neu(\Z_2^{xr}\times\Z_2^{xr^3}),t}\,,
\end{equation}
the idempotents can be computed by starting with the $\Z_2^x$ SSB phase II for $2\Vec_{D_8}$ of section \ref{sec:D8_Z2x_SSB2}  (which has the same $\Bphys$ as our current sandwich) and gauging $D_8$ on the symmetry boundary. The $\Z_2^{xr}\times\Z_2^{xr^3}$ symmetry was preserved in each vacuum, so after gauging $D_8$ we obtain a unique vacuum whose underlying TFT carries $\Z_{2}\times\Z_2$ Dijkgraaf-Witten theory with a 3-cocycle twist $t\in H^{3}(\Z_2\times\Z_2,U(1))$. We denote the 3d TFT as DW$(\Z_2\times\Z_2)_t$ and the lines generating its electric $\Z_2\times\Z_2$ 1-form symmetry as $D_1^{\wh{xr}}$ and $D_1^{\wh{r}^2}$.

The SymTFT bulk surfaces $\Q_2^{[r^2]}$ and $\Q_2^{[xr]}$ ending on both boundaries produce, after compactification, genuine lines $\cL^{r^2}\,,\;\cL^{xr}\,,\;\cL^{xr^3}$. From the SymTFT bulk linking, we know that $\cL^{xr}$ and $\cL^{xr^3}$ carry charge $-1$ under $D_1^{1_x}$ and $D_1^{1_{r}}$, charge $+1$ under $D_1^{1_{xr}}$ and $0$ under $D_1^E$. $\cL^{r^2}$ is charged only under $D_1^E$, with linking equal to $-2$.
The $2\Rep(D_8)$ symmetry generators are therefore realized in this phase as follows: 
\be
\ba
    D_1^{1_{xr}}&\simeq D_1^{1}\,,\quad \\
    D_1^{1_x}&\simeq D_1^{1_r}\simeq D_1^{\wh{xr}} \,,\\
    D_1^E&\simeq D_1^{\wh{r}^2}\oplus D_1^{\wh{xr}^3}\,,
\ea
\ee
so it is a $2\Rep(\Z_2\times\Z_2)$ SSB phase for the 1-form symmetry generated by $D_1^{1_x}$ and $D_1^{1_{r}}$.
\begin{tcolorbox}[
colback=white,
coltitle= black,
colbacktitle=ourcolorforheader,
colframe=black,
title= $\TwoRep(D_8)$: $\TwoRep(\Z_2\times\Z_2)$ SSB Phase II, 
sharp corners]
\be
\begin{split}
\begin{tikzpicture}
\begin{scope}[shift={(0,0)},scale=1] 
 \pgfmathsetmacro{\del}{-0.5}
\node at  (0,0)  {$ {\rm DW(\Z_2\times\Z_2)}_{t}$} ;
\end{scope}
\end{tikzpicture}    
\end{split}
\ee
The $\Rep(D_8)$ 1-form symmetry is realized as
\be
\ba
    D_1^{1_{xr}}\simeq D_1^{1}\,, \quad D_1^{1_x}\simeq D_1^{1_{r}}\simeq D_1^{\wh{xr}} \,,\quad 
    D_1^E\simeq D_1^{\wh{r}^2}\oplus D_1^{\wh{xr}^3}\,,
\ea
\ee
i.e. $\TwoRep(D_8)$ acts on this gapped phase via a projection $\TwoRep(D_8)\to \TwoRep(\Z_2\times\Z_2)$.\\
\end{tcolorbox}

\subsubsection{$\TwoRep(\Z_2\times\Z_2)$ SSB Phase III}
Taking the physical boundary to be
\begin{equation}
    \Bphys=\fB_{\Neu(\Z_2^{x}\times\Z_2^{xr^2}),t}\,,
\end{equation}
the idempotents can be computed by starting with the $\Z_2^{xr}$ SSB phase for $2\Vec_{D_8}$ of section \ref{sec:D8_Z2xr_SSB}  (which has the same $\Bphys$ as our current sandwich) and gauging $D_8$ on the symmetry boundary. The $\Z_2^{x}\times\Z_2^{r^2}$ symmetry was preserved in each vacuum, so after gauging $D_8$ we obtain a unique vacuum whose underlying TFT carries $\Z_{2}\times\Z_2$ Dijkgraaf-Witten theory with a 3-cocycle twist $t\in H^{3}(\Z_2\times\Z_2,U(1))$. We denote the 3d TFT as DW$(\Z_2\times\Z_2)_t$ and the lines generating its electric $\Z_2\times\Z_2$ 1-form symmetry as $D_1^{\wh{x}}$ and $D_1^{\wh{r}^2}$.

The SymTFT bulk surfaces $\Q_2^{[r^2]}$ and $\Q_2^{[x]}$ ending on both boundaries produce, after compactification, genuine lines $\cL^{r^2}\,,\;\cL^{x}\,,\;\cL^{xr^2}$. From the SymTFT bulk linking, we know that $\cL^{x}$ and $\cL^{xr^2}$ carry charge $-1$ under $D_1^{1_r}$ and $D_1^{1_{xr}}$, charge $+1$ under $D_1^{1_{x}}$ and $0$ under $D_1^E$. $\cL^{r^2}$ is charged only under $D_1^E$, with linking equal to $-2$.
The $2\Rep(D_8)$ symmetry generators are therefore realized in this phase as follows:
\be
\ba
    D_1^{1_{x}}&\simeq D_1^{1}\,,\quad \\
    D_1^{1_r}&\simeq D_1^{1_{xr}}\simeq D_1^{\wh{x}} \,,\\
    D_1^E&\simeq D_1^{\wh{r}^2}\oplus D_1^{\wh{xr}^2}\,,
\ea
\ee
so it is a $2\Rep(\Z_2\times\Z_2)$ SSB phase for the 1-form symmetry generated by $D_1^{1_{r}}$ and $D_1^{1_{xr}}$.
\begin{tcolorbox}[
colback=white,
coltitle= black,
colbacktitle=ourcolorforheader,
colframe=black,
title= $\TwoRep(D_8)$: $\TwoRep(\Z_2\times\Z_2)$ SSB Phase III, 
sharp corners]
\be
\begin{split}
\begin{tikzpicture}
\begin{scope}[shift={(0,0)},scale=1] 
 \pgfmathsetmacro{\del}{-0.5}
\node at  (0,0)  {$ {\rm DW(\Z_2\times\Z_2)}_{t}$} ;
\end{scope}
\end{tikzpicture}    
\end{split}
\ee
The $\Rep(D_8)$ 1-form symmetry is realized as
\be
\ba
    D_1^{1_{x}}\simeq D_1^{1}\,, \quad 
    D_1^{1_{r}}\simeq D_1^{1_{xr}}\simeq D_1^{\wh{x}} \,,\quad 
    D_1^E\simeq D_1^{\wh{r}^2}\oplus D_1^{\wh{xr}^2}\,,
\ea
\ee
i.e. $\TwoRep(D_8)$ acts on this gapped phase via a projection $\TwoRep(D_8)\to \TwoRep(\Z_2\times\Z_2)$.
\end{tcolorbox}

\subsubsection{$\TwoRep(\Z_2)$ SSB Phase I}
Let us now take the physical boundary to be
\begin{equation}
    \Bphys=\fB_{\Neu(\Z_2^{x}),s}\,.
\end{equation}
The idempotents can be computed by starting with the $\Z_4^r$ SSB phase I for $2\Vec_{D_8}$ of section \ref{sec:D8_Z4SSB1} (which has the same $\Bphys$ as our current sandwich) and gauging $D_8$ on the symmetry boundary. Since the $\Z_2^{x}$ symmetry was preserved in each vacuum, after gauging $D_8$ we obtain a unique vacuum whose underlying TFT carries $\Z_{2}$ Dijkgraaf-Witten theory with a 3-cocycle twist $s\in H^{3}(\Z_2,U(1))$, i.e. toric code or double semion.  We denote the 3d TFT as DW$(\Z_2)_s$, and the line generating its electric $\Z_2^{(1)}$ symmetry as $D_1^{\wh{x}}$.

The SymTFT bulk surface $\Q_2^{[x]}$ ending on both boundaries produces, after compactification, a genuine line $\cL^{x}$ that carries charge $-1$ under $D_1^{1_r}$ and $D_1^{1_{xr}}$, charge $+1$ under $D_1^{1_x}$ and $0$ under $D_1^E$. The $2\Rep(D_8)$ symmetry generators are therefore realized in this phase as follows:
\be
\ba
    D_1^{1_{x}}&\simeq D_1^{1}\,,\quad \\
    D_1^{1_r}&\simeq D_1^{1_{xr}}\simeq D_1^{\wh{x}} \,,\\
    D_1^E&\simeq D_1^{1}\oplus D_1^{\wh{x}}\,,
\ea
\ee
so it is a $2\Rep(\Z_2)$ SSB phase for the 1-form symmetry generated by $D_1^{1_{r}}$.

\begin{tcolorbox}[
colback=white,
coltitle= black,
colbacktitle=ourcolorforheader,
colframe=black,
title= $\TwoRep(D_8)$: $2\Rep(\Z_2)$ SSB phase I, 
sharp corners]
\be
\begin{split}
\begin{tikzpicture}
\begin{scope}[shift={(0,0)},scale=1] 
 \pgfmathsetmacro{\del}{-0.5}
\node at  (0,0)  {$ {\rm DW(\Z_2)}_{s}$} ;
\end{scope}
\end{tikzpicture}    
\end{split}
\ee
The $\Rep(D_8)$ 1-form symmetry is realized as
\be
\ba
   D_1^{1_{x}}&\simeq D_1^{1}\,,\quad 
    D_1^{1_r}\simeq D_1^{1_{xr}}\simeq D_1^{\wh{x}} \,,\quad
    D_1^E\simeq D_1^{1}\oplus D_1^{\wh{x}}\,,
\ea
\ee
i.e. $\TwoRep(D_8)$ acts on this gapped phase via a projection $\TwoRep(D_8)\to \TwoRep(\Z_2)$. 
\end{tcolorbox}

\subsubsection{$\TwoRep(\Z_2)$ SSB Phase II}
Let us now take the physical boundary to be
\begin{equation}
    \Bphys=\fB_{\Neu(\Z_2^{xr}),s}\,.
\end{equation}
The idempotents can be computed by starting with the $\Z_4^r$ SSB phase II for $2\Vec_{D_8}$ of section \ref{sec:D8_Z4SSB2} (which has the same $\Bphys$ as our current sandwich) and gauging $D_8$ on the symmetry boundary. Since the $\Z_2^{xr}$ symmetry was preserved in each vacuum, after gauging $D_8$ we obtain a unique vacuum whose underlying TFT carries $\Z_{2}$ Dijkgraaf-Witten theory with a 3-cocycle twist $s\in H^{3}(\Z_2,U(1))$, i.e. toric code or double semion.  We denote the 3d TFT as DW$(\Z_2)_s$, and the line generating its electric $\Z_2^{(1)}$ symmetry as $D_1^{\wh{xr}}$.

The SymTFT bulk surface $\Q_2^{[xr]}$ ending on both boundaries produces, after compactification, a genuine line $\cL^{xr}$ that carries charge $-1$ under $D_1^{1_r}$ and $D_1^{1_{x}}$, charge $+1$ under $D_1^{1_{xr}}$ and $0$ under $D_1^E$. The $2\Rep(D_8)$ symmetry generators are therefore realized in this phase as follows:
\be
\ba
    D_1^{1_{xr}}&\simeq D_1^{1}\,,\quad \\
    D_1^{1_r}&\simeq D_1^{1_{x}}\simeq D_1^{\wh{xr}} \,,\\
    D_1^E&\simeq D_1^{1}\oplus D_1^{\wh{xr}}\,,
\ea
\ee
so it is a $2\Rep(\Z_2)$ SSB phase for the 1-form symmetry generated by $D_1^{1_{r}}$.

\begin{tcolorbox}[
colback=white,
coltitle= black,
colbacktitle=ourcolorforheader,
colframe=black,
title= $\TwoRep(D_8)$: $2\Rep(\Z_2)$ SSB phase II, 
sharp corners]
\be
\begin{split}
\begin{tikzpicture}
\begin{scope}[shift={(0,0)},scale=1] 
 \pgfmathsetmacro{\del}{-0.5}
\node at  (0,0)  {$ {\rm DW(\Z_2)}_{s}$} ;
\end{scope}
\end{tikzpicture}    
\end{split}
\ee
The $\Rep(D_8)$ 1-form symmetry is realized as
\be
\ba
   D_1^{1_{xr}}&\simeq D_1^{1}\,,\quad 
    D_1^{1_r}\simeq D_1^{1_{x}}\simeq D_1^{\wh{x}} \,,\quad
    D_1^E\simeq D_1^{1}\oplus D_1^{\wh{xr}}\,,
\ea
\ee
i.e. $\TwoRep(D_8)$ acts on this gapped phase via a projection $\TwoRep(D_8)\to \TwoRep(\Z_2)$. 
\end{tcolorbox}

\subsubsection{$\TwoRep(D_8)$ SSB Phase}
Let us finally take the physical boundary to be 
\begin{equation}
    \Bphys=\fB_{\Neu(D_8),\tau}\,.
\end{equation}
This phase can be obtained by gauging the $D_8$ symmetry of the $D_8$ SPT phase of section \ref{sec:D8_D8-SPT}.
Since the IR TQFT contains the lines $\cL^{[g]}$, for all $D_8$ conjugacy classes, the full $\TwoRep(D_8)$ symmetry is spontaneously broken in this phase.
There is a unique vacuum since the IR TQFT contains only a single topological local operator which is the identity.
Finally the F-symbols of the charged lines $\cL^{[g]}$ are controlled by the 3-cocycle $\tau\in H^3(D_8,U(1))$.
This gapped phase is the $D_8$ Dijkgraaf-Witten theory with a $\tau$ 3-cocycle twist.
\begin{tcolorbox}[
colback=white,
coltitle= black,
colbacktitle=ourcolorforheader,
colframe=black,
title= $\TwoRep(D_8)$: $\TwoRep(D_8)$ SSB, 
sharp corners]
\be
\begin{split}
\begin{tikzpicture}
\begin{scope}[shift={(0,0)},scale=1] 
 \pgfmathsetmacro{\del}{-0.5}
\node at  (0,0)  {$ {\rm DW(D_8)}_{\tau}$} ;
\end{scope}
\end{tikzpicture}    
\end{split}
\ee
The $\Rep(D_8)$ 1-form symmetry is generated by the Wilson lines of the $D_8$ Dijkgraaf-Witten theory. 
Since the IR theory contains topological lines charged under the full symmetry category, this is the $\TwoRep(D_8)$ SSB phase.
\end{tcolorbox}

\section{Minimal and Non-Minimal SSB Phases for $\TwoRep(\Z_n^{(1)}\rtimes \Z_2^{(0)})$}

In the past sections, we encountered one particularly interesting gapped phase, which is the SSB Phase for $\cS=\TwoRep(\mathbb G^{(2)})$, where the 2-group symmetry is 
\be
\mathbb{G}^{(2)} = \Z_n^{(1)}\rtimes \Z_2^{(0)} \,,
\ee
for $n=3\,,4$, which we discussed in sections \ref{sec:apple3} and \ref{sec:carrot3}, and dubbed ``superstar phases".
We will generalize this now to higher $n$.
In this section, we provide an alternate route to this gapped phase via a direct gauging of a certain SSB phase with $\TwoVec(S_3)$ and $\TwoVec(D_8)$ respectively.
This allows for an immediate generalization to the case where $n>4$ as well as to non-minimal generalizations of this phase.

\subsection{Minimal SSB Phases}

 In general, the symmetry category $\TwoRep(\Z_n^{(1)}\rtimes \Z_2^{(0)})$ can be obtained from  $\TwoVec(\Z_n^{(0)}\rtimes \Z_2^{(0)})$ by gauging the non-normal $\Z_2^{(0)}$ sub-symmetry.
Specifically, the $\TwoRep(\Z_n^{(1)}\rtimes \Z_2^{(0)})$ SSB phase corresponds to the $\Z_{n}^{(0)}$ SSB phase with $\TwoVec(\Z_n^{(0)}\rtimes \Z_2^{(0)})$.
We now describe this $\Z_{n}^{(0)}$ SSB in $\TwoVec(D_{2n}^{(0)})$ symmetric systems for different $n$ and directly perform its gauging.

\begin{itemize}
    \item {$\boldsymbol{n=3}$:} Let us begin with the $\Z_3$ SSB phase realized in $S_3=D_{6}=\Z_3\rtimes Z_2$ symmetric systems.
This gapped phase has 3 indistinguishable vacua. 
These vacua are permuted by the $\Z_3$ symmetry generated by $D_2^a$ and each vacuum preserves one $\Z_2$ symmetry (generated by $D_2^b$, $D_2^{ab}$ or $D_2^{a^2b}$).
There are in fact two distinct $\Z_3$ SSB phases labeled by $t\in H^{3}(\Z_2\,,U(1))$ corresponding to the SPT nature of the preserved $\Z_{2}$ in each vacuum. 
Now lets consider gauging the $\Z_2^{b}$ symmetry.
Once we single out $\Z_2^b$, the vacua are in fact distinguishable. 
We depict the vacua as the vertices of a triangle:
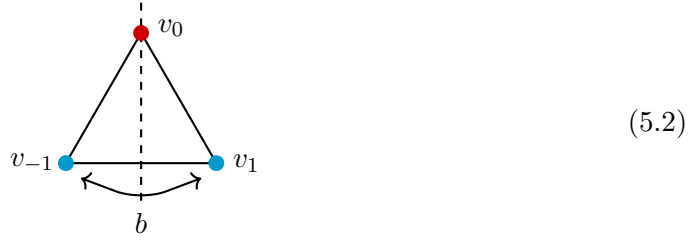
\be
\begin{split}
\begin{tikzpicture}
 \begin{scope}[shift={(0,0)},scale=1] 
\draw[dashed, thick] (1,-0.5)-- (1,2.23);
\draw[thick] (0,0) -- (2,0) -- (1,{sqrt(3)}) -- cycle;
\draw[<->, thick, rounded corners = 10pt] (0.2,-0.2)--(1,-.5)--(1.8,-0.2);
\fill[red!80!black] (1,{sqrt(3)}) circle (3pt);
\fill[cyan!80!black] (0,0) circle (3pt);
\fill[cyan!80!black] (2,0) circle (3pt);
\node at  (1,-0.8)  {$b$} ;
\node at  (1.4,1.8)  {$v_0$} ;
\node at  (-0.45,0)  {$v_{-1}$} ;
\node at  (2.4,0)  {$v_{1}$} ;
\end{scope}
\end{tikzpicture}    
\end{split}
\ee
The vacuum $v_0$ is invariant under $\Z_2^{b}$ while $v_1$ and $v_{-1}$ are exchanged under the $\Z_2^{b}$ action.
Therefore the $\Z_3$ SSB phase realizes the phase 
\begin{equation}
    {\rm Triv}_0 \oplus {\rm SSB}_{\pm 1}\,.
\end{equation}
Once we gauge $\Z_2^{b}$, the trivial vacuum $v_0$ maps to the vacuum where the dual $\Z_2$ 1-form symmetry is spontaneously broken or equivalently this vacuum realizes a $\Z_2$ topological order.
For $t=+1$, this is the toric code topological order while for $t=-1$, it is the double semion topological order.
Specifically the end of the $D_2^{b}$ surface (whose associator is determined by $t$) in the $v_0$ vacuum becomes a topological line that is charged under the dual $\Z_2$ 1-form symmetry.
Conversely, the vacua $v_{\pm 1}$ become a single vacuum $v_{[1]}$ in the gauged theory that realizes the trivial phase for the dual $\Z_2$ 1-form symmetry.
Finally the $\Z_3$ generators $D_2^{a}\oplus D_2^{a^2}$ combine into a non-invertible symmetry operator $D_2^{A}$ whose action can be inferred from its action on $v_{j}$.
\begin{equation}
    D_2^{A}:(v_0\,, v_{[1]})\longmapsto (v_{[1]}\,, 2v_0+v_{[1]})\,.
\end{equation}
This is precisely the symmetry action that was previously obtained from the SymTFT construction of gapped phases in section \ref{sec:apple3}.

\item $\boldsymbol{n=4}$: We now start with the $D_8$ symmetric gapped phase that realizes the $\Z_4$ SSB phase.
We present $D_8$ as
\begin{equation}
    D_8=\langle a\,, b \ | a^{4}=b^2=1\,, bab=a^3\rangle\,.  
\end{equation}
The $\Z_4$ SSB phase has 4 vacua that can be conveniently depicted as the vertices of a square as 
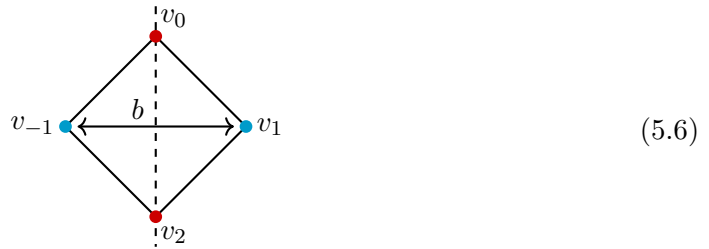
\be
\begin{split}
\begin{tikzpicture}
 \begin{scope}[shift={(0,0)},scale=0.8] 
\draw[dashed, thick] (1.5,2)-- (1.5,-2);
\draw[thick] (0,0) -- (1.5,1.5) -- (3,0) -- (1.5,-1.5) -- cycle;
\draw[<->, thick, rounded corners = 10pt] (0.2,0)--(2.8,0);
\fill[red!80!black] (1.5,1.5) circle (3pt);
\fill[red!80!black] (1.5,-1.5) circle (3pt);
\fill[cyan!80!black] (0,0) circle (3pt);
\fill[cyan!80!black] (3,0) circle (3pt);
\node at  (1.8,1.8)  {$v_0$} ;
\node at  (1.8,-1.8)  {$v_2$} ;
\node at  (-0.55,0)  {$v_{-1}$} ;
\node at  (3.4,0)  {$v_{1}$} ;
\node at  (1.2,0.3)  {$b$} ;
\end{scope}
\end{tikzpicture}    
\end{split}
\ee
The $D_8$ symmetry has a natural action on a square. 
The $\Z_4^{a}$ sub-symmetry rotates the square while the $\Z_2^{b}$ symmetry exchanges the blue vacua $v_{1}$ and $v_{-1}$ while leaving the red vacua invariant.
The red vacua can realize $\Z_2$ SPT labeled by $t\in H^3(\Z_2\,, U(1))$.
From the perspective of the $\Z_2^{b}$ sub-symmetry, this gapped phase is
\begin{equation}
    {\rm Triv}_0\oplus {\rm SSB}_{\pm 1}\oplus {\rm Triv}_{2}\,.
\end{equation}
Upon gauging $\Z_2^{b}$, one correspondingly obtains a a phase with the 3 vacua
\be
\begin{split}
\begin{tikzpicture}
 \begin{scope}[shift={(0,0)},scale=0.8] 
\fill[red!80!black] (0,0) circle (3pt);
\fill[cyan!80!black] (3,0) circle (3pt);
\fill[red!80!black] (6,0) circle (3pt);
\node at  (0.2,0.3)  {$v_0$} ;
\node at  (3.2,0.3)  {$v_{[1]}$} ;
\node at  (6.2,0.3)  {$v_2$} ;
\node at  (0,-0.6)  {\footnotesize $(\Z_2^{(1)} \ \rm{SSB})$} ;
\node at  (3,-0.6)  {\footnotesize $(\Z_2^{(1)} \ \rm{Triv})$} ;
\node at  (6,-0.6)  {\footnotesize $(\Z_2^{(1)} \ \rm{SSB})$} ;
\node at  (6.5,0)  {$,$} ;
\end{scope}
\end{tikzpicture}    
\end{split}
\ee
The vacua $v_0$ and $v_2$ are dual to the $\Z_2^{b}$ symmetric vacua and therefore spontaneously break the dual $\Z_2$ 1-form symmetry, realizing the $t$-$\Z_2$ topological gauge theory, i.e., Toric Code or Double Semion for $t=+1$ or $-1$ respectively.
While $v_{[1]}$ is dual the the $\Z_2^{b}$ SSB vacua $v_{\pm 1}$, therefore realizes a trivial phase for the dual $\Z_2$ 1-form symmetry.

The full symmetry category after gauging $\Z_2^{b}$ is $\TwoRep(\Z_4\rtimes \Z_2^{(0)})$.
In addition to the $\Z_2$ 1-form symmetry, this category contains invertible and a non-invertible 0-form generators (upto condensations), which descend from $D_2^{a^2}$ and $D_{2}^{a}\oplus D_2^{a^3}$ respectively. 
We will denote these as $D_2^{a^2}$ and $D_2^{A}$ respectively.
Their fusion rules are \cite{Bhardwaj:2022maz}
\begin{equation}
\begin{split}
    D_2^{a^2}\otimes D_2^{a^2} &= D_2^{\id} \,, \\
    D_2^{a^2}\otimes D_2^{A} &= D_2^{A} \,, \\
    D_2^{A}\otimes D_2^{A} &= \frac{D_2^{\id}}{D_1^{1}\oplus D_1^{\wh{b}}} \oplus \frac{D_2^{a^2}}{D_1^{1}\oplus D_1^{\wh{b}}} \,,
\end{split}
\end{equation}
where $D_1^{\wh{b}}$ is the dual $\Z_2$ 1-form symmetry generator.
The action of the 0-form symmetries in $\TwoRep(\Z_4\rtimes \Z_2^{(0)})$ can be obtained from the action of the lifted symmetries in $\TwoVec(D_8)$.
\begin{equation}
\begin{split}
    D_{2}^{a^2}:(v_0\,, v_{[1]}\,, v_{2})& \longmapsto (v_2\,, v_{[1]}\,, v_{0}) \,, \\ 
     D_{2}^{A}:(v_0\,, v_{[1]}\,, v_{2})& \longmapsto (v_{[1]}\,, 2v_0 + 2v_2\,, v_{[1]}) \,, \\
\end{split}    
\end{equation}
In summary, we obtain a phase with 3 vacua, 2 of which are topologically ordered while the third is topologically trivial and there is an action of a non-invertible 2-categorical symmetry that stabilizes this phase structure.

\item $\boldsymbol{n=2\ell+1}:$ To obtain the $\TwoRep(\Z_{2\ell+1}^{(1)}\rtimes \Z_2^{(0)})$ SSB phase, our starting point will be a $D_{4\ell+2}=\Z_{2\ell+1}\rtimes \Z_2$ 0-form symmetric theory that realizes the $\Z_{2\ell+1}$ SSB phase.
We present $D_{4\ell+2}$ as
\begin{equation}
    D_{4\ell +2}=\langle a\,, b \ | a^{2\ell +1}=b^2=1\,, bab=a^{2\ell}\rangle\,.  
\end{equation}
This phase has $2\ell +1$ vacua labeled as $v_{j}$ with $j\in \left\{-\ell\,,-\ell+1\,, \dots, \ell-1\,, \ell\right\}$ which are cyclically permuted by the action of $D_2^{a}$
The $\Z_2^{b}$ action exchanges $v_j$ and $v_{-j}$ and realizes a $t$ $\Z_2$ SPT in $v_0$.
For example for the case $\ell=2$, there is a $5$ vacua phase on which the $D_{10}$ symmetry acts as 
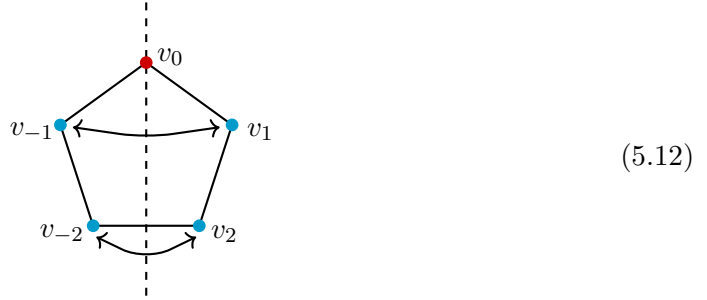
\be
\begin{split}
\begin{tikzpicture}
 \begin{scope}[shift={(0,0)},scale=0.8] 
 \pgfmathsetmacro{\del}{1.5}
\draw[dashed, thick] (0,2.5)-- (0,-2.5);
\draw[thick] (0,\del) -- ({\del*cos(162)},{\del*sin(162)}) -- ({\del*cos(234)},{\del*sin(234)}) -- ({\del*cos(306)},{\del*sin(306)}) -- ({\del*cos(18)},{\del*sin(18)}) -- cycle;
\draw[<->, thick, rounded corners = 10pt] (-1.22,.425)--(0,0.25)--(1.22,.425);
\fill[red!80!black] (0,\del) circle (3pt);
\fill[cyan!80!black] ({\del*cos(162)},{\del*sin(162)}) circle (3pt);
\fill[cyan!80!black] ({\del*cos(234)},{\del*sin(234)}) circle (3pt);
\fill[cyan!80!black] ({\del*cos(306)},{\del*sin(306)}) circle (3pt);
\fill[cyan!80!black] ({\del*cos(18)},{\del*sin(18)}) circle (3pt);
\draw[<->, thick, rounded corners = 10pt] (-.83,-1.4)--(0,-1.8)--(0.83,-1.4);
\node at  (0.4,1.6)  {$v_0$} ;
\node at  (1.9,0.35)  {$v_1$} ;
\node at  (-1.9,0.35)  {$v_{-1}$} ;
\node at  (-1.4,- 1.35)  {$v_{-2}$} ;
\node at  (1.3,- 1.35)  {$v_2$} ;
\end{scope}
\end{tikzpicture}    
\end{split}
\ee
In terms of the $\Z_2^b$ symmetry, the $\Z_{2\ell+1}$ SSB phase splits as
\begin{equation}
    {\rm Triv}_0\bigoplus_{j=1}^{\ell}{\rm SSB}_{\pm j}\,.
\end{equation}
Therefore upon gauging $\Z_2^{b}$ we obtain the phase with $\ell+1$ vacua summarized as
\begin{tcolorbox}[
colback=white,
coltitle= black,
colbacktitle=ourcolorforheader,
colframe=black,
title= $\TwoRep(\Z_{2\ell+1}\rtimes \Z_2)$: $\TwoRep(\Z_{2\ell+1}\rtimes \Z_2)$ SSB Phase, 
sharp corners]
The underlying 3d TFT is 
\begin{equation}
    \left[{\rm DW}(\Z_2)_t\right]_0 \ \bigoplus_{j=1}^{\ell}\ {\rm Triv}_{[j]} \,.
\end{equation}
which we depict as
\be
\begin{split}
\begin{tikzpicture}
 \begin{scope}[shift={(0,0)},scale=0.8] 
\fill[red!80!black] (0,0) circle (3pt);
\fill[cyan!80!black] (3,0) circle (3pt);
\node at  (4.8,0)  {$\dots \dots$} ;
\fill[cyan!80!black] (6.5,0) circle (3pt);
\node at  (0.2,0.3)  {$v_0$} ;
\node at  (3.2,0.3)  {$v_{[1]}$} ;
\node at  (6.7,0.3)  {$v_{[\ell]}$} ;
\node at  (0,-0.6)  {\footnotesize $(\Z_2^{(1)} \ \rm{SSB})$} ;
\node at  (3,-0.6)  {\footnotesize $(\Z_2^{(1)} \ \rm{Triv})$} ;
\node at  (6.5,-0.6)  {\footnotesize $(\Z_2^{(1)} \ \rm{Triv})$} ;
\end{scope}
\end{tikzpicture}    
\end{split}
\ee
The symmetry category $\TwoRep(\Z_{2\ell+1}^{(1)}\rtimes \Z_2^{(0)})$ contains $\ell$ non-invertible symmetry defects that descend from $D_2^{a^p}\oplus D_2^{a^{-p}}$ upong gauging $\Z_2^{b}$. 
We denote the symmetry operator thus obtained as $D_2^{A,p}$.
The fusion rules of these is 
\begin{equation}
    D_2^{A,p}\oplus D_2^{A,p'}=D_2^{A,p+p'}\oplus D_2^{A,p-p'}\,,
\end{equation}
where we use the identification
\begin{equation}
\begin{split}
    D_2^{A,0}& \equiv \frac{D_2^{\id}}{D_1^{1}\oplus D_1^{\wh{b}}}\,, \qquad
    D_2^{A,\ell+j}\equiv D_2^{A_{\ell+1-j}}\,,
\end{split}
\end{equation}
for $j=1,..,\ell$.
The symmetry action on the vacua takes the form
\begin{equation}
    D_{2}^{A_{p}}:
    \begin{pmatrix}
    v_{0} \\
    v_{[j]}
    \end{pmatrix}
    \longrightarrow 
    \begin{pmatrix}
        v_{[j]} \\
        v_{[j+p]}+v_{[j-p]}
    \end{pmatrix}\,,
\end{equation}
where we have made the identifications $v_{[0]}\equiv 2v_0$ and $v_{[\ell+j]}=v_{[\ell+1-j]}$.
\end{tcolorbox}

\item {$\boldsymbol{n=2\ell}$:} Now we construct the $\TwoRep(\Z_{2\ell}^{(1)}\times \Z_{2}^{(0)})$ SSB phase.
We start with the $D_{4\ell}$ symmetric gapped phase that is the $\Z_{2\ell}$ SSB Phase.
We present $D_{4\ell}$ as
\begin{equation}
    D_{4\ell }=\langle a\,, b \ | a^{2\ell }=b^2=1\,, bab=a^{2\ell-1}\rangle\,.  
\end{equation}
This phase has $2\ell$ vacua, which we label as 
\begin{equation}
    v_{j}\,, \qquad j=-\ell+1\,, \dots\,, 0\,, \dots\,, \ell\,.
\end{equation}
All these states are cyclically permuted under the action of $\Z_{2\ell}$, hence this is the $\Z_{2\ell}$ SSB phase.
Under the action of $\Z_2^b$, $v_0$ and $v_{\ell}$ are invariant while $v_{j}$ and $v_{-j}$ are interchanged.
For example, the case of $\ell=3$ has 6 vacua which realize the $D_{12}$ symmetry as 
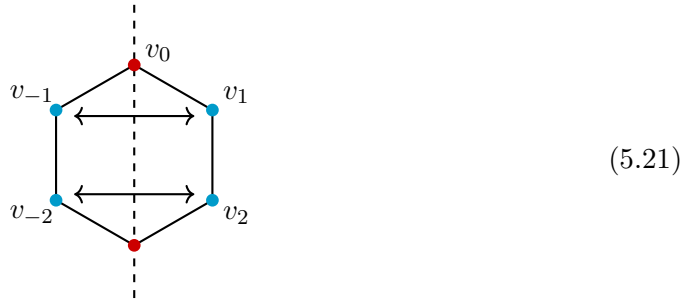
\be
\begin{split}
\begin{tikzpicture}
 \begin{scope}[shift={(0,0)},scale=0.8] 
 \pgfmathsetmacro{\del}{1.5}
\draw[dashed, thick] (0,2.5)-- (0,-2.5);
\draw[thick] (0,\del) -- ({\del*cos(150)},{\del*sin(150)}) -- ({\del*cos(210)},{\del*sin(210)}) -- ({\del*cos(270)},{\del*sin(270)}) -- ({\del*cos(330)},{\del*sin(330)}) --({\del*cos(30)},{\del*sin(30)})-- cycle;
\fill[red!80!black] (0,\del) circle (3pt);
\fill[red!80!black] ({\del*cos(270)},{\del*sin(270)}) circle (3pt);
\fill[cyan!80!black] ({\del*cos(150)},{\del*sin(150)}) circle (3pt);
\fill[cyan!80!black] ({\del*cos(210)},{\del*sin(210)}) circle (3pt);
\fill[cyan!80!black] ({\del*cos(330)},{\del*sin(330)}) circle (3pt);
\fill[cyan!80!black] ({\del*cos(30)},{\del*sin(30)}) circle (3pt);
\draw[<->, thick] (1,0.65)--(-1,0.65);
\draw[<->, thick] (1,-0.65)--(-1,-0.65);
\node at  (0.4,1.7)  {$v_0$} ;
\node at  (1.7,1)  {$v_1$} ;
\node at  (-1.7,1)  {$v_{-1}$} ;
\node at  (1.7,-1)  {$v_2$} ;
\node at  (-1.7,-1)  {$v_{-2}$} ;
\end{scope}
\end{tikzpicture}    
\end{split}
\ee
From the perspective of the $\Z_2^b$ symmetry, the ground states in this phase split as
\begin{equation}
    {\rm Triv}_0\oplus {\rm Triv}_{\ell}  \oplus \bigoplus_{j=1}^{\ell-1} {\rm Triv}_{\pm j}\,,
\end{equation}
Therefore after gauging the $\Z_2^b$ symmetry, one obtains a phase with $\ell+1$ vacua, two of which are topologically ordered and $\ell-1$ are topologically trivial. 
The phase structure can be summarized as
\begin{tcolorbox}[
colback=white,
coltitle= black,
colbacktitle=ourcolorforheader,
colframe=black,
title= $\TwoRep(\Z_{2\ell+1}\rtimes \Z_2)$: $\TwoRep(\Z_{2\ell+1}\rtimes \Z_2)$ SSB Phase, 
sharp corners]
The underlying 3d TFT is
\begin{equation}
    \left[{\rm DW}(\Z_2)_t\right]_0 \ \bigoplus_{j=1}^{\ell}\ {\rm Triv}_{[j]} \ \bigoplus \ \left[{\rm DW}(\Z_2)_t\right]_{\ell} \,.
\end{equation}
\be
\begin{split}
\begin{tikzpicture}
 \begin{scope}[shift={(-2,0)},scale=0.8] 
\fill[red!80!black] (0,0) circle (3pt);
\fill[cyan!80!black] (3,0) circle (3pt);
\node at  (4.8,0)  {$\dots \dots$} ;
\fill[cyan!80!black] (6.5,0) circle (3pt);
\fill[red!80!black] (9.5,0) circle (3pt);
\node at  (0.2,0.3)  {$v_0$} ;
\node at  (3.2,0.3)  {$v_{[1]}$} ;
\node at  (6.7,0.3)  {$v_{[\ell-1]}$} ;
\node at  (9.7,0.3)  {$v_{\ell}$} ;
\node at  (0,-0.6)  {\footnotesize $(\Z_2^{(1)} \ \rm{SSB})$} ;
\node at  (3,-0.6)  {\footnotesize $(\Z_2^{(1)} \ \rm{Triv})$} ;
\node at  (6.5,-0.6)  {\footnotesize $(\Z_2^{(1)} \ \rm{Triv})$} ;
\node at  (9.5,-0.6)  {\footnotesize $(\Z_2^{(1)} \ \rm{SSB})$} ;
\end{scope}
\end{tikzpicture}    
\end{split}
\ee
The symmetry structure of the gauged theory is $\TwoRep(\Z_{2\ell}^{(1)}\rtimes \Z_2^{(0)})$ which has a $\Z_2$ 1-form symmetry generated by $D_1^{\wh{b}}$, an $\Z_2$ 0-form symmetry generated by $D_2^{a^{\ell}}$ and $\ell-1$ non-invertible symmerty generators $D_2^{A_{p}}$ which descend from $D_2^{a^p}\oplus D_2^{a^{-p}}$ upon gauging $\Z_2^b$.
The fusion rules among these symmetry generators is
\begin{equation}
\begin{split}
    D_2^{A_{p}}\otimes D_2^{a^{\ell}}&=D_2^{A_{\ell-p}}\,, \\
    D_2^{A_{p}}\otimes D_2^{A_{p'}}&= D_2^{A_{p+p'}}\oplus D_2^{A_{p-p'}}\,, 
\end{split}
\end{equation}
where we use the identifications 
\begin{equation}
    D_2^{A_0}\equiv\frac{D_2^{\id}}{D_1^{\id}\oplus D_1^{\wh{b}}}\,, \quad 
    D_2^{A_\ell}\equiv\frac{D_2^{a^{\ell}}}{D_1^{\id}\oplus D_1^{\wh{b}}}\,, \quad
    D_2^{A_{\ell+j}}\equiv D_2^{A_{\ell-j}}\,, 
\end{equation}
Similarly, the action on the ground states is determined to be
\begin{equation}
\begin{split}
 D_2^{a^{\ell}}&:v_0 \longleftrightarrow v_{\ell}\,, \\
 &: v_{[j]} \longleftrightarrow v_{[\ell-j]}\,, \\
 D_2^{A_{p}}&:v_{0}\longrightarrow v_{[j]}\,, \\
 &: v_{[j]} \longrightarrow v_{[j+p]}+ v_{[j-p]}\,,
 \end{split}
\end{equation}
where we use the identifications $ v_{[0]}\equiv 2v_0$, $v_{[\ell]}=2v_{\ell}$ and $v_{[\ell +j]} = v_{[\ell-j]}$.
\end{tcolorbox}

\end{itemize}

\subsection{Non-minimal SSB Phases} The above discussion can be immediately generalized to non-minimal $\Rep(\Z_{n}^{(1)}\rtimes \Z_2^{(0)})$ SSB phases.
For this, before gauging $\Z_2^{b}$, one considers a $D_{2n}$ 0-form symmetric phase which realizes a non-minimal $\Z_{n}$ SPT.
This has $n$-vacua that are permuted by $\Z_n^{(0)}$.
Either one or two vacua (depending on whether $n$ is odd or even respectively) are invariant under $\Z_2^b$ while the remaining are organized into pairs that are mapped into each other under $\Z_2^{b}$.
To construct the non-minimal $\Rep(\Z_{n}^{(1)}\rtimes \Z_2^{(0)})$ phase, we consider each of these vacua to be topologically ordered.
I.e., each vacuum effectively realizes a 3d TFT whose underlying category of lines is some fixed MTC $\cM$.
The vacua that are $\Z_2^{b}$ invariant have a certain $\Z_2$ symmetry defined via choice of a $\Z_2$ extended braided fusion category $\cM^{\times}_{\Z_2}$.
Upon gauging $\Z_2^{b}$, these vacua map to vacua in which the dual $\Z_2^{(1)}$ symmetry has been spontaneouly broken.
These are concretely obtained via the $\Z_2$ equivariantization of $\cM_{\Z_2}^{\times}$.
The rest of the vacua realize topological orders $\cM$ after gauging $\Z_2^{b}$.




\subsection*{Acknowledgements}
We thank Lea Bottini, Weiguang Cao, Yuhan Gai, Sheng-jie Huang, Kansei Inamura, Daniel Pajer, Jingxiang Wu for discussions.
We thank Weiguang Cao, Linhao Li and Masahito Yamazaki for coordinating submission with their upcoming work \cite{Cao:2025qhg}.
The work of SSN and AW is supported by the UKRI Frontier Research Grant, underwriting the ERC Advanced Grant ``Generalized Symmetries in Quantum Field Theory and Quantum Gravity''.
The work of AT is funded by Villum Fonden Grant no. VIL60714.

%

\bibliography{GenSym}
\bibliographystyle{JHEP}

\end{document}